\documentclass[twoside,11pt,fleqn]{Latex/Classes/PhDthesisPSnPDF} %oneside

\usepackage{lineno}
\usepackage{amsbsy}
\usepackage{xspace}
\usepackage{wtmmPkg}
\usepackage{natbib}
\usepackage{multirow}
\usepackage{paralist}
\usepackage{fancyhdr} % for better header layout
\usepackage{sidecap}
\usepackage{rotating}
\usepackage{hyperref}
\usepackage{chngcntr}
\counterwithin{footnote}{chapter}

\DeclareRobustCommand{\ion}[2]{%
\relax\ifmmode
\ifx\testbx\f@series
{\mathbf{#1\,\mathsc{#2}}}\else
{\mathrm{#1\,\mathsc{#2}}}\fi
\else\textup{#1\,{\mdseries\textsc{#2}}}%
\fi}

\newcommand{\helix}{\textsf{{H\textsc{e}LI\textsc{x}$^{+}$}}}

\newcommand{\aap}{    {\it Astronomy \& Astrophysics}}

\newcommand{\apjl}{   {\it Astrophysical Journal Letters}}

\newcommand{\solphys}{{\it Solar Physics}}

\ifpdf  
    \pdfcatalog { /PageMode (/UseOutlines)
                  /OpenAction (fitbh)  }
\fi

\title{Fields and Flares: \\ Understanding the Complex \\ Magnetic Topologies of \\ Solar Active Regions}

\ifpdf
  \author{\href{mailto:somurray@tcd.ie}{Sophie A. Murray, B. A. (Mod.), M. Sc. }}
  \collegeordept{\href{http://www.physics.tcd.ie}{School of Physics}}
  \university{\href{http://www.tcd.ie}{University of Dublin, Trinity College}}

  \crest{
  \vspace{.61cm}
  \includegraphics[width=0.35\textwidth]{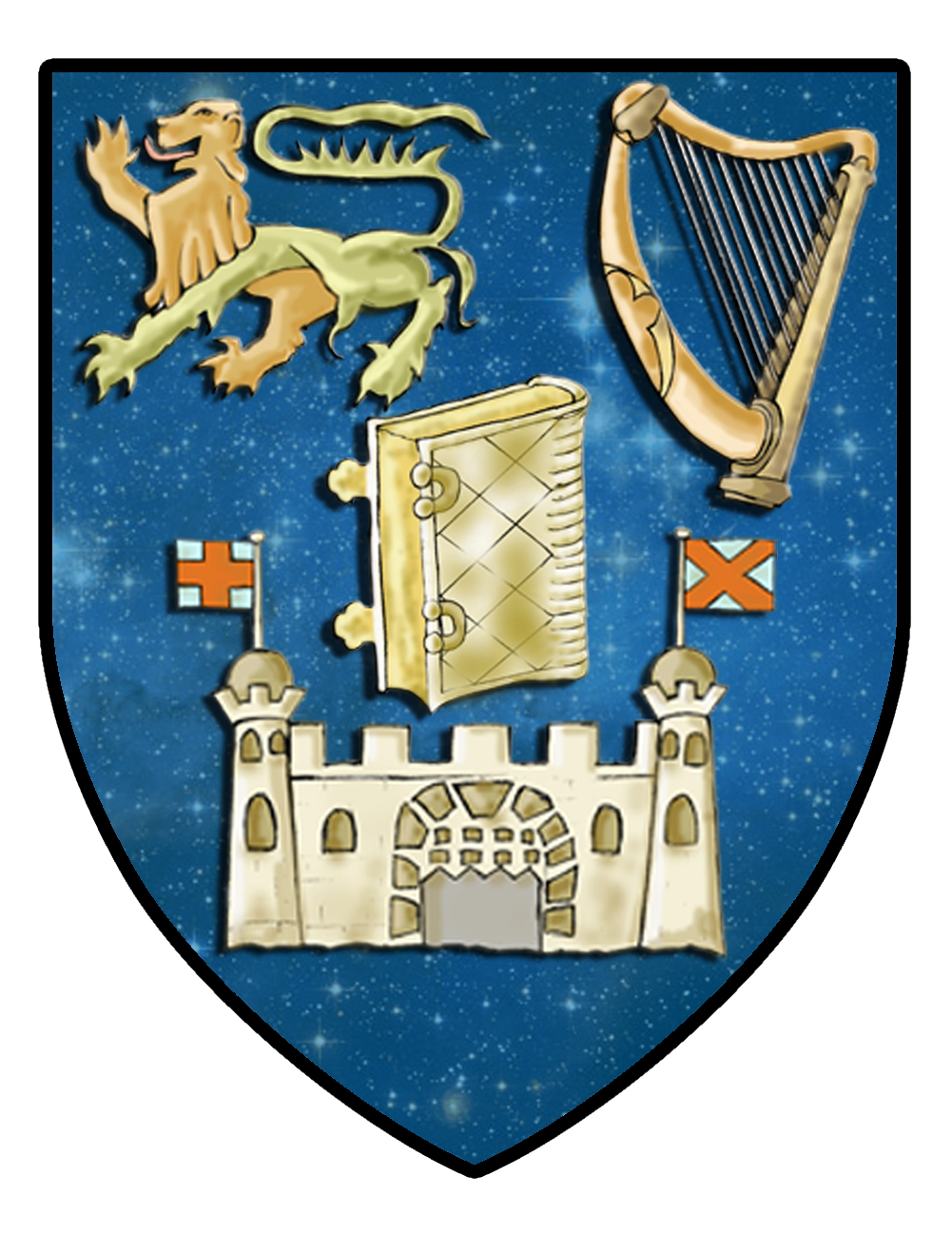}
   \vspace{1cm}
    }
  
\else
  \author{\\ Sophie A. Murray}
  \collegeordept{School of Physics}
  \university{University of Dublin, Trinity College}
  \crest{\includegraphics[width=6cm]{tcd_crest_asgi1.png}}
\fi
\degree{Philosophi\ae Doctor (PhD)}
\degreedate{2013 January}

\usepackage{setspace}
\begin{document}

%\language{english}

% sets line spacing
\renewcommand\baselinestretch{1.2}
\baselineskip=18pt plus1pt

%: ----------------------- generate cover page ------------------------

\maketitle  % command to print the title page with above variables

%
%%: ----------------------- cover page back side ------------------------
%% Your research institution may require reviewer names, etc.
%% This cover back side is required by Dresden Med Fac; uncomment if needed.

%\newpage
%\vspace{10mm}
%1. Reviewer: Name

%\vspace{10mm}
%2. Reviewer: 

%\vspace{20mm}
%Day of the defense:

%\vspace{20mm}
%\hspace{70mm}Signature from head of PhD committee:

%: ----------------------- abstract ------------------------

% Your institution may have specific regulations if you need an abstract and where it is to be placed in the document. The default here is just after title.

% The original template provides and abstractseparate environment, if your institution requires them to be separate. I think it's easier to print the abstract from the complete thesis by restricting printing to the relevant page.
% \begin{abstractseparate}
%   \input{Abstract/abstract}
% \end{abstractseparate}

%: ----------------------- tie in front matter ------------------------

\frontmatter
%: Declaration of originality

% Thesis statement of originality -------------------------------------

% Depending on the regulations of your faculty you may need a declaration like the one below. This specific one is from the medical faculty of the university of Dresden.

\begin{declaration}        %this creates the heading for the declaration page

I, Sophie A. Murray, hereby certify that I am the sole author of this thesis and that all the work presented in it, unless otherwise referenced, is entirely my own. I also declare that this work has not been submitted, in whole or in part, to any other university or college for any degree or other qualification. 

\textbf{Name:} Sophie A. Murray
 \begin{center}
 \textbf{Signature:}  ........................................		\textbf{Date:}  ..............
\end{center}
\vspace{5mm}

The thesis work was conducted from October 2008 to September 2012 under the supervision of Dr. Peter T. Gallagher and Dr. D. Shaun Bloomfield at Trinity College, University of Dublin.

\vspace{5mm}
In submitting this thesis to the University of Dublin I agree that the University Library may lend or copy the thesis upon request. 

\textbf{Name:} Sophie Murray
 \begin{center}
 \textbf{Signature:}  ........................................		\textbf{Date:}  ..............
\end{center}

\end{declaration}

% ----------------------------------------------------------------------

% Thesis Abstract -----------------------------------------------------

%\begin{abstractslong}    %uncommenting this line, gives a different abstract heading
\begin{abstracts}        %this creates the heading for the abstract page

Sunspots are regions of decreased brightness on the visible surface of the Sun (photosphere) that are associated with strong magnetic fields. They have been found to be locations associated with solar flares, which occur when energy stored in sunspot magnetic fields is suddenly released. The processes involved in flaring and the link between sunspot magnetic fields and flares is still not fully understood, and this thesis aims to gain a better understanding of these topics. The magnetic field evolution of a number of sunspot regions is examined using high spatial resolution data from the \emph{Hinode} spacecraft.

Photospheric magnetic field data is first investigated, and significant increases in negative vertical field strength, negative vertical current density, and field inclination angle towards the vertical are observed just hours before a flare occurs, which is on much shorter timescales than previously studied. 
These parameters then return to their pre-flare `quiet' state after the flare has ended. First observations of spatial changes in field inclination across a magnetic neutral line (generally believed to be a typical source region of flares) are also discovered. 
The changes in field inclination observed in this thesis confirm field configuration changes due to flares predicted by a number of previous works. 

3D magnetic field extrapolation methods are then used to study the coronal magnetic field, using the photospheric magnetic field data as a boundary condition. Significant geometrical differences are found to exist between different field configurations obtained from three types of extrapolation procedure (potential, linear force free, and non-linear force free). Magnetic energy and free magnetic energy are observed to increase significantly a few hours before a flare, and decrease afterwards, which is a similar trend to the photospheric field parameter changes observed. 
Evidence of partial Taylor relaxation is also detected after a flare, as predicted by several previous studies. 

The research presented in this thesis gives insight into photospheric and coronal magnetic field evolution of flaring regions. The magnetic field changes observed only hours before a flare could be useful for flare forecasting. Field changes observed due to the flare itself have confirmed currently proposed magnetic field topology changes due to flares.
The results outlined show that this particular field of research is vital in furthering our understanding of the magnetic nature of sunspots and its link to flare processes.

\end{abstracts}
%\end{abstractlongs}

% ---------------------------------------------------------------------- 

% Thesis Dedictation ---------------------------------------------------

\begin{dedication} %this creates the heading for the dedication page

\begin{center}
\emph{``Were it not for magnetic fields, the Sun would be as uninteresting as most astronomers seem to think it is.."}
\end{center}
\begin{flushright}\emph{- R.B Leighton, 1969}
\end{flushright}

\end{dedication}

% ----------------------------------------------------------------------
%Thesis Acknowledgements ------------------------------------------------

%\begin{acknowledgementslong} %uncommenting this line, gives a different acknowledgements heading
\begin{acknowledgements}      %this creates the heading for the acknowlegments
I must first thank my supervisors, Shaun Bloomfield and Peter Gallagher, for their help and guidance over the past four years. Peter, thank you for agreeing to supervise me a second time, your experience and advice has been vital on so many occasions. Shaun, thank you for all your wise suggestions and advice over those Monday afternoon meetings. Your keen eye for detail has been invaluable! 

Thank you also to all in the Astrophysics Research Group, particularly to the postgrads (past and present) that I shared the fourth-floor office with. 
I also thank the staff in the School of Physics, anyone who has helped me over the years I have spent in TCD.

I must extend my appreciation to colleagues outside of TCD who have given me advice throughout my PhD. Particular thanks must go to those who provided codes I have used during analysis. Thank you also to the AXA Research Fund, who agreed to fund my PhD.

Finally, thanks to my family and friends for putting up with me on this long stressful journey!
\end{acknowledgements}

\chapter{List of Publications}
\label{chapter:publications}

\textbf{\Large{\underline{Refereed}}}
\begin{enumerate}
 
 \item \textbf{Murray, S. A.}, Bloomfield, D. S., \& Gallagher, P. T (2012)\\
 ``Variation of Magnetic Field Inclination across a Magnetic Neutral Line during a Solar Flare", \\ \apjl, in prep

 \item Bloomfield, D. S., Gallagher, P. T, Carley, E., Higgins, P. A., Long, D. M., Maloney, S. A., \textbf{Murray, S. A.}, O'Flannagain, A., Perez-Suarez, D. Ryan, D., \& Zucca, P. (2013b)\\
 ``A Comprehensive Overview of the 2011 June 7 Solar Storm'', \\ \aap, submitted

  \item \textbf{Murray, S. A.}, Bloomfield, D. S., \& Gallagher, P. T (2013a)\\
 ``Evidence for Partial Taylor Relaxation from Changes in Magnetic Geometry and Energy during a Solar Flare'', \\ \aap, in press (DOI: 10.1051/0004-6361/201219964)

  \item \textbf{Murray, S. A.}, Bloomfield, D. S., \& Gallagher, P. T (2011)\\
 ``The Evolution of Sunspot Magnetic Fields Associated with a Solar Flare'', \\ \solphys, 277, 1, 45-57 

\end{enumerate}

%: ----------------------- contents ------------------------

\setcounter{secnumdepth}{3} % organisational level that receives a numbers
\setcounter{tocdepth}{3}    % print table of contents for level 3
\tableofcontents            % print the table of contents
% levels are: 0 - chapter, 1 - section, 2 - subsection, 3 - subsection

%: ----------------------- list of figures/tables ------------------------

\listoffigures	% print list of figures

\listoftables  % print list of tables

%: ----------------------- glossary ------------------------

% Tie in external source file for definitions: /0_frontmatter/glossary.tex
% Glossary entries can also be defined in the main text. See glossary.tex
\include{0_frontmatter/glossary} 

\begin{multicols}{2} % \begin{multicols}{#columns}[header text][space]
\begin{footnotesize} % scriptsize(7) < footnotesize(8) < small (9) < normal (10)

\printnomenclature[1.5cm] % [] = distance between entry and description
\label{nom} % target name for links to glossary

\end{footnotesize}
\end{multicols}

%: --------------------------------------------------------------
%:                  MAIN DOCUMENT SECTION
% --------------------------------------------------------------

% the main text starts here with the introduction, 1st chapter,...
\mainmatter

%\renewcommand{\chaptername}{} % uncomment to print only "1" not "Chapter 1"

%: ----------------------- subdocuments ------------------------

% Parts of the thesis are included below. Rename the files as required.
% But take care that the paths match. You can also change the order of appearance by moving the include commands.
\doublespacing
\pagestyle{fancy}
%\include{1_introduction/ch1}	% background information

% this file is called up by thesis.tex
% content in this file will be fed into the main document

%: ----------------------- introduction file header -----------------------
\chapter{Introduction}
\label{chapter:introduction}

% the code below specifies where the figures are stored
\ifpdf
    \graphicspath{{1_introduction/figures/PNG/}{1_introduction/figures/PDF/}{1_introduction/figures/}}
\else
    \graphicspath{{1_introduction/figures/EPS/}{1_introduction/figures/}}
\fi

%\hrule height 1mm
%\vspace{0.5mm}
%\hrule height 0.4mm 
%\noindent 
%\\ {\it This Chapter .........  }
%\\ 
%\hrule height 0.4mm
%\vspace{0.5mm}
%\hrule height 1mm 

\newpage

\begin{figure}[!t]
\centerline{\includegraphics[width=\textwidth]{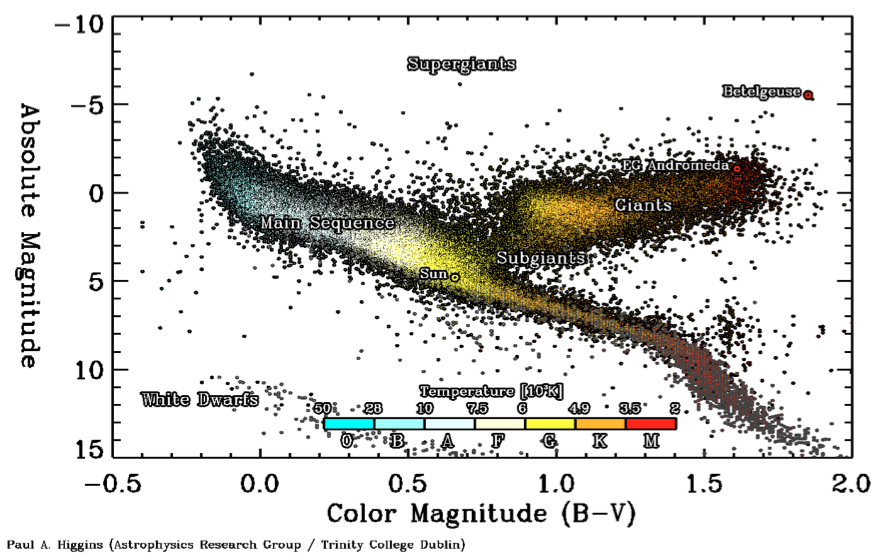}}
\caption[A Hertzsprung-Russell colour-magnitude diagram \citep{higgo12}.]{
A Hertzsprung-Russel colour-magnitude diagram. The colour bar indicates spectral class and temperature range. The Sun is a G2 spectral class star located on the main sequence. HIPPARCOS (black dots) and ÒGliese 1991Ó (dark gray dots) catalogues were used to create the plot, excluding binary systems \citep{higgo12}.}
\label{intro:hr}
\end{figure}

\noindent The Sun is a middle-aged G2V star currently situated on the main sequence of a Herzsprung-Russel (HR) diagram. This is a graphical representation of stars obtained by plotting absolute magnitude against spectral class (see Figure~\ref{intro:hr}). The Sun has a luminosity $L_\odot = 3.85 \times 10^{26}$~W, mass $M_\odot = 1.99 \times 10^{30}$~kg, and radius $R_\odot = 6.96 \times 10^{8}$~m \citep{phillips92}. It was formed by the gravitational collapse of an interstellar gas cloud. While on the main sequence, the Sun's energy is supplied by hydrogen fusion reactions in the core (see Section~\ref{intro:interior}). When this ceases (after $\sim10^{10}$~years), the Sun will progress to the red giant phase. Red giants are larger, more luminous stars, of spectral type K or M, with lower effective surface temperatures. After another $150\times 10^{6}$~years, the red giant sheds it outer layers to form a planetary nebula (a ring-shaped nebula formed by an expanding shell of gas around the aging star). The star will leave behind its core, becoming a compact object known as a white dwarf.

The Sun is the only star that we can observe at high angular resolution, and is therefore ideal for improving our understanding of Sun-like stars in the Universe. Events on the Sun also have a direct impact on life and technologies in the near-Earth environment. Solar eruptions can cause large amounts of energetic particles, plasma, and radiation across the entire electromagnetic spectrum, to hurtle towards the Earth. This can damage spacecraft instrumentation and electrical powergrids, disrupt radio and GPS communications, as well as divert polar airline flights and spacewalks (amongst other effects). As a consequence, solar physics is on the cutting edge of current physics research. The need to accurately predict these solar eruptions increases as our society grows more technologically dependent. This thesis examines regions on the Sun that are the source of these eruptions, as improving our understanding of the basic physics behind the eruption processes will aid prediction methods.

\begin{figure}[!t]
\centerline{\includegraphics[scale=0.55]{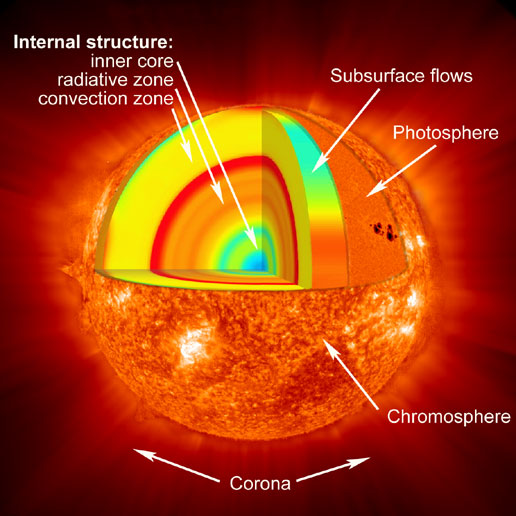}}
\caption[Cut of the layers of the Sun, \emph{courtesy of NASA}.]{Cut of the layers of the Sun, \emph{courtesy of NASA}. The solar core is the source of energy, where fusion heats plasma up to $\sim15 \times10^{6}$~K. Radiative diffusion transports the energy from the core in the radiative zone, out to 0.7$~R_\odot$. The convection zone is heated from the base of the tachocline, allowing convective currents to flow to the photosphere, the visible surface of the Sun. Temperatures rise from $\sim10^{3}$~K in the photosphere to $\sim10^{4}$~K in the chromosphere, then rise rapidly in the transition region to over 10$^{6}$~K in the solar corona.}
\label{intro:sunlayers}
\end{figure}

In this chapter, the basic theory behind the processes on the Sun relevant to this thesis will be discussed, beginning with a description of its structure. The Sun can be separated into distinct regimes, namely the interior, surface zones, atmosphere, and its extension, the solar wind. See Figure~\ref{intro:sunlayers} for an illustration of the various solar layers, which will be described in more detail in the following sections.

%%%%%%%%%%%%%%%%%%%%%%%%%%%%%%%%%%%%%%%%%%%%%%%%%%%%%%%%%%%%%%%%%%%%%%%%%%%%%%%%%%%%%%%%%%%%%%%%%%

\section{Internal Structure}
\label{intro:interior}
Nuclear fusion of hydrogen into helium accounts for the energy source in the solar core (out to 0.25~$R_\odot$, temperature of 15~MK, and density of $1.6 \times 10^5$~kg~m$^{-3}$). The dominant process is the proton-proton chain, which provides 99\% of the Sun's energy. The proton-proton chain consists of a series of reactions. First, deuterium ($^{2}$H) is formed from the collision of two protons (p), 
\begin{equation}
\label{intro:1a}
\mathrm{p~+~p~\to~^{2}H~+~e^{+}~+~\nu_{e}} \ ,
\end{equation}
where e${^{+}}$ is a positron, and $\nu_e$ is an electron neutrino. Alternatively, the chain can be started by a proton-electron-proton reaction,
\begin{equation}
\label{intro:1b}
\mathrm{p~+~e^{-}+~p~\to~^{2}H~+~\nu_{e}} \ ,
\end{equation}
where e$^{-}$ is an electron. The next steps are then,
\begin{equation}
\label{intro:2}
\mathrm{^{2}H~+~p~\to~^{3}He~+~\gamma} \ ,
\end{equation}
\begin{equation}
\label{intro:3}
\mathrm{^{3}He~+^{3}He~\to~^{4}He~+~2p} \ ,
\end{equation}
where $\gamma$ is a gamma ray, and $\mathrm{^{3}He}$ and $\mathrm{^{4}He}$ are helium isotopes with one and two neutrons respectively. Note that Equations~\ref{intro:1a} and \ref{intro:2}, or Equations~\ref{intro:1b} and \ref{intro:2}, must occur twice for each time Equation~\ref{intro:3} occurs. 

The other 1\% of the Sun's energy comes from the carbon-nitrogen-oxygen (CNO) cycle, more details of which can be found in \citet{phillips92}. For both the p-p chain and CNO cycle nuclear reaction sequences, the net result can be written,
\begin{equation}
\label{intro:pp}
\mathrm{4p~\to~^{4}He~+~2e^{+}~+~2\nu_e~+~energy} \ .
\end{equation}
The neutrinos escape from the Sun and carry away a small amount of energy. The energy, in $\gamma$ ray photons, positron annihilation, and particle kinetic energy, are all available to heat the Sun, i.e., to provide its luminosity.

Energy is transported via radiation from the core out to a radius of $\sim0.7~R_\odot$. The temperature in the radiative zone drops to $\sim5$~MK, with the radiation field closely approximated by a black body, for which the spectral radiance, $B_\nu$, can be described by the Planck function \citep{planck1900},
\begin{equation}
\label{intro:planck}
B_\nu(T)~=~\frac{2h\nu^{3}}{c^2}~\frac{1}{\exp(h\nu/k_BT)~-~1} \ ,
\end{equation}
where $B_\nu(T)$ is the intensity of radiation per unit frequency at a temperature $T$, $h$ is Planck's constant, $c$ is the speed of light, and $k_B$ is Boltzmann's constant.

At the top of the radiative zone, the temperature is low enough that elements such as iron are not fully ionised and radiative processes are less effective, and convective processes become more important.  Thus, from $\sim0.7~R_\odot$ out to the solar surface, energy is transported by convection. The Schwarzschild criterion \citep{schwarzchild1906} indicates when convection is likely to occur, when the radial gradient of the temperature is greater than the adiabatic temperature gradient,
\begin{equation}
\left | \left ( \frac{dT}{dr} \right )_{\mathrm{radiative}} \right |~>~\left | \left ( \frac{dT}{dr} \right )_{\mathrm{adiabatic}} \right | \  \ .
\end{equation}
Here, the adiabatic temperature describes that of a convecting cell, while the radiative temperature is that exterior to the cell. This criterion basically determines whether a rising (sinking) globule of gas will continue to rise (sink), or if it will return to its original depth. In the convection zone, heat from below can no longer be transmitted towards the surface by radiation alone, and heat is transported by material motion. As the gas rises it cools and begins to sink.  When it falls to the top of the radiative zone, it heats up and starts to rise once more.  This process repeats, creating convection cells and the visual effect of boiling on the Sun's surface (i.e., granulation). Temperatures of $\sim1 - 2$~K can be found in this zone, and at the top (on the surface) $T \sim5800$~K. It is worth noting that the tachocline is found at the base of the convection zone, specifically where the convection zone transitions into the stable radiative zone. This is where solar magnetic fields are regenerated via the solar dynamo on solar cycle time scales \citep{rempel03}, which will be discussed in more detail in Section~\ref{intro:solar_cycle}.

%%%%%%%%%%%%%%%%%%%%%%%%%%%%%%%%%%%%%%%%%%%%%%%%%%%%%%%%%%%%%%%%%%%%%%%%%%%%%%%%%%%%%%%%%%%%%%%%%%

\section{Solar Surface and Atmosphere}
\label{intro:atm}

The photosphere is the visible surface of the Sun, with a total density of $\sim10^{17}$~cm$^{-3}$ (neutral hydrogen, electron and helium densities), and a thickness of less than 500~km. It is usually defined as where the optical depth, $\tau$, equals 2/3 for radiation at a wavelength of 5000~\AA\ (visible light). Note that the `2/3' value can be found from the Eddington-Barbier expression, telling us that the solar surface flux is equal to $\pi$ times the source function (defined in Section~\ref{theory:rte_opt}) at an optical depth of 2/3 \citep{rutten03}. Optical depth expresses the quantity of light removed from a beam by scattering or absorption during its path through a medium, and can be defined as,
\begin{equation}
\frac{I}{I^o}~=~e^{-\tau}~~,
\label{intro:optical}
\end{equation}
where $I^o$ is the intensity of radiation at the source, and $I$ is the observed intensity after a given path. Optical depth will be discussed further in Section~\ref{theory:rte_opt}. The photosphere has a radiation spectrum similar to that of a blackbody at 5778~K. The strongest lines observed in the spectrum are Fraunhofer absorption lines due to the tenuous layers of the atmosphere above the solar surface.

The main features of interest to this thesis on the solar surface are sunspots, which are discussed in detail in Section~\ref{intro:sunspots}. As mentioned previously, the solar surface is covered by granulation, representing the tops of convective cells rising from the solar interior. There are two main cell sizes: granules of the order of $\sim100 -1000$~km across and with lifetimes of $\sim10$~minutes, and supergranules that are typically $\sim30,000$~km in diameter and have lifetimes of $\sim1 - 2$~days (\citealp{simon64}; \citealp{rieutord10}). The boundaries of supergranules contain magnetic field concentrations that give rise to the  magnetic network in the layer above the photosphere, known as the chromosphere. 

\begin{figure}
\centerline{\includegraphics[width=\textwidth]{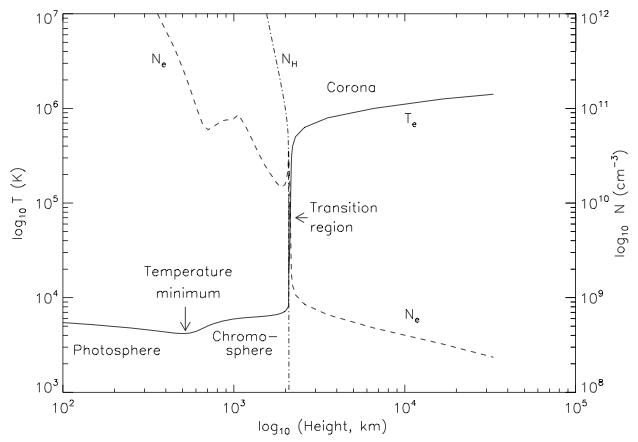}}
\caption[1D static model of solar temperature ($T\mathrm{_e}$; K) and density ($N\mathrm{_e}$ and $N\mathrm{_H}$; cm$^{-3}$) as a function of height.]{1D static model of solar electron temperature ($T\mathrm{_e}$; K), electron density ($N\mathrm{_e}$; cm$^{-3}$), and neutral hydrogen density ($N\mathrm{_H}$; cm$^{-3}$) as a function of height. Variation of $T\mathrm{_e}$, $N\mathrm{_e}$, and $N\mathrm{_H}$ are shown using logarithmic scales. \emph{Modified Figure by \citet{phillips08b}, based on one-dimensional model calculations of \citet{vernazza81}, \citet{fontenla88}, and \citet{gabriel76}}.}
\label{intro:temp}
\end{figure}

The temperature of the chromosphere first falls with height to 4400~K (see Figure~\ref{intro:temp}), at the temperature minimum. From here, the temperature rises with height to $\sim20,000$~K at the chromosphere, with a total density of $\sim10^{15}$~cm$^{-3}$, and a thickness of $\sim2,000$~km. The brightness of the photosphere tends to overwhelm the chromosphere in the optical continuum, however the hotter chromospheric temperatures mean strong H$\alpha$ emission is present. Plage regions are a typical example of a feature in the chromosphere. These are bright regions above the photosphere that are typically found near sunspots, of opposite polarity to the main spot. Other interesting features of note in the chromosphere are filaments, which are long, dark structures on the solar disk. If found on the solar limb, they are referred to as prominences. Hair-like structures, known as fibrils, and dynamic jets, known as spicules (plasma columns), are also observed in the chromosphere. Spicules can be found on the solar limb, and typically reach heights of $\sim$~3,000 -- 10,000 km above the solar surface. They are very short-lived, rising and falling over $\sim5-15$~minutes. 

A so-called `transition region' exists above the chromosphere, where, across a height of $\sim100$~km, temperatures increase from 0.01~MK to 1~MK, and densities decrease from $\sim10^{11}$~cm$^{-3}$ to $\sim10^{9}$~cm$^{-3}$. It marks a point where magnetic forces dominate over gravity, gas pressure and fluid motion (compared to the photosphere). This concept will be discussed in more detail in Section~\ref{mhd:motion}. Most observations of the transition region are obtained from UV and EUV wavelengths. This is because the extreme temperatures in this region result in emission of UV and EUV from carbon, oxygen, and silicon ions \citep{mariska93}.

\begin{figure}
\centerline{\includegraphics[width=\textwidth]{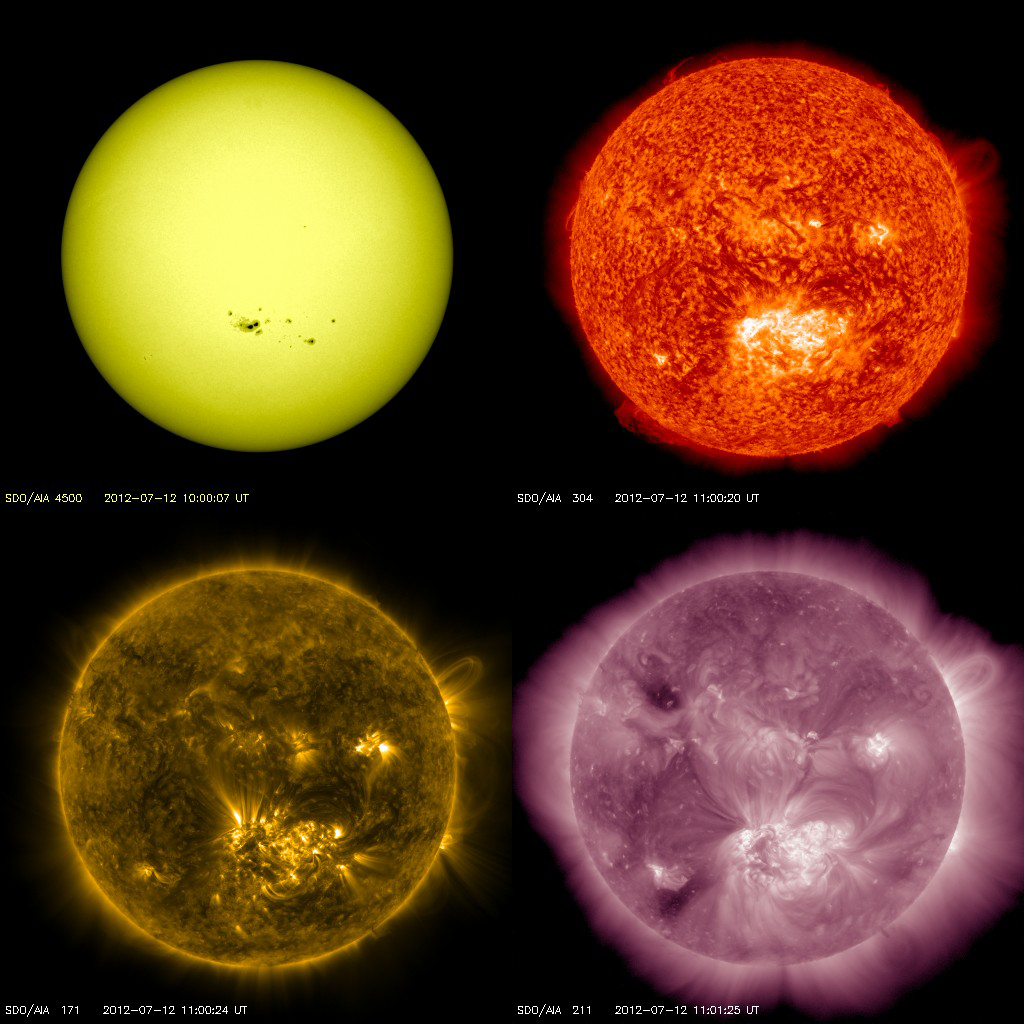}}
\caption[Full-disk images of the Sun at various wavelengths on 2012 July 12, observed mainly in the extreme ultraviolet by the AIA instrument onboard \emph{SDO}.]{Full-disk images of the Sun at various wavelengths on 2012 July 12, observed mainly in the extreme ultraviolet by the Atmospheric Imaging Assembly (AIA) instrument onboard \emph{SDO}. Upper left panel is a white light 4500~\AA\ image ($\sim5 \times 10^{3}$~K, photosphere), upper right panel at 304~\AA\ ($\sim5 \times 10^{4}$~K, chromosphere and transition region), lower left panel at 171~\AA\ ($\sim6.3 \times 10^{5}$~K, quiet corona and upper transition region), and lower right panel at 211~\AA\ ($\sim2 \times 10^{6}$~K, active region corona).}
\label{intro:aia}
\end{figure}

Figure~\ref{intro:temp} presents an illustration that summarises the changing temperature and density in the solar atmosphere. This is a 1D static model of the solar atmosphere, showing layers stratified into photosphere, chromosphere, transition region and corona. However, it must be noted that this stratification is a simplified view of the atmosphere; in reality the solar atmosphere is an inhomogeneous mix of these zones due to complex dynamic processes such as heated upflows, cooling downflows, field line motions and reconnections, hot plasma emission, acoustic waves, and shocks \citep{aschwanden05}. Figure~\ref{intro:aia} shows typical examples of observations of these solar atmospheric layers, using data from the \emph{Solar Dynamics Observatory} (\emph{SDO}) spacecraft. The 4500~\AA\ image shows the photospheric layer, with a number of dark sunspot regions near disk centre. The 304~\AA\ image is indicative of the chromosphere, 171~\AA\ image the quiet corona, and 2111~\AA\ image the active corona. All of these three wavelengths show brighter intensity in the sunspot regions, as well as some loops along the limb (particularly evident in the solar northwest).

Above the transition region is the corona, the uppermost part of the solar atmosphere, and an extremely hot and tenuous region. Total densities range from $\sim(1-2) \times 10^{8}$~cm$^{-3}$ at the base ($\sim$ 2,500 km above the photosphere), to $\leq 10^{6}$~c‡m$^{-3}$ for heights $\geq 1$~R$_\odot$ above the photosphere \citep{aschwanden05}. The total density here represents electron, proton, and helium densities, as plasma is fully ionised in the corona due to the high temperatures (compared to the presence of neutral hydrogen in the photosphere and chromosphere). The temperature is at $\sim2 \times 10^{6}$~K in the corona. Temperatures can be hotter in regions associated with sunspots\footnote[2]{In regions of increased magnetic field density, such as sunspots, temperatures of 2 -- 6 MK are observed, while temperatures of 1 -- 2 MK are observed across the quiet Sun.} (see Section~\ref{intro:sunspots}), where large complex loop structures exist. The amount of complexity of the magnetic field structures has been found to be related to the occurrence of solar flares (a topic of investigation in this thesis, see Section~\ref{intro:complexity} for a description). Note that there are two distinct magnetic zones in the corona with different magnetic topologies, closed and open magnetic field regions. In closed-field regions, closed-field lines connect back to the solar surface, and these regions are the source location of the slow solar wind\footnote[3]{Stream of charged particles, consisting mainly of electrons and protons.} ($\sim400$~km~s$^{-1}$). Open magnetic field regions pervade the solar north and south poles. Regions of open field known as coronal holes can generally be found at the poles, and can also be found intermittently closer to the equator. Open field regions connect the solar surface to interplanetary space, and are the source of the fast solar wind ($\sim800$~km~s$^{-1}$).

The solar wind is a constant out-stream of charged particles of plasma from the solar atmosphere, consisting mainly of electrons and protons at energies of $\sim1$ keV. It was predicted by \citet{parker58}, who assumed the wind is steady, isothermal, and spherically symmetric. He derived an equation of motion that reveals the existence of the solar wind,
\begin{equation}
\frac{1}{v}\frac{dv}{dr}(v^2 - \frac{2kT}{m}) = \frac{4kT}{mr} - \frac{GM_\odot}{r^2}
\label{eqn:parker}
\end{equation}
where $r$ is the distance from the centre of the star, $v$ is the solar wind speed, $k$ is Boltzmann's constant, and $G$ is the gravitational constant. The right hand side (RHS) of the equation was found to be negative for temperatures, T, observed in the lower corona. However, the gravitational force term decreases as $1/r^2$, so it will eventually become smaller than the first term on the RHS, which decreases only as $1/r$. Thus, there is a critical radius, $r_c$, beyond which the RHS changes sign and the left hand side (LHS) goes through zero,
\begin{equation}
r_c = \frac{GM_\odot m}{4kT}
\end{equation}

\begin{figure}
\centerline{\includegraphics[width=0.8\textwidth]{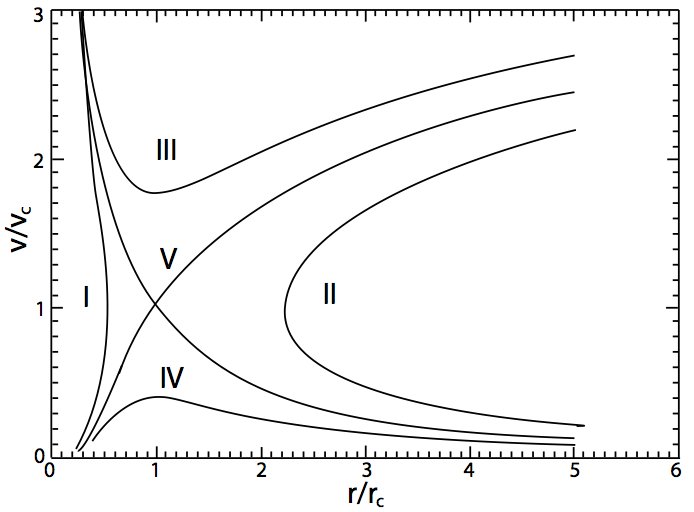}}
\caption{The five classes of solutions to the Parker solar wind model for a steady, isothermal, spherically symmetric outflow.}
\label{intro:wind}
\end{figure}

Analysis of these conditions for Parker's solar wind model shows that there exists five classes of solutions to Equation~\ref{eqn:parker}, which are shown in Figure~\ref{intro:wind}. Solutions I and II are excluded as they are double-valued (two values of velocity at the same height), and confined to small and large $r$ respectively. Solution II is also disconnected from the surface. Solution III is supersonic close to the Sun, which does not satisfy observations. Solution IV, known as the `solar breeze' solution, remains subsonic. Solution V is the considered the standard solar wind solution, starting subsonically near the Sun, and reaching supersonic speeds. This solution passes through the critical point, where $dv/dr$ is undefined at $v = v_c$ and $r = r_c$. At this point the coefficient of $dv/dr$ and the RHS of Equation~\ref{eqn:parker} vanish simultaneously. Although Solution V is generally considered the correct solution, it must be noted that it is only an approximation, as the assumptions of radial expansion and isothermality are not entirely correct in reality. 

Gas pressure dominates over magnetic pressure in the solar wind, and thus the solar wind drags out solar magnetic field lines. These field lines become wound up as a result of solar rotation, to form a Parker spiral. This is an Archimedean spiral described by the equation,
\begin{equation}
r - r_0~=~\frac{v}{\Omega} (\theta - \theta_0) \ ,
\end{equation}
where $r_0$ can be taken at the solar surface, $\Omega = 2.7 \times 10^{-6}$~rad s$^{-1}$ is the solar angular rotation rate, and $\theta$ is a longitude angle \citep{zirin98}. 
The resulting Archimedean spirals leave the Sun near-vertically to the surface at an angle of $0^\circ$, and cross the Earth orbit at an angle of $\sim 45^\circ$. The solar wind eventually terminates when it reaches the edge of the heliosphere. The termination shock is the location where the solar wind slows from supersonic to subsonic speeds, which was observed by the Voyager II mission in August 2007 \citep{burlaga08}. In this thesis, the extension of the solar magnetic field into interplanetary space is not of concern, but the structure of the magnetic field on the solar surface and in the solar atmosphere is. Active region magnetic fields are the main focus of investigation, which will be discussed in the following section.

%%%%%%%%%%%%%%%%%%%%%%%%%%%%%%%%%%%%%%%%%%%%%%%%%%%%%%%%%%%%%%%%%%%%%%%%%%%%%%%%%%%%%%%%%%%%%%%%%%%%%%%%%%%%%%%%%%%%%%%%%%%%%%%%%%%%%%%%%%%%%%%%%%%

%solar cycle
%the flux emergence
%then structure

\section{Sunspots}
\label{intro:sunspots}

\begin{figure}[!t]
\centerline{\includegraphics[width=\textwidth]{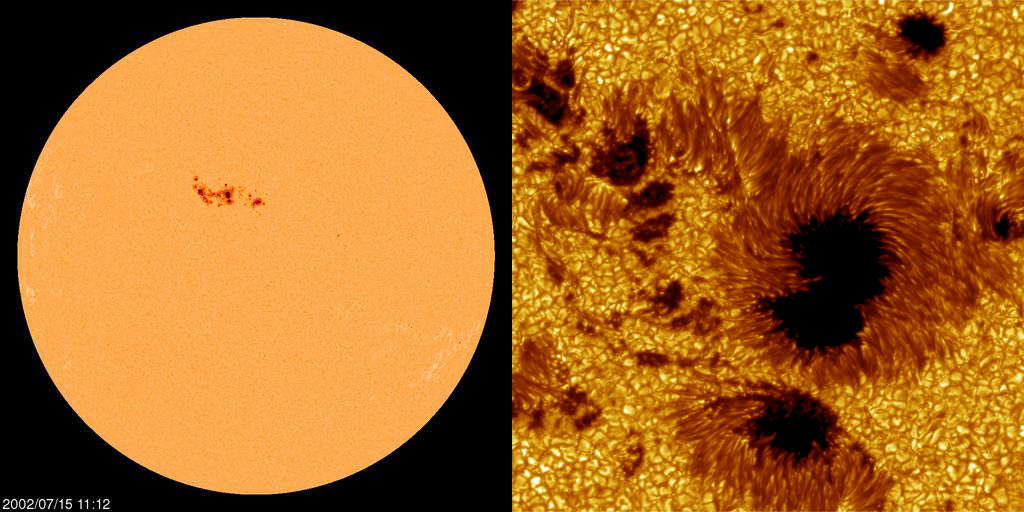}}
\caption[Typical active region on the solar disk (left), with a zoom-in showing the sunspots in the region in more detail (right).]{Left shows a full-disk \emph{SOHO}/MDI continuum image of NOAA AR 10030 on 2002 July 15. Right shows a zoom-in of this region using observations from the \emph{Swedish Solar Telescope}. \emph{Images are courtesy of SOHO (NASA \& ESA) and the Royal Swedish Academy of Sciences.}}
\label{intro:sst}
\end{figure}

Active regions (ARs) are localised volumes of the Sun's outer atmosphere where powerful and complex magnetic fields, emerging from subsurface layers, form loops that extend into the corona. They give rise to features such as sunspots, plage, fibrils and filaments in the photosphere and chromosphere \citep[see review by][]{solanki03}. Sunspots appear as dark zones in the photosphere, mainly located between latitudes of 60$^{\circ}$~N and 60$^{\circ}$~S. Figure~\ref{intro:sst} shows a typical AR (NOAA 10030\footnote[4]{The National Oceanic and Atmospheric Administration numbers ARs consecutively as they are observed on the Sun. An AR must be observed by two observatories before it is given a number, or before this if a flare is observed to occur in it. The consecutive numbering began on 1972 January 5.}) on the solar disk (left, \emph{SOHO}/MDI continuum image), with a zoom-in (right, \emph{Swedish Solar Telescope} image) showing sunspots in this AR. The dark interior of a sunspot is known as the umbra and the lighter area surrounding this is known as the penumbra\footnote[5]{Note that an umbra existing without a penumbra is known as a pore.}. This Section will discuss sunspots in detail, as they are the main focus of study in this thesis, beginning with their formation and the solar cycle, flux emergence, sunspot structure, and finally sunspot evolution.

%%%%%%%%%%%%%%%%%%%%%%%%%%%%%%%%%%%%%%%%%%%%%%%%%%%%%%%%%%%%%%%%%%%%%%%%%%%%%%%%%%%%%%%%%%%%%%%%%%

\subsection{Solar Cycle}
\label{intro:solar_cycle}

The $\alpha - \omega$ dynamo \citep{parker55a} describes the process by which the Sun's magnetic field is generated, and thus is a starting point for describing the formation of sunspots. The $\alpha$ effect describes the sub-surface magnetic field lines becoming distorted, twisted, and more complex in shape under the effect of the rotation of solar material. The $\omega$ effect describes the way in which magnetic fields are stretched out and wound around the Sun by solar differential rotation. The differential rotation rate has the general form,
\begin{equation}
\label{intro:diff_rot}
\omega(\Phi)~=~A + Bsin^2\Phi + Csin^4\Phi
\end{equation}
where $\omega(\Phi)$ is the angular velocity in degrees per day, $\Phi$ is the latitude, and A, B, and C are constants. The values of these constants differ depending on the techniques used to make the measurement, and the time period selected. \citet{snodgrass90} quote current accepted values as A~$\sim 14.7$~degrees/day, B~$\sim - 2.4$~degrees/day, and A~$\sim - 1.8$~degrees/day, obtained using cross-correlation measurements from magnetograph observations.

The magnetic polarity of sunspot pairs reverses and then returns to its original state in a $\sim$~22-year Hale cycle \citep{hale1925}, with the Sun's magnetic poles reversing every 11 years. A number of models have been put forward to explain the 22-year cycle, perhaps most notably the Babcock-Leighton model, which was first proposed by \citet{babcock61} and further elaborated by \citet{leighton64,leighton69}. This is illustrated in Figure~\ref{intro:alphaomega}, and consists of multiple stages \citep{vandriel08}:

\begin{figure}[!t]
\centerline{\includegraphics[scale=0.6]{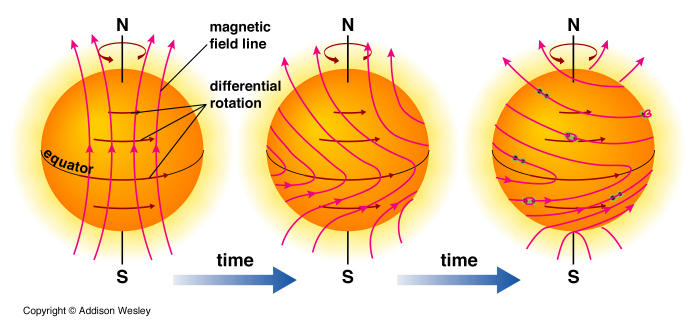}}
\caption[Illustration of the $\alpha$-$\omega$ effect that is the source of the solar magnetic field \citep{carroll06}.]{Illustration of the $\alpha$-$\omega$ effect that is the source of the solar magnetic field, beginning with a poloidal field on the left, winding up towards a toroidal field in the middle, and the emergence of complex active regions on the right \citep{carroll06}. }
\label{intro:alphaomega}
\end{figure}

\begin{itemize}
\item About three years before the onset of the new sunspot cycle, the new field to be involved is approximated by a dipole field symmetric about the rotation axis- a pure poloidal field (see left panel in Figure~\ref{intro:alphaomega}).
\item The originally poloidal field is pulled into a helical spiral in the activity belts, with the resulting field becoming toroidally amplified ($\omega$ effect, see middle panel in Figure~\ref{intro:alphaomega}). This is due to solar differential rotation, i.e., the rate of solar rotation is observed to be fastest at the equator, decreasing as latitude increases. Further amplification is reached by the $\alpha$-effect, i.e., twisting by the effect of the radial differential rotation shear (see right panel in Figure~\ref{intro:alphaomega}).
\item Each $\Omega$-shaped loop erupting through the surface (see Section~\ref{intro:emergence}) produces a bipolar active region with preceding and following magnetic polarities. The sense of positive and negative polarities is equivalent to Hale's polarity law, which states that active regions on opposite hemispheres have opposite leading magnetic polarities, alternating between successive sunspot cycles. With the $sin^2\Phi$ term in solar differential rotation (Equation~\ref{intro:diff_rot}) Sp\"{o}rerÕs Law is followed, i.e., the first active regions of the cycle are expected at higher latitudes $\pm \sim30^{\circ}$, and tend to emerge at progressively lower latitudes as a cycle progresses.
\item The general poloidal field is neutralised and reversed due to flux cancellation, with flux diffused out of following polarity spots which are closer to the poles (known as Joy's Law).
\item After 11 years, there is renewed winding up by differential rotation with the polar field reversed; the 22 year cycle restores it to the original polarity.
\end{itemize}
This model has its limitations; being heuristic, semi-quantative and kinematical, amongst others. Recent models have aimed to find fully dynamical solutions of the induction equation (defined in Chapter~\ref{chapter:theory}) together with the coupled mass, momentum and energy relations for the plasma (see the review by \citet{charbonneau10}). For example, a large focus is placed on the meridional circulation of the poloidal field in the advective dynamo model of \citet{choudhuri95}. However, the Babcock-Leighton model is sufficient to give a general overview of the source of the solar magnetic field. 

\begin{figure}[!t]
\centerline{\includegraphics[scale=0.32]{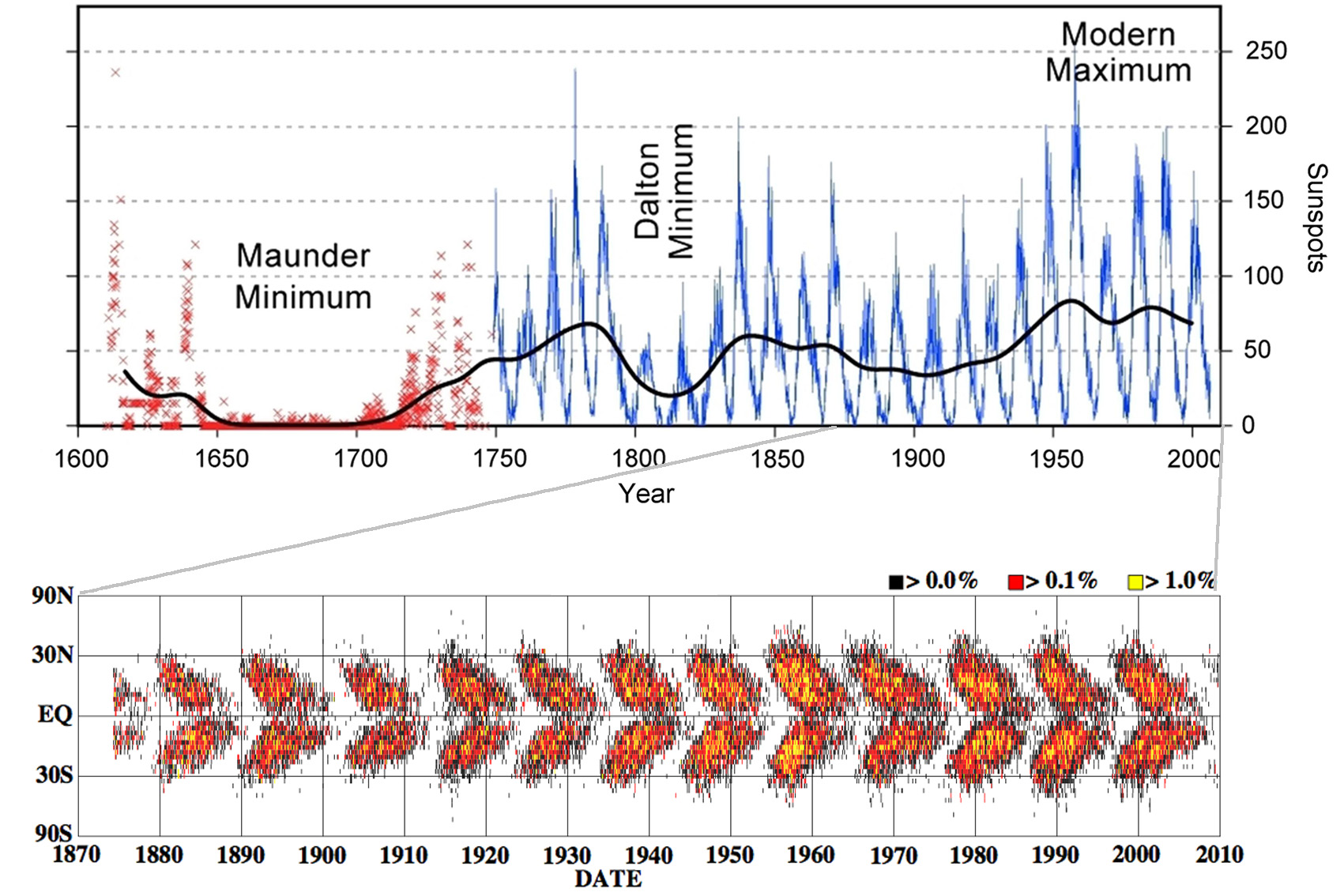}}
\caption[\emph{Upper panel}: 400 years of regular sunspot number observations, based on an average of measurements from multiple observatories worldwide. \emph{Lower panel}: Sunspot area in equal area latitude strips over 140 years of observations. \emph{Both images courtesy of NASA}.]{\emph{Upper panel}: 400 years of regular sunspot number observations, based on an average of measurements from multiple observatories worldwide. The sunspot records are shown in blue, with a polynomial fit in black. Prior to 1749 (shown in red) only sporadic observations of sunspots are available. \emph{Lower panel}: Sunspot area in equal area latitude strips (colours indicate \% of strip area as per legend) over 140 years of observations. \emph{Both images courtesy of NASA}. }
\label{intro:hale}
\end{figure}

\citet{schwabe1843} first discovered this 11-year solar cycle, with the magnetic polarities of sunspots reversing from one cycle to the next. The behaviour of the entire solar magnetic field is governed by this reversal. At the peak of the sunspot cycle (solar maximum), the greatest number of sunspots is observed, with increased solar activity. The opposite is the case at solar minimum, with very few visible spots. As the cycle progresses, more spots form closer to the equator, in a butterfly wing-like development (Sp\"{o}rers Law). Figure~\ref{intro:hale} illustrates features of the solar magnetic cycle. A number of periods of prolonged sunspot number minimum/maximum are marked in the top of the Figure, e.g, the Maunder Minimum from $\sim$ 1645 to 1715. The bottom of Figure~\ref{intro:hale} shows the butterfly-like development during the solar cycle. It is worth noting that the polarity of the foremost spots in one of the solar hemispheres is opposite to that in the other hemisphere (Hale's polarity law). 

%%%%%%%%%%%%%%%%%%%%%%%%%%%%%%%%%%%%%%%%%%%%%%%%%%%%%%%%%%%%%%%%%%%%%%%%%%%%%%%%%%%%%%%%%%%%%%%%%%

\subsection{Flux Emergence}
\label{intro:emergence}

The formation of sunspots originates in solar sub-surface magnetic field lines within the convection zone, which are distorted and become more complex in shape under the effect of the Sun's rotation. This is the $\alpha$ effect mentioned previously, describing a twisted geometry of a rising flux tube\footnote[6]{Magnetic flux tubes can be thought of as bundles of magnetic field lines.}. Since magnetic pressure adds to the gas pressure inside a magnetic flux tube \citep{parker55b}, local hydrostatic equilibrium requires,
\begin{equation}
P\mathrm{_{gas}^{external}}~=~P\mathrm{_{gas}^{internal}}+P\mathrm{_{mag}^{internal}}
\end{equation}
The pressure within the flux tube is the sum of the gas pressure $P\mathrm{_{gas}^{internal}}$ and the magnetic pressure $P\mathrm{_{gas}^{internal}}$; this is balanced by the external gas pressure $P\mathrm{_{gas}^{external}}$, while the magnetic field outside of a flux tube is assumed to be negligible. The continuous shearing of the magnetic field eventually causes a build-up of magnetic field in the azimuthal direction. The magnetic pressure ($P\mathrm{_{mag}^{internal}}$) associated with these field lines forces out the infused plasma in order to maintain a pressure balance with the surrounding plasma. Thus, $P\mathrm{_{gas}^{internal}} < P\mathrm{_{gas}^{external}}$ and buoyancy occurs. 

An $\Omega$-shaped loop of magnetic flux rises buoyantly from the convection zone, breaking through the photosphere. The emerging magnetic flux tube is described by \citet{thomasbook} as a closed-packed bundle of nearly vertical magnetic flux. The flux bundles are fragmented and twisted into separate strands as they reach the surface. The sunspot magnetic field is thus organised into a flux rope, a collection of twisted flux tubes. Upon emergence, the field lines become very dense and the small flux elements often coalesce to form pores on the surface. As more pores emerge, they grow and move towards each other, coalescing and thus forming a larger sunspot. 

\begin{figure}[!t]
\centerline{\includegraphics[width=0.6\textwidth]{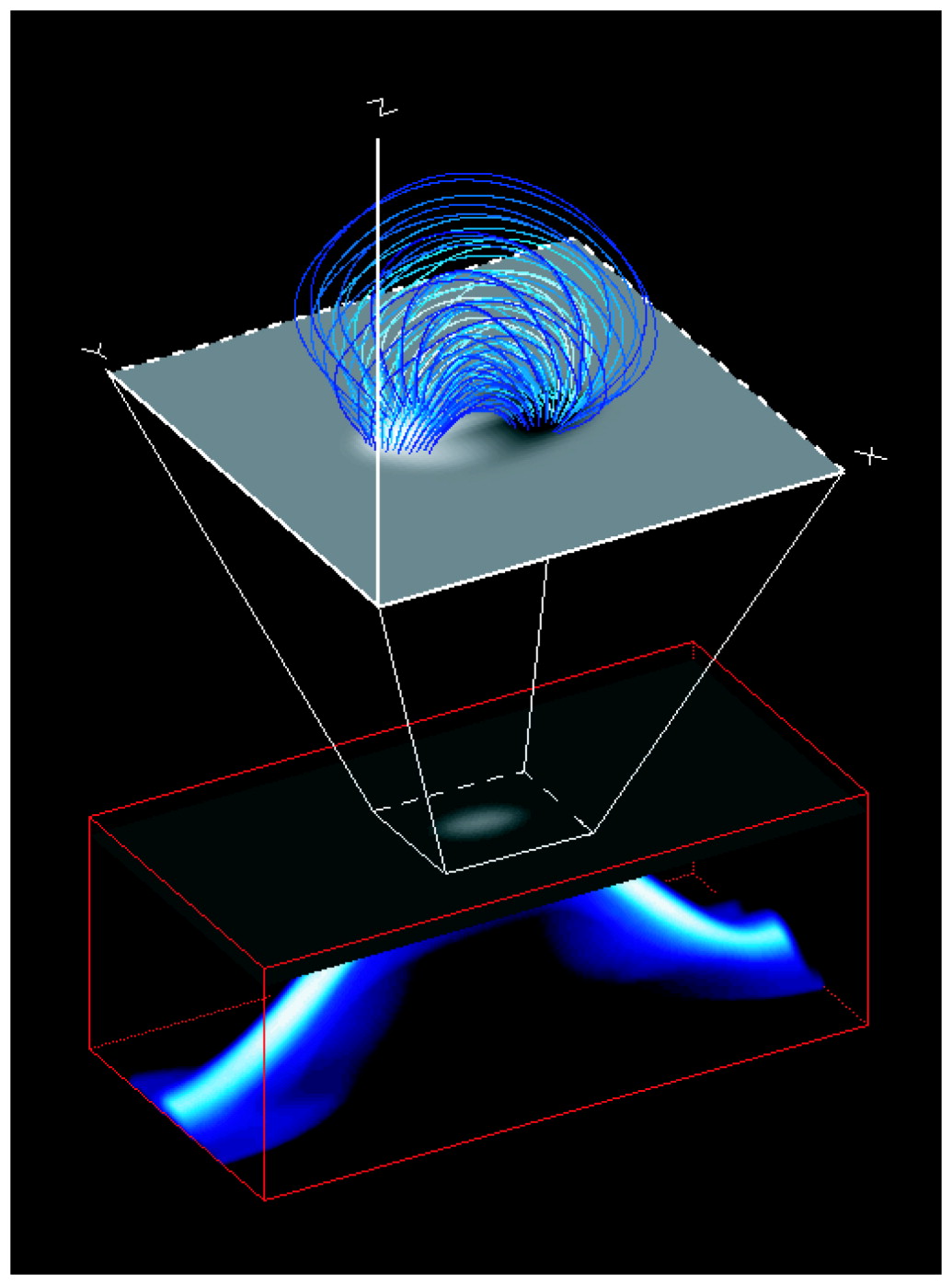}}
\caption[Snapshot of a simulation of the model corona being driven by an emerging $\Omega$ loop  \citep{abbett03}.]{Snapshot of a simulation of the model corona being driven by an emerging $\Omega$ loop. A buoyant flux rope is shown in the lower rectangular box. The vertical component of the magnetic field along the photospheric lower boundary is shown as a grey-scale image (white areas represent positive polarity and black areas represent negative polarity). Magnetic field lines are traced in blue in the atmosphere above \citep{abbett03}.}
\label{intro:flux_emerg}
\end{figure}

Although the structure of flux ropes near the photosphere has been well-studied, the deeper structure of sunspot regions and the processes involved in flux emergence remain poorly understood. There has been much previous work using simulations to try to accurately model an emerging sunspot. For example, much progress has been made by the \emph{Solar Multidisciplinary University Research Initiative} (MURI) consortium; an example of their work can be found in Figure~\ref{intro:flux_emerg}. This is a reproduction of Figure 1 of \citet{abbett03}, who present a combined subsurface-atmospheric model of an emerging sunspot in 3D.  

It is worth noting that, as a flux-tube reaches the photosphere, the magnetic pressure is greater than the gas pressure in the tube reaching into the atmosphere above. This causes the magnetic fields to fan apart, forming a magnetic field structure that will be discussed more thoroughly in the next section.

%%%%%%%%%%%%%%%%%%%%%%%%%%%%%%%%%%%%%%%%%%%%%%%%%%%%%%%%%%%%%%%%%%%%%%%%%%%%%%%%%%%%%%%%%%%%%%%%%%

\subsection{Field Structure}
\label{intro:structure}

\begin{figure}[!t]
\centerline{\includegraphics[scale=0.4]{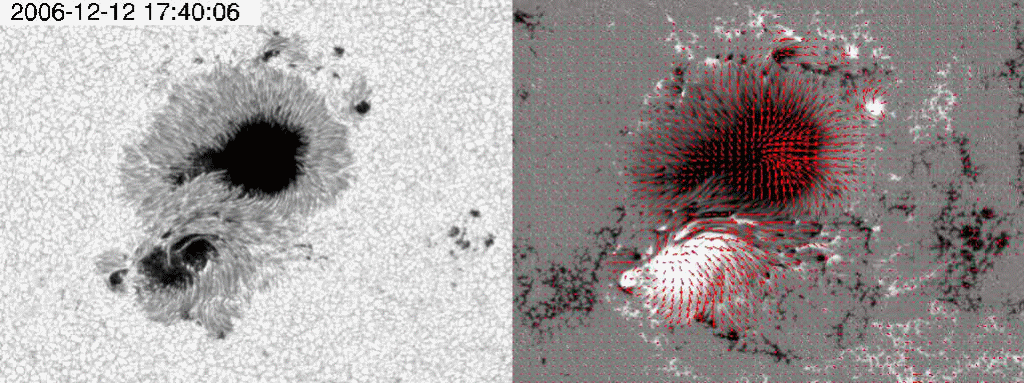}}
\caption[Sunspot observations taken with the Solar Optical Telescope (SOT) onboard the \emph{Hinode} spacecraft \citep{benz08}.]{Sunspot observations taken with the Solar Optical Telescope (SOT) onboard the \emph{Hinode} spacecraft. The left panel shows a white light image, with dark umbra and pores ($T\mathrm{_{eff}^{umbra}} \sim4500$~K), surrounding penumbra ($T\mathrm{_{eff}^{penumbra}} \sim5500$~K), and granules in the quiet outlying area ($T\mathrm{_{eff}^{granule}} \sim5800$~K). The right panel shows a vector magnetogram of the same region, black indicating negative polarity (into surface) and white positive polarity (out of surface). The direction of the magnetic field is shown by the red arrows, with its strength indicated by the length of the arrow \citep{benz08}.}
\label{intro:sunspot_structure}
\end{figure}

Sunspot magnetic fields are from 100 to 5000 times more intense than the surrounding quiet field, i.e., thousands of gauss (G) versus tens to hundreds of G, where 1~G$~= 10^{-4}$~Tesla. This inhibits convection and hence heat transport. This lowers the temperature to $\sim4000$~K, and the sunspots appear darker than the rest of the surface. The opacity also decreases with decreased temperature and density in the sunspot, and deeper geometrical levels in the spot can be seen than in the surrounding photosphere. A typical sunspot structure is shown in Figure~\ref{intro:sunspot_structure}, including typical temperatures for the various features, from observations by the \emph{Hinode} spacecraft (see Chapter~\ref{chapter:instrumentation}). It is worth noting that as a sunspot approaches the limb, the width of the penumbrae on the diskward side decreases at a greater rate than the side of the spot near the limb \citep[known as the Wilson effect;][]{brayloughhead58}. 

\begin{figure}[!t]
\centerline{\includegraphics[scale=0.4]{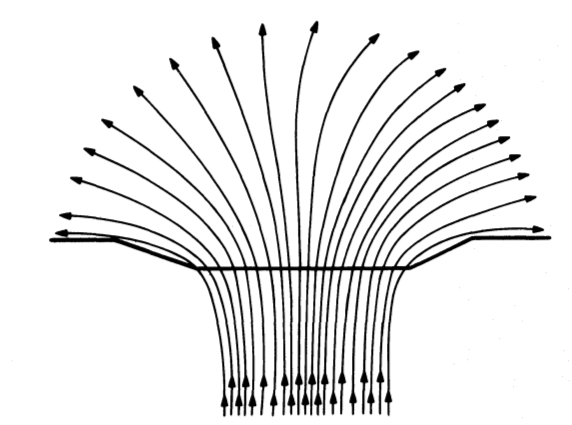}}
\caption[Sketch of the Parker sunspot model, showing the conventional idea of the magnetic field configuration of a sunspot \citep{parker79}.]{Sketch of the Parker sunspot model, showing the conventional idea of the magnetic field configuration of a sunspot. The heavy line represents the visible surface of the Sun \citep{parker79}.}
\label{intro:parkersunspot}
\end{figure}

The magnetic field of a sunspot is almost vertical (i.e., radial) at the centre of the spot, with umbral field strength usually between $2000 - 4000$~G.  The inclination of the field from vertical increases with increasing radius, while field strength decreases, with the field at $\sim60 - 70^{\circ}$ inclination at the edge of the sunspot and field strength dropping below $\sim1000$~G. The field strength also decreases with height \citep{solanki03}. See Figure~\ref{intro:parkersunspot} for a sketch of the typical magnetic field configuration of a sunspot. This shows the fan-like structure of field lines mentioned Section~\ref{intro:emergence}. \citet{beckers69} found that the radial variation of $B$ could be best represented by $B(r) = B_{0}/(1+ r^2)$, where $B_{0}$ is the central umbral field strength and $r$ the fractional radius, so that the field falls to half its central value at the edge of the spot. 

\begin{figure}[!t]
\centerline{\includegraphics[width=0.75\textwidth]{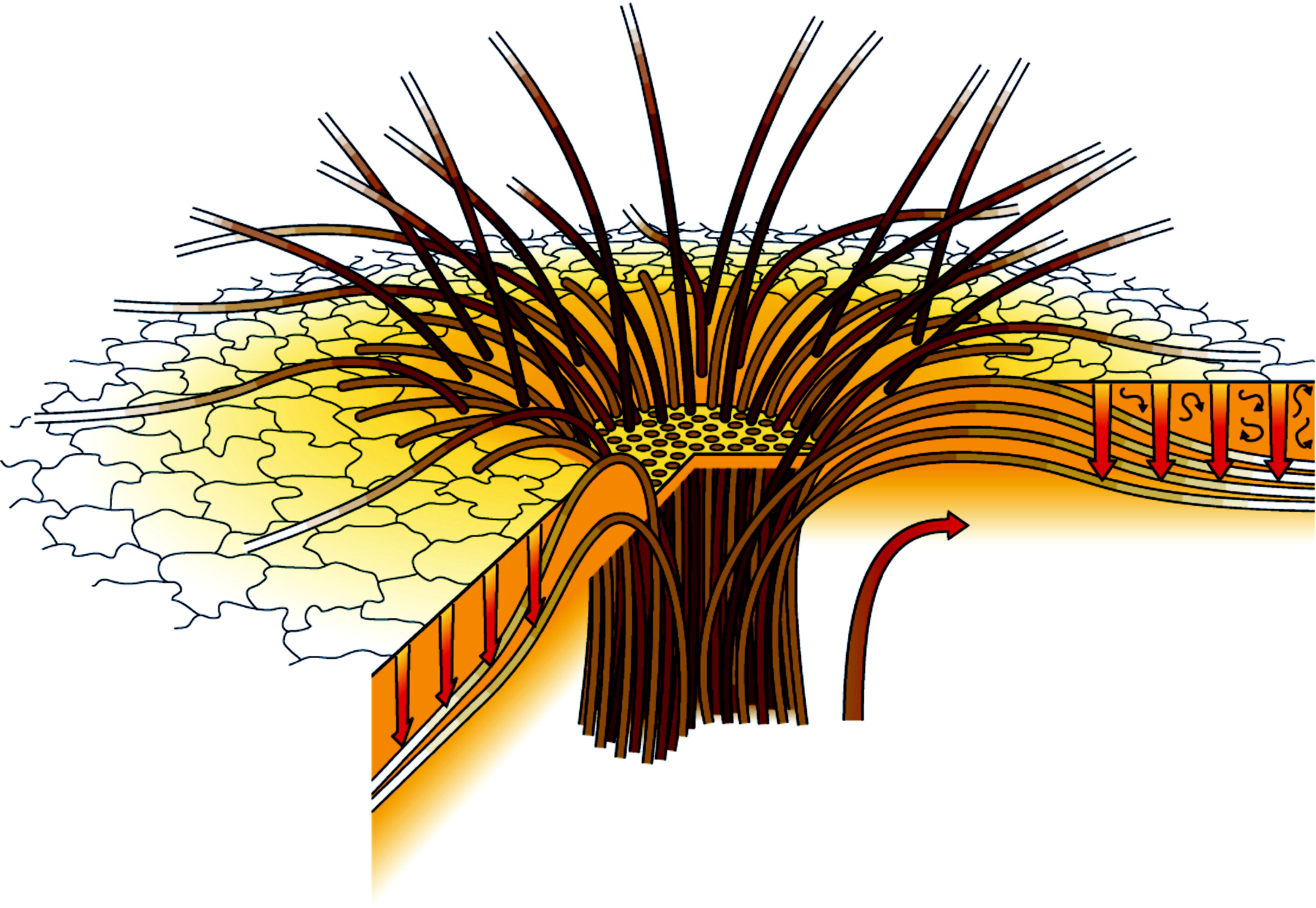}}
\caption[Sketch of an `interlocking-comb' structure \citep{thomas02}.]{Sketch of an `interlocking-comb' structure, in which granular convection plays a key role in submerging the returning penumbral flux tubes and establishing the structure of the penumbral magnetic field \citep{thomas02}.}
\label{intro:fine}
\end{figure}

\citeauthor{beckers69} also noted that the inclination of the magnetic field in the penumbra varies azimuthally. Figure~\ref{intro:fine} reproduces Figure 1 of \citet{thomas02}, showing a more detailed sketch of the `interlocking-comb' structure\footnote[7]{This term was coined in a study by \citet{thomas92}, however this form of structure has also been called by various other names: `spines' \citep{lites93}, `fluted' \citep{title93}, and `uncombed' \citep{bellotrubio03} are some examples.} of the magnetic field in a sunspot penumbra. Bright radial filaments, where the magnetic field is more inclined towards the vertical (known as `spines'), alternate with dark filaments in which the field is nearly horizontal. \citet{langhans05} found the inclination of the bright components with respect to the vertical to be $\sim35^{\circ}$ in the inner penumbra, increasing to $\sim60^{\circ}$ towards the outer boundary. The inclination of the dark component increases outwards from approximately 40$^{\circ}$ in the inner penumbra, and are nearly horizontal in the middle penumbra. Within the dark filaments, some magnetic flux tubes extend radially outward beyond the penumbra along an elevated magnetic canopy, while other `returning' flux tubes dive back below the surface (known as `sea serpents'). The sunspot is surrounded by a layer of small-scale granular convection (squiggly arrows in Figure~\ref{intro:fine}) embedded in the radial outflow. The outflow is associated with a long-lived supergranule (large curved arrow). The submerged parts of the returning flux tubes are held down by the granular convection (vertical arrows). 

\begin{figure}
\centerline{\includegraphics[width=\textwidth]{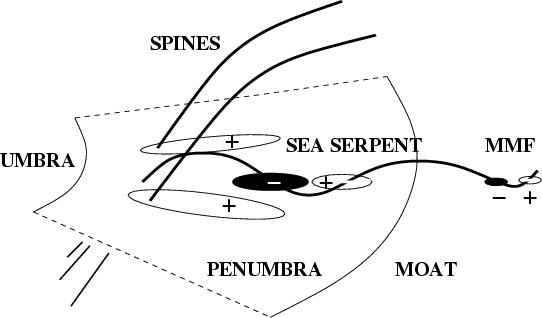}}
\caption[Sketch of a moving magnetic feature in a moat surrounding a penumbra \citep{sainzdalda08}.]{Sketch of a penumbra showing `sea-serpent' field lines in between more vertical `spine' fields, with a MMF at the moat. The ovals indicate the polarity of the different structures when observed at disk center \citep{sainzdalda08}.}
\label{intro:mmf}
\end{figure}

Even smaller features exist in sunspot regions, and moving magnetic features (MMFs) are of particular interest in this thesis. They were first detected by \citet{sheeley69} as small bright features moving radially outwards from sunspots. \citeauthor{sheeley69} postulated that these features could be the manifestation of magnetic field erupting through the surface. Magnetic field measurements by \citet{harvey73} later confirmed that these bright features represented bipolar magnetic field concentrations. More recent observations have showed that MMFs are extensions of the sunspot penumbral field (\citealp{ravindra06}; \citealp{sainzdalda08}), their orientation correlating well with the sign and amount of twist in the sunspot field \citep[having a U-shaped configuration according to][]{lim12}. Figure~\ref{intro:mmf} reproduces Figure 3 from \citet{sainzdalda08} showing a sketch of a penumbra with MMFs. As mentioned previously, the inclined fields of the penumbra resemble `sea serpents', flanked by more vertical field lines representing penumbral `spines'. The `sea serpents' propagate across the penumbra and reach the moat, where they become bipolar MMFs. Note a sunspot moat is an annular region of typical radius 10 -- 30 Mm \citep{sheeley69}, cleared of stationary magnetic flux, that surrounds sunspots as a moat surrounds a castle. MMFs emerge and move outward at approximately the moat flow speed of  $\sim0.5 -  1$~kms$^{-1}$ (surface outflows beyond the penumbra) to form the outer boundary of the moat. However, the small spatial scale ($< 1"$) of MMFs means much of the details of their structure remains unknown. These features are investigated in more detail in Chapter~\ref{chapter:paper1} , where high-resolution photospheric magnetic field observations are used to study the evolution of small-scale magnetic field structures near a sunspot. 

\citet{thomas02} and \citet{schlichenmaier02} (among others) proposed that MMFs are the continuation of the penumbral fields that harbour the Evershed flow, later confirmed by the observations of \citet{sainzdalda08}. The Evershed flow is a horizontal outward flow of plasma across the photospheric surface of the penumbra, from the inner border with the umbra towards the outer edge. It was first discovered by \citet{evershed1909}. The speed of the flow tends to vary from $\sim1$~km$^{-1}$ at the border between the umbra and the penumbra to a maximum of approximately double this value in the middle of the penumbra. The flow speed then falls off to zero at the outer edge of the penumbra. In the chromosphere and transition region the flow reverses to an inflow (with generally higher velocities than the outflow), termed the inverse Evershed flow.

\begin{figure}
\centerline{\includegraphics[width=\textwidth]{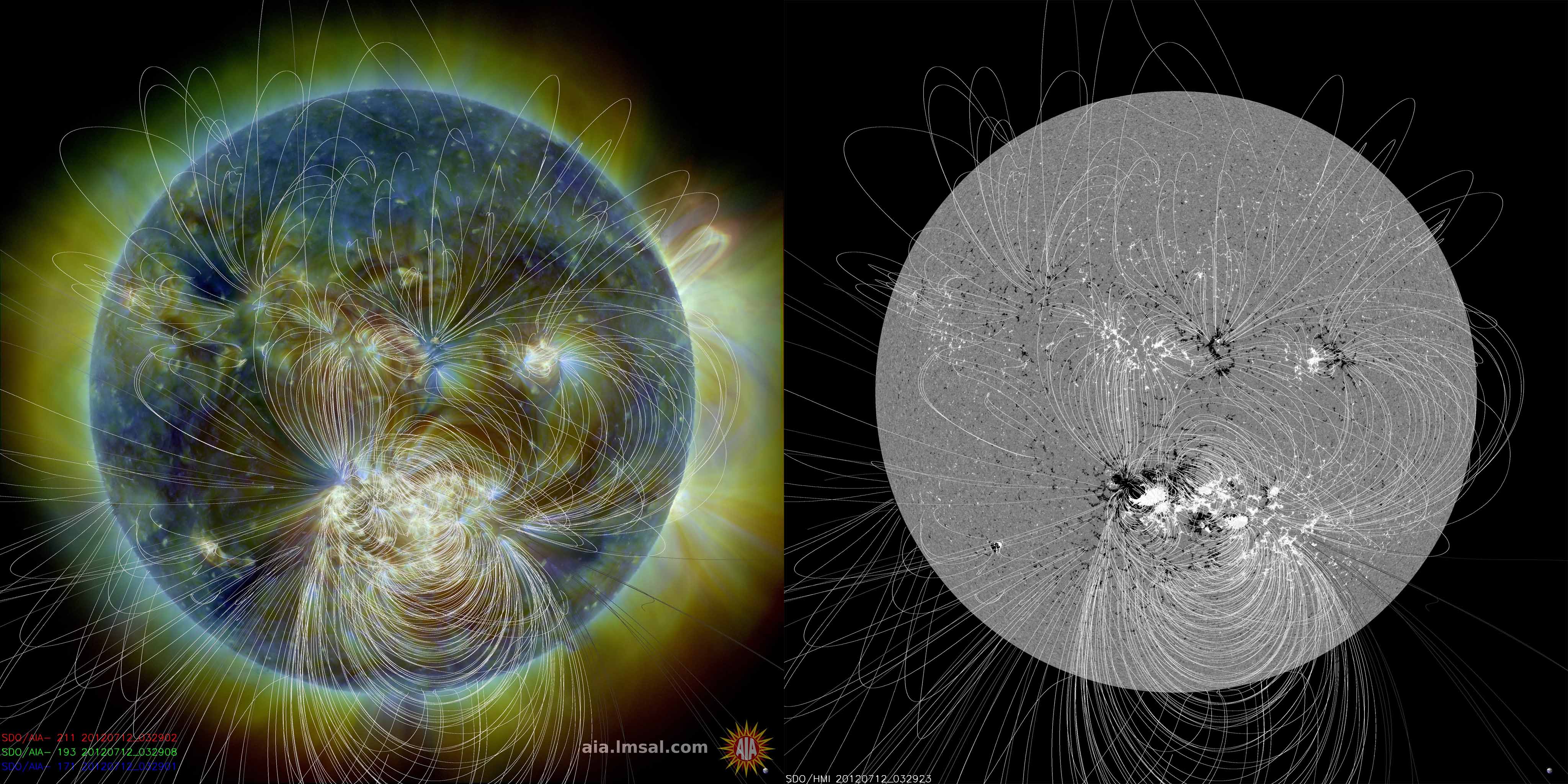}}
\caption[Example of extrapolations obtained using observations from the $SDO$ spacecraft.]{Example of extrapolations obtained using observations from the $SDO$ spacecraft. Multiple coronal wavelengths observations are shown on the left from AIA (211~\AA, 193~\AA, and 171~\AA), and a magnetogram from HMI on the right. The white lines show the field line traces obtained by the PFSS model. The enormous complex NOAA region 11520 at disk centre (Hale classification $\beta\gamma\delta$, or McIntosh $Fkc$ classification) was the source of an X1.4-class flare (see Table~\ref{intro:goes}) at 15:30~UT on 2012 July 12.}
\label{intro:hmi}
\end{figure}

Although much work has been done to try to understand the structure of sunspots, the intricacies of sunspot fine structure is still not fully resolved \citep{thomas04}. Studying the 3D magnetic field topology of ARs is important in improving our understanding of sunspot structure. Early work focused on simple extrapolation methods to study the full solar global field. For example, Figure~\ref{intro:hmi} shows a typical global potential field extrapolation (explained in Chapter~\ref{chapter:theory}) obtained using the \emph{Potential Field Source Surface} model \citep[PFSS, ][]{schrijverderosa03} with observations from $SDO$. The observations correspond to the same time as those used for Figure~\ref{intro:aia}. Recently, more advanced techniques have been used to study smaller FOVs, investigating the topology of sunspot magnetic fields in greater detail (\citealp{regnierpriest07aa}; \citealp{derosa09}; \citealp{conlon10}). 3D extrapolations will be used in Chapter~\ref{chapter:3D} in order to investigate the magnetic topology of small flux elements in a sunspot region.

%%%%%%%%%%%%%%%%%%%%%%%%%%%%%%%%%%%%%%%%%%%%%%%%%%%%%%%%%%%%%%%%%%%%%%%%%%%%%%%%%%%%%%%%%%%%%%%%%%%%%%

\subsection{Classification Schemes}

\begin{figure}[!t]
\centerline{\includegraphics[width=\textwidth]{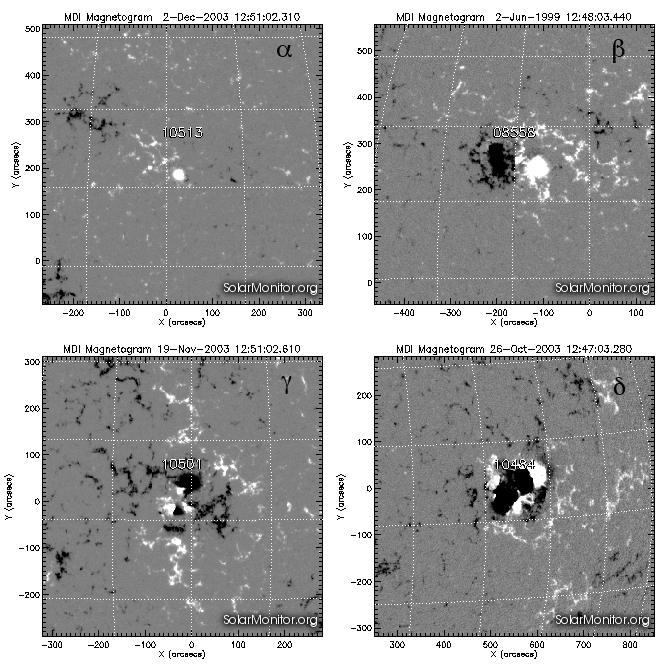}}
\caption[Examples of $\alpha$, $\beta$, $\gamma$, and $\delta$ regions.]{Examples of $\alpha$ (upper left panel), $\beta$ (upper right), $\gamma$ (lower left), and $\delta$ (lower right) regions, as defined by the Modified Mount Wilson Classification Scheme. \emph{Images are courtesty of \url{SolarMonitor.org}}.}
\label{intro:classes}
\end{figure}

It is worth mentioning how sunspots are traditionally classified. \citet{hale1919} determined a classification scheme for sunspots, with unipolar spots designated $\alpha$, bipolar spots $\beta$, and multipolar spots $\gamma$. \citet{kunzel60} added a $\delta$ classification for more complex regions that are generally associated with flaring (as will be discussed in Section~\ref{intro:flares}). Mount Wilson Observatory developed this classification further after examining regular measurements of sunspot polarities, as detailed in Table~\ref{intro:class}, and this classification scheme will be referred to throughout this thesis. Examples of $\alpha$, $\beta$, $\gamma$, and $\delta$ regions are shown in Figure~\ref{intro:classes}. Also, the sunspot shown in Figure~\ref{intro:sunspot_structure} is a $\beta\gamma\delta$ class region.

\begin{table}[!t]
\caption{The Modified Mount Wilson sunspot classification scheme \citep{brayloughhead64}.} 
\centering 
\begin{tabular}{c l} 
\vspace{0.1cm} \\
\hline\hline
Class & Description\\
\hline
$\alpha$					&	Unipolar sunspot group.				\\ 
$\beta$					&	Sunspot group having both positive and negative magnetic polarities  \\
						&	(bipolar), with a simple and distinct division between the polarities.			\\ 
$\gamma$				&	Complex sunspot group in which the positive and negative polarities are \\
						&	so irregularly distributed as to prevent classification as a bipolar group.	\\ 
$\beta\gamma$			&	Sunspot group that is bipolar but sufficiently complex that no single, \\
						&	continuous line can be drawn between spots of opposite polarities. 	\\ 
$\delta$					&	Sunspot group in which the umbrae of the positive and negative polarities \\
						&	are within 2 degrees of one another and within the same penumbra.	\\ 
$\beta\delta$				&	Sunspot group of $\beta$ class but containing one (or more) $\delta$ spots.	\\ 
$\beta\gamma\delta$		&	Sunspot group of $\beta\gamma$ class but containing one (or more) $\delta$ spots.	\\
$\gamma\delta$			&	Sunspot group of $\gamma$ class but containing one (or more) $\delta$ spots.	\\
\hline
\hline
\vspace{0.2cm} \\
\end{tabular} 
\label{intro:class} 
\end{table} 

\begin{figure}[!t]
\centerline{\includegraphics[scale=0.75]{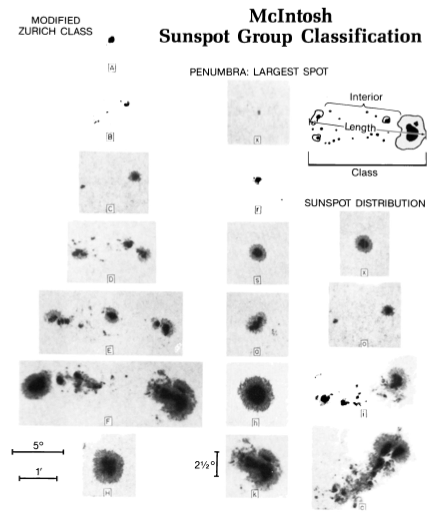}}
\caption[The three-component McIntosh classification, with examples from each category \citep{McIntosh90}.]{The three-component McIntosh classification, with examples from each category (as described in \citet{McIntosh90}). The left column shows the modified Z\"{u}rich class, the middle column primarily describes the penumbra, and the right column indicates the degree of compactness in the interior of the group.}
\label{intro:mcintosh}
\end{figure}

\citet{McIntosh90} developed a new classification of sunspots based on their likelihood of flaring, which is shown in more detail in Figure~\ref{intro:mcintosh}. The general form is Zpc, where Z is a modified Z\"{u}rich class\footnote[8]{The original Z\"{u}rich classification scheme was developed by \citet{waldmeier38}.} (left column of Figure~\ref{intro:mcintosh}, classifying sunspot group), p is the type of principal sunspot (middle column, primarily describing the penumbra), and c is the degree of compactness in the interior of the group. \citeauthor{McIntosh90} found that \emph{Fkc} class spots are much more likely to flare: \emph{F} being defined as a `bipolar group with penumbra on spots at both ends of the group', \emph{k} meaning a large and asymmetric penumbra, and \emph{c} suggesting a compact sunspot distribution. Sunspots will generally change classification at various stages of a their life cycle, often beginning with a simple structure and becoming more complex. Sunspot evolution will be described in more detail in the next section.

%%%%%%%%%%%%%%%%%%%%%%%%%%%%%%%%%%%%%%%%%%%%%%%%%%%%%%%%%%%%%%%%%%%%%%%%%%%%%%%%%%%%%%%%%%%%%%%%%%%%%%

\subsection{Field Evolution}

Flux emergence was described in detail in Section~\ref{intro:emergence}, however there are multiple stages of sunspot evolution that can be examined as an AR crosses the solar disk. Considering that ARs emerge in generally $3-5$ days, and spend $70-94$\% of their life decaying \citep{harvey93}, studying the evolution of ARs is a necessary step in improving the understanding of AR magnetic field structure. 

After an AR emerges, the region will generally continue to grow, and often an increase in complexity of the region becomes apparent. Flare activity may increase as complexity increases, a concept that will be discussed in more detail in Section~\ref{intro:complexity}. \citet{vandriel00} studied the long-term evolution of an AR with photospheric magnetic field and X-ray observations, and found solar flares occurring in the first three rotations only (the evolution lasted several months). However, coronal mass ejections (CMEs) were found to occur from emergence phase through to the decay phase of the AR (relating to removal of excess shear, which will be defined in Section~\ref{intro:complexity}).

Once ARs reach maximum maturity, they begin to decay; sunspots diminish in size and flux, then pores and small spots disappear \citep{vandriel98}. Flaring activity is strongly diminished. The decay-phase time scale of an AR differs depending on the type of region, but if there are multiple sunspots the leading spot can generally last for a number of months, with trailing spots splitting first. \citet{schrijver94} note an average time scale of disappearance to be $\sim$~4 months for solar cycle maximum, and $\sim 10$ months at solar minimum. Plage areas generally expand and dissolve into the chromospheric network. As the magnetic structure simplifies, a bipolar nature generally becomes more obvious. Following the loss of bipolar structures, remnants of the two main polarities may be tracked as monopolar structures, which often merge with other AR remnants. Larger composite unipolar regions often drift towards the poles, and can give rise to coronal holes \citep{vandriel08}.

\begin{sidewaysfigure}
\centerline{\includegraphics[width=0.9\textwidth]{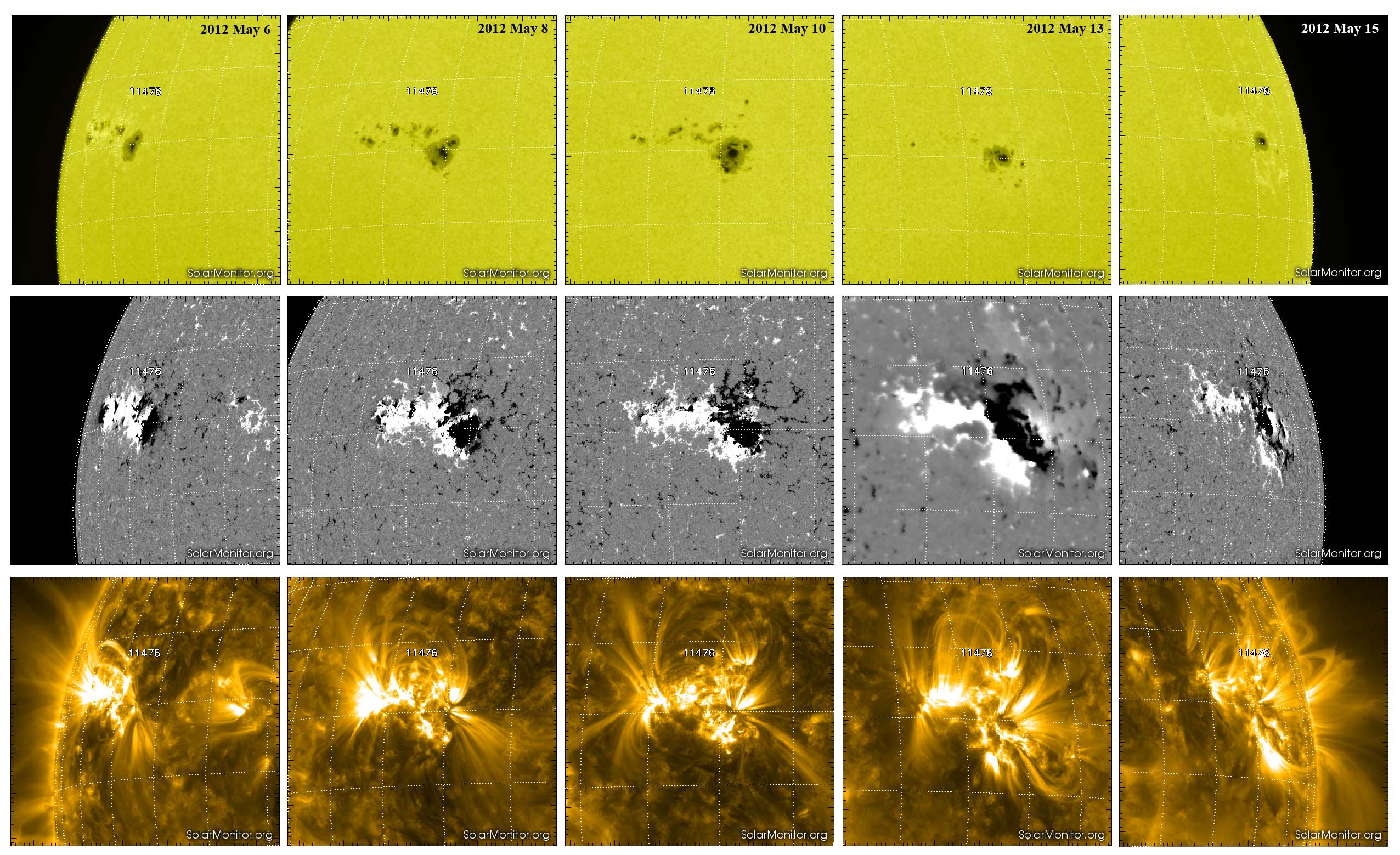}}
\caption[The evolution of NOAA AR 11476 from 5--17 May 2012 in 4500~\AA\ images (upper row), LOS magnetograms (middle row), and 171~\AA\ images (lower row).]{The evolution of NOAA AR 11476 from 5--17 May 2012 in 4500~\AA\ images (upper row, \emph{SDO}/AIA observations), LOS magnetic field (middle row, \emph{SDO}/HMI observations), and 171~\AA\ images (lower, \emph{SDO}/AIA observations). Note that the magnetogram on May 13 (fourth column, middle row) was taken from the SOLIS instrument (chromospheric magnetic field), as no HMI data was available that day. Left to right columns show observations from 2012 May 6 (at Solar X~$= - 751"$, Solar Y~$= 183"$), May 8 ($- 525"$, $210"$), May 10 ($- 211"$, $215"$), May 13 ($495"$, $219"$), and May 15 ($813"$, $169"$) respectively. \emph{Images courtesy of} \url{SolarMonitor.org}.}
\label{intro:evolu}
\end{sidewaysfigure}

Figure~\ref{intro:evolu} shows the evolution of an AR as observed by the \emph{SDO} spacecraft. The evolution is shown with LOS photospheric magnetograms from the HMI instrument, and 4500~\AA\ (temperature minimum/photosphere) and 171~\AA\ (quiet corona, upper transition region) wavelength images taken by the AIA instrument. The AR moved across the solar disk over a period of 13 days. It first emerged on the eastern limb on 2012 May 5, however was not given a classification until May 6. It was designated NOAA 11476, and determined to be a $\beta$ class region (see first column of Figure~\ref{intro:evolu}). By May 7 it had evolved into a $\beta\gamma$ region, and was the source of multiple medium-magnitude flares, which became more frequent on May 8 (second column of Figure~\ref{intro:evolu}). The AR was designated $\beta\gamma\delta$ on May 9, i.e., the most complex classification likely to flare. The region was a source of several flares larger in magnitude than previously observed. 
More larger flares occurred on May 10 as the AR neared disk centre (third column of Figure~\ref{intro:evolu}), with the largest magnitude flare (GOES M5.7, which will be defined in Section~\ref{intro:flares}) beginning at 04:11~UT on May 11. Only medium-magnitude flares occurred from the May 13 onwards, and the AR showed clear signs of decay, with trailing elements disappearing (fourth column of Figure~\ref{intro:evolu}). As the AR neared the western limb on May 15 it became a $\beta\gamma$ classification AR (fifth column of Figure~\ref{intro:evolu}), still with some flaring, but of smaller magnitude and less frequent. The AR was last observed on the western limb on May 17, when it was designated a $\beta$-class region.

\citet{deng99} also tracked an AR over 5 days as it traversed the solar disk. \citeauthor{deng99} found that increasing AR area and complexity, as well as emergence of new flux, led to a large solar flare occurring. With such large changes observed in the magnetic field structure over the course of an AR's life, it is clear that studying sunspot magnetic field evolution is a useful tool for improving our understanding of sunspot structure. Also, clear trends in flare activity over the course of the evolution period indicate a worthy area of investigation. This thesis will investigate the link between sunspot field evolution and flaring processes, but over much shorter timescales than months of a full sunspot life cycle. Magnetic field evolution will be examined over hours leading up to and after a solar flare.

%%%%%%%%%%%%%%%%%%%%%%%%%%%%%%%%%%%%%%%%%%%%%%%%%%%%%%%%%%%%%%%%%%%%%%%%%%%%%%%%%%%%%%%%%%%%%%%%%%%%%%%%%%%%%%%%%%%%%%%%%%%%%%%%%%%%%%%%%%%%%%%%%%%

\section{Solar Flares}
\label{intro:flares}

Sunspots decay and fragment by the motion of supergranulation and solar rotation. This, along with the emergence of new magnetic field, can lead to very complex and stressed magnetic fields. Solar flares occur when the energy stored in the sunspot magnetic fields is suddenly released, converting magnetic energy to kinetic energy of near-relativistic particles, mass motions, and radiation emitted across the entire electromagnetic spectrum. Large solar energetic eruptive events can include CMEs as well as flares \citep{green01,svestka01,zhang01}, but flares are the particular phenomena of interest in this thesis\footnote[9]{It is worth noting that whenever `eruptions' are mentioned in the text, CMEs are also being referred to.}. In the soft X-ray range, flares are classified as A-, B-, C-, M- or X- class according to the peak flux measured near Earth by the \emph{GOES} spacecraft over $1-8$~\AA\ (in Watts~m$^{-2}$). Each class has a peak flux ten times greater than the preceding one, with X-class flares having a peak flux of order 10$^{-4}$~Wm$^{-2}$. Table~\ref{intro:goes} outlines this in more detail. The radiated energy of a large flare may be as high as 10$^{32}$~erg (10$^{25}$~J).

%%%%%%%%%%%%%%%%%%%%%%%%%%%%%%%%%%%%%%%%%%%%%%%%%%%%%%%%%%%%%%%%%%%%%%%%%%%%%%%%%%%%%%%%%%%%%%%%%%

 \subsection{Standard Model}

The basic physics that govern flares is still not well understood despite many years of observations, but the `standard model' is widely accepted. The model suggests three main phases of the flare: pre-flare, impulsive, and decay phase. In the pre-flare phase there is a build up of stored magnetic energy. This phase will be discussed in detail throughout the thesis, however the energy can be built up via, e.g., twisting or shearing\footnote[10]{Magnetic shear angle is defined as the angle between the measured transverse field and calculated potential field.} of the field lines (see Section~\ref{intro:trigger}).

\begin{table}
\caption{GOES classification scheme for solar flares.} 
\centering 
\begin{tabular}{c c} 
\vspace{0.1cm} \\
\hline\hline
GOES class & Peak flux\\
			& 	[$W~m^{-2}]$  	\\
\hline
A			&		10$^{-8}$		\\ 
B			&		10$^{-7}$		\\ 
C			&		10$^{-6}$		\\ 
M			&		10$^{-5}$	 	\\ 
X			&		10$^{-4}$		\\ 
\hline
\hline
\end{tabular} 
\label{intro:goes} 
\end{table} 

Plasma is heated to high temperatures in the impulsive phase, with strong particle acceleration and a rapid upflow of heated material. The impulsive phase is clearly seen in hard X-rays (HXR) and radio, but intense emission is also observed in optical, UV and EUV. It is generally believed that the impulsive phase is driven by magnetic reconnection, where the stored energy that has built up is released, which accelerates coronal particles.

\begin{figure}[!t]
\centerline{\includegraphics[scale=0.8]{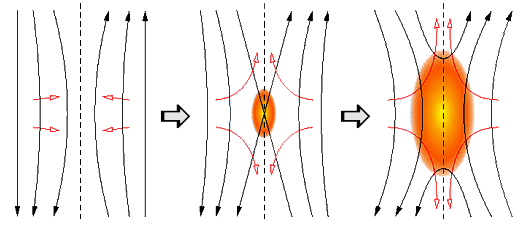}}
\caption[Basic 2D model of the reconnection process, where two oppositely directed regions of magnetic flux converge.]{Basic 2D model of the reconnection process, where two oppositely directed regions of magnetic flux converge. A diffusion region forms at the boundary where magnetic field is zero, creating horizontal outflows. \\
 ($\mathrm{http://www.aldebaran.cz/astrofyzika/plazma/reconnection\_en.html}$).}
\label{intro:reconnection}
\end{figure}

The so-called CSHKP model is a standard model of large-scale magnetic reconnection to explain solar flares and CMEs, named after pioneers in this line of research (\citealp{carmichael64}; \citealp{sturrock66}; \citealp{hirayama74}; \citealp{kopp76}). Reconnection is believed to occur when the shearing and twisting of magnetic field lines pushes the coronal magnetic field to an unstable state, one which wishes to reach a preferred lower energy state \citep{aschwanden05}. The driving force behind all reconnection models is a scenario where two regions of oppositely directed magnetic flux converge (see Figure~\ref{intro:reconnection}). Usually solar plasma is assumed to be perfectly conducting and so the magnetic fields are said to be `frozen-in' to the plasma; see Section \ref{mhd:flux_freezing} for a more detailed description. However, here the magnetic field across the boundary (diffusion zone) is zero and the frozen-in condition breaks down. This allows plasma to diffuse and reconnect, resulting in a lower energy topology. 

Note that the impulsive phase of a flare is often described in more detail by the `thick-target model', originally developed by \citet{brown71}. In this model, once the eruption has begun a reconnection `jet' of fast moving material collides with the soft X-ray (SXR) loop below, producing an MHD fast shock. This shock produces a HXR loop-top source and further acceleration. Electrons and ions stream down the legs of the loop, and produce HXRs by Bremsstrahlung radiation when they meet the chromosphere. Sometimes chromospheric material is heated so rapidly that energy cannot be radiated away, and pressure gradients can build up in the plasma. The plasma thus expands to fill the SXR loops. This process is known as chromospheric evaporation. As the eruption progresses, more and more field lines reconnect, producing an arcade of loops seen in SXR. This occurs over a timescale of $\sim$~tens of minutes \citep{phillips08}. It is worth noting that in recent times a number of issues have been raised regarding the way the energetics and plasma physics are described in the the thick target model (see, e.g., \citet{brown90}). The model faced problems with high electron beam densities, leading to a reworking of the theory to include local re-accelerations of fast electrons throughout the HXR source itself (rather than just streaming down to the chromosphere) \citep{brown09}.

\begin{figure}[!t]
\centerline{\includegraphics[scale=0.55]{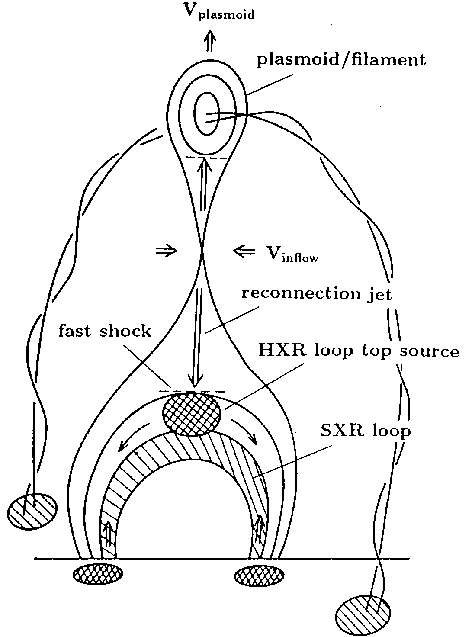}}
\caption[Standard solar flare model, as illustrated by \citet{shibata99}.]{Standard solar flare model, as illustrated by \citet{shibata99}.
}
\label{intro:flare}
\end{figure}

The decay phase of the standard model can be simply described as when the plasma cools down. A summary of the process of flaring outlined by the standard model is shown via an illustration in Figure~\ref{intro:flare}. 

%%%%%%%%%%%%%%%%%%%%%%%%%%%%%%%%%%%%%%%%%%%%%%%%%%%%%%%%%%%%%%%%%%%%%%%%%%%%%%%%%%%%%%%%%%%%%%%%%%%%%%%%%%%%%%%%%%%%%%%%%%%%%%%%%%%%%%%%%%%%%%%%%%%%%%%%%%%%%%%%%%%%%%%%%%%%%%%%%%%%%%%%%%%%%%%%%%%%

\label{intro:complexity}
\begin{figure}[!t]
\centerline{\includegraphics[width=\textwidth]{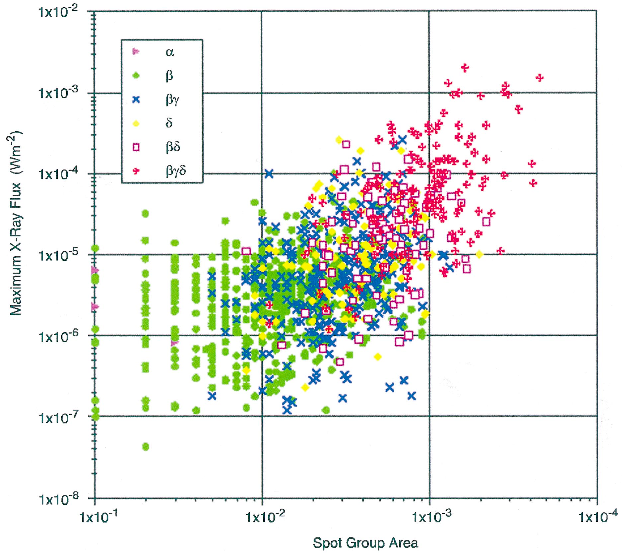}}
\caption[Figure 2 of \citet{sammis00}, showing peak flare intensities (Wm$^{-2}$) for the largest flare of each of the 2,789 sunspot groups studied as a function of peak area in disk fraction.]{Figure 2 of \citet{sammis00}, showing peak flare intensities (Wm$^{-2}$) for the largest flare of each of the 2,789 sunspot groups studied as a function of peak area in disk fraction. Each magnetic Hale class is plotted separately (colours indicated in the legend). Note that regions producing no flares are omitted.}
\label{intro:sammis}
\end{figure}

This thesis will explore the connection between sunspot magnetic field structure and solar flares. Early studies of solar flare activity found a close relationship between flares and sunspots, and was the reason for the extra $\delta$ classification scheme developed by \citet{kunzel60} for the Hale clasification scheme. \citet{zirin87} examined eighteen years of observations from Big Bear Solar Observatory, and found a link between the existence of $\delta$ configuration sunspot groups and very large flares. Later statistical studies confirmed this early research, for example a statistical study by \citet{sammis00} described relationships between active region size, peak $GOES$ flare flux, and magnetic classification. Their Figure 2 is reproduced in Figure~\ref{intro:sammis}, showing that bigger flare events (in terms of peak flux) occur in more complex Hale magnetic classification sunspot groups ($\beta\gamma\delta$ producing the biggest). It must be noted that the values plotted are largest quantities at any time in the AR's lifetimes (not necessarily at flare peak). The tendency of larger flares occurring in more complex regions could also be seen in the AR evolution illustrated in Figure~\ref{intro:evolu}, as the AR produced its largest-magnitude flare at the height of its complexity. Thus the degree of complexity of active regions has been found to be linked to the likelihood of flaring and maximum flare magnitude: larger, more numerous flares occur near larger and more complex sunspot groups. The following subsections will explore the relationships between sunspots and flares further, reviewing previous work in the field.

%%%%%%%%%%%%%%%%%%%%%%%%%%%%%%%%%%%%%%%%%%%%%%%%%%%%%%%%%%%%%%%%%%%%%%%%%%%%%%%%%%%%%%%%%%%%%%%%%%

\subsection{Trigger Mechanisms}
\label{intro:trigger}
Much previous work has expanded upon the link between sunspot complexity and flaring, to focus on possible flare trigger mechanisms. \cite{hagyard84} found that points of flare onset are where both magnetic shear and field strength are the strongest, suggesting that a flare is triggered when the magnetic shear stress exceeds some critical amount. Shearing is taken to mean that the field is aligned almost parallel to the magnetic neutral line (NL, where field falls to zero) rather than perpendicular to it, as would be observed in a potential field configuration \citep{schmieder96}. Many early theoretical studies suggested a link between both the emergence of new flux and the shearing and twisting of field lines with the flare trigger mechanism \citep[see][for a review]{rust94}. 

Flux emergence has been found to play a significant role, with \citet{feynman95} showing that eruptions of filaments are most likely to occur when new flux emerges nearby in an orientation favourable for reconnection. More recently, \citet{wallace10} found the emergence of small-scale flux near a magnetic NL to be a trigger to a B-class flare, with pre-flare flows observed along two loop systems in the corona $\sim40$~minutes before the flare began. The earliest indication of activity in the event studied by \citeauthor{wallace10} was a rise in non-thermal velocity $\sim$ 1 hour before flare start. 
An increase in non-thermal velocity was suggested by \citet{harra10} as an indicator of turbulent changes in an AR prior to a flare that are related to the flare trigger mechanism. Computational simulations have also been used to investigate flux emergence as a flare trigger, for example \citet{kusano12} used 3D magnetohydrodynamic simulations to investigate small-scale flux emergence of opposite polarity as possible flare triggers.

Aside from non-thermal velocity, a number of other pre-flare indicators have been discovered across multiple wavelengths (see review by \citet{benz08}). \citet{desjardins03} found changes in H$\alpha$ observations up to $\sim 3$~hours before the start of 11 eruptive flares, a phenomenon they called `moving blueshift events'. \citeauthor{desjardins03} concluded that reconnection in the chromosphere or low corona plays an important role in establishing the conditions that lead to solar flare eruptions. Nonthermal emission was observed in microwaves and hard X-Rays by \citet{asai06} during the pre-flare phase of an X-class flare, from $\sim 30$~minutes before flare start. A faint ejection associated with the flare was also observed in EUV images. \citet{joshi11} also observed significant pre-flare activities for $\sim 9$ minutes before the onset of a flare impulsive phase, in the form of an initiation phase observed at EUV/UV wavelengths, followed by an X-ray pre-cursor phase. Although multi-wavelength pre-flare observations in the chromopshere and corona have been well-studied, currently no observations of magnetic field changes on such short timescales before flaring have been found. In this thesis, the magnitude and orientation of the photospheric magnetic field will be investigated over a number hours before flare start.

As analytical methods have improved, and more high-resolution magnetic field data has become available, studies have begun to focus more on increasing twist and magnetic helicity as important flare triggers
(e.g., \citealp{harra09}; \citealp{georgoulis11}; \citealp{li11a}). Note that magnetic helicity is a measure of magnetic topological complexity, e.g., twists and kinks of field lines \citep[see][]{canfield98}. However, early theoretical work hinted at the importance of these factors, with \citet{hood79} estimating a value of twist of 2.5$\pi$~radians as a critical threshold after which a flux tube will become linearly unstable to kinking. \citet{PevtsovCanfieldZirin96} confirmed this theory with X-ray and vector magnetic field observations of NOAA AR 7154, finding twist in the AR to exceed this threshold before an M-class flare occurred. The use of sophisticated 3D magnetic field extrapolations can now improve upon earlier studies, giving a more accurate depiction of field line topology \citep{chandra09,jing12}. 3D extrapolations will be used in Chapter~\ref{chapter:3D} to investigate possible flare triggers for the event studied there.

%%%%%%%%%%%%%%%%%%%%%%%%%%%%%%%%%%%%%%%%%%%%%%%%%%%%%%%%%%%%%%%%%%%%%%%%%%%%%%%%%%%%%%%%%%%%%%%%%%

\subsection{Flaring Process}
\label{intro:flaring_process}
\begin{figure} [!t]
\centerline{\includegraphics[width=\textwidth]{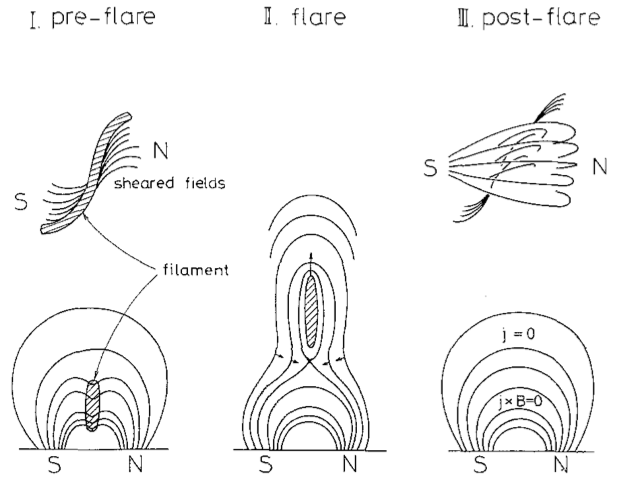}}
\caption[Depiction of a magnetic field configuration during flaring by \citet{tanaka86}.]{Depiction of a magnetic field configuration during flaring by \citet{tanaka86}. The configuration is shown at pre-flare (left column), flare (middle column), and post-flare (right column) states. Upper row shows a view from above the configuration, and the lower row shows a side-on view.}
\label{paper1:tanaka86}
\end{figure}

Analysing the magnetic field configuration of ARs is an important step in understanding the flaring process. \citet{tanaka86} depicts a possible evolution of large-scale fields in a flare, shown in Figure~\ref{paper1:tanaka86}. The depiction includes an ensemble of sheared fields containing large currents and a filament located above the NL in the pre-flare state. \citeauthor{tanaka86} suggests a flare may be triggered by a filament eruption at this NL location. More sophisticated flare models were later developed, e.g., \citet{antiochos98} described a `breakout' model for large eruptive flares, beginning with newly-emerged, highly-sheared field held down by an overlying un-sheared field. Reconnection takes place between the un-sheared overlying flux and flux in additional neighboring systems. This reconnection transfers un-sheared flux to the neighboring flux systems, thereby removing the overlying field and the restraining pull. Hence, reconnection allows the innermost core field to `break out' to infinity, without opening the overlying field itself and thus violating an upper limit of free energy known as the Aly-Sturrock limit \citep{aly91,sturrock91}. This limit states that an open state of magnetic field contains the largest energy of any possible field configuration. \citeauthor{antiochos98} note that, while a bipolar AR does not have the necessary complexity, a $\delta$ sunspot has the correct topology for this model.

\begin{figure} [!t]
\centerline{\includegraphics[width=\textwidth]{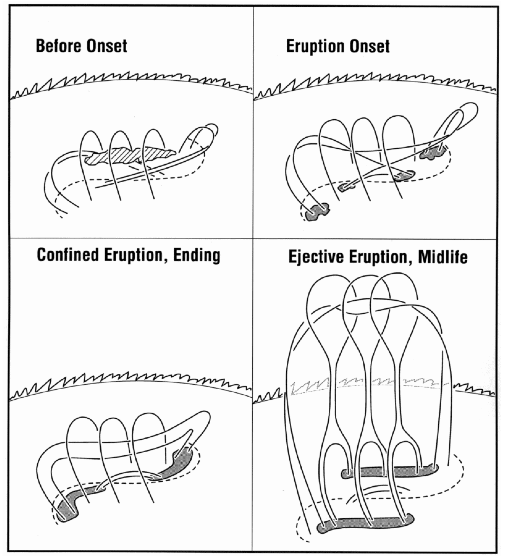}}
\caption[Flare model of \citet{moore01} showing the magnetic field configuration before, during, and after the onset of an explosion that is unleashed by internal tether-cutting reconnection.]{Flare model of \citet{moore01} showing the magnetic field configuration before, during, and after the onset of an explosion that is unleashed by internal tether-cutting reconnection. The dashed curve indicates the photospheric NL; the arc in the background of the panels defines the chromospheric limb; the dashed area in the upper left panel shows a filament of chromospheric temperature plasma; the dark gray areas in the other panels are ribbons of chromospheric flare emission (at the feet of reconnected field lines).}
\label{paper1:moore01}
\end{figure}

In the `tether-cutting' scenario of \citet{moore01}, as shown in Figure~\ref{paper1:moore01}, there is strong shear at low altitudes along the NL before the flare occurs (upper left panel). The upper right panel of Figure~\ref{paper1:moore01} shows reconnection via `tether-cutting' occurring in the region, and lower left panel shows the completion of the early reconnection. Finally, the lower right panel of Figure~\ref{paper1:moore01} shows a rising plasmoid distending the outer field lines, which continue the process of reconnection to form the expanding flare. In the \citeauthor{moore01} model the plasmoid eventually escapes as a CME. Note the main difference between this and the `breakout model' is that reconnection occurs above the unstable structure in the `breakout' model, whereas it occurs below it in the `tether-cutting' scenario. 

Field topology studies have been used to place constraints on theoretical models, for example \citet{mandrini06} reviewed a number of flaring active-region topologies, finding that magnetic reconnection can occur in a greater variety of magnetic configurations than traditionally thought. The reader is referred to the reviews of \citet{priestforbes02} and \citet{Schrijver09}, and references therein, for more recent developments in eruptive event models.

%%%%%%%%%%%%%%%%%%%%%%%%%%%%%%%%%%%%%%%%%%%%%%%%%%%%%%%%%%%%%%%%%%%%%%%%%%%%%%%%%%%%%%%%%%%%%%%%%%

\subsection{Flaring Locations}
\label{intro:flaring_locations}
Although sunspot regions are closely associated with solar flares, it is rare that flares occur in the sunspot umbra itself \citep{moore85}. The magnetic neutral line (NL, where the field falls to zero) is often found to be a region of flare activity, a locus across which the line-of-sight (LOS) field component changes sign. The review of \citet{moore85} noted that the most common flare-productive field configuration is characterised by strong shear across the NL. For example, \citet{hagyard84} found flare onset to occur across a NL that exhibited strong field strength and magnetic shear, as mentioned in Section~\ref{intro:trigger}. This location will be investigated further in Chapter~\ref{chapter:NL} for its importance to flaring.

It has been found that in the lead up to a flare occuring, field lines rooted to the NL run nearly parallel rather than perpendicular to it \citep[see, e.g, the tether-cutting scenario of][]{moore01}. The field near such a NL is far from potential and holds ample free magnetic energy for flares. Magnetic energy changes over the course of a flare observation period will be investigated in Chapter~\ref{chapter:3D}. It is worth noting that there is some debate over what overlays a NL; sheared magnetic arcades or helical magnetic flux ropes. However, \citet{georgoulis10} examined the magnetic pre-flare configuration near the NL and suggested instead that numerous small-scale magnetic reconnections, constantly triggered in the NL area, can lead to effective transformation of mutual helicity (i.e, crossing of field lines) to self magnetic helicity (i.e., twist and writhe) that, ultimately, may force the magnetic structure above NLs to erupt to be relieved from its excess helicity.

%%%%%%%%%%%%%%%%%%%%%%%%%%%%%%%%%%%%%%%%%%%%%%%%%%%%%%%%%%%%%%%%%%%%%%%%%%%%%%%%%%%%%%%%%%%%%%%%%%

\subsection{Observed Magnetic Field Changes}
Numerous observational studies have confirmed the importance of emergence and shearing to flare phenomena. \citet{zirin93} investigated flux emergence and sunspot group motions, which resulted in complicated flow patterns leading to flaring. \citet{wang94} used vector magnetograms to observe magnetic shear in five X-class solar flares; in all cases increasing along a substantial portion of the magnetic NL. They suggested flux emergence being key to eruption, but the increase in shear persisted much longer after the flare rather than decreasing as per model predictions, and no theoretical explanation was given. Recent evidence has furthered the idea that emerging-flux regions and magnetic helicity are crucial to the pre-flare state (e.g., \citealp{liuzhang01}; \citealp{wang02}; \citealp{chandra09}).

However, these parameters are not the only ones of interest when studying the links between AR magnetic fields and flaring. \citet{wang02} studied LOS magnetic field observations during three X-class flares, and found changes in magnetic flux and transverse field due to the flare. \citet{sudolharvey05} also used LOS magnetgrams to observe abrupt and permanent changes in the LOS field after 15 X-class flares. Sunspot magnetic field inclination has also been found to change dramatically due to solar flares (\citealp{liu05}; \citealp{li09}; \citealp{gosain12}), as well as current density (\citealp{dela93}; \citealp{su09}; \citealp{canou10}). These parameters will be investigated further in Chapters~\ref{chapter:paper1} and \ref{chapter:NL}. 
It is clear that significant changes in various magnetic field parameters occur due to solar flares, but most previous work has focused on investigating changes due to the flare itself. The research in this thesis will examine the full evolution of ARs associated with flare events (over shorter timescales than previously studied) looking for changes leading up to the flare as well as afterwards. 

Early investigations were limited by the lack of high-resolution solar vector magnetic field data, mainly using only LOS observations of the magnetic field. \citet{wang06} listed predictions of results from future flare observations, when high-resolution vector magnetograms from \emph{Hinode} and \emph{Solar Dynamics Observatory} would become available,
\begin{itemize}
\item Transverse magnetic fields at a flaring NL will increase rapidly following flares.  
\item The unbalanced flux change will be more prominent when the regions are closer to the limb due to the enhanced projection effect there.
\item Evershed flow will decrease in the outer boundary of a $\delta$-configuration, as outward-inclined fields will become more vertical. 
\item In the initial phase of the flare, two flare footpoints may move closer before they start the usual separation motion.  
\item As a consequence of the reconnection, some current will be able to be measured near the photosphere, and therefore, an increase of the magnetic shear near the flaring NLs may be detected.
\end{itemize}
This list is not exhaustive (other indicators will be discussed throughout this thesis in detail), but it summarises a selection of current theories and observations well. The research described in this thesis uses this newly available high-resolution data to test currently proposed theories, by examining magnetic field parameters in active regions for any significant changes before and after a flare. Photospheric magnetic field evolution is examined in Chapters~\ref{chapter:paper1} and \ref{chapter:NL}, while 3D magnetic field extrapolation methods are used in Chapter~\ref{chapter:3D} to study the coronal magnetic field.

%%%%%%%%%%%%%%%%%%%%%%%%%%%%%%%%%%%%%%%%%%%%%%%%%%%%%%%%%%%%%%%%%%%%%%%%%%%%%%%%%%%%%%%%%%%%%%%%%%%%%%%%%%%%%%%%%%%%%%%%%%%%%%%%%%%%%%%%%%%%%%%%%%%%%%%%%%%%%%%%%%%%%%%%%%%%%%%%%%%%%%%%%%%%%%%%%%%%%%%%%%%%%%%%%%%%%%%%%%%%%%%%%%%%%%%%%%%%%%%%%%%%%

\section{Outline of Thesis}

The research contained in this thesis examines the evolution of a number of ARs undergoing flaring using various analysis techniques. Examination of the magnetic field conditions in an AR before a flare occurs could produce useful indicators for flare forecasting. Studying differences in AR topology between pre- and post- flare states helps to test the validity of currently proposed changes in magnetic topology during solar flares. Through examination of the evolution of a sunspot, changes in the topology observed, if any, can give an insight into how and when a flare might occur from this kind of region. 

In this chapter, the introductory theory behind the research project has been presented, starting with some background to the Sun, and then focusing on sunspots and solar flares in greater detail. In Chapter~\ref{chapter:theory}, some basic theory needed to understand the analysis techniques throughout this thesis are introduced. Fundamentals of magnetohydrodynamics are described to introduce 3D magnetic field extrapolations, and an introduction to radiative transfer is also presented. The instruments used to obtain data for analysis in this thesis are described in Chapter~\ref{chapter:instrumentation}. Specifically instruments onboard the \emph{Hinode} spacecraft are described, including the software methods used to process the raw data obtained. The techniques used for analysis in Chapters~\ref{chapter:paper1}, ~\ref{chapter:NL}, and ~\ref{chapter:3D} are also introduced in Chapter~\ref{chapter:instrumentation}, including some background theory not previously described in Chapter~\ref{chapter:theory}. 

In Chapter~\ref{chapter:paper1}, the results of examining photospheric magnetic field information are presented, from small flux elements of a sunspot region undergoing flaring. Data from the \emph{Hinode} spacecraft are used to observe the AR magnetic field before and after a B-class flare. Distinctive magnetic field changes are observed leading up to and after the flare in an area of chromospheric flare brightening. In Chapter~\ref{chapter:NL}, a magnetic neutral line location is examined with \emph{Hinode} observations during a period of observation in which a C-class flare occurred. Both temporal and spatial changes are observed across the magnetic NL during the observation period.

While 2D photospheric magnetic field observations are studied in Chapters~\ref{chapter:paper1} and ~\ref{chapter:NL}, the 3D coronal field is examined in the third research chapter, using magnetic field extrapolation methods. In Chapter~\ref{chapter:3D}, the results of the 3D extrapolations are described, using the observations of the B-class flare event previously examined in Chapter~\ref{chapter:paper1}. Magnetic geometry and energy changes are observed both before and after the flare in the region of flare brightening. Finally, the main conclusions of the research presented in this thesis are described in Chapter~\ref{chapter:concs}, with some directions for future work.
	% background information

\chapter{Theory}
\label{chapter:theory}
% the code below specifies where the figures are stored
\ifpdf
    \graphicspath{{3/figures/PNG/}{3/figures/PDF/}{3/figures/}}
\else
    \graphicspath{{3/figures/EPS/}{3/figures/}}
\fi

% ----------------------------------------------------------------------
%: ----------------------- introduction content ----------------------- 
% ----------------------------------------------------------------------

\hrule height 1mm
\vspace{0.5mm}
\hrule height 0.4mm 
\noindent 
\\ {\it In this chapter, the theory needed to understand the various analysis techniques used in this thesis is presented, beginning with a background to magnetohydrodynamics. Fundamental magnetohydrodynamic equations are outlined, and the basic theory behind 3D magnetic field extrapolations is described. Radiative transfer is then discussed, including the various assumptions that must be made in order to obtain a model of the solar atmosphere.}
\\ 
\hrule height 0.4mm
\vspace{0.5mm}
\hrule height 1mm 

\newpage

\section{Magnetohydrodynamics}

The Sun is composed of material called plasma, which is defined as a state similar to gas in which a certain fraction of atoms are ionised. In the Sun, this ionisation is caused by the extreme temperatures and pressures that exist, and matter in and around the Sun can be treated as electrically charged fluid. Magnetohydrodynamics (MHD) describes the flow of electrically conducting fluid in the presence of electromagnetic (EM) fields. Its equations govern the magnetised plasma of the Sun. It is useful to include a description of basic MHD in studies of AR magnetic fields in order to understand the dynamics of solar plasma, thus this section introduces some fundamental theory needed to understand this complex topic.

\subsection{Maxwell's Equations}

Maxwell's equations \citep{maxwell1861} are a set of four fundamental equations governing electromagnetism (i.e., the behavior of electric and magnetic fields). For time varying fields they are defined (in differential microscopic form\footnote[1]
{There are numerous forms of these equations, including integral and macroscopic form, but the form defined here is what will be referred to throughout the rest of this thesis.}) as,
\begin{eqnarray}
\nabla  \times \textbf{B} & = & \mu_0 \textbf{j} ~ + \frac{1}{c^2}\frac{\partial \textbf{E}}{\partial t} \ , \label{mhd:amperes_law} \\ 
\nabla  \cdot \textbf{B} & = & 0 \ , \label{mhd:monopoles} \\
\nabla \times \textbf{E} & = & - \frac{\partial \textbf{B}}{\partial t} \ , \label{mhd:faradays_law} \\
\nabla \cdot \textbf{E} & = & \frac{1}{\epsilon }\rho \ , \label{mhd:gauss_law}
\end{eqnarray}
where \textbf{B} is the magnetic field, $\mu_0$ is the magnetic permeability of free space, \textbf{j} is the current density, $c$ is the speed of light in a vacuum, \textbf{E} is the electric field, $\epsilon$ is the permittivity of free space, and $\rho$ is the charge density. 

Equation~\ref{mhd:amperes_law} is generally referred to as Ampere's Law, which means that either currents or time-varying electric fields may produce magnetic fields. If it is assumed that the typical plasma velocity, \textbf{v}, is much less than $c$, the second term on the RHS of Equation~\ref{mhd:amperes_law} can be ignored (MHD approximation). Thus Ampere's Law becomes,
\begin{equation}
\label{mhd:amperes_law_reduced}
\nabla\times\mathbf{B}~=~\mu_{0}\mathbf{j} \ .
\end{equation}
Equation~\ref{mhd:monopoles} is often described as Gauss's Law for magnetic fields (no magnetic monopoles). Faraday's Law is defined in Equation~\ref{mhd:faradays_law}, meaning that time-varying magnetic fields can give rise to electric fields. Finally, Equation~\ref{mhd:gauss_law} is known as Poisson's Law (or Gauss's Law for electric fields), and means that electric charges may give rise to electric fields.

Ohm's Law couples the plasma velocity to Maxwell's equations, and is written as,
\begin{equation}
\label{mhd:ohms_law}
\mathbf{E'}~=~\mathbf{E}+\mathbf{v}\times \mathbf{B}~=~\mathbf{j}/\sigma \ ,
\end{equation} 
where $\sigma$ is the plasma conductivity. It expresses that the electric field in the frame moving with the plasma (\textbf{E$'$}) is proportional to the current, or that the moving plasma in the presence of magnetic field \textbf{B} is subject to an electric field $\textbf{v} \times \textbf{B}$ (in addition to \textbf{E}).

%%%%%%%%%%%%%%%%%%%%%%%%%%%%%%%%%%%%%%%%%%%%%%%%%%%%%%%%%%%%%%%%%%%%%%%%%%%%%%%%%%%%%%%%%%%%%%%%%%

\subsection{Induction Equation}

Ampere's Law (Equation~\ref{mhd:amperes_law_reduced}) and Ohm's Law (Equation~\ref{mhd:ohms_law}) can be combined to give,
\begin{equation}
\textbf{E}~=~-\textbf{v} \times \textbf{B}~+~\eta \nabla \times \textbf{B} \ ,
\end{equation}
where $\eta = {1}/{\mu_0 \sigma}$ is the magnetic diffusivity. Taking the curl of both sides, and substituting for Faraday's Law (Equation~\ref{mhd:faradays_law}) gives,
\begin{equation}
\label{mhd:temp}
- \frac{\partial \mathbf{B}}{\partial t}~=~- \nabla \times (- \textbf{v} \times \textbf{B}~+~\eta \nabla \times \textbf{B}) \ .
\end{equation}
Since $\nabla \times (\nabla \times \textbf{B})= \nabla (\nabla \cdot \textbf{B})~-~\nabla^{2}\textbf{B}$, Equation~\ref{mhd:temp} can be rewritten using Equation~\ref{mhd:monopoles} to derive the induction equation for the solar magnetic field,
\begin{equation}
\label{induction}
\frac{\partial \mathbf{B}}{\partial t}~=~\nabla \times (\mathbf{v} \times \mathbf{B}) + \eta \nabla ^{2}\mathbf{B}  \ .
\end{equation}
The first term of the RHS of the equation describes advection, and the second term diffusion. 

The magnetic Reynold's number defines the ratio of the advection and diffusion terms of the induction equation, such that,
\begin{equation}
R_m~=~\frac{\nabla \times ( \textbf{v} \times \textbf{B})}{\eta \nabla^{2} \textbf{B}} \ .
\end{equation}
Replacing vector quantities by their magnitudes, and considering $\nabla \sim l^{-1}$, where $l$ is a typical length scale, the equation can be written as,
\begin{equation}
R_m~=~\frac{vB/l}{\eta B/l^{2}}~=~\frac{l v}{\eta} \ ,
\end{equation}
where $v$ is a typical velocity. For example in a sunspot $l \sim10^7$~m (typical radius), $v \sim10^3$~m~s$^{-1}$ (super granular motion at surface), and $\eta \sim10^3$~m$^2$~s$^{-1}$ (medium-sized sunspot at the photosphere; \citep{aschwanden05}). This gives a value of $R_m \sim10^7$. For the solar corona, $R_m \sim10^8$ (since $l \sim10^5$~m and $\eta \sim1$~m$^2$~s$^{-1}$).

If $R_m \gg 1$ (as in a sunspot) then the advection term dominates and the induction equation becomes $\partial \mathbf{B}/\partial t = \nabla \times (\mathbf{v} \times \mathbf{B})$. This is for ideal MHD, i.e., in the limit of perfect electrical conductivity ($\sigma \rightarrow \infty$, $\eta \rightarrow 0$). The field is said to be `frozen-in' (see Section~\ref{mhd:flux_freezing}), with the field lines in this perfectly conducting plasma behaving as if they move with the plasma. If the field is very strong, plasma is constrained to flow along it, and if the field is weak, it is advected with the plasma.

If $R_m \ll 1$ the diffusion term dominates and the induction equation becomes $\partial \mathbf{B}/\partial t = \eta \nabla ^{2}\mathbf{B}$. Magnetic field irregularites will then diffuse away over a time scale of $\sim l^{2}/\eta$. For a sunspot, this yields a diffusion times of $10^{11}$~seconds, thus it is not the main method of particle motion here (unless there are short length scales or large gradients in the field). Note for flares, the diffusion time is $\sim100$~seconds, which gives a lengthscale of 10~m. However, the finest instrument resolution is currently a few hundred km, leading to the need for better resolution instruments to fully analyse the solar flare process. 

%%%%%%%%%%%%%%%%%%%%%%%%%%%%%%%%%%%%%%%%%%%%%%%%%%%%%%%%%%%%%%%%%%%%%%%%%%%%%%%%%%%%%%%%%%%%%%%%%%

\subsection{Magnetic Flux Freezing}
\label{mhd:flux_freezing}
Conservation of magnetic flux in a perfectly conducting fluid is one of the most fundamental conservation laws of MHD. Also known as Alfv$\acute{\mathrm{e}}$n's Theorem, it implies that magnetic field lines are ÔfrozenÕ into the fluid, so that the field lines and the plasma move together. The idea of MHD is that magnetic fields can induce currents in a moving conductive fluid, which create forces on the fluid, and also change the magnetic field itself. In ideal MHD, the fluid is a perfect conductor, and applies to partially ionised plasmas which are strongly collisional and have little or no resistivity.

A more detailed description of this theorem begins by considering a gas parcel threaded with magnetic field. The flux anchored to the gas parcel, $\Phi _{B} \equiv \int \mathbf{B}.\mathrm{d}\mathbf{S}$ will change in two ways:
\begin{equation}
\frac{\mathrm{d}\Phi _{B}}{\mathrm{d}t}~=~\frac{\mathrm{d}}{\mathrm{d}t}\int\mathbf{B}.\mathrm{d}\mathbf{S}~=~\int \frac{\partial \mathbf{B}}{\partial t}.\mathrm{d}\mathbf{S} + \int \mathbf{B}.\mathbf{v}\times \mathrm{d}\mathbf{s}
\label{1}
\end{equation}
where $\mathbf{s} = \partial \mathbf{S}$. One part of the change of flux anchored to a surface $\mathrm{d}\mathbf{S}$ comes from a temporal change in the magnetic flux density over the surface (first term on RHS), while the other part comes from a change of surface boundary due to gas motion (second term on RHS; $\mathrm{d}\mathbf{S} = \mathbf{v} \times \mathrm{d}\mathbf{s}$).

Stoke's theorem, $\int \nabla \times \mathbf{A} . \mathrm{d}\mathbf{S}= \int \mathbf{A} . \mathrm{d} \mathbf{s}$, can be used to reduce the last part of Equation~\ref{1} to $\int \mathbf{B}\times \mathbf{v}.\mathrm{d}\mathbf{s}=\int\mathbf{\nabla}(\mathbf{B} \times \mathbf{v}).\mathrm{d}\mathbf{S}$. This gives
\begin{equation}
\frac{\mathrm{d}\Phi _{B}}{dt}~=~\int\left [ \frac{\partial \mathbf{B}}{\partial t} - \nabla(\mathbf{v}\times\mathbf{B})\right ].\mathrm{d}\mathbf{S}~=~0
\label{2}
\end{equation}
which means the magnetic flux anchored to a gas parcel does not change in the ideal MHD regime, it remains constant in time for any arbitrary contour. Thus magnetic field lines must move with the plasma, i.e., they are `frozen' into the perfectly conducting fluid. 

Note that in Equation~\ref{2}, ${\partial \textbf{B}}/{\partial t}$ can be written in terms of Faraday's Law (i.e, Equation~\ref{mhd:faradays_law}). Alfv$\acute{\mathrm{e}}$n's theorem is thus a direct result of Faraday's law applied to a medium of infinite electrical conductivity. Motions along the field lines do not change the field but motions transverse to the field carry the field with them, i.e., the field is dragged with the plasma or vice-versa. One important consequence of this theorem is that the topology of the field lines is preserved, and, in particular, that crossing field lines cannot reconnect. This presents a puzzle in many situations where ideal MHD is supposed to hold to a very good first-order approximation, such as in many astrophysical systems, when the conditions for ideal MHD break down, allowing magnetic reconnection that releases the stored energy from the magnetic field.

\subsection{Equation of Motion}
\label{mhd:motion}

The equation of motion (momentum equation), $\textbf{F}=m\textbf{a}$, is described by,
\begin{equation}
\label{mhd:momentum}
\rho \frac{D\textbf{v}}{Dt}~=~-\nabla P~+~\textbf{j} \times \textbf{B}~+~\rho \textbf{g}
\end{equation}
with the external forces $\textbf{F}$ on the RHS indicating the gradient of the gas pressure, Lorentz force, and gravitational force onto the plasma, respectively. Note on the LHS of the equation of motion that $D/Dt$ is the convective time derivative,
\begin{equation}
\frac{D}{Dt}~=~\frac{\partial}{\partial t}~+~\textbf{v} \cdot \nabla
\end{equation}

The Lorentz term of the equation of motion can be expanded upon. Using Ampere's law (Equation~\ref{mhd:amperes_law}) with the \textbf{j} $\times$ \textbf{B} term of Equation~\ref{mhd:momentum}, and the vector identity $\nabla(\frac{1}{2} \textbf{B} \cdot \textbf{B})=\textbf{B} \times(\nabla \times \textbf{B})~+~(\textbf{B} \cdot \nabla) \textbf{B}$, gives
\begin{equation}
\textbf{j} \times \textbf{B}~=~\frac{1}{\mu_0} (\nabla \times \textbf{B}) \times \textbf{B}~=~-~\nabla \left ( \frac{B^2}{2\mu_0} \right ) ~+~\frac{(\textbf{B} \cdot \nabla)\textbf{B}}{\mu_0}
\label{maxwell_stress}
\end{equation}
where the first term on the RHS is the magnetic pressure force, and the second term on the RHS is the magnetic tension force. The magnitude of the plasma pressure, p\footnote[2]{p=2$nk_\mathrm{b}T$, where n is the number of hydrogen atoms, k$_\mathrm{b}$ is Boltzmann's constant and $T$ is the temperature. The factor of 2 is included as both electrons and protons contribute.}, and magnetic pressure, $B^2/2\mu_0$ are compared by the plasma $\beta$ parameter, which is defined as,
\begin{equation}
\beta~=~\frac{2\mu_0p}{B^2}
\end{equation}
If $\beta \gg 1$, the gas pressure dominates (e.g., in the solar photosphere) and the influence of the magnetic field is negligible (with plasma motions dominating over the magnetic field forces). If $\beta \ll 1$, the magnetic pressure dominates (e.g., in the solar corona) and the magnetic field is often assumed to be force free for the purposes of extrapolation procedures (see below). Figure~\ref{mhd:beta}\footnote[3]{Note that the units for $\beta$ in the plot are CGS units (rather than SI), with magnetic pressure defined as $B^2/8\pi$ and gas pressure as $2nk_\mathrm{b}T$. The CGS version of magnetic pressure will be used for calculations in Chapter~\ref{chapter:3D}.} illustrates the change in $\beta$ value throughout the solar atmosphere as depicted by \citet{gary01}. After a high $\beta$ plasma in the photosphere, the $\beta$ value decreases to low values through the chromosphere and corona, where the magnetic field structures are observed to suspend plasma in loops and filaments. In the extended upper atmosphere $\beta$ rises again, and the magnetic field is advected out with the solar wind plasma flow to ultimately form the Parker spiral (as described in Section~\ref{intro:atm}).

%%%%%%%%%%%%%%%%%%%%%%%%%%%%%%%%%%%%%%%%%%%%%%%%%%%%%%%%%%%%%%%%%%%%%%%%%%%%%%%%%%%%%%%%%%%%%%%%%%%%%%%%%%%%%%%%%%%%%%%%%%%%%%%%%%%%%%%%%%%%%%%%%%%

\section{3D Magnetic Field Extrapolations}
\label{mhd:extraps}
In Chapter~\ref{chapter:3D} the evolution of the 3D coronal magnetic field in an active region is investigated using three types of extrapolation procedure: potential, linear force free (LFF), and non-linear force free (NLFF). This section aims to discuss the theory and techniques behind these procedures.

\begin{figure}[!t]
\centerline{\includegraphics[width=\textwidth]{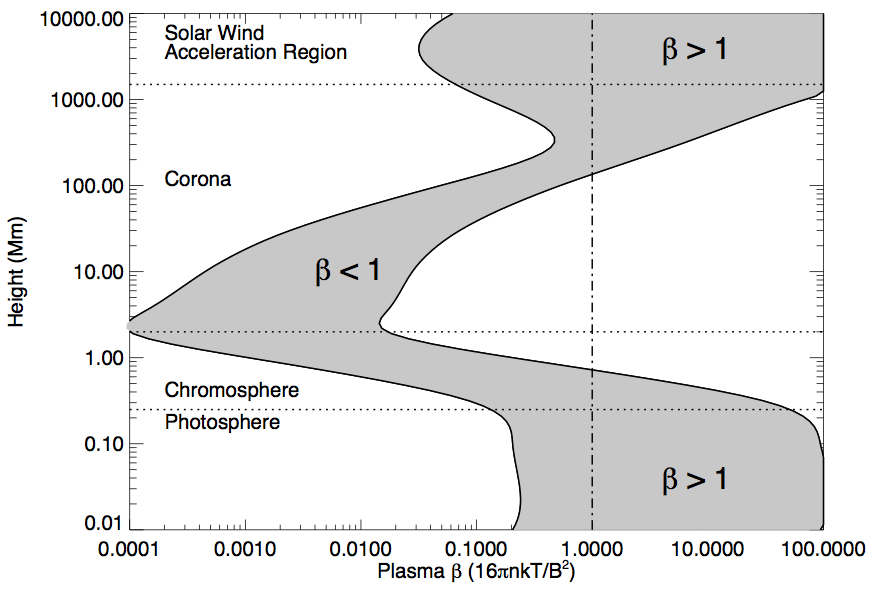}}
\caption[Plasma $\beta$ as a function of height over an AR \citep{gary01}.]{Plasma $\beta$ as a function of height over an AR (magnetic field strengths between 150 G for a plage region, and 2500 G for a sunspot). The dotted lines divide the photosphere ($\beta > 1$), chromosphere and corona ($\beta < 1$), and the solar wind ($\beta > 1$) \citep{gary01}.}
\label{mhd:beta}
\end{figure}

First the special condition of magnetohydrostatic equilibrium is applied to the equation of motion (Equation~\ref{mhd:momentum}). Flows are neglected, so that $\textbf{v} = 0$, and it is assumed there is no time variation, so that $\partial/\partial t = 0$. Hence, the equation of motion becomes,
\begin{equation}
\label{mhd:momentum_temp}
-\nabla P~+~\textbf{j} \times \textbf{B}~+~\rho \textbf{g}~=~0
\end{equation}

The corona is considered to be generally force free \citep{gold60}, dominated by the relatively stable magnetic field in a low-$\beta$ plasma. The gas pressure term in Equation~\ref{mhd:momentum_temp} is negligible compared to the Lorentz term, and gravity can also be considered negligible high in the upper solar atmosphere. Equation~\ref{mhd:momentum_temp} thus reduces to,
\begin{equation}
\label{3D:force_free_eqn}
\textbf{j} \times \textbf{B}~=~0
\end{equation}
This is known as the force-free approximation, which all three types of 3D extrapolation mentioned above assume \citep{gary01}. The approximation results in the current being parallel vectorially to the magnetic field, with a proportionality factor $\alpha$ termed the force-free field parameter, and is a scalar function of position (i.e., $\alpha$ is a spatially varying function to be determined). There are three general forms of the force-free relation, 
\begin{eqnarray}
\textbf{j} &=& 0, \label{potential} \\
\textbf{j} &=& \alpha \textbf{B}, \label{lff} \\
\textbf{j} &=& \alpha(x,y,z) \textbf{B}. \label{nlff}
\end{eqnarray}

A potential field configuration is defined as one containing no currents, resulting in the case of Eqn.~\ref{potential} where $\alpha=0$. When $\alpha$ is non-zero but constant throughout a given volume the field configuration is referred to as LFF (Eqn.~\ref{lff}). Finally, when $\alpha$ is allowed to vary spatially (i.e., differing from field line to field line, but constant along one field line) the field configuration is referred to as NLFF (Eqn.~\ref{nlff}). This specific case allows for the existence of both potential and non-potential fields within the given volume. The following subsections will examine these three forms in more detail.

%%%%%%%%%%%%%%%%%%%%%%%%%%%%%%%%%%%%%%%%%%%%%%%%%%%%%%%%%%%%%%%%%%%%%%%%%%%%%%%%%%%%%%%%%%%%%%%%%%

\subsection{Potential and Linear Force-Free Fields}
\label{mhd:pot_lff}
\begin{figure}[!t]
\centerline{\includegraphics[scale=.6]{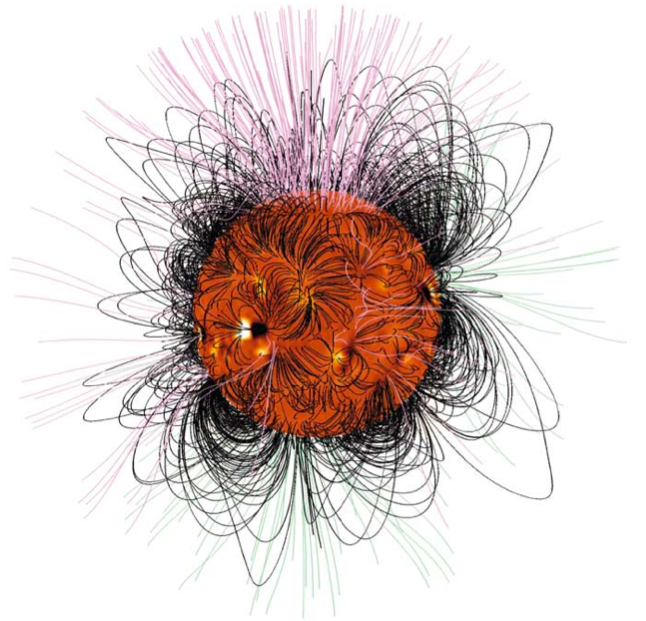}}
\caption[Low-resolution equivalent of a full-disk magnetogram, with computed field lines from a potential field extrapolation overlayed \citep{schrijverderosa03}.]{Low-resolution equivalent of a full-disk magnetogram, with computed field lines from a potential field extrapolation overlayed: black indicates closed field lines, and green and pink indicate open field lines (for field pointing away from or towards the Sun, respectively). The computed field lines start from a set of uniformly distributed points on the solar surface \citep{schrijverderosa03}.}
\label{mhd:pot_example}
\end{figure}

For a current-free potential field, assuming the force-free approximation, Ampere's Law reduces to $\nabla \times \textbf{B} = 0$. The most general solution to this is
\begin{equation}
\textbf{B}~=~\nabla \phi
\end{equation}
where $\phi(x,y,z)$ is the scalar magnetic potential. Substituting this into Equation~\ref{mhd:monopoles} gives $\nabla^2 \phi =0$, showing that potential magnetic fields satisfy Laplace's equations. Green's functions are often used to solve a potential magnetic field, first proposed by \citet{schmidt64}, and further developed by \citet{sakurai82}. An example of a typical global potential field extrapolation is shown in Figure~\ref{mhd:pot_example}, as previously discussed in Section~\ref{mhd:comparison}.

\begin{figure}[!t]
\centerline{\includegraphics[width=\textwidth]{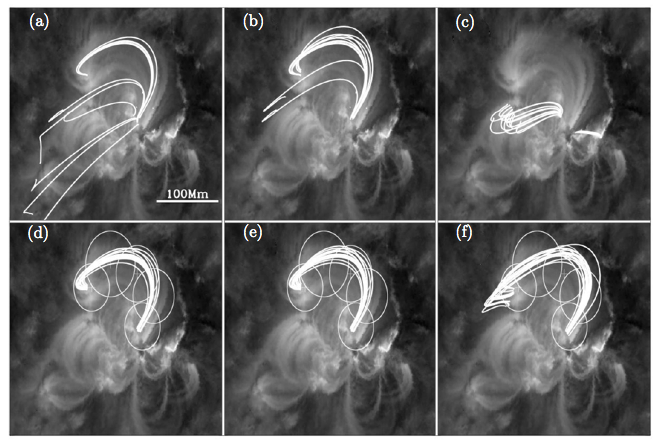}}
\caption[LFF extrapolated field lines with a value of $\alpha = 6.25 \times 10^{-3}$~Mm$^{-1}$, overlayed on STEREO A EUVI images of NOAA 10956. \citep{conlon10}.]{LFF extrapolated field lines with a value of $\alpha = 6.25 \times 10^{-3}$~Mm$^{-1}$, overlayed on STEREO A EUVI images of NOAA 10956. The field lines start from a user-defined footpoint, comparing the ability of three cost functions to recover the geometry of the active region loop \citep{conlon10}.}
\label{mhd:lff_example}
\end{figure}

If the magnetic field is not potential, then using Ampere's Law we obtain,
\begin{equation}
\label{mhd:lff_temp}
\nabla \times \textbf{B}~=~\alpha \textbf{B}
\end{equation}
Using the vector identity $\nabla \cdot (\nabla \times \textbf{B}) = 0$ with Equation~\ref{mhd:lff_temp} gives,
\begin{eqnarray}
\nabla \cdot (\nabla \times \textbf{B}) & = & \nabla \cdot (\alpha \textbf{B}) \nonumber \\
& = & \alpha \nabla \cdot \textbf{B}~+~\textbf{B} \cdot \nabla \alpha \nonumber \\
							& = & 0 \,
\end{eqnarray}
However, Equation~\ref{mhd:monopoles} shows the first term on the RHS is zero, thus
\begin{equation}
\textbf{B} \cdot \nabla \alpha~=~0
\end{equation}
so that $\alpha$ is constant along each field line, although it may vary from field line to field line in the case of a NLFF field. Note that if $\alpha = 0$, the magnetic field is potential. Figure~\ref{mhd:lff_example} shows a typical example of a LFF extrapolation. \citet{gary89} outlines typical methods for LFF extrapolations, discussing their limitations and usefulness (see Section~\ref{mhd:comparison} for a comparison with potential and NLFF cases). 

%%%%%%%%%%%%%%%%%%%%%%%%%%%%%%%%%%%%%%%%%%%%%%%%%%%%%%%%%%%%%%%%%%%%%%%%%%%%%%%%%%%%%%%%%%%%%%%%%%

\subsection{Non-Linear Force-Free Fields}
\label{mhd:nlff}
\begin{figure}
\centerline{\includegraphics[width=0.9\textwidth]{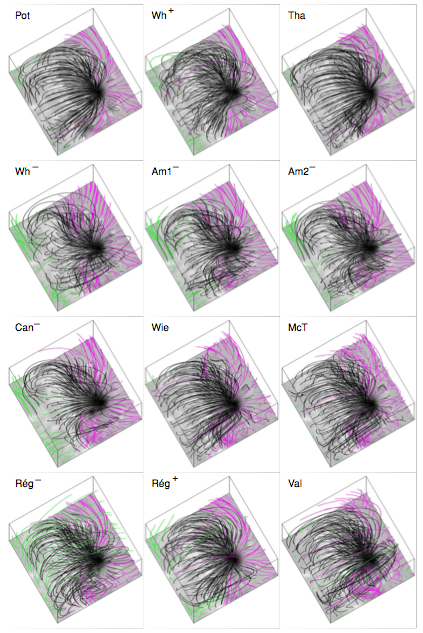}}
\caption[Representative field lines in the central portion of the active region for each NLFFF model listed in Table 1 of \citet{derosa09}.]{Representative field lines in the central portion of the active region for each NLFFF model listed in Table 1 of \citet{derosa09}. Black field lines indicate (closed) lines that intersect the lower boundary twice, and red and green field lines represent field lines that leave the box through either the sides or top, with colour indicative of polarity.}
\label{mhd:derosa09}
\end{figure}

If $\alpha$ is constant along a field line, then $\nabla \times \textbf{B} = \alpha(\textbf{r})\textbf{B}$, where $\textbf{r}=(x,y,z)$. Taking the curl of this gives,
\begin{eqnarray}
\nabla \times (\nabla \times \textbf{B}) & = & \nabla \times (\alpha(\textbf{r}) \textbf{B}) \nonumber \\
& = & \alpha(\textbf{r}) \nabla \times \textbf{B}~+~\nabla \alpha(\textbf{r}) \times \textbf{B} \nonumber \\
& = & \alpha(\textbf{r})^2 \textbf{B}~+~\nabla \alpha(\textbf{r}) \times \textbf{B}
\end{eqnarray}
Using the vector identity $\nabla \times (\nabla \times \textbf{B})=\nabla(\nabla \cdot \textbf{B})~-~\nabla^2 \textbf{B}$ with Equation~\ref{mhd:monopoles} and the above equation gives,
\begin{eqnarray}
\nabla^2\textbf{B}~+~\alpha(\textbf{r})^2\textbf{B} & = & \textbf{B} \times \nabla \alpha(\textbf{r}) \label{mhd:nlff1} \\
\textbf{B} \cdot \nabla \alpha(\textbf{r}) & = & 0 \label{mhd:nlff23} 
\end{eqnarray}
Thus there are two coupled equations for \textbf{B} and $\alpha(\textbf{r})$ for the case of a NLFF field, which need to be solved together. It is worth noting that for the LFF case, Equation~\ref{mhd:nlff1} reduces to $\nabla^2\textbf{B}~+~\alpha^2\textbf{B}=0$, known as the Helmholtz equation. There are several different approaches to a NLFF solution currently available, e.g., Grad-Rubin style current-field iteration procedures, boundary integral methods, magnetofrictional methods, and optimization approaches \citep[see the review by ][]{schrijver06}. Figure~\ref{mhd:derosa09} shows examples of the various types of NLFF extrapolation procedures, featured in the review of \citet{derosa09}.

%%%%%%%%%%%%%%%%%%%%%%%%%%%%%%%%%%%%%%%%%%%%%%%%%%%%%%%%%%%%%%%%%%%%%%%%%%%%%%%%%%%%%%%%%%%%%%%%%%

\subsection{Comparison}
\label{mhd:comparison}
\begin{figure}[!t]
\centerline{\includegraphics[width=0.9\textwidth]{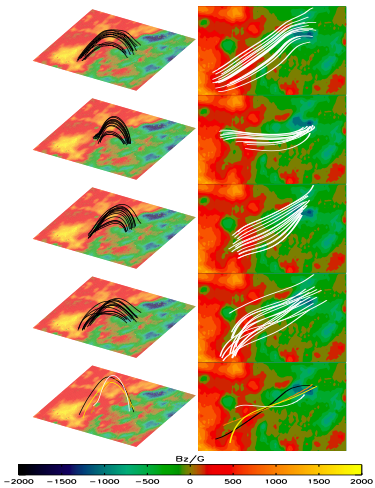}}
\caption[Observed and extrapolated fields compared by \citet{Wiegelmann05}.]{Observed and extrapolated fields compared by \citet{Wiegelmann05}. The upper row shows the original observed chromospheric loops, second row the potential field extrapolation, third row the LFF field extrapolation, and fourth row the NLFF field extrapolation. The bottom row shows one loop for comparison of all extrapolation types (black line: original, white: potential field, orange: LFF field, yellow: NLFF field.)}
\label{mhd:wiegelmann05}
\end{figure}

Potential field extrapolations can be thought of as the first-order approximation to the coronal magnetic topology. They are often used to represent solar global magnetic fields, which are dominated by simple dipolar configurations with few currents \citep[][see Figure~\ref{mhd:pot_example}]{liu08}. In contrast, LFF field extrapolations have been considered sufficient to represent the large-scale, current-carrying coronal magnetic fields present on whole active region size scales \citep[][see Figure~\ref{mhd:lff_example}]{gary89, wheatland99}. However, NLFF field extrapolations intrinsically contain more information on the complex nature of solar magnetic fields and more accurately represent coronal magnetic structures on size scales smaller than whole active regions. For example, \citet{Wiegelmann05} compared field extrapolations from photospheric measurements to observed chromospheric magnetic fields, finding the potential solution provided no agreement with any observed field lines, the LFF solution provided 35\% agreement, while the NLFF solution provided 64\% agreement. Figure~\ref{mhd:wiegelmann05} shows Figure 1 of \citet{Wiegelmann05}, which illustrates these differences.

It should be noted that there are still inconsistencies between existing NLFF extrapolation methods, particularly regarding the treatment of boundary conditions \citep[see, e.g., the comparison paper of][as in Figure~\ref{mhd:derosa09}]{derosa09}. However, for the purposes of this thesis the NLFF extrapolation is considered the most accurate representation of sunspot magnetic field lines. 

%%%%%%%%%%%%%%%%%%%%%%%%%%%%%%%%%%%%%%%%%%%%%%%%%%%%%%%%%%%%%%%%%%%%%%%%%%%%%%%%%%%%%%%%%%%%%%%%%%

\section{Radiative Transfer}
\label{radtransfer}
Although analysis of the 3D structure of ARs is the ultimate goal of this thesis, first magnetic field data must be obtained to be used as inputs to the extrapolation procedures described above. The input vector magnetic field information is often obtained using radiative transfer techniques. Radiative transfer is described as the physical phenomenon of energy transfer in the form of EM radiation. The convection found below the solar surface rarely carries a significant fraction of energy flux into the photosphere, and the dominant mechanism of energy transport through surface layers of a Sun-like star is radiation \citep{Gray}. Radiative transfer theory is used to create model atmospheres that can be compared to observations. This section serves to introduce the reader to the basic theory behind radiative transfer, which will be built upon to describe specific analysis techniques in Section~\ref{instr:data}.

\begin{figure}[!t]
\centerline{\includegraphics[width=\textwidth]{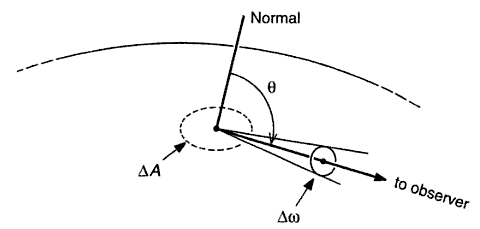}}
\caption[An illustration of the geometrical portion of the definition of specific intensity \citep{Gray}.]{An illustration of the geometrical portion of the definition of specific intensity. Note here that $\theta$ indicates the inclination angle, which is denoted by $\gamma$ in this thesis \citep{Gray}.}
\label{rad_trans:specificintensity}
\end{figure}

The fundamental quantity which describes a field of radiation is the specific intensity, $I_\nu$. Consider light in a particular frequency range passing through a small area, $dA$, into a solid angle, $d\omega$, over a particular time, with the light beam inclined at an angle $\theta$ to the direction perpendicular to the area (normal axis in Figure~\ref{rad_trans:specificintensity}). $I_\nu$ can be defined as the amount of energy passing through unit area (perpendicular to the beam of radiation) per unit time per unit frequency interval into unit solid angle. In other words,
\begin{equation}
dE~=~I_\nu~cos\theta~dA~dt~d\nu~d\omega \ ,
\end{equation}
where $dE$ is the amount of energy (in units of erg) crossing the area $dA$. See Figure~\ref{rad_trans:specificintensity} for an illustrative example. The units of the spectral intensity are thus W~m$^{-2}$~sr$^{-1}$~Hz$^{-1}$. Note that if the source emits radiation equally in all directions, the radiation is said to be isotropic (as in the solar interior). In the solar atmosphere there is much more radiation coming from the direction of the Sun's centre than from the outside direction, i.e., the radiation is anisotropic \citep{phillips08b}.

\subsection{Radiative Transfer Equation}
\label{theory:rte_opt}

The propagation of radiation through a medium can be affected by emission, absorption and scattering processes, and specific intensity plays a part in the definition of these processes. As a beam of radiation travels, it loses energy to the atmosphere by absorption, gains energy by atmospheric emission, and redistributes energy by scattering. 

\begin{figure}[!t]
\centerline{\includegraphics[width=0.75\textwidth]{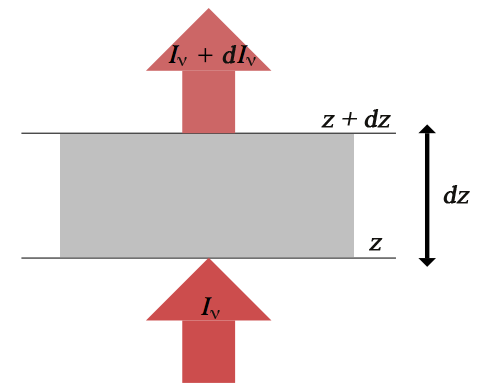}}
\caption[Radiative transfer in a medium. \emph{Modified figure originally by \citet{bannister07}}.]{Radiative transfer in a medium (gray rectangular slab). Radiation enters the medium at height $z$, and leaves it at height $z + dz$. \emph{Modified figure originally by \citet{bannister07}}.}
\label{theory:rte_fig}
\end{figure}

Consider a beam of radiation passing through a medium of thickness $dz$ (it enters the medium at point $z$, and leaves at $z + dz$), which may add to or subtract from the radiation, as shown in Figure~\ref{theory:rte_fig}. The intensity of the beam, $I_\nu$, is increased by processes of emission, and decreased by processes of absorption (including scattering). Let $j_\nu$ be the emission coefficient at frequency $\nu$, such that the amount of energy added to the beam $dE_\nu$ is given by,
\begin{equation}
dE_\nu~=~j_\nu~dA~dt~d\nu~d\omega~dz \ .
\end{equation}
Similarly, the absorption coefficient $\kappa_\nu$ is defined as the amount of energy subtracted from the beam,
\begin{equation}
dE_\nu~=~-~\kappa_\nu~I_\nu~dA~dt~d_\nu~d\omega~dz \ .
\end{equation}
The change in intensity over length $dz$ of the material is thus,
\begin{equation}
dI_\nu~=~[j_\nu~-~\kappa_\nu~I_\nu]dz \ .
\end{equation}
Rearranging this equation,
\begin{equation}
\label{rad_trans:RTE1}
\frac{dI_{\nu}}{dz}~=~j_{\nu}-\kappa_{\nu}I_{\nu} \ .
\end{equation}
This is known as the radiative transfer equation (RTE), where $z$ is the direction of propagation (\citealp{rutten03}; \citealp{phillips08b}). In general, the intensity decreases as the beam travels, due to the absorption of photons being greater than the emission of photons. 

There are two physical processes contributing to $\kappa_\nu$: absorption where a photon is destroyed and energy `thermalised', and scattering where a photon is deviated in direction and removed from the solid angle. Similarly with $j_\nu$, there is emission with creation of photons, and scattering of photons into the direction being considered. The source function, $S_{\nu}$, describes the ratio of the emission coefficient to the absorption coefficient, such that $S_{\nu} \equiv j_{\nu} / \kappa{_\nu}$. It is a measure of how photons in a light beam are removed and replaced by new photons by the material it passes through. Equation~\ref{rad_trans:RTE1} can thus be written,
\begin{equation}
\label{rad_trans:RTE2}
\frac{dI_{\nu}}{dz}~=~-\kappa_\nu(I_{\nu}-S_\nu)~~~.
\end{equation}
The source function $S_{\nu}$ here can be thought of as the specific intensity `emitted' at some `point' in a hot gas, and has the same units as $I_\nu$.

The RTE is commonly written in terms of the optical depth, $\tau$, which is a measure of transparency, expressing the quantity of light removed from a beam by scattering or absorption during its passage through the medium. For $\tau<1$, the plasma is optically thin (transparent), and if $\tau>1$ it is optically thick (opaque). 
The optical depth can be defined as $d\tau_{\nu}=-\kappa_{\nu}dz$, and Equation~\ref{rad_trans:RTE2} can then be re-written as,
\begin{equation}
\label{rad_trans:RTE3}
\frac{dI_{\nu}}{d\tau}~=~I_{\nu}-S_\nu~~~. 
\end{equation}
Also note that integrating $d\tau_{\nu}=-\kappa_{\nu}dz$ gives the optical depth as $\tau_{\nu}=-\int\kappa_{\nu}dz$.

%%%%%%%%%%%%%%%%%%%%%%%%%%%%%%%%%%%%%%%%%%%%%%%%%%%%%%%%%%%%%%%%%%%%%%%%%%%%%%%%%%%%%%%%%%%%%%%%%%%%%%%%%%%%%%%%%%%%%%%%%%%%%%%%%%%%%%%%%%%%%%%%%%%

\subsection{Local Thermodynamic Equilibrium}
A particularly useful simplification of the RTE occurs under the conditions of local thermodynamic equilibrium (LTE). In LTE, intensive parameters are varying in space and time, but are varying so slowly that, for any point, one can assume thermodynamic equilibrium (i.e., constant temperature) in some neighborhood about that point. Radiation passing through a blackbody volume of gas suffers no change with time, i.e., there is as much radiation absorbed as emitted. So at thermal equilibrium, the ratio of emissivity to absorptivity is a universal function only of radiative wavelength and temperature, i.e, a perfect blackbody emissive power (this is known as Kirchoff's Law). Therefore, $S_\nu = B_\nu(T)$, where $B_{\nu}(T)$ is the Planck function, i.e., a blackbody emits radiation in thermal equilibrium according to Planck's Law (defined in Equation~\ref{intro:planck}). Thus, $j_{\nu}=\kappa_{\nu} B_{\nu}(T)$. The RTE in Equation~\ref{rad_trans:RTE2} can then be written as,
\begin{equation}
\label{rad_trans:LTE}
\frac{dI_{\nu}}{dz}~=~-\kappa_{\nu}(I_{\nu}-B_{\nu}(T)) ~~~,
\end{equation}
or in terms of optical depth from Equation~\ref{rad_trans:RTE3},
\begin{equation}
\frac{dI_{\nu}}{d\tau}~=~I_{\nu}-B_{\nu}(T)~~~. 
\end{equation}
Generally, LTE is often only applied to massive particles. In a radiating gas, the photons being emitted and absorbed by the gas need not be in thermodynamic equilibrium with each other or with the massive particles of the gas in order for LTE to exist. LTE enables calculations of thermodynamic properties in terms of temperature, density and composition as they change from the solar centre to the surface. In the photosphere the radiation varies from place to place, but sufficiently slowly that the emitting material and radiation are close to thermodynamic equilibrium.

A number of inversion techniques use model atmospheres that assume LTE to make calculations simpler (using photospheric observations for comparison), including the Milne-Eddington approximation, which is described in Section~\ref{rad_trans:m-e}. However first the photospheric measurements must be obtained from instruments observing the Sun. The instruments used in this thesis will be described in the next chapter, as well as describing the analysis techniques used on the observations obtained. 

	% background information

%\include{1_introduction/rad_transfer}	% spectropolarimetry

% this file is called up by thesis.tex
% content in this file will be fed into the main document

%: ----------------------- name of chapter  -------------------------
\chapter{Instrumentation and Analysis Techniques} % top level followed by section, subsection
\label{chapter:instrumentation}

\ifpdf
    \graphicspath{{2/figures/PNG/}{2/figures/PDF/}{2/figures/}}
\else
    \graphicspath{{2/figures/EPS/}{2/figures/}}
\fi

\hrule height 1mm
\vspace{0.5mm}
\hrule height 0.4mm 
\noindent 
\\ {\it In this chapter, details of the instruments and data analysis techniques used in this thesis are presented, including preparation of the data obtained from them. First some background to the physics used in the measurements is given, as well as a discussion of early observations. Instruments onboard the Hinode spacecraft are the main source of data for this thesis. The Spectropolarimeter onboard Hinode's Solar Optical Telescope is described in detail, followed by a brief outline of the Broadband Filter Instrument. Data analysis techniques are then described, including details of the atmospheric inversion code, field azimuth disambiguation, coordinate transformations, and 3D magnetic field extrapolation methods.
}
\\ 
\hrule height 0.4mm
\vspace{0.5mm}
\hrule height 1mm 

\newpage

Before the dawn of satellite instrumentation, decades of ground-based longitudinal magnetic measurements could not provide information on the true complexity of active-region magnetic topology, as these did not provide information on the 3D field. Even with good seeing and image stabilisation methods through adaptive optics, atmospheric distortions cause instruments to fall short of resolving sunspot fine structure in the photosphere \citep[angular resolution of $0.1" \sim72$~km or better is needed;][]{lites02}. The polarimetric signals from ground-based instruments must also be accumulated over at least a number of seconds in order to achieve the required precision. High spatial resolution observations of the solar magnetic field from spacecraft can now provide this information.

\section{Solar Magnetic Field Observations}
\subsection{Spectropolarimetry}
 
Spectropolarimetry is used for the investigation of the magnetic field structure of sunspots in this thesis. Its basis comes from spectroscopy and polarimetry. Spectroscopy aims at measuring the flux distribution of a light source as a function of the radiation wavelength (spectra). Emission and absorption lines result from radiative transitions through absorption, emission, recombination and de-excitation of photons, with different elements producing different spectra according to their differing atomic structure. The Sun radiates light across a broad range of the EM spectrum. In the infra-red and visible wavelengths, the solar spectrum is a continuous background crossed by absorption lines, with the greatest intensity found in the visible range. At wavelengths $<1600$~\AA\ (extreme ultraviolet), the solar spectrum is in the form of emission lines with a relatively weak continuum (broadband EM radiation).

\begin{figure}[!t]
\centerline{\includegraphics[scale=0.7]{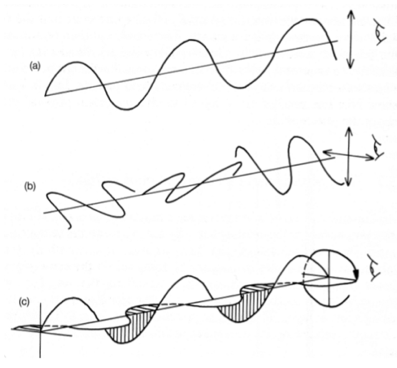}}
\caption[Depiction of the various types of polarisation of light \citep{phillips92}.]{Depiction of the various types of polarisation of light \citep{phillips92}. (a) Linearly polarised light, a confinement of the electric magnetic field vector to a given plane along the direction of propagation. (b) Unpolarised light, segments of light waves occur along line of sight with randomly oriented planes. (c) Circularly polarised light, when two linearly polarised rays are a quarter of a wavelength out of phase, so that the vector sum of the wave motions rotates as the wave propagates.}
\label{rad_trans:polarisation}
\end{figure}

Polarimetry measures the degree to which radiation from a light source is polarised, as well as the polarisation state of the corresponding light. Polarisation is a property of an EM wave describing the orientation of the electric field oscillations. Unpolarised light is observed when segments of light waves occur along the LOS with randomly oriented planes. Linearly polarised light is a confinement of the EM field vector to a given plane along the direction of propagation. Circularly polarised light occurs, for example, when two linearly polarised rays are a quarter of a wavelength out of phase, so that the vector sum of the wave motions rotates as the wave propagates. Figure~\ref{rad_trans:polarisation} depicts the various types of polarised light, with polarisation being a property of waves describing the orientation of their oscillations.

\label{rad_transfer:stokes}
The Stokes parameters are quantities used to describe the polarisation state of light \citep{stokes}. The four parameters \emph{I}, \emph{Q}, \emph{U}, and \emph{V} completely describe polarisation: \emph{I} is the total unpolarised intensity; \emph{Q} and \emph{U} are the components of linear polarisation (where the frames of \emph{Q} and \emph{U} are rotated by 45$^{\circ}$); \emph{V} is the degree of circular polarisation. In other words,
\begin{eqnarray}
I & = & \circlearrowright + \circlearrowleft \mathrm{or} \leftrightarrow + \updownarrow \mathrm{etc} \ , \\
Q & = & \updownarrow - \leftrightarrow \ ,  \\
U & = & \nwarrow - \nearrow \ , \\
V & = & \circlearrowright - \circlearrowleft \ ,
\end{eqnarray}
where $\circlearrowright$ and $\circlearrowleft$ represent right and left circularly polarised light, $\leftrightarrow$ and $\updownarrow$ represent linearly polarised light in the $0^\circ$ and $90^\circ$ directions, and $\nearrow$ and $\nwarrow$ represent polarisation in the $45^\circ$ and $135^\circ$ directions. In a spectral line, the state of polarisation varies as a function of wavelength, so the Stokes parameters are a function of wavelength, i.e., \emph{I}($\lambda$), \emph{Q}($\lambda$), \emph{U}($\lambda$), \emph{V}($\lambda$).

\begin{figure}[!t]
\centerline{\includegraphics[width=\textwidth]{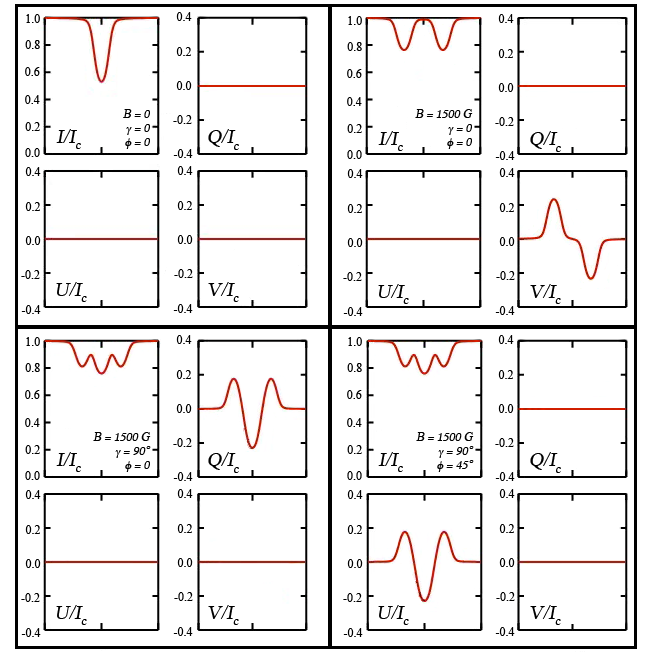}}
\caption[The change of the synthetic emergent Stokes \emph{I}, \emph{Q}, \emph{U}, and \emph{V} profiles when the magnetic field present in the solar plasma varies.]{The change of the synthetic emergent Stokes \emph{I}, \emph{Q}, \emph{U}, and \emph{V} profiles when the magnetic field present in the solar plasma varies. This is depicted in terms of changes in the magnetic field strength ($B$), inclination of the magnetic field vector with respect to the observer's LOS ($\gamma$), and azimuth of the magnetic field vector in the plane perpendicular to the observer's LOS ($\phi$). Modified from a movie originally by \citet{borrero11}.}
\label{instr:stokesiquv}
\end{figure}

Figure~\ref{instr:stokesiquv} shows examples of Stokes \emph{I}, \emph{Q}, \emph{U}, and \emph{V} profiles  under various circumstances. In the upper left panel, magnetic field strength, inclination angle, and azimuthal angle are all zero. This leads to only a Stokes \emph{I} profile observed. When the magnetic field strength, $B$, is increased along the LOS only (i.e. inclination, $\gamma$, and azimuthal, $\phi$, angles are still both zero), the \emph{V} profile is also observed. The upper right panel shows this for a field strength of 1500~G, with a typical two-lobe structure for Stokes \emph{V} and the \emph{I} profile beginning to split due to the presence of the magnetic field. When field inclination is at $90^{\circ}$ (and still a 1500~G field), as in the lower left panel, Stokes \emph{V} disappears, however a typical Stokes \emph{Q} profile is now observed. Finally, when the azimuthal angle is at $45^{\circ}$ (with field strength at 1500~G and inclination at $90^{\circ}$), the Stokes \emph{Q} profile disappears and only \emph{U} profile is observed along with the total intensity \emph{I}. The two-lobe structure of the \emph{V} profile, and three-lobe structures of the \emph{Q} and \emph{U} profiles can be explained by the Zeeman effect, which will be discussed in the next section.

%%%%%%%%%%%%%%%%%%%%%%%%%%%%%%%%%%%%%%%%%%%%%%%%%%%%%%%%%%%%%%%%%%%%%%%%%%%%%%%%%%%%%%%%%%%%%%%%%%%%%%%%%%%%%%%%%%%%%%%%%%%%%%%%%%%%%%%%%%%%%%%%%%%

\subsection{Zeeman Effect}
\label{rad_transfer:zeeman_effect}
\begin{figure}[!t]
\centerline{\includegraphics[scale=0.6,trim=5mm 0mm 0mm 0mm,clip]{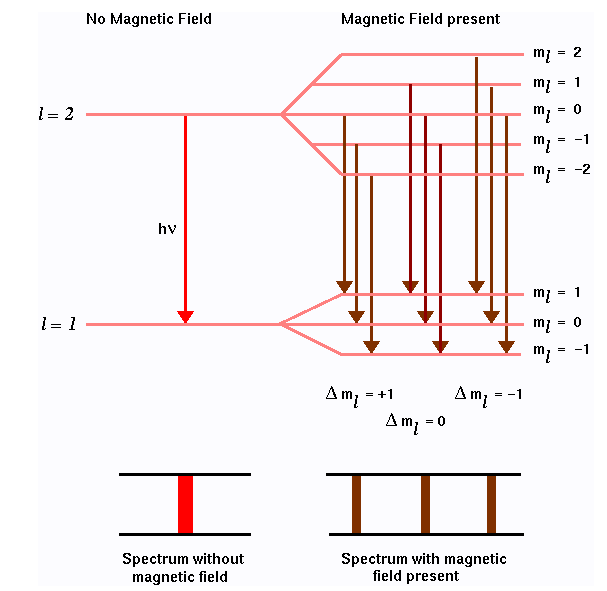}}
\caption[Depiction of a normal Zeeman triplet, with the splitting of the line into three components due to the presence of a magnetic field \citep{majumdar00}.]{Depiction of a normal Zeeman triplet, with the splitting of the line into three components due to the presence of a magnetic field. Transitions with $\Delta m_l=0$ are called $\pi$ transitions, and transitions with $\Delta m_l = \pm1$ are called $\sigma$ transitions \citep{majumdar00}.
}
\label{rad_trans:zeeman_triplet}
\end{figure}

The strong magnetic fields of sunspots can result in large splittings of atomic levels. The Zeeman effect can be defined as the splitting of a spectral line into several components in the presence of a static magnetic field. Atomic and molecular energy levels split into sublevels, which are characterised by their magnetic quantum number, $M$ \citep{condon35}. More specifically, the presence of the magnetic field breaks degeneracy and an atomic level of total angular momentum \emph{J} is split into (2\emph{J} + 1) magnetic sublevels, where $J = L + S$ ($L$ is orbital angular momentum and $S$ is spin angular momentum). For the normal Zeeman effect\footnote[1]{The anomalous Zeeman effect appears for transitions where the net spin of the electrons is non-zero, but is not of concern here.} (zero spin), the magnetic quantum number selection rule simplifies the available transitions into three allowed states: $\pi$ ($\Delta m=0$; linear), $\sigma_r$ ($\Delta m = -1$; left circular), and $\sigma_b$ ($\Delta m = +1$; right circular). See Figure~\ref{rad_trans:zeeman_triplet} for an example of a typical normal Zeeman triplet showing this splitting. 

\begin{figure}[!t]
\centerline{\includegraphics[scale=0.5]{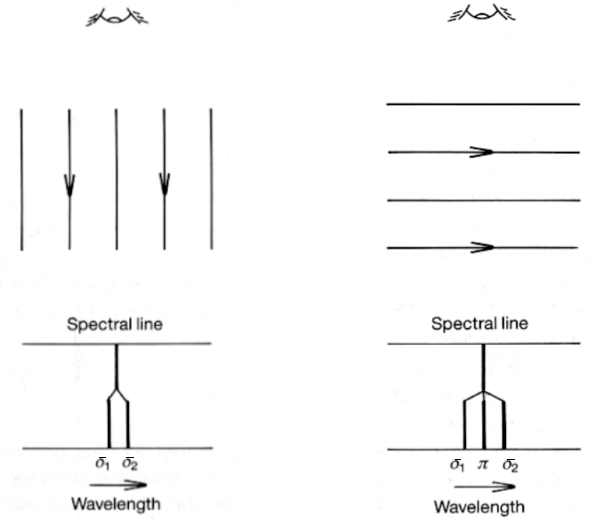}}
\caption[Illustration of the Zeeman effect for an absorption line, \emph{courtesy of L. van Driel-Gesztelyi}.]{Illustration of the Zeeman effect for an absorption line. The left column depicts the longitudinal Zeeman effect, and right column the transverse Zeeman effect. \emph{Figure courtesy of L. van Driel-Gesztelyi}.}
\label{rad_trans:zeeman_abs}
\end{figure}

The lines corresponding to Zeeman splitting thus exhibit polarisation effects. Figure~\ref{rad_trans:zeeman_abs} depicts the longitudinal and transverse Zeeman effect for an absorption line. For $B$ parallel to the LOS (longitudinal Zeeman effect), only the two shifted $\sigma$ components are present, which have opposite sense of circular polarisation. The $\pi$ component is not observed because its axis of linear polarisation is directed along the LOS. For \emph{B} perpendicular to the LOS (transverse Zeeman effect), all three components are seen as linearly polarised: the unshifted $\pi$ component, and both shifted $\sigma$ components. The $\sigma$ components appear linearly polarised from viewing their circular polaristion side-on. Note that this description is valid for absorption lines; for emission lines the sense of circular polarisation is reversed and `parallel' is replaced with `perpendicular'.

The splitting that occurs (i.e, distance between the Zeeman sub-levels) is proportional to the Land$\acute{\mathrm{e}}$ $g$ factor and the field strength. The presence of a solar magnetic field can thus be deduced from the Zeeman splitting of spectral lines for a sufficiently strong field \citep{bueno06}. The wavelength shift involved can be represented in terms of the Land$\acute{\mathrm{e}}$ $g$ factor,
\begin{equation}
\label{landeg}
g~=~1+\frac{J(J+1)-L(L+1)+S(S+1)}{2J(J+1)}~~~,
\end{equation}
where \emph{L}, \emph{S} and \emph{J} are the quantum numbers characterising the orbital, spin and total angular momentum, respectively. The wavelength displacement of the spectrum is thus,
\begin{equation}
\label{wavelengthdisplacement}
\lambda-\lambda_0~=~\frac{e}{4 \pi m_\mathrm{e} c}g^{*}\lambda^2B~~~,
\end{equation}
where $\lambda_0$ is the original wavelength, $m_{\mathrm{e}}$ and \emph{e} are the mass and charge on the electron, \emph{c} is the speed of light , $g^*$ is the Land$\acute{\mathrm{e}}$ $g$ factor for the transition\footnote[2]{$g^{*}=g_{\mathrm{u}}M_{\mathrm{u}}-g_{\mathrm{l}}M_{\mathrm{l}}$, where \emph{M} the additional magnetic quantum number and subscripts `u' and `l' denote the upper and lower levels of the transition}, and $B$ is the magnetic field strength \citep{thomasbook}.

\begin{figure}[!t]
\centerline{\includegraphics[scale=0.6]{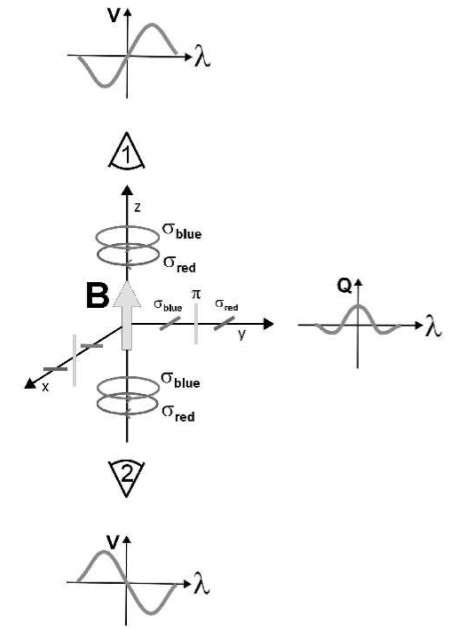}}
\caption[Resultant polarisation states of light through the Zeeman effect \citep{bueno03}.]{Resultant polarisation states of light through the Zeeman effect. The Stokes \emph{V}($\lambda$) profile changes its sign for opposite orientations of the magnetic field. The Stokes \emph{Q}($\lambda$) profile is invariant to a reversal of the magnetic field direction, while it reverses sign when the transverse field is rotated by $\pm 90^{\circ}$ \citep{bueno03}. }
\label{rad_trans:illustrativezeeman}
\end{figure}

Relating the Stokes parameters back to the Zeeman effect, as in Figure~\ref{rad_trans:zeeman_abs}, for the transverse effect, $B_{\perp}\propto \sqrt{Q^{2}+U^{2}}$, and for the longitudinal effect, $B_{\mathrm{||}}\propto V$. See Figure~\ref{rad_trans:illustrativezeeman} for an illustrative example. Realistically, magnetic fields will not always be directed parallel or perpendicular to the observer's LOS. This results in the detection of linearly and circularly polarised components, where the relative degrees of $Q$ and $U$ provide information on the orientation of the magnetic field azimuthal angle from a chosen reference frame, $\phi$, and their relation to $V$ yields information on the inclination angle from the LOS, $\gamma$, such that \citep{auer77}, 
\begin{equation}
\phi~=~\mathrm{arctan}\left ( \frac{U}{Q} \right )~~~,
\label{azieqn}
\end{equation}
\begin{equation}
\gamma~=~\mathrm{arccos}\left ( \frac{V}{\sqrt{Q^{2}+U^{2}+V^{2}}} \right )~~.~~~~~~~~~~~~~~
\label{inceqn}
\end{equation}
Note that this is simply for a good initial guess of the field orientation. Inversion techniques are needed to obtain more accurate field information, which will be described in Section~\ref{rad_trans:helixmethod}.

%%%%%%%%%%%%%%%%%%%%%%%%%%%%%%%%%%%%%%%%%%%%%%%%%%%%%%%%%%%%%%%%%%%%%%%%%%%%%%%%%%%%%%%%%%%%%%%%

\subsection{Early Observations}

\begin{figure}[!t]
\centerline{\includegraphics[width=0.9\textwidth]{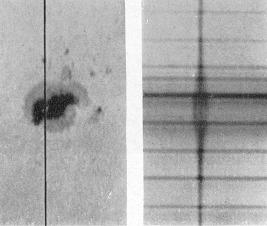}}
\caption{Observations of the magnetically-induced Zeeman splitting in the spectrum of a sunspot obtained by \citet{hale1919}.}
\label{instr:hale1919}
\end{figure}

Although sunspot observations trace back as far as Chinese astronomers around 28 BC, \cite{hale08} is generally credited with the first direct measurement of the magnetic field within a sunspot. By measuring the Zeeman splitting in magnetically sensitive $6000 - 6200$~\AA\ lines in the spectra of sunspots, and detecting the polarisation of the split spectral components, \citeauthor{hale08} provided the first quantitative demonstration that sunspots are regions of strong magnetic fields (he inferred the field strength to be $\sim$~3000~G). Figure~\ref{instr:hale1919} reproduces the observations obtained by \citet{hale1919}, showing Zeeman splitting in a sunspot due to its strong magnetic field.

The Babcock magnetograph \citep{babcock53} paved the way for later magnetic field observations. It is operated on the same principle \citeauthor{hale08} used to measure the Zeeman effect, by alternatively comparing right circularly polarised and left circularly polarised light from a particular location on the Sun. However, while \citeauthor{hale08} used only one slit to disperse the spectrum, \citeauthor{babcock53} used two slits to determine the line shifts. In the standard Babcock design, the right circularly polarised and left circularly polarised components are separated using an electro-optic crystal. Passing the correct voltage through the crystal produces a retardation of $\lambda$/4, which converts the circularly polarised components into two linearly polarised beams at right angles to each other. A linear polaroid then allows only one beam to pass through. A retardation in the opposite sense can be achieved by applying an opposite voltage to the crystal, thus allowing both circularly polarised components to be calculated. The difference in intensity of the signal in each of these states is then used to infer the LOS magnetic field strength.

The work of \citet{babcock53} paved the way for modern magnetographs that use tuneable Michelson-Morley interferometers in place of slits to simultaneously allow the spectral analysis of every point on the Sun. However, the instruments described in this section only measure the LOS field. As mentioned in Chapter~\ref{chapter:introduction}, many previous studies of AR magnetic fields used LOS field observations only, as the instrumentation available was not advanced enough. Now, instruments onboard spacecraft such as \emph{Hinode} can measure the vector magnetic field with high spectral and spatial resolution. It is advantageous to have full vector magnetic field measurements as field orientation and topology can be determined. Also, higher resolution is important when examining small spatial-scale magnetic field changes (as is investigated in Chapter~\ref{chapter:paper1}). LOS field information is mostly only useful for studies of large spatial- and temporal-scale evolution of the Sun. The measurements used for analysis in this thesis were obtained from a spectropolarimeter onboard \emph{Hinode}, and will be described in more detail in the following Section.

%%%%%%%%%%%%%%%%%%%%%%%%%%%%%%%%%%%%%%%%%%%%%%%%%%%%%%%%%%%%%%%%%%%%%%%%%%%%%%%%%%%%%%%%%%%%%%%%

\subsection{\emph{Hinode}}

\label{instr:hinode}

\emph{Hinode}, formerly known as \emph{Solar-B}, is a Japan Aerospace Exploration Agency solar mission, with collaboration from the United States as well as the United Kingdom \citep{Kosugi07}. It was launched from the Uchinoura Space Center on 2006 September 22, on the final flight of the M-V rocket. The satellite maneuvered into a quasi-circular, sun-synchronous orbit over the day/night terminator, allowing near-continuous observation of the Sun. 

\begin{figure}[!t]
\centerline{\includegraphics[scale=0.7]{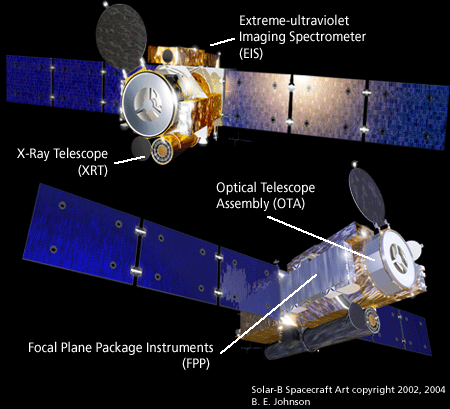}}
\caption{Illustration of the main components of \emph{Hinode} (EIS, XRT, OTA, FPP) from a front and side view of the craft. \emph{Figure is courtesy of NASA.}}
\label{instr:hinode_spacecraft}
\end{figure}

The spacecraft payload consists of an X-ray telescope (XRT), Extreme Ultraviolet Imaging Spectrometer and the Solar Optical Telescope (SOT). See Figure~\ref{instr:hinode_spacecraft} for a depiction of the spacecraft and its instruments. The X-Ray Telescope (XRT) uses grazing incidence optics to image the solar corona's hottest components, with full-Sun image capturing when pointed at the solar disk. The EUV Imaging Spectrometer (EIS) obtains spatially resolved spectra in two wavelength bands, $170-212$~\AA\ and $246-292$~\AA, which is used to identify physical processes involved in heating the solar corona. The Solar Optical Telescope (SOT) is used as a primary data source in this research, and provides measurements of solar vector magnetic fields on spatial scales of $\sim200 - 300$ km over a FOV large enough to contain small active regions \citep{Tsuneta08}.

\begin{figure}[!t]
\centerline{\includegraphics[width=\textwidth]{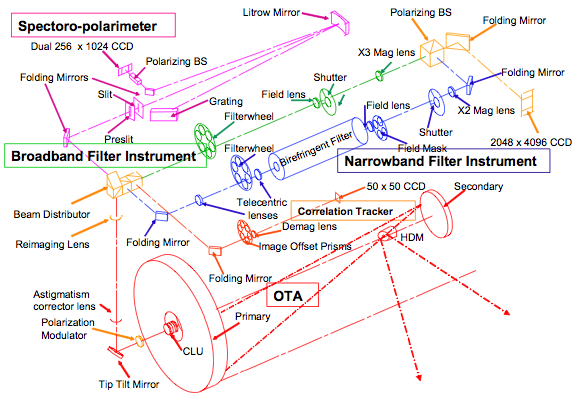}}
\caption[Optical schematic of the Optical Telescope Assembly and Focal Plane Package instruments \citep{Tsuneta08}.]{Optical schematic of the Optical Telescope Assembly and Focal Plane Package instruments \citep{Tsuneta08}. The OTA consists of the primary and secondary mirrors, the collimator lens unit, polarisation modulator and the tip-tilt mirror.}
\label{instr:sot_schematic}
\end{figure}

SOT is composed of two main parts (see Figure~\ref{instr:sot_schematic} for a schematic): the Optical Telescope Assembly \citep[OTA; ][]{Suematsu08}, which holds the mirrors and other optics of the instrument, and the Focal Plane Package (FPP). The OTA is an aplanatic Gregorian telescope, with an aperture of 0.5~m. It holds the collimating lens unit (CLU), the polarisation modulator (PMU), a tip-tilt mirror (CTM) behind the primary mirror, and a heat dump mirror. With the CLU and CTM, the OTA provides a pointing-stabilised parallel beam to the FPP \citep{Kosugi07}. The FPP was built by Lockheed Martin Solar and Astrophysics Laboratory in Palo Alto, California. It consists of the Broadband Filter Imager (BFI), the Narrowband Filter Imager (NFI), and the Spectropolarimeter (SP), along with a correlation tracker (CT). The CT is a high-speed charge-coupled device (CCD) camera used to sense jitter of solar features on the focal plane. It obtains a displacement error by the tracking of solar granulation, and this displacement is fed back to the CTM in the OTA for correction. SOT stabilisation is described further by \citet{shimizu08}. The SP on the FPP uses polarising optics to measure magnetic fields in the solar photosphere, and is the primary instrument used in this research.

%%%%%%%%%%%%%%%%%%%%%%%%%%%%%%%%%%%%%%%%%%%%%%%%%%%%%%%%%%%%%%%%%%%%%%%%%%%%%%%%%%%%%%%%%%%%%%%%%%

\subsubsection{Spectropolarimeter}
\label{chapter:sot-sp}
The SOT-SP is a modified Littrow spectrometer that operates in a synchronous mode with the PMU. It records photospheric vector magnetic fields with a polarimetric accuracy better than 10$^{-3}$ of the continuum intensity \citep[$I_c$;][]{Tsuneta08}. The instrument is based on the design of the Advanced Stokes Polarimeter (ASP), which was constructed by the National Center for Atmospheric Research at the National Solar Observatory, in collaboration with the High Altitude Observatory \citep{Elmore92}.

Spectra of the Fe \textsc{i} 6301.5~\AA\ and 6302.5~\AA\ spectral lines and nearby continuum are recorded through a $0.16" \times 164"$~slit (oriented North-South on the heliographic disk), with a spectral sampling of 21.5~m\AA\ (i.e., Doppler velocity resolution $\sim 1.02$~km s$^{-1}$). These Fe \textsc{i} lines are photospheric lines of the same multiplet with different sensitivity to the Zeeman effect by having different Land$\acute{\mathrm{e}}$ $g$ factors (1.67 and 2.5 respectively) and, as such, provide a way to measure or infer the magnetic field experienced within that emitting portion of the solar atmosphere. See Table~\ref{instr:atomic} for some atomic information on these lines \citep{bard91,nave94}. 

\emph{Hinode} observes a sunspot by exposing the SP slit for a certain time duration (depending on the mapping mode) whilst recording different polarisation states. Light passes through the slit onto a diffraction grating to disperse the light, which is imaged onto two CCDs. It then moves the slit in the horizontal direction across the sunspot, repeating the process multiple times to scan the entire sunspot. The slit maps a finite area, up to the full 320" wide FOV. Note the largest observable FOV is $320" \times 164"$.

\begin{table}[!t]
\caption[Table of atomic information for the Fe \textsc{i} spectral line transition used by \emph{Hinode}/SOT-SP.]{Table of atomic information for the Fe \textsc{i} spectral lines used by \emph{Hinode}/SOT-SP, including lower and upper levels of the transitions. Note that s, p, and d represent the sharp, principal, and diffuse orbitals, respectively. The term symbol has the general form $^{2S+1}{L_J}$, where $S$, $L$ and $J$ represent the total spin, orbital, and angular momentum quantum numbers, respectively. }
\centering
\begin{tabular}{ l l l l}
\vspace{0.1cm} \\
\hline\hline
&  & 6301.51 \AA\ &  6302.49 \AA\  \\
\hline
Lower Level & Energy ($\mathrm{E_i}$)  & 3.65 eV &  3.69  eV  \\
& Configuration & 3d$^6$($^5$D)4s4p($^3$P) & 3d$^6$($^5$D)4s4p($^3$P) \\
& Term &  z$^5$P$_2$ & z$^5$P$_1$ \\
& $S$  & 2 & 2 \\
& $L$  & 1 & 1 \\
& $J$ & 2 & 1 \\
\hline
Upper Level  & Energy ($\mathrm{E_k}$)  & 5.62eV & 5.65 eV  \\
& Configuration &   3d$^6$($^5$D)4s ($^6$D)5s &   3d$^6$($^5$D)4s ($^6$D)5s \\
& Term &  e$^5$D$_2$ & e$^5$D$_0$ \\
& $S$ & 2 & 2 \\
& $L$ & 2 & 2 \\
& $J$  & 2 & 0 \\
  \hline\hline
\end{tabular}
\label{instr:atomic}
\end{table}

\begin{figure}[!t]
\centerline{\includegraphics[scale=0.7]{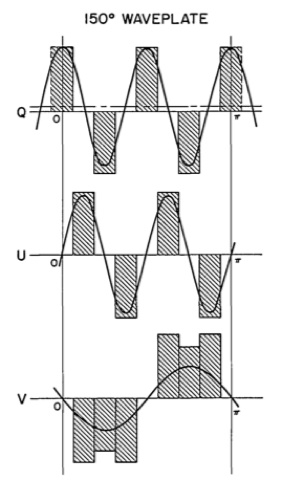}}
\caption[Ilustration of the periods of sampling within the ASP waveplate rotation cycle \citep{Lites87}.]{The periods of sampling within the ASP waveplate rotation cycle are shown. Onboard \emph{Hinode}, the linear polarisations $Q$ and $U$, and circular polarisation $V$, are converted into sinusoidal variations of intensity by the polarising beam splitters. $Q$ differs from $U$ by $22.5^{\circ}$. Demodulation is produced by sampling the intensity 16 times for revolution of the PMU, and $I$, $Q$, $U$, and $V$ are obtained by adding/subtracting each sample into four memories corresponding to the four Stokes parameters.\citep{Tsuneta08, Lites87}}
\label{instr:waveplate}
\end{figure}

The PMU is a rotating quarter waveplate with a rate of $T =$ 1.6~seconds per revolution, located after the CLU. It is used to modulate the intensity on the CCDs according to the polarisation of incident light. Spectra are exposed and readout continuously, 16 times per rotation of the PMU. The polarising beam splitter is located just before the cameras and is a dual-beam system, whereby the beam passes through the waveplate before splitting into two orthogonal states of linear polarisation. This eliminates polarisation cross-talk, i.e., to mitigate polarisation noise induced by residual image motions during the sampling \citep{ichimoto05}. The two beams are then imaged onto two separate detectors and the difference in these signals is sampled at discrete intervals of the waveplate orientation \citep{Lites87}. See Figure~\ref{instr:waveplate} for an illustration of the waveplate operation of the ASP, on which the SP is based.

Stokes parameters $Q$, $U$, and $V$ (see Section~\ref{rad_transfer:stokes} for description) are converted into sinusoidal variations of intensity by the polarising beam splitter, and are encoded as harmonic variations of intensity at periods proportional to $T/4$, $T/4$ and $T/2$ respectively. The signal vector $Q$ differs in phase from the signal $U$ by 22.5$^{\circ}$ (relative to the rotational phase of the waveplate; see Figure~\ref{instr:waveplate}). Demodulation of this signal is done by sampling the intensity 16 times per revolution of the PMU waveplate, with a fixed exposure time of 0.1~s. The CCD records these raw spectra, where each sample is either added or subtracted into four on-board memories to obtain the Stokes \emph{I}, \emph{Q}, \emph{U}, and \emph{V} spectra \citep{Tsuneta08}. A single SP `observable' is thus a set of \emph{I}, \emph{Q}, \emph{U}, and \emph{V} spectra in each of the two orthogonal polarisation states, where orthogonal here is taken to mean the two states have their polarisation vectors anti-parallel. The Stokes $I$, $Q$, $U$, $V$ spectra it creates are used to derive vector magnetogram maps of the solar surface. For more details on obtaining the data product of the SOT-SP from the incident Stokes vector to the telescope, and details on polarisation calibration, see \citet{ichimoto08}.

The SP has four modes of operation: normal, fast, dynamic and deep magnetogram. The normal map mode takes 83 minutes to scan a $160"$ wide area. It produces a polarimetric accuracy of 0.1$\%$ of the continuum intensity ($I_{\mathrm{c}}$) and a spatial sampling of $0.16"$. It accumulates data for 3 rotations of the PMU (i.e., taking 4.8~s per slit position). The scanning size can be reduced to speed up the cadence, which is useful for examining smaller features. The fast map mode, which is the source of data for this thesis, provides 30~minute cadence for $160"$ wide scanning, but with $0.32"$ sampling (2 rotations of PMU, one at each of two slit positions which are co-added). Thus the effective pixel size is $0.32" \times 0.32"$. The dynamic map mode provides higher cadence and $0.16"$ sampling, but with lower polarimetric accuracy (1 full rotation of PMU). In deep magnetogram mode, photons are accumulated over multiple rotations of the PMU in order to achieve high polarimetric accuracy in quiet solar regions, at the cost of lower time resolution.

%%%%%%%%%%%%%%%%%%%%%%%%%%%%%%%%%%%%%%%%%%%%%%%%%%%%%%%%%%%%%%%%%%%%%%%%%%%%%%%%%%%%%%%%%%%%%%%%%%

\subsubsection{Data preparation}
\label{instr:hinode:prep}

In terms of onboard data processing, two types of compression are performed sequentially: pixel-by-pixel bit compression and then image compression. First, look-up tables are used to perform 16-to-12 bit compression. Then a 12 bit JPEG lossy compression is employed for image compression of SP data, with the Stokes vector data compressed to $\sim1.5$~bits/pixel. The telemetry is received by a ground station at the Uchinoura Space Center in Japan, and the master archive of data there is mirrored into data centres across the globe. For more information on the data archive, see \citet{matsuzaki07} and references therein. All \emph{Hinode} data used for this thesis were obtained from the \emph{Hinode} Science Data Center Europe\footnote[3]{\url{http://sdc.uio.no/search/API}}. SOT data are available as FITS files, as first developed by \citet{wells81}, and updated by \citet{hanisch01}. The Level 0 data are in the format of SP 4D arrays of [$y$, $\lambda$, $O_{s}$, $S$], where $y$ is the spatial dimension along the slit, $\lambda$ the wavelength, $O_s$ the orthogonal state and $S$ is the Stokes vector ($I$, $Q$, $U$, $V$). 

The raw Level 0 data are calibrated using an IDL routine called \textsf{sp$\_$prep.pro}, from the \emph{Hinode}/SOT tree within the SolarSoft library \citep{Freeland98}. The \textsf{sp$\_$prep} routine makes two passes through the data. The first determines the thermal shifts (in both offset and dispersion) in the spectral dimension across successive slit positions ($x$), which is mainly caused by internal heating of equipment during normal operation. Thermal drift can also occur due to changes in external ambient temperature, with day-to-night solar heating variation on each Earth orbit. Thermal noise arises from thermal energy in the CCD material, creating electron-hole pairs in the absence of illumination. Exposures known as `darks' are made with no illumination falling on the CCD. 

The second pass corrects the thermal variations and merges the two orthogonal polarisation states to create a 3D Level 1 data array, [$y$, $\lambda$, $S$]. Correction must also be made for the fact that a CCD output is an analogue signal, i.e., a fluctuating voltage. To make sure the signal is always greater than the reference voltage, a DC offset is added. Other corrections in the prep routine include cosmic ray removal (despiking), and flat-fielding. Flat-field correction stems from the fact that the CCD material varies from pixel-to-pixel, and the CCD may have surface flaws. This can been seen by exposing the CCD to a uniform light source known as a `flat field', and is corrected for by dividing the data by the flat field. Bad pixels are also removed during the calibration process.

\begin{figure}[!t]
\centerline{\includegraphics[width=\textwidth]{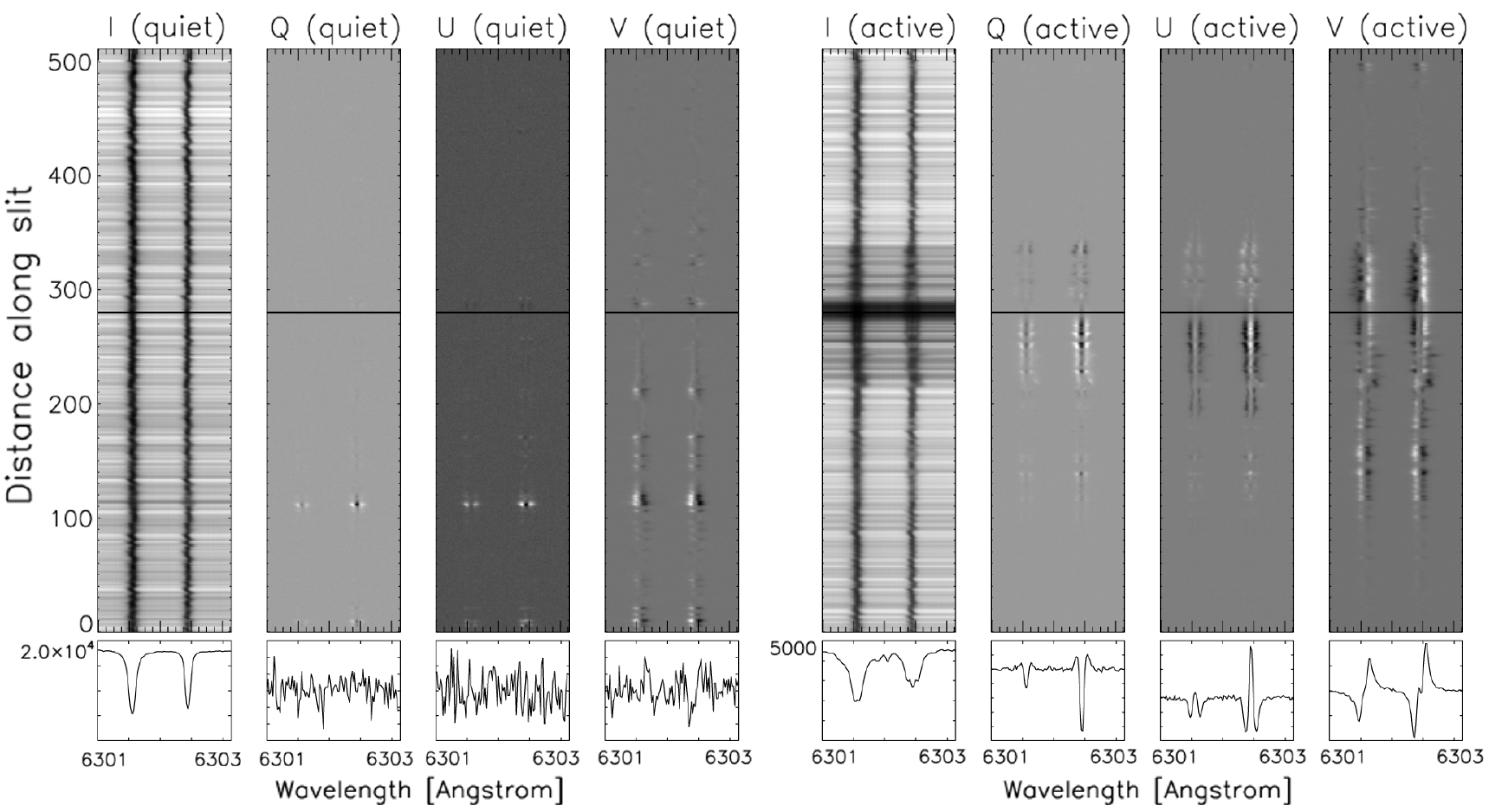}}
\caption[\emph{Top:} Stokes \emph{I}, \emph{Q}, \emph{U}, and \emph{V} spectra of the Fe \textsc{i} 6301.5~\AA\ and 6302.5~\AA\ spectral lines from the first SOT-SP scan of the event studied in Section~\ref{section:paper1}. \emph{Left:} Spectra from quiet region slice of the scan. \emph{Right:} Spectra from central umbral (active) region slice of the scan. \emph{Bottom:} Corresponding $I$, $Q$, $U$, and $V$ profiles for both of the regions at pixel 280 along the slit.]{\emph{Top:} Stokes \emph{I}, \emph{Q}, \emph{U}, and \emph{V} spectra of the Fe \textsc{i} 6301.5~\AA\ and 6302.5~\AA\ spectral lines from the first SOT-SP scan of the event studied in Section~\ref{section:paper1}. \emph{Left:} Spectra from quiet region slice of the scan. \emph{Right:} Spectra from central umbral (active) region slice of the scan. The colour scale is black corresponding to lowest intensity, and white the brightest. The wavelength scale is in pixels. \emph{Bottom:} Corresponding $I$, $Q$, $U$, and $V$ profiles for both of the regions at pixel 280 along the slit.}
\label{instr:iquvactivequiet}
\end{figure}

After running \textsf{sp$\_$prep.pro}, the resulting Level 1 data are confirmed to be correctly calibrated by examination of the spectra resulting from the routine. Figure~\ref{instr:iquvactivequiet} shows example spectra for quiet and AR slit positions. The data used is from the first SOT-SP scan of an event studied in Chapter~\ref{section:paper1}. In the AR Stokes \emph{I} spectrum (panel 5 of the Figure), as the sunspot is crossed there is a black area (low signal) which signifies the umbra with strong magnetic field, and line splitting is visible in the surrounding penumbral region. The \emph{Q} and \emph{U} spectra (panels 6 and 7) show the transverse Zeeman effect, with two lobes in the wings, which are of opposite sign to the central lobe. The AR Stokes \emph{V} plot (panel 8) shows the typical signature of the longitudinal Zeeman effect, with one positive and one negative lobe for each spectral line. 

The corresponding 1D Stokes spectral profiles are located beneath their 2D spectra in Figure~\ref{instr:iquvactivequiet}, with the black horizontal line through the top plots showing the slice at pixel 280 on the y-axis where the lower plots were taken. The profiles further confirm the typical characteristics of each region, with a decreased intensity and less uniform Stokes \emph{I} profile (panel 5) for the AR compared to that of the quiet region (panel 1), simply noise for the \emph{Q}, \emph{U}, and {V} plots of the quiet region (panels 2, 3, and 4), and typical strong field profiles for the AR \emph{Q}, \emph{U}, and \emph{V} profiles (panels 6, 7, and 8).

%%%%%%%%%%%%%%%%%%%%%%%%%%%%%%%%%%%%%%%%%%%%%%%%%%%%%%%%%%%%%%%%%%%%%%%%%%%%%%%%%%%%%%%%%%%%%%%%%%%%%%%%%%%%%%%%%%%%%%%%%%%%%%%%%%%%%%%%%%%%%%%%%%%%%%%%%%%%%%%%%%%%%%%%%%%%%%%%%%%%%%%%%%%%%%%%%%%%

\subsubsection{Broadband Filter Instrument}
\label{instr:sot_bfi}
Broadband filtergrams are the only type of observable made by the BFI, recording diffraction-limited images over a range of wavelengths from 3883.5~\AA\ to 6684~\AA\ . The BFI consists of six interference filters mounted in a user-controlled filterwheel, producing photometric images in six different bands (CN band, Ca \textsc{ii} H line, G band, and three continuum bands). It has the highest spatial resolution available from SOT, with 0.0541$"$ pixel sampling. A cadence between 8~s and 64~s is typical (depending on data size), covering a FOV of $218 \times 109"$. No onboard processing is performed in the FPP for the BFI, and data compression is the same as for the SP. Filtergram data may be compressed to less than 3 bits/pixel by the JPEG algorithm. Note that the BFI shares a $4096\times2048$~pixel CCD with the NFI. The raw data are calibrated using the \textsf{fg$\_$prep.pro} routine, similar to the routine used for SP. The routine removes cosmic rays (optional) and bad pixels, as well as flat-fielding and subtracting dark current and DC offset.

\begin{figure}[!t]
\centerline{\includegraphics[width=0.85\textwidth]{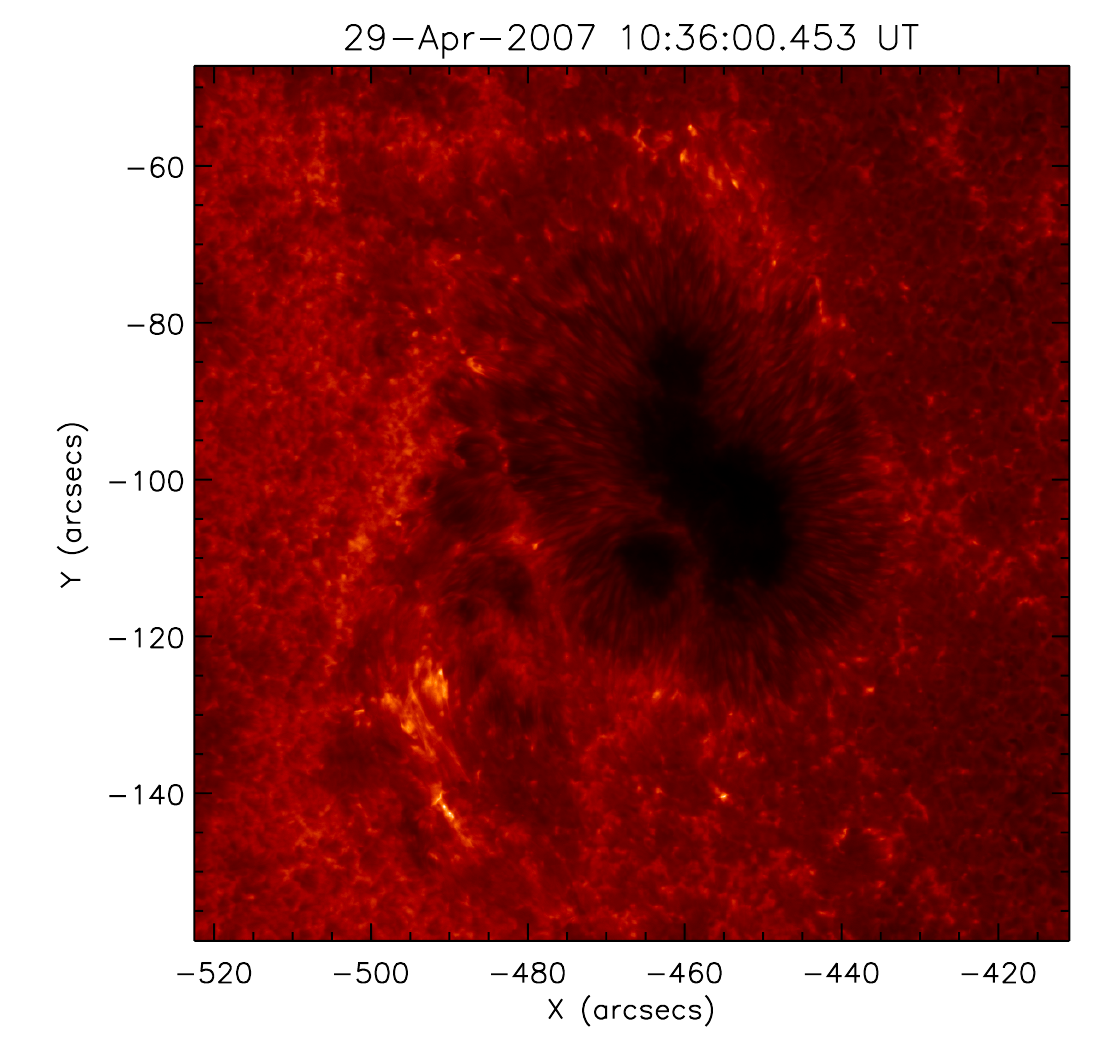}}
\caption[Ca \textsc{ii} H image at 10:36 UT on 2007 April 29, obtained from \emph{Hinode}/SOT-BFI (after data prepping using \textsf{fg$\_$prep.pro}).]{Ca \textsc{ii} H image at 10:36 UT on 2007 April 29, obtained from \emph{Hinode}/SOT-BFI (after data prepping using \textsf{fg$\_$prep.pro}). The greatest amount of brightening can be seen to the solar south east of the main sunspot umbra, corresponding to the chromospheric flare brightening of a GOES B1.0 event.}
\label{instr:caiih}
\end{figure}

The wavelength of interest for this thesis is Ca \textsc{ii} H at 3968.5~\AA, which allows study of chromospheric structure. Brightness indicates the strength of heating in the chromosphere, which coincides with magnetic field concentration in the photosphere. Information on flare brightening in Chapters~\ref{chapter:paper1},~\ref{chapter:NL}, and~\ref{chapter:3D} is obtained from the Ca \textsc{ii} H filter on SOT/BFI. Figure~\ref{instr:caiih} shows an example of a typical Level 1 image, taken from the time of flaring for the B-class flare event studied in Section~\ref{section:paper1}.

In terms of using both SP and BFI data together for analysis of an event, it is worth noting the work of \citet{centeno09} on the alignment between the two instruments. In terms of slit orientation, \citeauthor{centeno09} found that the SP spectrograph slit is not oriented vertically with respect to a column of pixels in a G band BFI image. Note that the G band is centered at 4305~\AA, with a filter width of 8.3~\AA. SP images have to be rotated by $0.26^\circ$ counter-clockwise with respect to the BFI images in order to correct for the inclination of the slit. Vertical drift must also be corrected for, as the SP slit shifts towards the south with respect to the BFI image by 0.0023~pixels for each scanning step, producing a skew effect on the image. \citeauthor{centeno09} also gives details of correcting pixel scale along the slit and scanning directions. These alignment issues are accounted for when overlaying SP and BFI images in Chapters~\ref{chapter:paper1} and \ref{chapter:NL}.

%%%%%%%%%%%%%%%%%%%%%%%%%%%%%%%%%%%%%%%%%%%%%%%%%%%%%%%%%%%%%%%%%%%%%%%%%%%%%%%%%%%%%%%%%%%%%%%%%%%%%%%%%%%%%%%%%%%%%%%%%%%%%%%%%%%%%%%%%%%%%%%%%%%%%%%%%%%%%%%%%%%%%%%%%%%%%%%%%%%%%%%%%%%%%%%%%%%%

\section{Data Analysis Techniques} 
\label{instr:data}

\subsection{Radiative Transfer of Stokes Profiles}

The theory of radiative transfer described in Chapter~\ref{chapter:theory} can be applied to fit a model solar atmopshere to the observed Stokes profiles. The measured Stokes \emph{I}, \emph{Q}, \emph{U}, and \emph{V} profiles obtained with the spectropolarimeter described above can be compared with modelled profiles in order to determine the magnetic field vector at the source of the spectral line. The RTE defined in Equation~\ref{rad_trans:RTE2} can be written in a matrix form in terms of the Stokes parameters and the optical depth,
\begin{equation}
\label{rad_trans:RTE3_tau}
\frac{d\mathbf{I}}{d\tau}~=~\mathbf{K}(\mathbf{I}-\mathbf{S})~~~. 
\end{equation}
where \textbf{I} = ($I$, $Q$, $U$, $V$) is the Stokes vector, \textbf{K} is a propagation matrix (constant, described below) and \textbf{S} = ($S_{\nu}$, 0, 0, 0) is the source function vector. The source function vector is only acting on the Stokes $I$ component here. If LTE is assumed, the source function vector can be written as \textbf{S} = ($B_{\nu}$($T$), 0, 0, 0) from Equation~\ref{rad_trans:LTE}. 

\begin{figure}[!t]
\centerline{\includegraphics[scale=0.75]{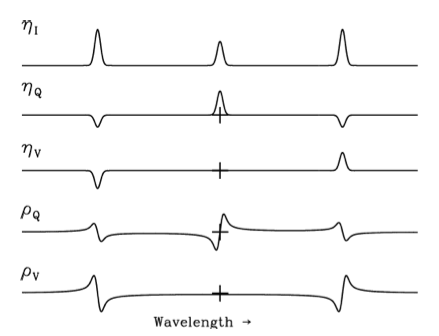}}
\caption{Coefficients of the propagation matrix for the normal Zeeman effect (\emph{courtesy of J. Sanchez Almeida}).}
\label{rad_trans:matrix}
\end{figure}

It is often useful to expand the matrix form of Equation~\ref{rad_trans:RTE3_tau} in order to show that the Stokes parameters obey the RTE, such that \citep{deltorobook},
\begin{equation}
\label{rad_trans:RTEmatrix}
\frac{d}{d\tau}
\left( 
\begin{array}{c}
I \\ 
Q \\ 
U \\ 
V
\end{array}\right)~=~
\left(\begin{array}{cccc}
\eta_{I} & \eta_{Q} & \eta_{U} & \eta_{V}\\ 
\eta_{Q} & \eta_{I} & \rho_{V} & -\rho_{U}\\ 
\eta_{U} & -\rho_{V} & \eta_{I} & \rho_{Q}\\ 
\eta_{V} & \rho_{U} & -\rho_{Q} & \eta_{I}
\end{array}\right)
\left(\begin{array}{c}
I-S_{\nu}\\ 
Q\\ 
U\\ 
V
\end{array}
\right)~~~.
\end{equation}
This equation shows the RTE for Stokes parameters, where the propagation matrix (\textbf{K}) elements are made up of the various absorption and disperson profiles characteristic of the medium and the geometry relevant to the problem. The diagonal $\eta$ elements describe absorption, where energy from all the polarisation states is withdrawn by the medium, hence the Stokes parameters evolve the same way (i.e., all reduced by a factor $\eta_I$). The remaining $\eta$ elements describe dichroism, where some of the polarised components of the beam are extinguished more than others (i.e., the contribution of $I-S_\nu$ to the resultant $I$, $Q$, $U$, $V$ components varies by the factors $\eta_I$, $\eta_Q$, $\eta_U$, and $\eta_V$, respectively). The $\rho$ elements describe dispersion, where phase shifts that take place during propagation switch between states of linear polarisation (i.e., $Q \rightarrow U$ and vice-versa) and switch between states of linear and circular polarisation (i.e., $V \rightarrow Q$ or $U$ and vice-versa). Figure~\ref{rad_trans:matrix} illustrates the propagation matrix components for a Zeeman triplet. The `+' in the Figure indicates the rest wavelength and location of the unshifted $\pi$ component, with outer lobes indicating shifted components due to Zeeman splitting. For example, $\sigma_r$ and $\sigma_b$ states are observed in the $\eta_V$ coefficient, and $\sigma_r$, $\sigma_b$, and $\pi$ states are observed in the $\eta_Q$ coefficient (the $\eta_U$ coefficient would be reversed).

%%%%%%%%%%%%%%%%%%%%%%%%%%%%%%%%%%%%%%%%%%%%%%%%%%%%%%%%%%%%%%%%%%%%%%%%%%%%%%%%%%%%%%%%%%%%%%%%%%%%%%%%%%%%%%%%%%%%%%%%%%%%%%%%%%%%%%%%%%%%%%%%%%%

\subsubsection{Milne-Eddington Approximation}
\label{rad_trans:m-e}
Proper interpretations of the Stokes vector requires a good knowledge of the atmosphere. The problem is often simplified by using a model atmosphere: a set of input parameters which describes the solar medium from which the Stokes spectrum is coming \citep{deltoro10}. It is constrained by the expected physical scenario, i.e., the set of simplifying assumptions concerning the atmosphere in which the line is formed. The measurements of the spectrum and polarisation states of the light must be transformed into magnetic field information using an inversion code. For the inversion technique described in Section~\ref{rad_trans:helixmethod}, a Milne-Eddington atmosphere is used. The physical assumptions in a Milne-Eddington atmosphere are as follows \citep{Landi92}:
\noindent
\begin{itemize}
\item the magnetic field vector is constant with $\tau$;
\item all the parameters specifying the absorption coefficient profiles (e.g., Doppler broadening, damping constant) are constant with $\tau$; 
\item there are no velocity field gradients (no macroturbulence); 
\item the ratio between the line and continuum absorption coefficients, $\eta_{o} = \kappa_{l}/\kappa_{c}$, is also constant with $\tau$; 
\item and $S_{\nu}$ varies linearly with $\tau$, $S(\tau) = S_{\mathrm{o}}+ S_{1}\tau$.
\end{itemize}

The inversion technique uses the optical depth form of the RTE as per Equation~\ref{rad_trans:RTE3_tau}, and in a Milne-Eddington atmosphere the formal solution becomes $\mathbf{I}=\int_{0}^{\infty }e^{-\mathbf{K}\tau}\mathbf{K}(\mathbf{S}_0+\mathbf{S}_1\tau)\mathrm{d}\tau$, which can be integrated analytically by parts to yield $\mathbf{I}=\mathbf{S}_0+\mathbf{K}^{-1}\mathbf{S}_1$. The explicit expression of all four Stokes profiles is then,
\begin{eqnarray}
I & = & S_{0}+\Delta ^{-1}\eta_{I}(\eta_{I}^{2}+\rho_{Q}^{2}+\rho_{U}^2+\rho_{V}^{2})S_1~, \\
Q & = & -\Delta ^{-1}[\eta_{I}^2\eta_Q+\eta_I(\eta_V\rho_U-\eta_U\rho_V)+\rho_U(\eta_Q\rho_Q+\eta_U\rho_U+\eta_V\rho_V)]S_1~, \\
U & = & -\Delta ^{-1}[\eta_{I}^2\eta_U+\eta_I(\eta_Q\rho_V-\eta_V\rho_Q)+\rho_U(\eta_Q\rho_Q+\eta_U\rho_U+\eta_V\rho_V)]S_1~, \\
V & = & -\Delta ^{-1}[\eta_{I}^2\eta_V+\eta_I(\eta_U\rho_Q-\eta_Q\rho_U)+\rho_V(\eta_Q\rho_Q+\eta_U\rho_U+\eta_V\rho_V)]S_1~,
\end{eqnarray}
where the source function depends linearly on optical depth as per the assumption above, and $\Delta$ is the determinant of the propagation matrix, $\mathbf{K}$,
\begin{equation}
\Delta~=~\eta_I^2(\eta_I^2-\eta_Q^2-\eta_U^2-\eta_V^2+\rho_Q^2+\rho_U^2+\rho_V^2)-(\eta_Q\rho_Q+\eta_U\rho_U+\eta_V\rho_V)^2~.
\end{equation}

The parameters $\eta_{I,Q,U,V}$ and $\rho_{I,Q,U,V}$ of the propagation matrix above depend on a number of parameters in the Milne-Eddington assumption, namely, $\Phi$ and $\Psi$, which are the absorption and dispersion profiles respectively, the inclination angle, $\gamma$, and the azimuthal angle, $\phi$, such that,
\begin{eqnarray}
\eta_{I} & = & 1+\frac{\eta_0}{2}\left \{ \Phi_{\mathrm{p}}\mathrm{sin}^{2}\gamma+\frac{1}{2}[\Phi_\mathrm{b}+\Phi_\mathrm{r}](1+\mathrm{cos}^2\gamma)\right \}~, \\
\eta_{Q} & = & \frac{\eta_0}{2} \left\{ \Phi_{\mathrm{p}}-\frac{1}{2}[\Phi_{\mathrm{b}} + \Phi_\mathrm{r}]\right \} \mathrm{sin}^2\gamma \mathrm{cos}2\phi~, \\
\eta_{U} & = & \frac{\eta_0}{2} \left\{ \Phi_{\mathrm{p}}-\frac{1}{2}[\Phi_{\mathrm{b}} + \Phi_\mathrm{r}]\right \} \mathrm{sin}^2\gamma \mathrm{sin}2\phi~, \\ 
\eta_{V} & = & \frac{\eta_0}{2}[\Phi_{\mathrm{r}}-\Phi_{\mathrm{b}}]\mathrm{cos}\gamma~, \\
\rho_Q & = & \frac{\eta_0}{2}\left \{ \Psi_\mathrm{p} - \frac{1}{2}\left [ \psi_\mathrm{b} + \psi_\mathrm{r} \right ]\right \}\mathrm{sin}^2\gamma \mathrm{cos}2\phi~, \\
\rho_U & = & \frac{\eta_0}{2}\left \{ \Psi_\mathrm{p} - \frac{1}{2}\left [ \psi_\mathrm{b} + \psi_\mathrm{r} \right ]\right \}\mathrm{sin}^2\gamma \mathrm{sin}2\phi~, \\
\rho_V & = & \frac{\eta_0}{2}\left[ \Psi_\mathrm{r}-\psi_\mathrm{b}\right]\mathrm{cos}\gamma~,
\end{eqnarray}
where subscripts `$\mathrm{p}$', `$\mathrm{r}$' and `$\mathrm{b}$' denote the `principal', `red' (left circ.) and `blue' (right circ.) components, respectively. Also note that the absorption and dispersion profiles are given by Voigt\footnote[4]{Voigt profiles result from the convolution of two broadening mechanisms, one of which alone would produce a Gaussian profile, and the other would produce a Lorentzian profile.}~and Faraday-Voigt functions respectively. These equations are collectively known as the Unno-Rachowsky solutions for \emph{I}, \emph{Q}, \emph{U}, and \emph{V} (see \citealp{Unno56}; \citealp{Rachkowsky67}), giving Stokes profiles as a function of wavelength, $\lambda$, and a number of magnetic parameters (which vary depending on the atmospheric inversion code used). The solution to the RTE in a Milne-Eddington atmosphere is used by inversion techniques to derive the magnetic field vector from observed Stokes profiles through a best fit procedure. 

%%%%%%%%%%%%%%%%%%%%%%%%%%%%%%%%%%%%%%%%%%%%%%%%%%%%%%%%%%%%%%%%%%%%%%%%%%%%%%%%%%%%%%%%%%%%%%%%

\subsection{\helix\ Inversion Code}
\label{rad_trans:helixmethod}
\label{32}

In this thesis, Stokes profiles obtained from spectropolarimetric observations onboard the \emph{Hinode} spacecraft are used as inputs to an atmospheric inversion code. Inversion techniques deduce physical properties of a magnetic atmosphere upon the interpretation of the polarisation it produces. A Stokes inversion finds a magnetic field vector solution and other associated properties of the magnetised atmosphere from information contained in the Stokes profiles. It computes synthetic Stokes profiles and tries to find a match between these and the measured profiles. This is achieved by varying the parameters used in the synthetic profile generation, with the number of iterations set by the user.

The inversion code used for this report is the He-Line Information Extractor (\helix)\footnote[5]{The `He' in \helix refers to the helium triplet at 10830\AA\ used to fit chromospheric data, and is not used in this thesis}, which is a flexible inversion code for the Radiative Transfer Equation, created by \citet{Lagg04} at the Max Planck Institute for Solar System Research. The Fortran 90 version of the code was used, with Message Passing Interface (MPI) supported to make use of multiple CPU cores. \helix\ fits the observed Stokes profiles with synthetic ones obtained from an analytic solution of the Unno-Rachkovsky equations in a Milne-Eddington atmosphere. These synthetic profiles are functions of a number of magnetic parameters detailed in Table~\ref{rad_trans:table:inputparams}.

Atmospheric parameters are obtained using a genetic algorithm-based general purpose optimisation subroutine called \textsf{PIKAIA}, developed by \citet{Charbonneau95}. A genetic algorithm is a search technique by iterating the generation, selection and mutation of solutions. It is a subclass of Evolutionary Computing and is based on Darwin's Theory of Evolution \citep{darwin1859}. It uses, in a computational setting, the biological notion of evolution by means of Ônatural selectionÕ. As random mutations occur, any beneficial mutations (evaluated by goodness of fit) are preserved to aid survival and hence passed onto the next generation. Over time the beneficial mutations accumulate and the result is an entirely different organism. \textsf{PIKAIA} is very reliable, reaching a global minimum of the merit function with higher reliability than classical Levenberg-Marquardt algorithms \citep[non-linear least-square fitting;][]{levenberg44,marquardt63}, which are commonly used with other inversion techniques and are highly dependent on the initial parameter values.

Before using the inversion code, the input Stokes profiles must be normalised to background continuum intensity using an IDL routine designed specifically for \helix\ called \textsf{make$\_$ccx.pro}. The routine performs a Voigt function fit to the spectral lines that \helix\ uses, to determine the wavelength calibration. Note that for this thesis \helix\ was chosen to use the photospheric Fe \textsc{i} 6301.5~\AA\ and 6302.5~\AA\ spectral lines, which are the spectral lines the Spectropolarimeter onboard the \emph{Hinode} spacecraft uses for measurements (see Section~\ref{chapter:sot-sp}). For the calibration, the routine uses an average over `the most quiet' profiles along one slit position. It also determines the local continuum value, the average continuum value along the slit and the average continuum value for the whole image. For more details on the calibration process, see \citet{helixmanual}.

Numerous input parameters have to be modified before the inversion is run. For \emph{Hinode} data (see Section~\ref{instr:hinode}), a model atmosphere is chosen in which the source function at $\tau$$=$0 is a free parameter. Table~\ref{rad_trans:table:inputparams} lists the parameter ranges used in the inversion. A polarisation threshold is chosen below which the magnetic field would be treated as zero (minimum magnetic signal). This is calculated from Level 1 data (described in Section~\ref{instr:hinode:prep}), by finding the average values of \emph{I}, \emph{Q}, \emph{U}, and \emph{V} for a quiet region on the scan and calculating the total polaristion,
\begin{equation}
\label{rad_transfer:polarisation}
P~=~\frac{\sqrt{Q^{2}+U^{2}+V^{2}}}{I_c}~~~.
\end{equation}
A quiet-Sun value of \emph{P} is taken as a threshold in this thesis, which will be described in more detail in Section~\ref{paper1:observations}.

\begin{table}[!t]
\caption{Input parameters used for \helix\ inversion, where the source function at $\tau$$=$0 is a free parameter. Note that the unit for the damping constant is the velocity of light, $c$.}
\centering
\begin{tabular}{ l l c c c}
\vspace{0.1cm} \\
\hline\hline
Parameter & Description &  Minimum &  Maximum & Unit\\
\hline
$B$ & Magnetic field strength &  0 & 3500 & G \\
$\phi$ & Azimuthal angle of B-vector &  - 90 & 90 & $^\circ$\\
$\gamma$ & Inclination angle of B-vector  &  0 &180 & $^\circ$\\
$v_{\mathrm{los}}$ & Line of sight velocity & - 4 & 4  & $\mathrm{km~s}^{-1}$\\
$v_{\mathrm{Dopp}}$ & Doppler broadening &  0.01  & 0.05 & \AA\ \\
 &  &  0.48  & 2.38 & km s$^{-1}$ \\
$\xi_{\mathrm{0}}$ & Amplitude of prop. matrix components &  10 & 100 & \ldots \\
$S_{\mathrm{0}}$ & Source (Planck) function at $\tau$ $=$ 0 &  0 &1 & \ldots \\
$S_{\mathrm{grad}}$ & Gradient of source function &  0.5 & 2.0 & \ldots\\ 
$\Gamma$ & Damping constant & 0.01 & 0.29
& \emph{c} \\
$f$ & Filling factor for this component. & 0.01 & 0.99 & \ldots\\ 
  \hline\hline
\end{tabular}
\label{rad_trans:table:inputparams}
\end{table}

The observed Stokes profiles may be a combination of more than one magnetic field component, e.g., in areas of unresolved complex field structure such as the penumbra. There may also be light scattered from bright granules near the sunspot because the telescope point spread function is greater than the pixel sampling. The scattering adds another, non-magnetic, component to the Stokes profile. To take heed of this, only one magnetic atmosphere component is used, with a local straylight component included. The average local straylight profile is calculated from the Stokes $I$ profiles surrounding the pixel for which the inversion is to be carried out, within a specified radius of 6 pixels. The averaging is performed with a Gaussian weighting centered on the pixel to be inverted, with a full-width-half-maximum of 4 pixels. The parameters are defined in all of the scans as in Table~\ref{rad_trans:table:inputparams} and the inversion run for 500 iterations, with one repetition per pixel and equal weighting across all of the \emph{I}, \emph{Q}, \emph{U}, and \emph{V} profiles.

Once the inversion has completed, a \textsf{.sav} file of the resulting magnetic parameters is created and the maps of these parameters can be displayed using a purposely made IDL routine for \helix\ called \textsf{xdisplay.pro}. When inverting a whole scan of an observation, some pixels are sometimes not converted to the optimal parameters. Speckles are small-scale non-smooth pixel-to-pixel variations. A routine called \textsf{despeckle.pro} is used to check for speckles in the parameter specified, allowing for a defined threshold deviation of surrounding pixels. If the deviation is larger than this threshold, then the pixel is recalculated using the original input-file parameters. If the recalculation gave a better fit, the original result is replaced. 

\begin{figure}
\centerline{\includegraphics[width=\textwidth]{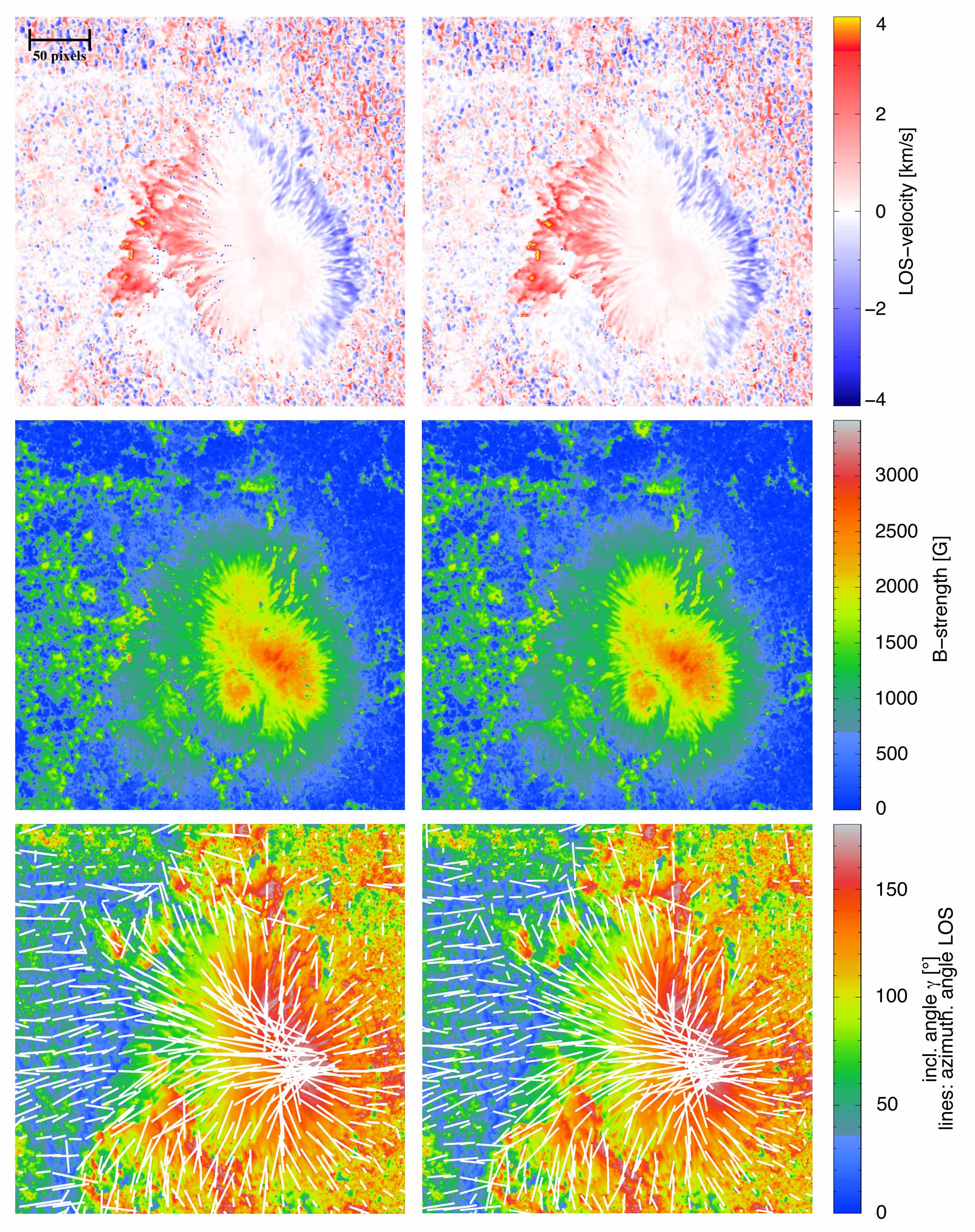}}
\caption[Examples of vector magnetic field components obtained from the \helix code before (left column) and after (right column) despeckling.]{Examples of vector magnetic field components obtained from the \helix code before (left column) and after (right column) despeckling. The maps are of a sunspot umbra on 2007 April 29. Upper to lower rows show line of sight velocity [kms$^{-1}$], magnetic field strength [G], and inclination angle [$^\circ$] with LOS azimuthal angle overplotted (white lines).}
\label{rad_trans:despeckle}
\end{figure}

Figure~\ref{rad_trans:despeckle} shows an example of the improvement the routine makes to various vector magnetic field parameters. Note that the maps shown are of a region analysed in Chapters~\ref{chapter:paper1} and \ref{chapter:3D}. The speckling is most obvious in the LOS velocity map (upper left panel). The variation across the sunspot in the LOS velocity maps from red (positive velocity, away from observer) to blue (negative velocity, towards observer) also demonstrates the Evershed flow. The flow speed varies from around $\pm~1$~km~s$^{-1}$ at the border between the umbra and penumbra, to around $\pm~3$~km~s$^{-1}$ in the middle of the penumbra, falling off to zero at the outer edge of the penumbra. Granulation is apparent in the quiet Sun, indicated by blue (upflowing) granules and red (downflowing) intergranular lanes, as expected for convective cells. The magnetic field strength map (middle row) shows typically stronger field in the umbra (red), decreasing radially with distance from this point. The inclination angle map also shows a typical sunspot field configuration, with vertical field in the umbral region, becoming more horizontal in the penumbra. The lines indicate the LOS azimuthal angle, however no directional arrows are included due to an inherent ambiguity in the azimuthal angle. This ambiguity will be discussed further in Section~\ref{rad_trans:ambiguityresolution}.

\begin{sidewaysfigure}[!]
\centerline{\includegraphics[width=\textwidth]{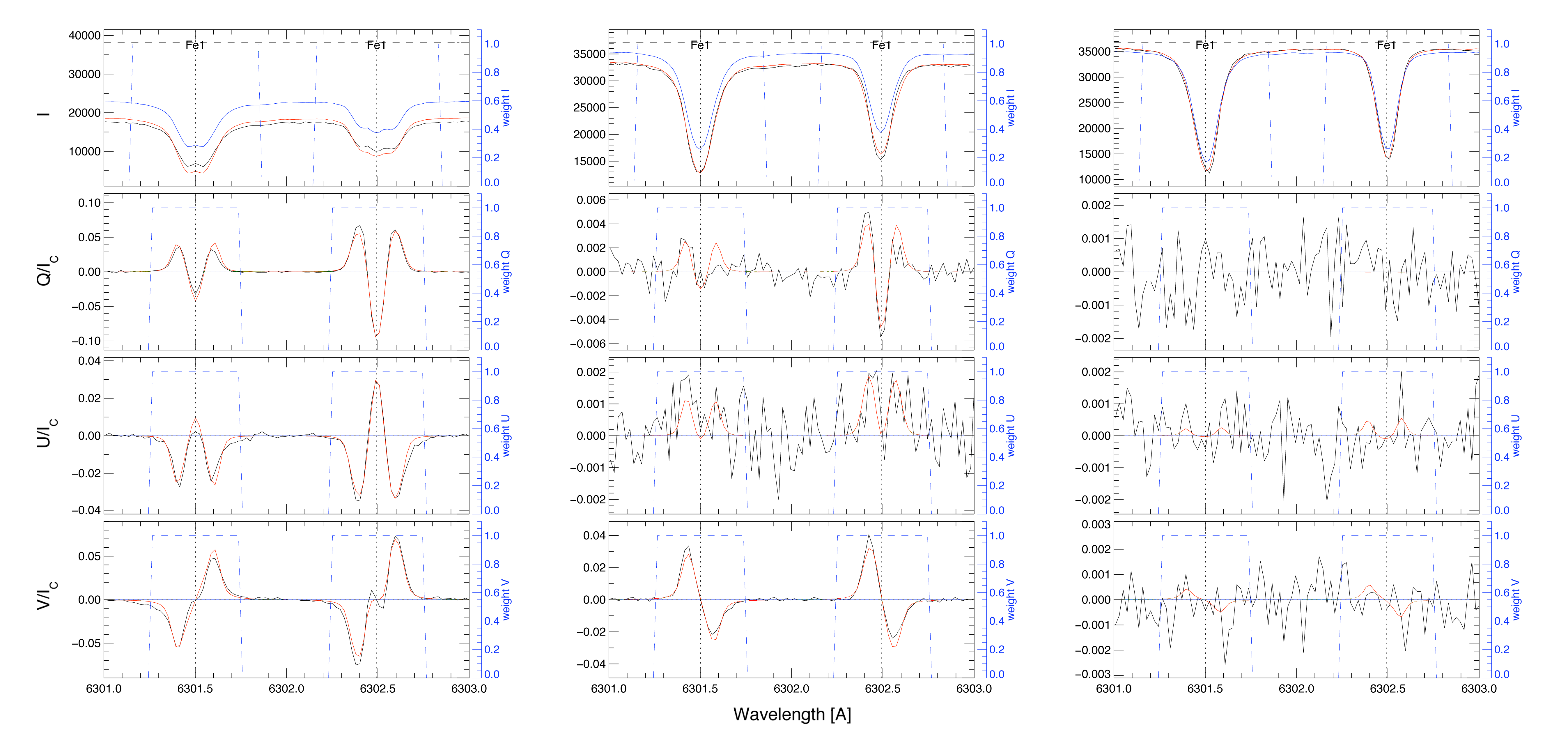}}
\caption[\emph{Left}: $I$, $Q$, $U$, and $V$ Stokes profiles for a pixel location in the umbra of the sunspot. \emph{Middle}: Stokes profiles for a weaker field pixel location in the plage region of the map. \emph{Right}: Stokes profiles at a pixel location in the quiet region of the map.]{\emph{Left}: $I$, $Q$, $U$, and $V$ Stokes profiles for a pixel location in the umbra of the sunspot, with a field strength of 1997~G. \emph{Middle}: Stokes profiles for a weaker field pixel location in the plage region of the map, with a field strength of 884~G. \emph{Right}: Stokes profiles at a pixel location in the quiet region of the map, with field strength of 146~G. In all panels, black depicts the observations, red the fitted model, and blue the unpolarised non-magnetic straylight component.}
\label{rad_trans:profile}
\end{sidewaysfigure}

Stokes profiles can be viewed for each pixel of the $N_x \times N_y$~pixel$^2$ (where $N$ is the number of pixels in the x or y direction) scans resulting from the code. Figure~\ref{rad_trans:profile} shows an example of the profiles that are obtained. Note that the profiles shown are from the same sunspot shown in Figure~\ref{rad_trans:despeckle}. Comparing the red fitted lines to the observations (black lines), agreement was reached to a good degree in the umbral region of the scan, shown in the first panel. Note that in general, an observed Stokes spectrum can be expressed as,
\begin{equation}
 I_{\mathrm{obs}}~=~f I_{\mathrm{mag}} + (1 - f ) I_{\mathrm{non-mag}}~~,
 \end{equation}
where $I_{\mathrm{obs}}$ stands for the observed intensity, $I_{\mathrm{mag}}$ the magnetic component, $I_{\mathrm{non-mag}}$ the non-magnetic straylight contribution, and $f$ the magnetic filling factor (with a value between 0 and 1). For this inversion code however a linear equation does not simply result, as the intensity at any point will become a convolution. In Figure~\ref{rad_trans:profile}, the umbral intensity in the first panel is lower than that of the quiet region in the third panel, with strong signals for $Q/I_{\mathrm{c}}$, $U/I_{\mathrm{c}}$ and $V/I_{\mathrm{c}}$ profiles in both Fe lines. The quiet-Sun region in the third panel shows a clear Stokes $I/I_{\mathrm{c}}$ profile, and noise in the $Q/I_{\mathrm{c}}$, $U/I_{\mathrm{c}}$ and $V/I_{\mathrm{c}}$ profiles. However, the red lines indicate an attempt to fit the noise, albeit of a small amplitude. This is to be expected from a region with a small magnetic contribution. The second panel shows little in the $Q/I_{\mathrm{c}}$ and $U/I_{\mathrm{c}}$ plage profiles, but more in $V/I_{\mathrm{c}}$, indicating that the field is pointing primarily towards the observer.

%%%%%%%%%%%%%%%%%%%%%%%%%%%%%%%%%%%%%%%%%%%%%%%%%%%%%%%%%%%%%%%%%%%%%%%%%%%%%%%%%%%%%%%%%%%%%%%%%%

\subsection{Azimuthal 180$^{\circ}$ Ambiguity}
\label{rad_trans:ambiguityresolution}

\begin{figure}
\centerline{\includegraphics[scale=0.65]{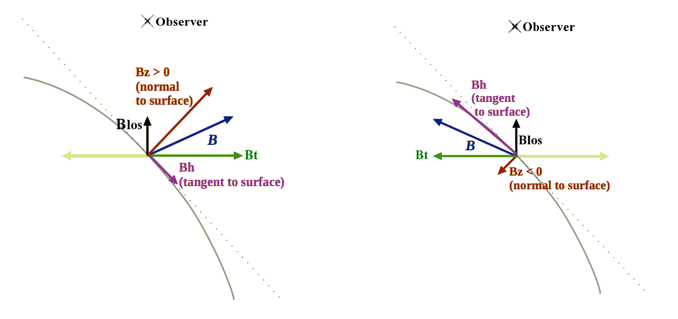}}
\caption{The direction of the transverse component of \emph{B}, $B_{t}$, is ambiguous and the direction choice influences the direction of the $|B|$ vector as well as radial component $B_{z}$ (\emph{modified image originally by K.D. Leka}).}
\label{azimuthalangles}
\end{figure}

The \helix\ inversion code gives magnetic field parameters, but the analysis is not complete until resolution of the $180^{\circ}$ ambiguity in the resulting azimuthal angle is obtained. Note that for an inversion procedure, the \emph{I}, \emph{Q}, \emph{U}, and \emph{V} spectra give the `observer's frame' magnetic field vector magnitude, $|B|$, azimuthal angle, $\phi$, and inclination angle, $\gamma$, but the LOS parallel field and the LOS transverse field can also be specified by,
\begin{eqnarray}
B_{\parallel} & = & \left | B \right |f\mathrm{cos}(\gamma)~~~, \label{b_long} \\
B_{\perp} & = & \left | B \right |f\mathrm{sin}(\gamma)~~~. \label{b_trans} 
\end{eqnarray}
The inclination angle relative to the LOS defines a cone of possibility, where the azimuthal angle defines a plane in which the vector lies, such that there are two points on the $\gamma$ cone which satisfy the same $\phi$. Thus there is a $180^{\circ}$ difference between two equally likely values of the direction of the magnetic field transverse to the LOS. See Figure~\ref{azimuthalangles} for illustration of this. 

The \helix\ \textsf{xdisplay} widget application has a built-in ambiguity resolution. Locations on the map must be manually clicked where $0^{\circ}$ or $180^{\circ}$ (e.g., umbra and opposite polarity location), and perhaps where horizontal (e.g., penumbrae). A routine called \textsf{los2solar.pro} then searches for a smooth solution in the solar $B_{\mathrm{x}}$, $B_{\mathrm{y}}$ and $B_{\mathrm{z}}$ maps. It is a very basic procedure and user-dependant, hence only suitable for hints as to what the real solution may be. Multiple ambiguity codes were tested on the inversion results instead, namely \textsf{NPFC2} created by Manolis Georgoulis \citep{Georgoulis05} and \textsf{AMBIG} by K.D. Leka et. al \citep{Leka09}. Both scored highly in the \citet{Metcalfetal06} and \citet{Lekaetal09} reviews on solar ambiguity angle resolution procedures. For the purposes of this thesis research (i.e, transformation of the disambiguation to solar normal reference frame), it was decided that \textsf{AMBIG} was the best choice. \textsf{AMBIG} gave a better azimuthal solution, with \textsf{NPFC2} being quite `patchy', and \textsf{AMBIG} also gave a more flexible output format for further analysis.

\subsubsection{Disambiguation Solution}
\textsf{AMBIG} is an updated version of the Minimum Energy Algorithm by \citet{Metcalf94}. The routine simultaneously minimises the magnetic field divergence, $\nabla \cdot \mathbf{B}$, and electric current density, $\mathbf{J}_\mathrm{z}$. The divergence is used to give a physically meaningful solution, while the current acts as a smoothing constraint. A potential field extrapolation (Section~\ref{mhd:pot_lff}) is used to find vertical derivatives of the magnetic field, in order to calculate a value for the divergence. The ambiguity is resolved by minimising the absolute value of the resulting divergence over the field of view (FOV). 

The calculations of $\mathbf{J}_\mathrm{z}$ and $\nabla \cdot \mathbf{B}$ are not local, so a simulated annealing algorithm is used to globally minimise the sum of the electric current density and magnetic field divergence. The idea of simulated annealing comes from metallurgy, where increasing of the size of crystals and reduction of defects is obtained by heating and controlled cooling of a material. Hence, the system is allowed to `cool' slowly into a state of minimum `energy', where the `temperature' in this case is a parameter that controls how fast the algorithm settles into the minimum state, and `energy' refers to the quantity being minimised, i.e., $ \sum_{pixels} (|\mathbf{J}_\mathrm{z}| + |\nabla \cdot \mathbf{B}|)$. At each `temperature' in the routine, many random orientations of the field are attempted, each adding 180$^{\circ}$ to a randomly selected pixel's azimuth. Any orientation that reduces the `energy' is accepted, and one that increases it isn't necessarily rejected, but given a probability of acceptance based on a Boltzmann probability distribution. More details about the process can be found in \citet{Leka09}. As the `temperature' falls slowly, the weaker field regions are in turn resolved. At a `temperature' which is low enough that the accepted re-orientations are all in regions where the magnetic field is noise, the important regions of the magnetogram have been resolved. 

After the ambiguity has been resolved based on the minimum-energy state found, pixels below an appropriate noise threshold in the transverse field strength are revisited using an iterative acute angle to nearest neighbors method \citep{Canfield93}. This is done because the minimum energy state is not considered the correct ambiguity solution in the presence of noise \citep{Lekaetal09}. An appropriate threshold generally ranges between 200 $\leq B_{\perp} \leq$ 400~G for lower resolution \emph{SOHO}/MDI data. For the higher-resolution \emph{Hinode} data used in this thesis, a value of 150~G was selected. Note that acute angle methods compare the observed field to an extrapolated model field. The 180$^\circ$ azimuthal ambiguity is resolved by requiring that some component (i.e., image-plane transverse or heliographic-plane horizontal; see Section~\ref{rad_trans:coords}) of the observed field and the extrapolated field make an acute angle, i.e., $-90^\circ \le \Delta\theta \le 90^\circ$, where $\Delta\theta$ is the angle between the observed and extrapolated components.

Note that the resulting \helix\ magnetic field vector magnitude, inclination angle, and azimuthal angle are used as inputs to the \textsf{AMBIG} code. Alternatively, \helix\ magnetic field vector magnitude, inclination angle, and magnetic filling factor can be used to calculate the longitudinal and transverse field, as per Equations~\ref{b_long} and \ref{b_trans}, for input. Table~\ref{rad_trans:AMBIGinput} outlines the input values needed for the code.

\newpage

\begin{table}[!t]
\caption{Description of input parameters used in the \textsf{AMBIG} $180^\circ$ ambiguity resolution code.}
\centering
\begin{tabular}{ l l l l ccc}
\vspace{0.1cm} \\
\hline\hline
Parameter & Description & Unit\\
\hline
n$_\mathrm{x}$, n$_\mathrm{y}$ &  Dimension of the input arrays.  			& pixels \\
x$_\mathrm{pix}$, y$_\mathrm{pix}$  & Pixel dimensions [horizontal,vertical]. 	& arcseconds \\ 
b$_\mathrm{0}$ &   Latitude of centre of solar disk (solar b angle).			& radians \\
p			&  Position angle of northern extremity of solar				& radians \\
			& rotation axis (solar p angle).							& \\
lat			& Latitude of the centre of the FOV.						& radians \\
cmd			& Central meridian angle of the centre of the FOV. 			& radians \\
B$_{\mathrm{l}}$ & $\mathrm{n_x \times n_y}$ array of the LOS field	 	& Gauss \\
 				& (or of the magnitude of the field).					& \\
B$_{\mathrm{t}}$ & $\mathrm{n_x \times n_y}$ array of the transverse field component 	& Gauss \\
 				& (or of the inclination angle of the field).				& (degrees)\\
B$_{\mathrm{a}}$ & $\mathrm{n_x \times n_y}$ array of the azimuthal angle, 	& degrees \\
 				& containing the ambiguity						& \\
\hline\hline
\end{tabular}
\label{rad_trans:AMBIGinput}
\end{table}

%%%%%%%%%%%%%%%%%%%%%%%%%%%%%%%%%%%%%%%%%%%%%%%%%%%%%%%%%%%%%%%%%%%%%%%%%%%%%%%%%%%%%%%%%%%%%%%%%%%%%%%%%%%%%%%%%%%%%%%%%%%%%%%%%%%%%%%%%%%%%%%%%%%

\subsection{Coordinate Systems}
\label{rad_trans:coords}
The resulting disambiguated inversion results are in the image plane of the $Hinode$ spacecraft. For thorough analysis, the results must be converted to heliographic planar coordinates. The first step is to obtain the image plane LOS $B_{x}^\mathrm{i}$, $B_{y}^\mathrm{i}$, and $B_{z}^\mathrm{i}$ parameters, which are defined as,
\begin{eqnarray}
B_{x}^\mathrm{i} & = & B_{\perp}cos(\phi)~~~\ , \\ 
B_{y}^\mathrm{i} & = & B_{\perp}sin(\phi)~~~\ , \\
B_{z}^\mathrm{i} & = & B_{\parallel}~~~~~~~~~~~~\ ,
\end{eqnarray}
where $B_{\perp}$ and $B_{\parallel}$ are the perpendicular and parallel field components as specified in Equations~\ref{b_long} and \ref{b_trans}, and $\phi$ is the disambiguated azimuthal angle.

\begin{figure}
\centerline{\includegraphics[scale=0.65]{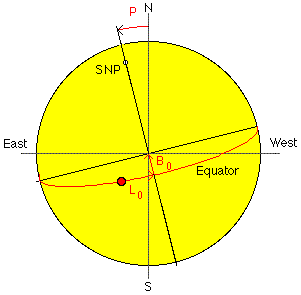}}
\caption[Illustration of the $P$, $B_0$ and $L_0$ angles, as used for conversion to heliographic planar coordinates, \emph{courtesy of J\"urgen Giesen (2012)}.]{Illustration of the $P$, $B_0$ and $L_0$ angles, as used for conversion to heliographic planar coordinates. The solar $P$ angle has a range of $\pm~26.31^\circ$, and $B_0$ a range of $\pm~7.23^\circ$ (correcting for the tilt of the ecliptic with respect to the solar equatorial plane). The value of $L_0$ is determined with reference to a system of fixed longitudes rotating on the Sun at a rate of 13.2$^\circ$/day. N and S indicate the poles of the ecliptic, and SNP the solar north pole. \emph{Figure courtesy of J\"urgen Giesen (2012).}}
\label{rad_trans:angles}
\end{figure}

These are then converted to the solar surface normal reference frame (i.e, obtaining $B_x^\mathrm{h}$, $B_y^\mathrm{h}$, and $B_z^\mathrm{h}$) by using the method outlined by \citet{gary90}. The orthogonal magnetic-field components in the observers (i.e., image plane, superscript ``i'') frame and solar surface normal (i.e., heliographic plane, superscript ``h'') frame are related by,
\begin{eqnarray}
B_x^\mathrm{h} & = & a_{11}B_{x}^\mathrm{i}+a_{12}B_{y}^\mathrm{i}+a_{13}B_{z}^\mathrm{i} \ , \label{gary1} \\
B_{y}^\mathrm{h} & = & a_{21}B_{x}^\mathrm{i}+a_{22}B_{y}^\mathrm{i}+a_{23}B_{z}^\mathrm{i} \ , \label{gary2} \\
B_{z}^\mathrm{h} & = & a_{31}B_{x}^\mathrm{i}+a_{32}B_{y}^\mathrm{i}+a_{33}B_{z}^\mathrm{i} \ , \label{gary3} 
\end{eqnarray}
where coefficients $a_{\mathrm{ij}}$ are defined in Equation 1 of \citet{gary90} as,
\begin{eqnarray}
a_{11} & = & -~sinB_{0}~sinP~sin(L-L_0)~+~cosP~cos(L-L_0)~ \ , \\
a_{12} & = & +~sinB_{0}~cosP~sin(L-L_0)~+~sinP~cos(L-L_0)\ , \\
a_{13} & = & -~cosB_{0}~sin(L-L_0)\ , \\
a_{21} & = & -~sinB~[sinB_0~sinP~cos(L-L_0)~+~cosP~sin(L-L_0)] \nonumber \\
&& -\:cosB~cosB_0~sinP \ , \\
a_{22} & = & +~sinB~[sinB_0~cosP~cos(L-L_0)~-~sinP~sin(L-L_0)] \nonumber \\
&& +\:cosB~cosB_0~cosP \ , \\
a_{23} & = & -~cosB_0~sinB~cos(L-L_0)~+~sinB_0~cosB \ , \\
a_{31} & = & +~cosB~[sinB_0~sinP~cos(L-L_0)~+~cosP~sin(L-L_0)] \nonumber \\
&& -\:sinB~[cosB_0~sinP] \ , \\
a_{32} & = & -~cosB~[sinB_0~cosP~cos(L-L_0)~-~sinP~sin(L-L_0)] \nonumber \\
&& +\:sinB~[cosB_0~cosP] \ , \\
a_{33} & = & +~cosB~cosB_0~cos(L-L_0)~+~sinB~sinB_0 \ .
\end{eqnarray}
$B$ and $L$ are arrays of the heliographic latitudes and longitudes of the image points, $B_0$ is the latitude of the central point of solar disk (angle over which the Sun is tilted towards or away from the observer), $L_0$ is the longitude of the central point of the solar disk, and $P$ is the position angle of the northern extremity of the solar axis of rotation (measured eastward from the north point of the disk). It is worth noting $B_0$ and $P$ correspond to the solar b$_0$ and p angles mentioned in Table~\ref{rad_trans:AMBIGinput}. These angles are illustrated in Figure~\ref{rad_trans:angles}.

In the image frame, $B_{z}^\mathrm{i}$ is the component along the LOS, and ($B_{x}^\mathrm{i}$, $B_{y}^\mathrm{i}$) define the plane of the image. In the heliographic frame, $B_{z}^\mathrm{h}$ is the component normal to the solar surface, and ($B_{x}^\mathrm{h}$, $B_{y}^\mathrm{h}$) lie in the plane tangential to the solar surface at the centre of the FOV. In terms of the field vector, $B_{x}^\mathrm{h} = \left | \mathbf{B} \right |\mathrm{sin}(\gamma^{\prime})\mathrm{cos}(\phi^{\prime})$, $B_{y}^\mathrm{h}=\left|\mathbf{B}\right|\mathrm{sin}(\gamma^{\prime})\mathrm{sin}(\phi^{\prime})$, and $B_{z}^\mathrm{h} = \left | \mathbf{B} \right |\mathrm{cos}(\gamma^{\prime})$. Here, $\left | \mathbf{B} \right |$ is the absolute magnetic field strength, $\gamma^{\prime}$ is the inclination angle from the solar normal direction, and $\phi^{\prime}$ is the azimuthal angle in the ($B_{x}^\mathrm{h}$, $B_{y}^\mathrm{h}$) plane measured counter-clockwise from solar west.

\begin{sidewaysfigure}
\centerline{\includegraphics[scale=1.35]{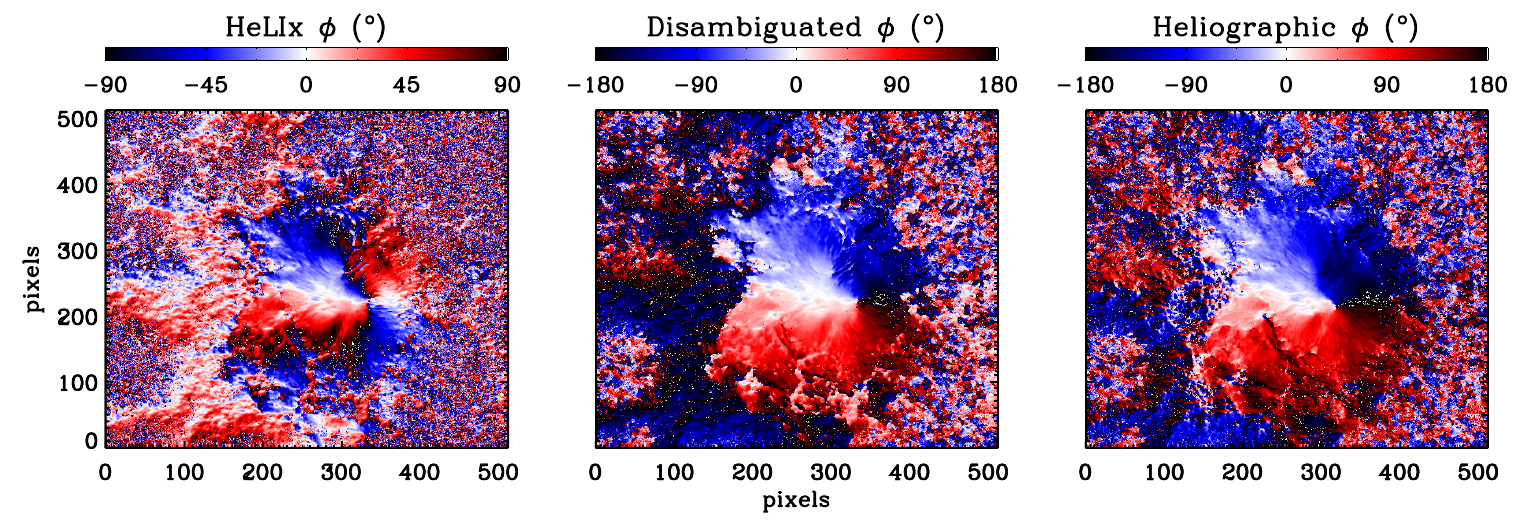}}
\caption[Various examples of an azimuthal angle scan from 2007 April 29.]{Various examples of an azimuthal angle scan from 2007 April 29. The left panel shows a resulting azimuthal angle scan from the \helix code, the middle panel the results of the \textsf{AMBIG} disambiguation procedure, and the right panel shows the results of converting to the heliographic planar coordinate system.}
\label{rad_trans:azi}
\end{sidewaysfigure}

As an example of the analysis process that has been described in this Chapter, Figure~\ref{rad_trans:azi} illustrates the various results for azimuthal angle after application of the described techniques. A typical azimuthal angle scan obtained from the \helix code is shown on the left, followed by a disambiguated version using the \textsf{AMBIG} code. Finally, on the right the azimuthal angle map is shown after conversion to the solar surface reference frame.

%%%%%%%%%%%%%%%%%%%%%%%%%%%%%%%%%%%%%%%%%%%%%%%%%%%%%%%%%%%%%%%%%%%%%%%%%%%%%%%%%%%%%%%%%%%%%%%%%%%%%%
%%%%%%%%%%%%%%%%%%%%%%%%%%%%%%%%%%%%%%%%%%%%%%%%%%%%%%%%%%%%%%%%%%%%%%%%%%%%%%%%%%%%%%%%%%%%%%%%%%%%%%

\subsection{Extrapolation Methods}

Once the photospheric vector magnetic field data have been correctly processed, they can be used as inputs to 3D magnetic field extrapolations, which were described in Section~\ref{mhd:extraps}. This Section describes the methods used to obtain the 3D coronal magnetic field in this thesis, the results of which are analysed in detail in Chapter~\ref{chapter:3D}.

\subsubsection{Potential and LFF Methods}

The method of \citet{seehafer78} is used to calculate the potential and LFF fields in this thesis, implemented in the {\tt LINFF} code developed by T. Wiegelmann (2011, private communication). To obtain the potential extrapolation using {\tt LINFF}, $\alpha$ is simply set to zero. The \citet{seehafer78} method gives the magnetic field components in terms of a Fourier series. The input FOV, covering a domain given by $0 \leq x \leq L_x$ and $0 \leq y \leq L_y$, is artificially extended to a region covering $-L_x \leq x \leq L_x$ and $-L_y \leq y \leq L_y$, by assuming $B_z(-x,y)=B_z(x,-y)=-B_z(x,y) $. The magnetic field is given by,
\begin{eqnarray}
B_x & = & \sum_{m,n=1}^{\infty}~\frac{C_{mn}}{\lambda_{mn}}~\exp(-r_{mn}z)~\cdot~\left[ \alpha~\xi_{y}^{n}~sin\xi_{x}^{m}~cos\xi_{y}^{n}~-~r_{mn}~~\xi_{x}^{m}~cos\xi_{x}^{m}~sin\xi_{y}^{n} \right] \nonumber , \\
B_y & = & \sum_{m,n=1}^{\infty}~\frac{C_{mn}}{\lambda_{mn}}~\exp(-r_{mn}z)~\cdot~\left[ \alpha~\xi_{x}^{n}~cos\xi_{x}^{m}~sin\xi_{y}^{n}~-~r_{mn}~\xi_{y}^{m}~sin\xi_{x}^{m}~cos\xi_{y}^{n} \right] \nonumber , \\
B_z & = & \sum_{m,n=1}^{\infty}~C_{mn}~\exp(-r_{mn}z)~\cdot~sin\xi_{x}^{m}~ sin\xi_{y}^{n}~ \ , 
\end{eqnarray}
where 
\begin{eqnarray}
\lambda_{mn} & = & \pi^2(\frac{m^2}{L_{x}^2}~+~\frac{n^2}{L_{y}^2}) \nonumber  \ , \\ %eigenvalues
r_{mn} & = & \sqrt{\lambda_{mn}~-~\alpha^2}  \nonumber  \ , \\
\xi_{i}^{m} & = & \frac{\pi m i}{L_i}  \nonumber  \ , \\
\xi_{i}^{n} & = & \frac{\pi n i}{L_i}  \nonumber  \ ,
\end{eqnarray}
such that \emph{i = (x, y, z)}, and the $C_{mn}$ coefficients are obtained by comparing the expression of $B_z$ for $z=0$ with a fast fourier transform of the input data. More details can be found in \citet{seehafer78}. A maximum value of $\alpha^2$ is given for $L_x$ and $L_y$ as,
\begin{equation}
\alpha^{2}_{max}~=~\pi^2 \left( \frac{1}{L_{x}^{2}}~+~\frac{1}{L_{y}^{2}} \right)
\end{equation}
which is defined because the numerical method must cut-off the Fourier series at some $m_{max}$ and $n_{max}$. The force-free parameter $\alpha$ is normalised by defining a harmonic mean, L,
\begin{equation}
\frac{1}{L^2}~=~\frac{1}{2} \left ( \frac{1}{L_{x}^{2}}~+~\frac{1}{L_{y}^{2}} \right) \,
\end{equation}
such that when $L_x=L_y$, $L=L_x=L_y$. This normalisation gives values for $\alpha$ in the range $-\sqrt{2\pi} < \alpha < \sqrt{2\pi}$. A fourth-order Runge-Kutta scheme\footnote[5]{RK4 is a method of numerically integrating ordinary differential equations by using a trial step at the midpoint of an interval, in order to cancel out lower-order error terms.} (RK4) with step-size control is used in the calculation of the magnetic field lines. The research discussed in Chapter~\ref{chapter:3D} uses both the potential and linear force-free versions of the {\tt LINFF} code to obtain information on the 3D coronal magnetic field, as well as a NLFF extrapolation procedure that will be described in the following sub-section. 

%%%%%%%%%%%%%%%%%%%%%%%%%%%%%%%%%%%%%%%%%%%%%%%%%%%%%%%%%%%%%%%%%%%%%%%%%%%%%%%%%%%%%%%%%%%%%%%%%%%%%%

\subsubsection{NLFF Method}
The chosen form of NLFF extrapolation to be used for analysis in this thesis is the weighted optimization method originally proposed by \citet{wheatland00} and implemented by \citet{wiegelmann04}. This is currently one of the most accurate NLFF procedures available, as demonstrated in the \citet{schrijver06}, \citet{metcalf08}, and \citet{derosa09} method reviews. It is worth noting that this NLFF code directly minimises the force-balance equation, which avoids the explicit computation of $\alpha$. In the optimization method, a function $M$ is defined as,
\begin{equation}
\label{mhd:l}
M~=~\int_{V}\ w(x,y,z)~B^2~(\Omega_{a}^{2}+\Omega^{2}_{b})~d^3x \,
\end{equation}
where $w(x,y,z)$ is a weighting function, $ \mathbf{\Omega}_{a} = B^{-2} \left [ (\nabla \times \textbf{B}) \right ]$, and $ \mathbf{\Omega}_{b} = B^{-2} \left [ (\nabla \cdot \textbf{B})~\textbf{B} \right ]$. An iterative scheme is used to minimise $M$ with respect to an iteration parameter $t$. The iteration equation for the magnetic field within the computational box is given by,
\begin{equation}
\label{mhd:iteration_equation}
\frac{1}{2} \frac{dM}{dt}~=~-\int_{V} \frac{\partial \textbf{B}}{\partial t} \cdot \textbf{\~F} d^3x \,
\end{equation}
where 
\begin{eqnarray}
\textbf{\~F} & = & w\textbf{F}+(\mathbf{\Omega}_{a} \times \textbf{B}) \times \nabla w~+~(\mathbf{\Omega}_{b} \cdot \textbf{B}) \nabla w \\
\textbf{F} & = & \nabla \times (\mathbf{\Omega}_{a} \times \textbf{B})~-~\mathbf{\Omega}_{a} \times (\nabla \times \textbf{B})~+~\nabla(\mathbf{\Omega}_{b} \cdot \textbf{B})~-~\mathbf{\Omega}_{b}(\nabla \cdot \textbf{B}) \nonumber \\
&& +\:(\mathbf{\Omega}_{a}^{2}~+~\mathbf{\Omega}_{b}^{2})\textbf{B} \ .
\end{eqnarray}
Assuming the magnetic field is described on the boundaries of a computational box, the field is iterated with
\begin{equation}
\label{mhd:f}
\frac{\partial \textbf{B}}{\partial t}~=~\textbf{\~F} \ .
\end{equation}
The continuous form of this equation ensures a monotonically decreasing $M$.

The method begins by computing a potential field in the computational box using the \citet{seehafer78} method. The bottom boundary is defined by the input vector magnetogram, with the lateral and top boundaries described using the potential field results. The function $M$ in Equation~\ref{mhd:l} is minimised with the help of Equation~\ref{mhd:f}, with an iteration step $dt$. The code checks if $M(t~+~dt) < M(t)$ after each time step, and the iteration step is repeated with $dt$ reduced by a factor of 2 each time until this condition is fulfilled. After each successful iteration step, $dt$ is increased by a factor of 1.01. Once $M$ becomes stationary the iterations cease, with `stationary' assumed if $(\partial M/\partial t)/M < 1.0 \times 10^{-4}$ for 100 consecutive iteration steps.

It is necessary to preprocess photospheric magnetic field data before its use as a boundary condition for 3D NLFF extrapolations. The procedure outlined by \citet{wiegelmann06} is used to deal with the inconsistency between the force-free assumption of NLFF models and the non force-free nature of the photospheric magnetic field, while also removing certain noise issues (such as uncertainties in the transverse field components). A function defined as $Y=\mu_1Y_1+\mu_2Y_2+\mu_3Y_3+\mu_4Y_4$ is minimised, where,
\begin{eqnarray}
Y_1 & = & \left [ \left ( \sum_{p} B_x B_z \right )^2+\left ( \sum_{p}B_yB_z \right )^2+\left ( \sum_{p}\zeta \right )^2 \right ] ~ \ , \\
Y_2 & = & \left [ \left ( \sum_{p} x \zeta  \right )^2+ \left ( \sum_{p} y \zeta \right )^2+ \left ( \sum_{p} yB_xB_z~-~xB_yB_z \right ) ^2 \right ] ~  \ , \\ 
Y_3 & = & \left [ \sum_{p} \left( B_x - B_{x}^{\mathrm{obs}} \right)^2+ \sum_{p} \left( B_y - B_{y}^{\mathrm{obs}} \right) ^2+ \sum_{p} \left( B_z - B_{z}^{\mathrm{obs}} \right ) ^2 \right ]  ~ \ , \\
Y_4 & = & \left[ \sum_{p} \left( \Delta B_x \right)^2+ \left( \Delta B_y \right)^2+ \left( \Delta B_z \right)^2 \right] ~ \ .
\end{eqnarray}
Note that $\zeta = B_{z}^{2}-B_{x}^{2}-B_{y}^{2}$, and $\sum_{p}$ is a summation over all grid nodes $p$ of the bottom surface grid. Also note that the $Y_n$ terms are weighted by the $\mu_n$ factors. Term $n=1$ corresponds to the force-balance condition, term $n=2$ to the torque-free condition, term $n=3$ ensures the optimised boundary condition agrees with the measured photospheric data, and term $n=4$ controls the smoothing. The $\Delta$ in the $n=4$ term designates a 2D Laplace operator that is used for smoothing. 

The force and torque parameters are defined in Section 2 of \citet{wiegelmann06} as,
\begin{eqnarray}
\epsilon_{force} & = & \frac{\left | \int_{S}B_{x}B_{z} \right |~+~\left | \int_{S}B_{y}B_{z} \right |~+~\left | \int_{S}(B_{x}^{2}~+~B_{y}^{2})~-~B_{z}^{2} \right |}{\int_{S}(B_{x}^{2}~+~B_{y}^{2}~+~B_{z}^{2})} \label{mhd:force} \\
\epsilon_{torque} & = & \frac{\left | \int_{S} x ((B_{x}^{2}+B_{y}^{2})-B_{z}^{2}) \right |+\left | \int_{S} y((B_{x}^{2}+B_{y}^{2})-B_{z}^{2} \right |+\left | \int_{S}yB_xB_z-xB_yB_z \right |} {\int_{S}\sqrt{x^2~+~y^2}(B_{x}^{2}~+~B_{y}^{2}~+~B_{z}^{2})} \label{mhd:torque} \nonumber \\ 
\end{eqnarray}
and also the flux balance is defined as,
\begin{equation}
\epsilon_{flux}~=~\frac{\int_{S}B_z}{{\int_{S}\left | B_z \right |}} ,
\label{mhd:flux}
\end{equation}
The integrals in $\epsilon_{force}$ and $\epsilon_{torque}$ correspond to the Maxwell stress tensor (see Equation~\ref{maxwell_stress}) and its first moment, respectively \citep[][see \citet{wiegelmann06} for a fuller description]{aly89}. The force and torque parameters are normalised by the magnetic pressure term of the Maxwell stress tensor \citep{tiwari12}. For perfectly force-free consistent boundary conditions these three quantities are zero, but for practical computations it is sufficient that they are $\ll$ 1. The preprocessing procedure aims to minimise $Y$ so that all $Y_n$ terms are made small simultaneously. Preprocessing results in a data set consistent with the assumption of a force-free coronal magnetic field, as close as possible to the measured data set within the noise level.

%%%%%%%

		% instrumenation

% this file is called up by thesis.tex
% content in this file will be fed into the main document

%: ----------------------- name of chapter  -------------------------
\chapter{Photospheric Magnetic Field Evolution}
\label{chapter:paper1}

\ifpdf
    \graphicspath{{4/figures/PNG/}{4/figures/PDF/}{4/figures/}}
\else
    \graphicspath{{4/figures/EPS/}{4/figures/}}
\fi

\hrule height 1mm
\vspace{0.5mm}
\hrule height 0.4mm 
\noindent 
\\ {\it As mentioned in Section~\ref{intro:complexity}, there is evidence that flaring is related to changes in the topology or complexity of an active region's magnetic field, but to date the pre-cursors to flaring are still not understood. In this chapter, research undertaken on a sunspot region over a period of $\sim12$~hours leading up to and after a B-class flare is described. The observations and data analysis techniques used are briefly discussed in Section~\ref{paper1:observations}. The main results are presented in Section~\ref{paper1:results}, in particular the changes in vertical and horizontal field in Section~\ref{paper1:vertical}, field orientation in Section~\ref{paper1:inclination}, and derived low-order 3D magnetic properties in Section~\ref{paper1:divergence}. Finally, the main conclusions are discussed in Section~\ref{paper1:conclusions}. This research described in this chapter has been published in Murray, S.~A., et al.,  \solphys, 2012, 277, 1, $45-57$.
\\}

\hrule height 1mm
\vspace{0.5mm}
\hrule height 0.4mm 

\newpage

\label{section:paper1}

\section{Introduction}
\label{paper1:introduction}
Active regions in the solar atmosphere have complex magnetic fields that emerge from subsurface layers to form loops which extend into the corona. When active regions undergo external forcing, the system may destabilise and produce a solar flare, where energy stored in sunspot magnetic fields is suddenly released as energetic particles and radiation across the entire solar spectrum (Section~\ref{intro:flares}; \citealp{rust92}; \citealp{conlon08}). As outlined in Section~\ref{intro:reconnection}, the initial impulsive phase of the flare is generally believed to be driven by magnetic reconnection, which leads to a change in the topology of the magnetic field, and energy stored in the field is released, accelerating coronal particles \citep{aschwanden05}.  The storage of magnetic energy in active regions is indicated by the degree of non-potentiality of sunspot magnetic fields \citep{regnierpriest07aa}. The processes leading up to reconnection and energy release are still not fully understood, and studying the links between solar flares and topology changes in active-region magnetic fields is an important step in understanding the pre-flare configuration and the process of energy release (\citealp{hewett08}; \citealp{conlon10}). 

As mentioned in Section~\ref{intro:flaring_process}, \citet{tanaka86} suggested that sheared fields along the NL that contain large currents are likely locations for flaring (see Figure~\ref{paper1:tanaka86}). \citet{canfield91} explored the importance of strong currents further, finding that sites of significant energetic-electron precipitation into the chromosphere are at the edges of regions of strong vertical current rather than within them. \citet{metcalf94a} and \citet{li97} subsequently found that flares do not necessarily coincide spatially with the locations of strong vertical current. 

LOS magnetic-field observations have shown that photospheric fields change rapidly during large solar flares (\citealp{sudolharvey05}; \citealp{petriesudol09}). Other studies use improved 3D extrapolation techniques to analyse the topology further, increasing our understanding of eruptions in the solar corona (\citealp{regniercanfield06}; \citealp{georgoulis07}). This will be discussed further in Chapter~\ref{chapter:3D}. The wide range of previous work has shown that observing active-region magnetic fields around the time of flaring can be very beneficial, as magnetic-field properties have been found to be viable flare-forecasting tools \citep{gallagher02}. However, the LOS magnetic field alone cannot provide complete information on the changing magnetic field.

High spatial resolution observations of the solar magnetic-field vector can now provide more in-depth information on the true topological complexities. Vector magnetic field observations are analysed in this chapter to examine how sunspot magnetic fields evolve leading up to and after flare activity. In particular, differences in the magnetic-field vector between pre- and post-flare states are examined in the vicinity of a chromospheric flare brightening. Studying the evolution of the magnetic field before the flare with these improved observations could outline some new flare precursors that may be of use in flare forecasting, perhaps in terms of how soon a flare could be expected after certain conditions are met. Any changes observed after the flare compared to the pre-flare conditions should also give insight into how a flare might occur from this kind of region, testing the validity of currently proposed changes in magnetic topology during solar flares \citep[e.g.,][]{PevtsovCanfieldZirin96}. 

%%%%%%%%%%%%%%%%%%%%%%%%%%%%%%%%%%%%%%%%%%%%%%%%%%%%%%%%%%%%%%%%%%%%%%%%%%%%%%%%%%%%%%%%%%%%%%%%%%%%%%%%%%%%%%%%%%%%%%%%%%%%%%%%%%%%%%%%%%%%%%%%%%%%%%%%%%%%%%%%%%

\section{Observations and Data Analysis}
\label{paper1:observations}

\begin{figure}[!t]
\centerline{\includegraphics[width=0.8\textwidth]{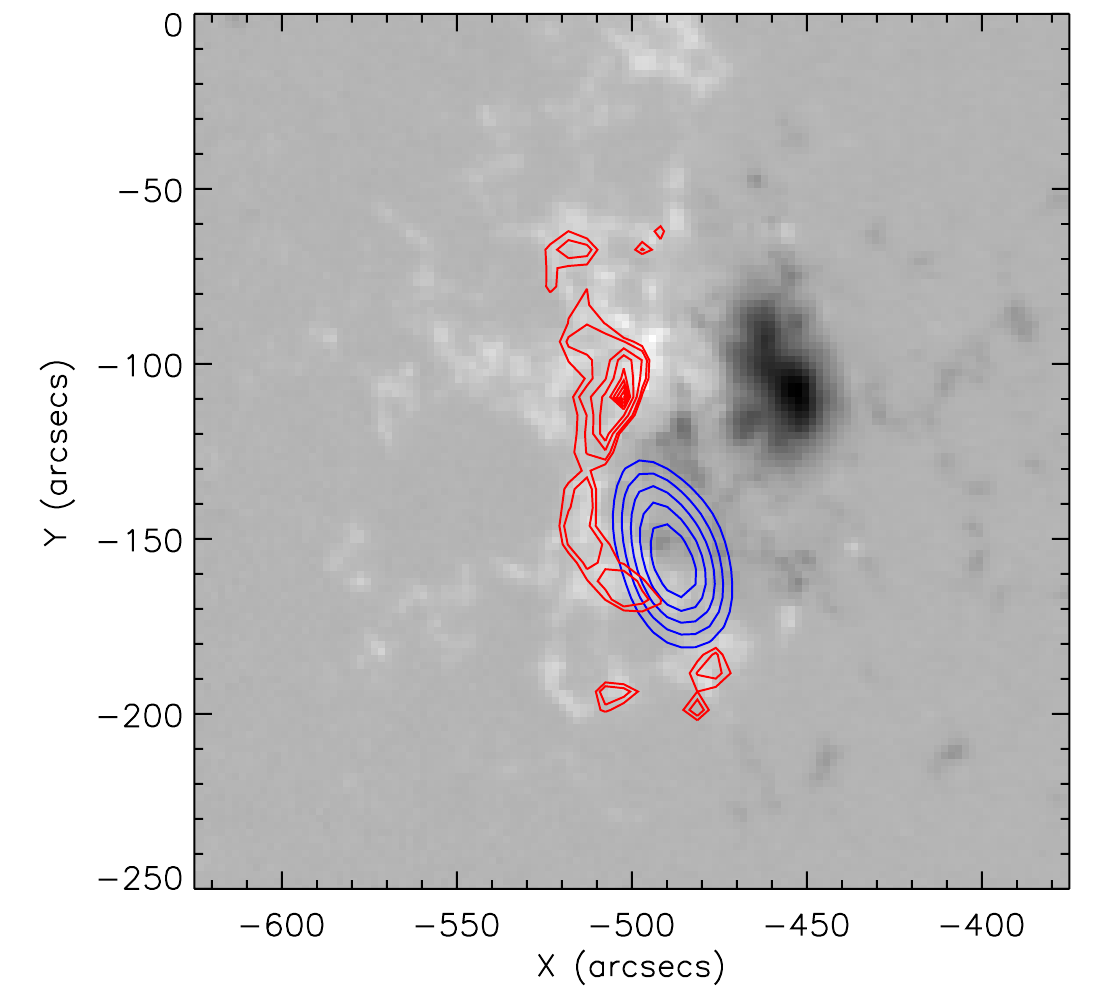}}
\caption[\emph{SOHO}/MDI magnetogram of the AR on 2007 April 29, with reconstructed RHESSI image and 195~\AA\ EUV peak contours overlayed.]{\emph{SOHO}/MDI magnetogram of the AR at 10:36~UT on 2007 April 29. Blue contours show the peak intensity of a reconstructed RHESSI image in the $6 - 12$~keV energy range. The RHESSI image was created using the \textsc{CLEAN} algorithm \citep[see][]{hurford02}. Red contours show the peak intensity of a SOHO/EIT 195~\AA\ image.}
\label{paper1:rhessi}
\end{figure}

Active region NOAA 10953\footnote{\url{http://www.solarmonitor.org/region.php?date=20070426&region=10953}}~crossed the solar disk from 2007 April 26 to 2007 May 9. Previous studies of this region have found evidence of twisting, e.g, \citet{canou10} examined the magnetic structure of the region on 2007 April 30. Their reconstructed magnetic configurations exhibited twisted flux ropes along the southern part of the NL, similar to observations by \citet{okamoto09} of twisted flux ropes emerging from below the photosphere. Here, observations of the main AR on 2007 April 29 are examined. The simple structured active region comprised of a negative-polarity leading sunspot and opposite-polarity trailing plage, with an `S-shaped' filament visible over this time. In addition, this region was the source of a low-magnitude X-ray flare (GOES B1.0): beginning at 10:34 UT; peaking at 10:37~UT; ending at 10:40~UT. Observations recorded by SOT-SP onboard \textit{Hinode} were used, see Table~\ref{paper1:table_obs} for a list of the scan start and end times and pointing information. Note that the flare location outlined in Table~\ref{paper1:table_obs} corresponds to a reconstructed \emph{Reuven Ramaty High Energy Solar Spectroscopic Imager} (RHESSI) spacecraft image peak, obtained using the RHESSI Quicklook Browser Interface\footnote[2]{\url{http://sprg.ssl.berkeley.edu/~tohban/browser/?show=grth+qlpcr}}. Figure~\ref{paper1:rhessi} shows the peak intensity contours of the reconstructed RHESSI image overlayed on a \emph{SOHO}/MDI magnetogram of the AR at 10:36~UT. Also included is the peak emission of a 195~\AA\ image at the time of flare peak, obtained from the Extreme Ultraviolet Imaging Telescope (EIT) on SOHO. Note that the low resolution of the RHESSI image ($\sim 2"$) means that the contours may depict blurred brightening from both footpoints and loop tops. The EUV brightening is located near the filament structure to the solar east and southeast of the main spot.

\begin{table}
\centering
\caption{Summary of SOT-SP scan times on 2007 April 29.}
\begin{tabular}{ c c c c c }
\vspace{0.1cm} \\
\hline\hline
  Scan Number & Begin Time 	& End Time 	& Centre of FOV \\
  			& (UT)			&  (UT)		&    (Solar X, Solar Y)           \\
  \hline
  1                        & 00:17     	& 00:49	   	&  -549$''$, -99$''$\\ 
  2                        & 03:30     	& 04:02	   	&  -525$''$, -98$''$\\ 
  3 			& 08:00     	& 08:32	  	&  -491$''$, -96$''$\\ 
Flare	& 10:34	 	& 10:40		&   -486$''$, -160$''$\\ 
  4 			& 11:27	 	& 11:59	 	&  -464$''$, -95$''$\\
  \hline\hline
  \end{tabular}
  \label{paper1:table_obs}
\end{table}

\begin{figure}[!t]
\centerline{\includegraphics[width=0.8\textwidth]{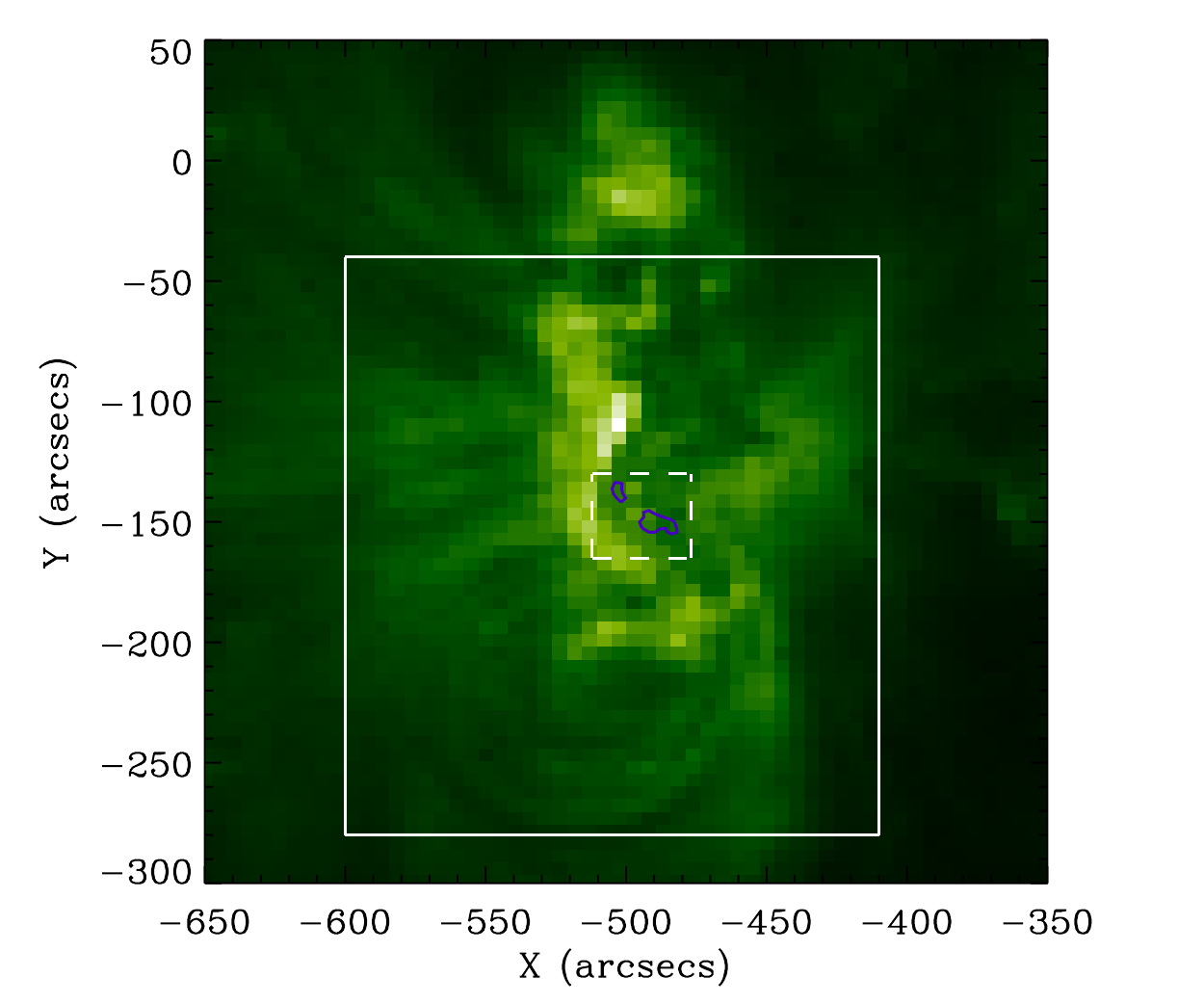}}

\

\centerline{\includegraphics[width=0.95\textwidth]{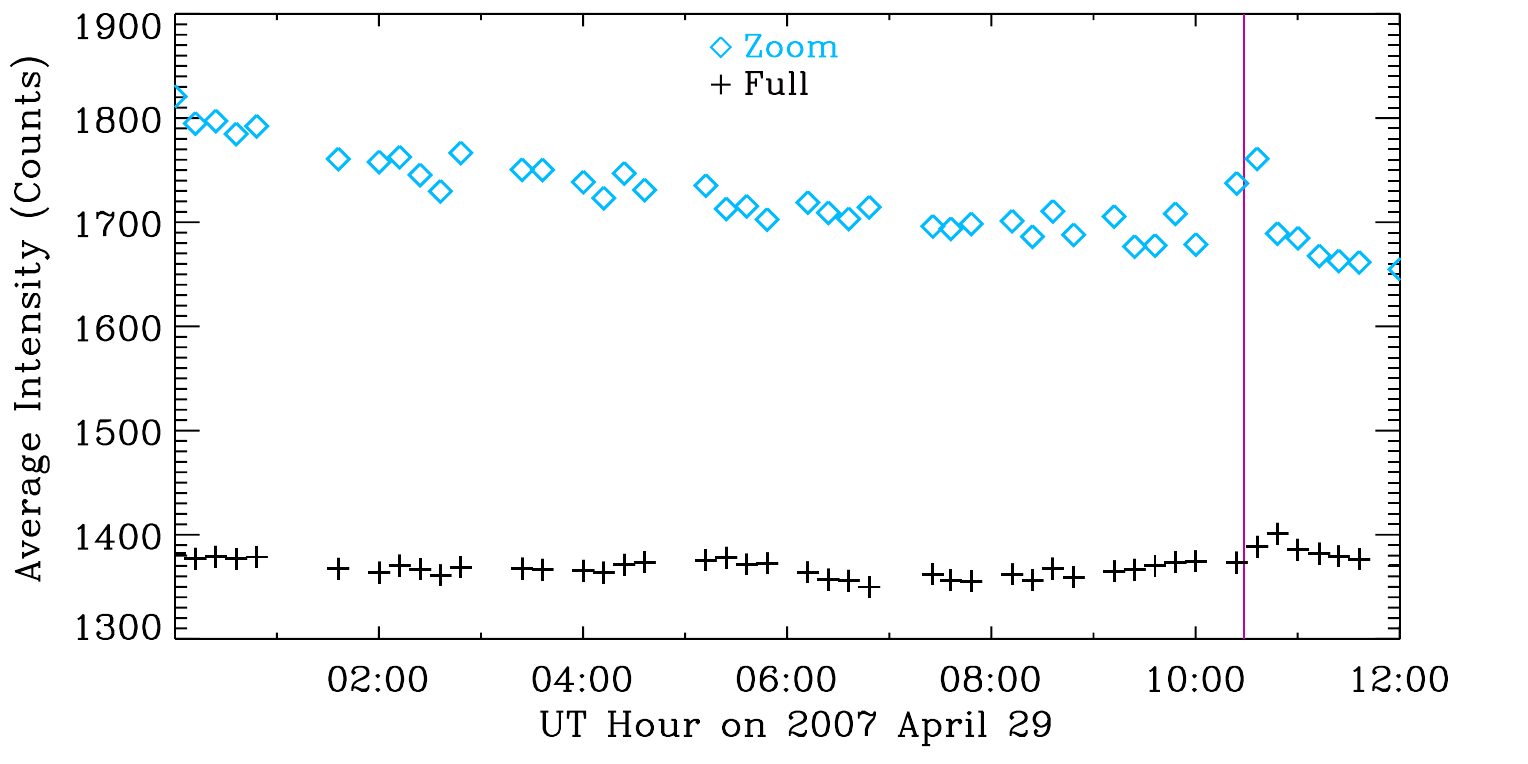}}
\caption[Upper panel shows SOHO/EIT 195\AA\ image of NOAA 10953 at the time of flare peak, while lower panel shows corresponding lightcurves for the AR over a 12-hour period of observation.]{Upper panel shows SOHO/EIT 195~\AA\ image of NOAA 10953 at the time of flare peak, with the solid white box outlining the whole AR FOV, and dashed white box the zoomed-in FOV. Purple contours indicate SOHO/MDI LOS magnetic field strength at $- 200$~G. Lower panel shows corresponding average intensity lightcurves for the full FOV (black crosses) and zoomed-in FOV (blue diamonds). The purple vertical line indicates the time of flare peak.}
\label{paper1:lightcurve}
\end{figure}

Four scans from the SOT-SP were used, with a scan duration of $\sim32$~minutes each. The temporal scan coverage was a critical reason for choosing this event, i.e., three scans before the flare and one immediately after (Table~\ref{paper1:table_obs}). Using multiple scans prior to the flare enables the non-flare related evolution of the magnetic-field properties to be analysed in detail, with changes over the flare able to be compared to this background evolution. No other flares occurred during the entire time period of observation, preventing the contamination of any of the scans. To confirm no other flares occurred in this AR during the period of observation, lightcurves were plotted from EUV 195~\AA\ data obtained with SOHO/EIT, the results of which are shown in Figure~\ref{paper1:lightcurve}. The upper panel shows a 195~\AA\ image from 10:36~UT (near flare peak), outlining two FOVs from which the lightcurves in the lower panel were obtained. Average values of 195~\AA\ intensity are shown for the whole AR FOV (black crosses), as well as a zoomed-in FOV (blue diamonds) that will be used for analysis later. An increase in average intensity is observed at the time of flaring (purple vertical line), with no other significant increases observed throughout the 12-hour period.

SOT-SP recorded the Stokes $I$, $Q$, $U$, and $V$ profiles of the Fe \textsc{i} 6301.5~\AA~and 6302.5~\AA~lines simultaneously using fast-map mode, as described in Section~\ref{chapter:sot-sp}. The raw SOT-SP data were calibrated using \textsf{sp$\_$prep.pro} from the \textit{Hinode}/SOT tree within the IDL \textsf{SolarSoft} library (see Section~\ref{instr:hinode:prep}). The resulting Stokes $I$, $Q$, $U$, and $V$ profiles were inverted using \helix\ in order to derive the magnetic-field vector, as previously discussed in Section~\ref{rad_trans:helixmethod}. Note that the model atmosphere used in fitting the observed profiles comprised of one magnetic component with a local straylight component included. A polarisation value was chosen below which the magnetic field would be treated as zero (minimum magnetic signal), such that regions with values below this were not inverted. This was calculated from the \emph{Hinode} Level 1 data (see Section~\ref{instr:hinode:prep}), by finding the average values of \emph{I}, \emph{Q}, \emph{U}, and \emph{V} for a quiet region in the scan and calculating the total polarisation as per Equation~\ref{rad_transfer:polarisation}. This quiet-Sun value of \emph{P} was taken as the threshold, calculated separately for each of the four SOT-SP scans. An average value of $P \sim3.5 \times 10 ^{-3}$~$I_c$ was found (i.e., units of continuum intensity), however Table~\ref{paper1:polarisation} outlines the exact values. Note that a value of $P$ for the sunspot umbral region was also calculated for comparison purposes to check the thresholds were reasonable, also shown in Table~\ref{paper1:polarisation}.

The \textsf{AMBIG} routine (Section~\ref{rad_trans:ambiguityresolution}) was used to remove the  $180^{\circ}$ ambiguity in the LOS azimuthal angle. As previously mentioned, the routine simultaneously minimises the magnetic field divergence, $\nabla \cdot \mathbf{B}$, and vertical electric current density, $J_{z}$, for pixels above a certain noise threshold in transverse field strength. In this research a value of 150~G was chosen, whereby pixels with values below this level are determined using an iterative acute-angle-to-nearest-neighbors method.
The resulting LOS inversion results were converted to the solar surface normal reference frame using the method described in Section~\ref{rad_trans:coords}, where the orthogonal magnetic-field components in the observers frame and solar surface normal frame are related by Equations~\ref{gary1},~\ref{gary2}, and~\ref{gary3}. Note that Figure~\ref{rad_trans:azi} showed a typical example of the data product of this analysis procedure for this event.

\begin{table}[!]
\centering
\caption[Values of total polarisation in quiet and active (umbra) regions for all four SOT-SP scans.]{Values of total polarisation in quiet and active (umbra) regions for all four SOT-SP scans. For a slice along the third scan 320$^\mathrm{{th}}$ pixel on the x axis, the 220$^\mathrm{{th}}$ and 490$^\mathrm{{th}}$ pixels along the y axis were chosen for the active and quiet region samples, respectively. }
\begin{tabular}{ c c c c c }
\vspace{0.1cm} \\
\hline\hline
  Scan Number 	& $P_{quiet}$ 		& $P_{active}$ 	 \\
  				& $\times 10 ^{-3}$~$I_c$			&  $\times 10 ^{-3}$~$I_c$			           \\
  \hline
  1                        & 3.27    		& 67.46	   	\\ 
  2                        & 4.13     		& 75.63	   	\\ 
  3 				& 2.17     		& 68.06	  	\\ 
  4 				& 4.58	 		& 67.87	 	\\
  \hline\hline
  \end{tabular}
  \label{paper1:polarisation}
\end{table}

\begin{figure} 
\centerline{\includegraphics[scale=1.05]{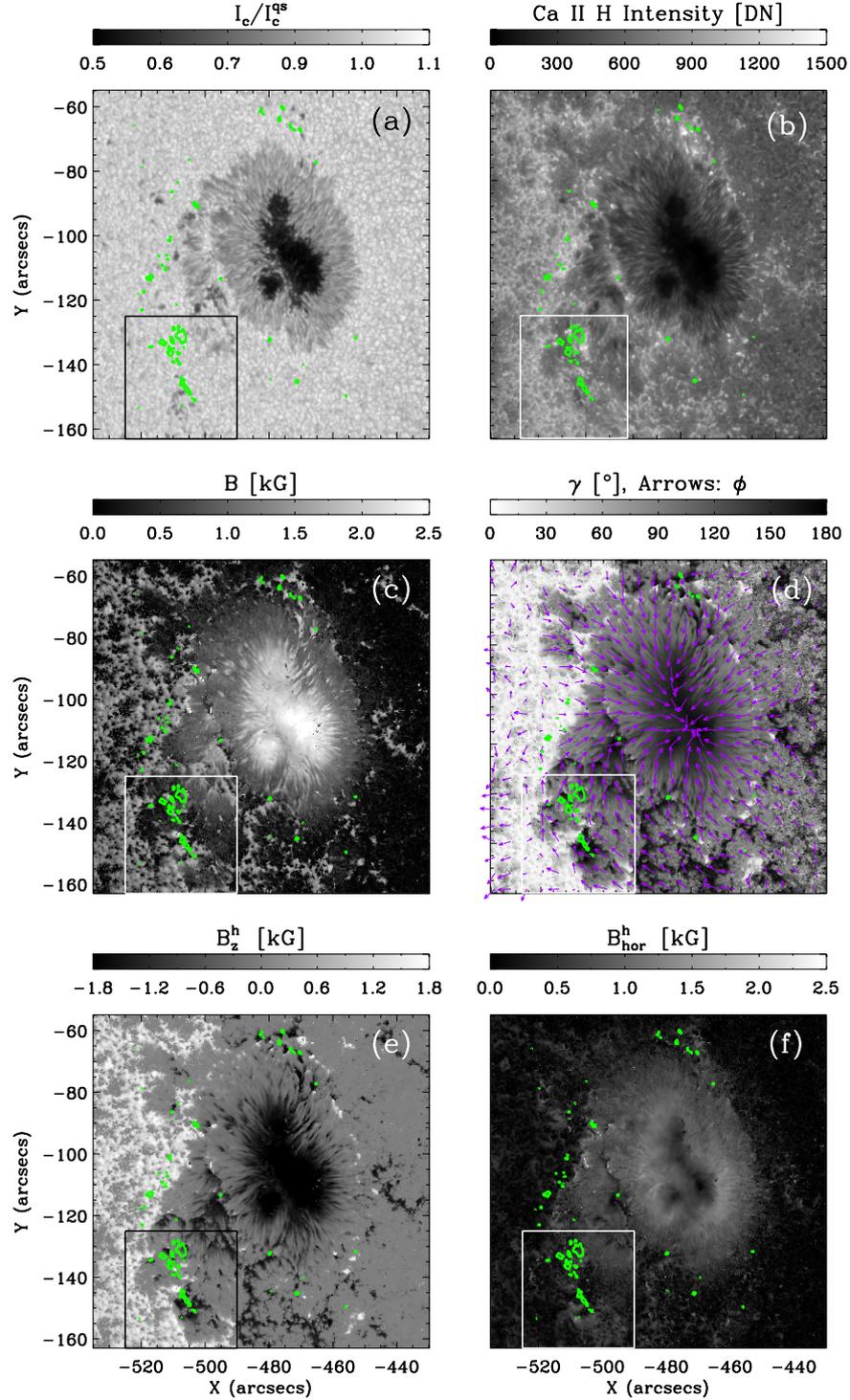}}
\caption[$108''\times108''$ FOV images showing the active region pre-flare state (08:00\,--\,08:32~UT): (a) continuum intensity; (b) Ca \textsc{ii} H intensity (08:16 UT); (c) absolute field strength; (d) inclination angle, with transverse magnetic field vectors overlaid as arrows (magenta); (e) vertical field strength; (f) horizontal field strength.]{$108''\times108''$ FOV images showing the active region pre-flare state (08:00\,--\,08:32~UT): (a) continuum intensity; (b) Ca \textsc{ii} H intensity (08:16 UT); (c) absolute field strength; (d) inclination angle, with transverse magnetic field vectors overlaid as arrows (magenta); (e) vertical field strength; (f) horizontal field strength.  Green contours in all panels outline the significant Ca \textsc{ii} H flare brightening (at the 1250~DN level) observed at 10:37~UT. The sub-region selected for further analysis in Figure~\ref{paper1:figure_2} is indicated by the box in all panels. }
\label{paper1:figure_1}
\end{figure}

The scans taken are $\sim$~three\,--\,four hours apart so it was necessary to correct for changes in scan pointing. To solve this, all scans were differentially rotated and their continuum intensity co-aligned to that of the third scan. Examples of observations from the third scan (i.e., immediately preceding the flare) are shown in Figure~\ref{paper1:figure_1}, including \textit{Hinode}/SOT-SP continuum intensity (Figure~\ref{paper1:figure_1}a) and resulting magnetic field parameters from the \textsf{H\textsc{e}LI\textsc{x}$^{+}$} code after disambiguation and transformation to the solar normal reference frame: absolute magnetic-field strength (Figure~\ref{paper1:figure_1}c); inclination angle with azimuthal-angle vectors overlayed (Figure~\ref{paper1:figure_1}d); vertical field strength, $B_{z}^\mathrm{h}$ (Figure~\ref{paper1:figure_1}e); horizontal field strength, $B_\mathrm{hor}^\mathrm{h}=[(B_{x}^\mathrm{h})^2 +(B_{y}^\mathrm{h})^2]^{1/2}$ (Figure~\ref{paper1:figure_1}f). 

\emph{Hinode}/SOT-BFI Ca \textsc{ii} H line images (3968~\AA, Section~\ref{instr:sot_bfi}) were also obtained close to the flare peak time, with a FOV of $108''\times108''$ (1024 $\times$ 1024~pixels$^2$). Figure~\ref{instr:caiih} in Chapter~\ref{chapter:instrumentation} showed a Ca \textsc{ii} H image at the time of flare peak. Figure~\ref{paper1:figure_1}b shows a Ca \textsc{ii} H image at the time of the third scan,  as well as contours of significant brightening at the time of the flare peak at 10:37~UT (1250~DN) overlaid on all other images. The brightening seems to be mostly located along the NL dividing the sunspot and plage regions in the solar east of the scan.  The location containing the most significant chromospheric flare brightening is found to the solar south east (SE) of the main sunspot, located near the trailing plage NL. A $35''\times40''$ box was chosen from this region for analysis, and is highlighted in Figure~\ref{paper1:figure_1}. 

The sub-region was divided into two specific regions of interest, ROI 1 and ROI 2, defined by thresholding the signed field magnitude (i.e., $|\mathbf{B}|$ times $-1$ or $+1$ for fields pointing in or out of the solar surface, respectively). ROI 1 was thresholded at $-800$~G and ROI 2 at $-1000$~G.  Both of these regions are small flux elements of the same polarity as the main sunspot, and are located SE of the main spot. They both lie close to the NL with the positive polarity plage (see Figure~\ref{paper1:figure_1}e). These two ROIs will be the focus for the study of magnetic field parameters in this Chapter.

%%%%%%%%%%%%%%%%%%%%%%%%%%%%%%%%%%%%%%%%%%%%%%%%%%%%%%%%%%%%%%%%%%%%%%%%%%%%%%%%%%%%%%%%%%%%%%%%%%%%%%%%%%%%%%%%%%%%%%%%%%%%%%%%%%%%%%%%%%%%%%%%%%%%%%%%%%%%%%%%%%

\section{Results}
\label{paper1:results}

\begin{sidewaysfigure}
\centerline{\includegraphics[width=0.63\textwidth,angle=90,clip]{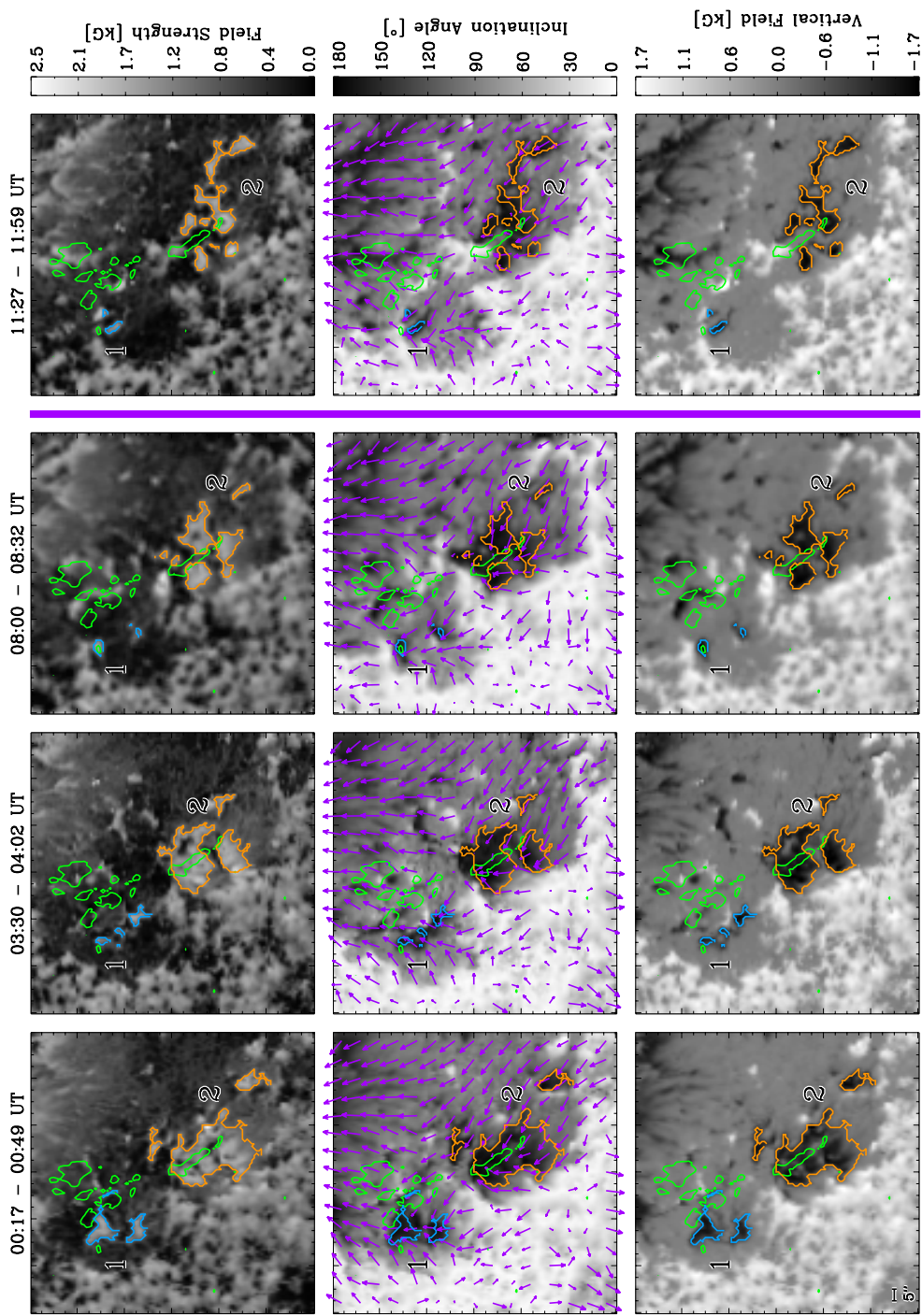}}
\caption[Evolution of the sub-region outlined in Figure~\ref{paper1:figure_1} at increasing scan times from left to right. Top to bottom: absolute magnetic field strength; inclination angle, with transverse magnetic field vectors overlaid as arrows (magenta); vertical field strength.]{Evolution of the sub-region outlined in Figure~\ref{paper1:figure_1} at increasing scan times from left to right. Top to bottom: absolute magnetic field strength; inclination angle, with transverse magnetic field vectors overlaid as arrows (magenta); vertical field strength. Two regions of interest, ROI 1 and ROI 2, are numbered and defined by blue and orange contours, respectively. Green contours in all panels outline the significant Ca \textsc{ii} H flare brightening (at the 1250~DN level) observed at 10:37~UT, as per Figure~\ref{paper1:figure_1}. The time of flaring is indicated by a magenta vertical line between the third and fourth scans. Note the $5"$ scale in the bottom left corner corresponds to $\sim 15.6$~pixels (or $\sim 3.65$~Mm).}
\label{paper1:figure_2}
\end{sidewaysfigure}

Figure~\ref{paper1:figure_2} shows the temporal evolution of the magnetic field in the chosen sub-region over the four scans. ROI 1 fragments significantly from the first to the third pre-flare scans, and almost completely disappears after the flare. ROI 2 also fragments, but changes less than ROI 1. The chromospheric flare brightenings are located over and north west (NW) of ROI 1, and directly over ROI 2. The brightening is indicated by overlaid green contours in Figure~\ref{paper1:figure_2}.

\begin{figure}[!t]
\centerline{\includegraphics[width=1.0\textwidth,clip]{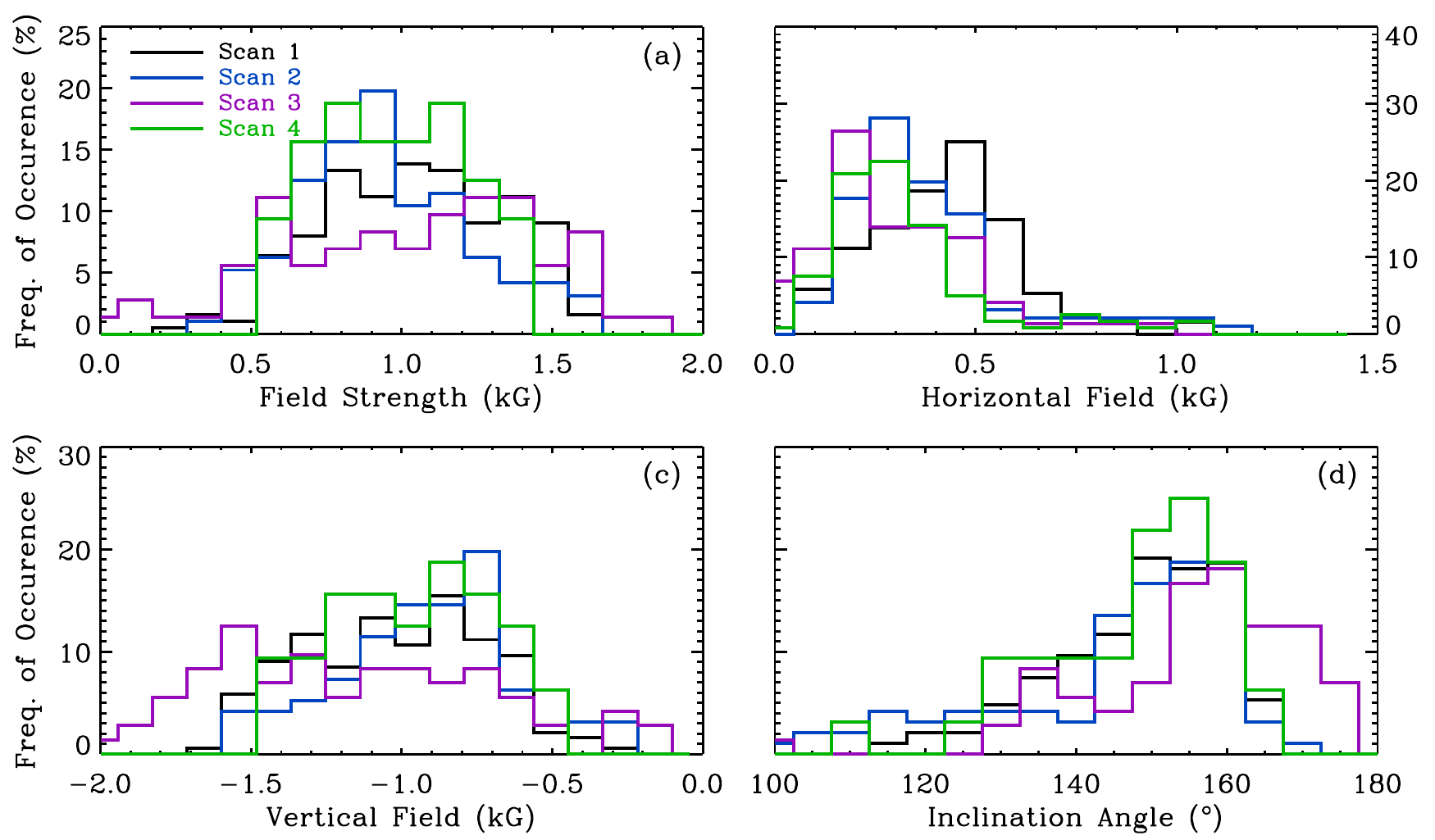}}
\caption[Distribution of (a) field strength, (b) horizontal field, (c) vertical field, and (d) inclination angle from all ROI 1 pixels defined in Figure~\ref{paper1:figure_2}.]{Distribution of (a) field strength, (b) horizontal field, (c) vertical field, and (d) inclination angle from all ROI 1 pixels defined in Figure~\ref{paper1:figure_2}. Y-axes show percentage occurrences. Each scan distribution is coloured as per the legend in panel (a). A bin size of 100~G was used for (a), (b) and (c), and of $5^\circ$ for (d).}
\label{paper1:hist_roi1}
\end{figure}

The parameters depicted in Figure~\ref{paper1:figure_2} were separately analysed in detail for both ROIs. Distributions of field strength, horizontal field, vertical field, and inclination angle were obtained for all pixels within the thresholded contours shown in Figure~\ref{paper1:figure_2}. Histograms of these distributions from ROI 1 are shown for each scan in Figure~\ref{paper1:hist_roi1}, where each frequency distribution is normalised to a total of 100\%. This is required as the number of pixels varies in the contoured ROIs from scan to scan. The field strength and horizontal field distributions for all scans vary little across all four scans. No significant variation is also observed in the vertical field and inclination angle distributions for scans 1 and 2. However, the vertical field strength distribution in the third pre-flare scan shifts towards larger negative values. A shift towards greater inclination angle values is also observed in the third scan distribution, i.e., the field becomes more vertical (note that increasing values of inclination once beyond $90^\circ$ indicates the field becomes more vertical). The distributions of negative vertical field strength and inclination angle in scan 4 (i.e, after the flare has ended) appear to be similar to those of scans 1 and 2.

\begin{figure}[!t]
\centerline{\includegraphics[width=1.0\textwidth,clip]{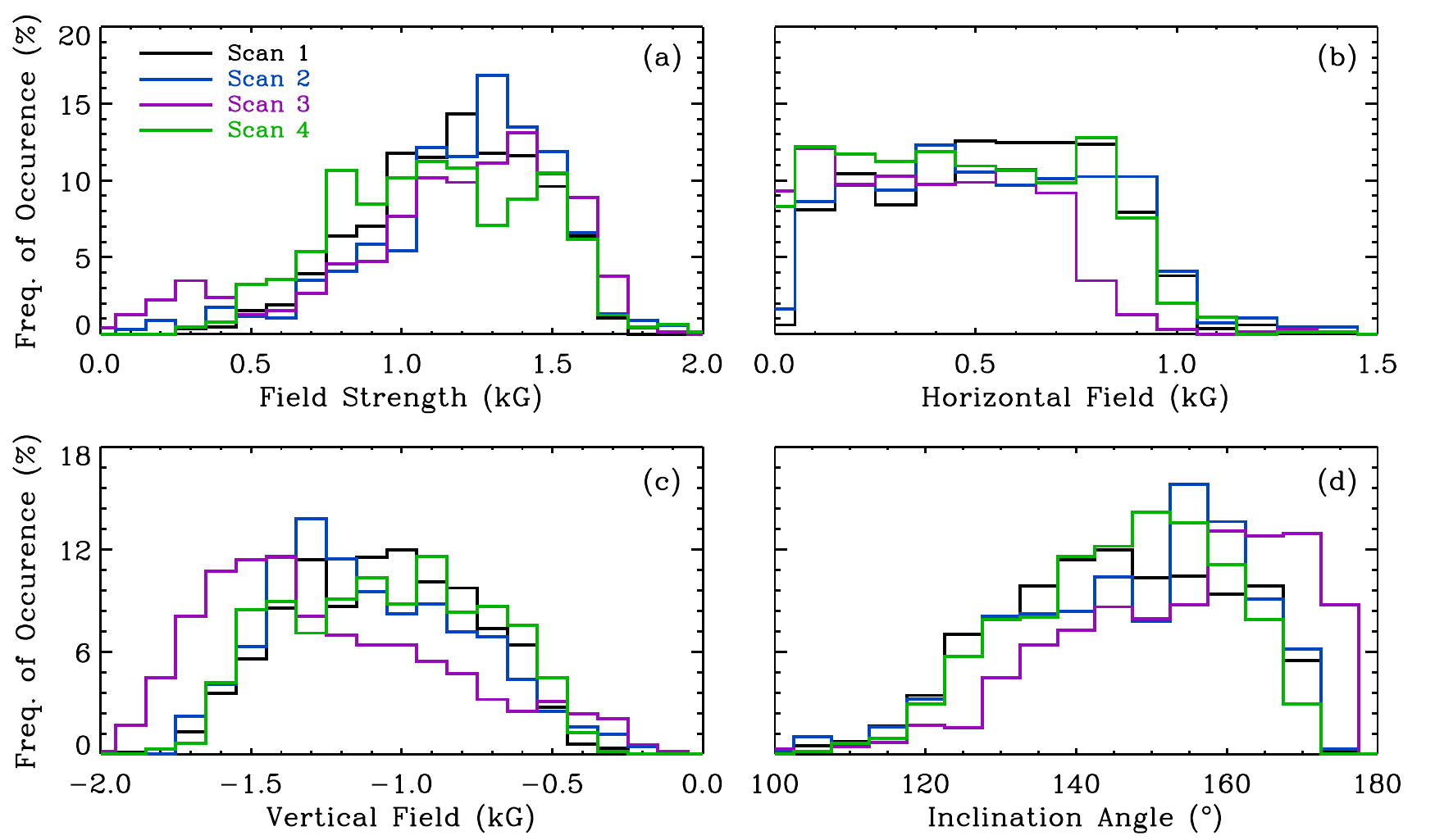}}
\caption[Distribution of (a) field strength, (b) horizontal field, (c) vertical field, and (d) inclination angle from all ROI 2 pixels defined in Figure~\ref{paper1:figure_2}.]{Distribution of (a) field strength, (b) horizontal field, (c) vertical field, and (d) inclination angle from all ROI 2 pixels defined in Figure~\ref{paper1:figure_2}. Y-axes show percentage occurrences. Each scan distribution is coloured as per the legend in panel (a). A bin size of 100~G was used for (a), (b) and (c), and of $5^\circ$ for (d).}
\label{paper1:hist_roi2}
\end{figure}

Figure~\ref{paper1:hist_roi2} shows the distributions obtained for all pixels within ROI 2. Similar changes are observed to those found in Figure~\ref{paper1:hist_roi1}. No significant changes are observed in the field strength and horizontal field strength distributions, except slight tendencies towards lower values in scan 3 for the horizontal field strength and again lower values in scan 4 for the field strength. Again, the vertical field distribution shows larger negative values in the third pre-flare scan compared to scans 1 and 2, with the inclination angle distribution showing the field distribution shifting towards a more vertical configuration. After the flare, both of these distributions again return to a shape resembling the distributions of scans 1 and 2. It is worth noting that the ROIs are at different locations at different scan times, but it is promising to see similar covered parameter ranges. 

There seems to be a clear indication of significant changes in vertical field strength and field inclination before and after the B-class flare. In order to investigate these changes more thoroughly, it is useful to obtain a gauge of the magnitude of changes observed. Thus the median and standard deviation of the values examined in Figures~\ref{paper1:hist_roi1} and ~\ref{paper1:hist_roi2} were extracted from all pixels within the ROI contour in each individual scan. Median values were used rather than other averaging methods due to their ease of interpretation and relative insensitivity to outlying values. The investigation of the field structure will be discussed in the following sub-sections: the vector field components (Section~\ref{paper1:vertical}), and the field-orientation angles (Section~\ref{paper1:inclination}). Signatures of magnetic non-potentiality will also be examined in Section~\ref{paper1:divergence}.

%%%%%%%%%%%%%%%%%%%%%%%%%%%%%%%%%%%%%%%%%%%%%%%%%%%%%%%%%%%%%%%

\subsection{Vector Field Components}
\label{paper1:vertical}

\begin{figure} 
\centerline{\includegraphics[width=1.0\textwidth,clip]{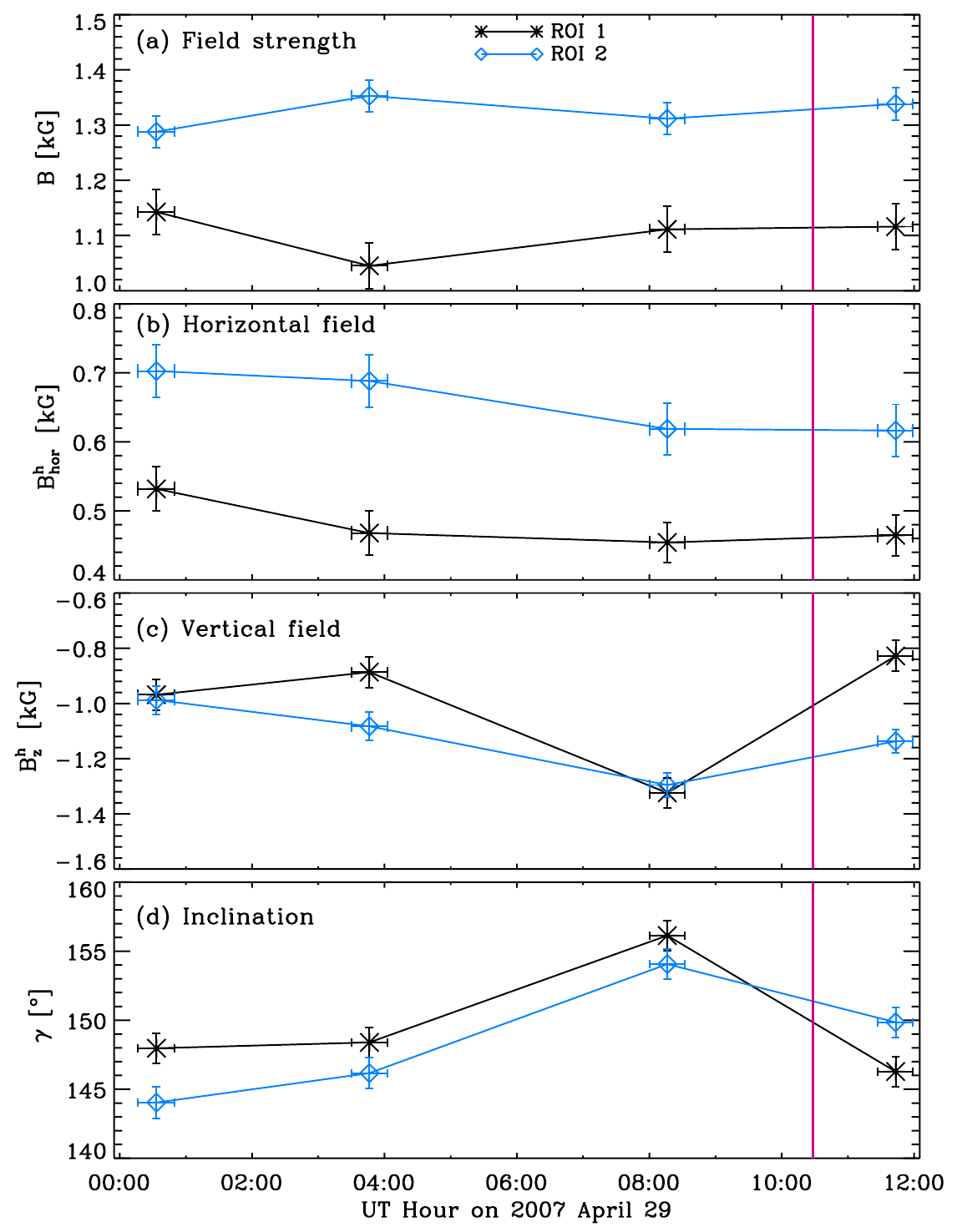}}
\caption[Temporal variation in the median values of: (a) absolute magnetic-field strength; (b) horizontal field strength; (c) vertical field strength; (d) inclination angle.]{Temporal variation in the median values of: (a) absolute magnetic-field strength; (b) horizontal field strength; (c) vertical field strength; (d) inclination angle. Values for ROI 1 are plotted with black asterisks, and ROI 2 with blue diamonds. Vertical bars indicate the standard deviation of the property within the ROI, while horizontal bars delimit the scan duration. The vertical line between the third and fourth scans marks the flare peak time. With values of inclination being beyond $90^{\circ}$, increasing values indicate the field becoming more vertical.}
\label{paper1:figure_3}
\end{figure}

Changes in ROI median values of the field magnitude, vertical field, and horizontal field were calculated in each scan (i.e., values from all pixels in the thresholded contours of a ROI). Figure~\ref{paper1:figure_3} depicts time lines of these ROI median values, with vertical bars representing the ROI standard deviation and horizontal bars depicting the scan duration. The magnetic field strength in Figure~\ref{paper1:figure_3}a varies little over all the scans within 1-$\sigma$ errors, with only a slight decrease in the second scan for ROI 1. The horizontal field strength, given in Figure~\ref{paper1:figure_3}b, shows only a slightly decreasing trend over the scans. The main source of interest here again comes from the observed changes in vertical field strength.

The vertical field median value also marginally changes within the spread of ROI values between the first two scans, as can be seen in Figure~\ref{paper1:figure_3}c. However, substantial variations are found between both the second and the third scans, as well as the third and fourth scans. An increase in vertical field magnitude is found between the second and third scans, increasing by $\sim440$~G for ROI 1 and $\sim210$~G for ROI 2. After the flare (i.e., some time between the third and fourth scans) $B_{z}^\mathrm{h}$ decreases by $\sim500$~G for ROI 1 and $\sim160$~G for ROI 2. It is likely that the changes prior to the flare are linked to the energy storage mechanism in the ROIs, while the changes over the course of the flare are due to the energy release. However, it is unclear from the median field magnitude measurements how the field structure is changing before and after the flare. Thus, field orientation was investigated further.

%%%%%%%%%%%%%%%%%%%%%%%%%%%%%%%%%%%%%%%%%%%%%%%%%%%%%%%%%%%%%%%

\subsection{Field Orientation}
\label{paper1:inclination}
The median inclination angle was also extracted from both ROIs and is included in Figure~\ref{paper1:figure_3}d. A similar trend in inclination evolution is seen to the vertical-field evolution, as would be expected given the relative stability of $|B|$. Again no changes of significance are found between the first two scans, with large changes observed between the second and third scans and after the flare. An increase in inclination is found in the third scan, with field becoming more vertical by  $\sim8^{\circ}$ for both ROI 1 and ROI 2. After the flare, inclination decreases (i.e., becomes more horizontal) by $\sim10 ^{\circ}$ for ROI 1 and $\sim4^{\circ}$ for ROI 2. These results show that the field in both ROIs becomes more vertical $\sim$~6.5\,--\,2.5 hours before the GOES B1.0 flare and more horizontal within $\sim$~one hour after the flare has ended. It is interesting to note that the location of the field change is localised near the NL with the plage region, in a negative polarity region to the SE of the sunspot.
 
To put the changes in field parameters observed over the scans into context, it is worth estimating where the field lines in ROI 1 and ROI 2 are connected to by examining the direction of the transverse magnetic field vectors (overlaid on the inclination scans in Figure~\ref{paper1:figure_2}). It is noted that the true connectivity cannot be determined from 2D results alone and the necessary 3D extrapolations of the region will be investigated in Chapter~\ref{chapter:3D}. As a first guess towards the possible connectivity, the field in ROI 1 seems to be generally in a northerly direction in scan 1 and scan 2, becoming increasingly more NE in scan 3 and scan 4. In ROI 2, the field is pointing in a general NE direction in the first scan, pointing in an increasingly more easterly direction as time progresses, before finally returning to a more NE direction after the flare. The purple azimuthal arrows in the middle row of Figure~\ref{paper1:figure_2} illustrate this well. It seems that the plage region SE of the ROIs extends towards the NW (i.e., between the ROIs) as the scans progress, before `pinching off' after the flare, and thus separating from the rest of the positive plage region. 

It is difficult to determine by eye exactly where the field may be connected to over the scans, especially if relying on median values of small groups of pixels. It is surmised that a region of plage NE of ROI 1 is a likely connection point. The fourth scan in Figure~\ref{paper1:figure_2} also indicates a possible connection between ROI 2 and the portion of intersecting plage that first extends between the ROIs before `pinching off' after the flare. Studying the 3D extrapolated field over the ROIs is necessary to fully understand the evolution (which will be examined in Chapter~\ref{chapter:3D}).

%%%%%%%%%%%%%%%%%%%%%%%%%%%%%%%%%%%%%%%%%%%%%%%%%%%%%%%%%%%%%%%
\begin{figure}[!t]
\centerline{\includegraphics[width=\textwidth,clip]{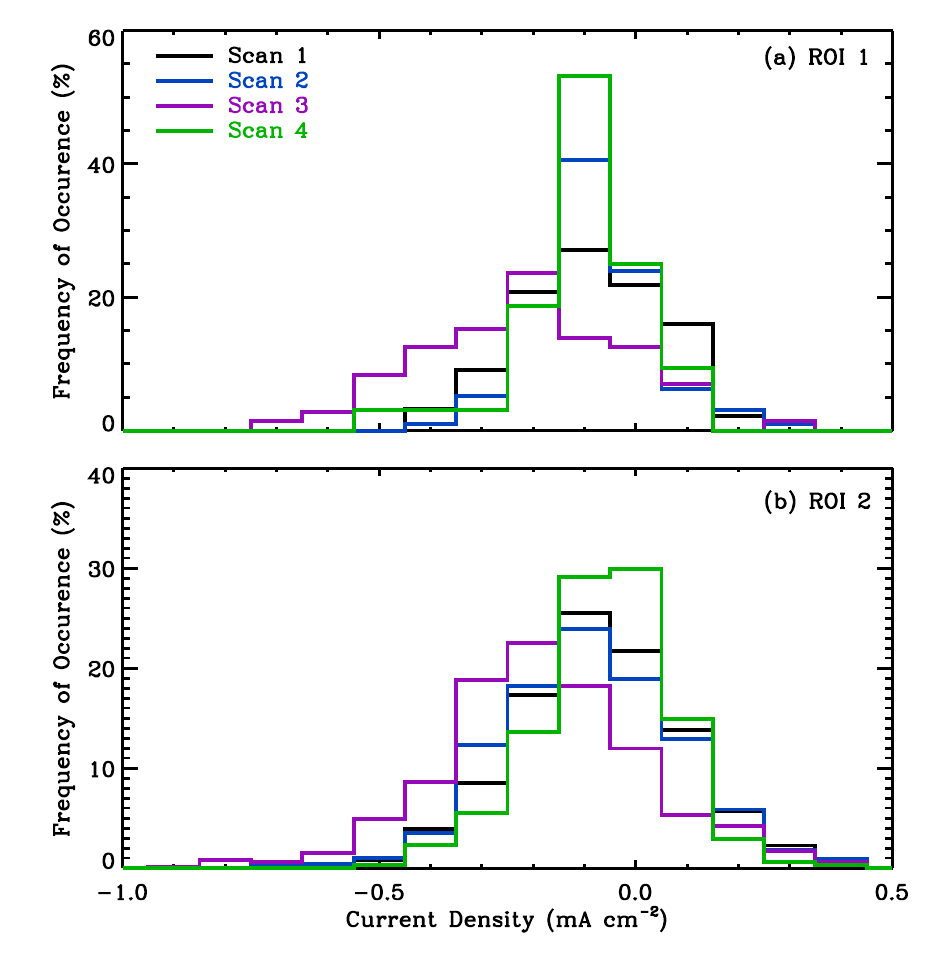}}
\caption[Distribution of vertical current density from all pixels within (a) ROI 1 and (b) ROI 2, as defined in Figure~\ref{paper1:figure_2}.]{Distribution of vertical current density from all pixels within (a) ROI 1 and (b) ROI 2, as defined in Figure~\ref{paper1:figure_2}. Y-axes show percentage occurrences. Each scan distribution is coloured as per the legend in panel (a).}
\label{paper1:hist_jz}
\end{figure}

\subsection{Signatures of Non-Potentiality}
\label{paper1:divergence}
Non-potential fields harbour currents, therefore signatures of currents supported by the surface vector field are indicators of non-potentiality. Here, the vertical current density is calculated by the method of \citet{crouch08}, as implemented in the \textsf{AMBIG} code. 
Figure~\ref{paper1:hist_jz} shows distributions of the values from all pixels within both ROIs for each scan. The distribution for the first two scans in both ROIs are quite similar, with the most significant difference observed in the third pre-flare scan. The distributions after the flare (i.e., scan 4) for both ROIs are also similar to the first two scans, with slightly higher frequency peaks of the distributions located around similar values. The distribution of vertical current density for both ROIs in the third pre-flare scan extends towards larger negative values than the other distributions. This suggests an increase in negative vertical current density before the flare, with a return to previous `quiet' pre-flare values afterwards.

\begin{figure}[!t]
\centerline{\includegraphics[width=1.0\textwidth,clip]{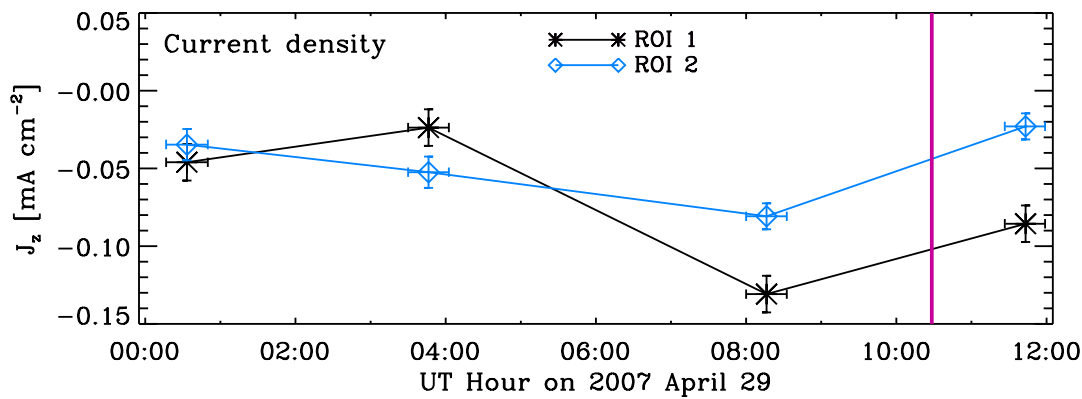}}
\caption[Temporal variation in the median values of vertical current density.]{Temporal variation in the median values of vertical current density. Values for ROI 1 are plotted with black asterisks, and ROI 2 with blue diamonds. Vertical bars indicate the standard deviation of the property within the ROI, while horizontal bars delimit the scan duration. The vertical line between the third and fourth scans marks the flare peak time.}
\label{paper1:figure_4}
\end{figure}

Median values of all pixels within the contours for each ROI are presented in Figure~\ref{paper1:figure_4}, with vertical bars again showing the ROI standard deviation. A familiar trend is observed between the first and second scans (i.e., no change within the spread of values in either ROI). Negative vertical current density increases in magnitude in the pre-flare state from the second to third scans by $\sim0.11$~$\mathrm{mA} \, \mathrm{cm^{-2}}$ for ROI 1 and by $\sim0.03$~$\mathrm{mA} \, \mathrm{cm^{-2}}$ for ROI 2. The magnitude subsequently decreases by $\sim0.07$~$ \, \mathrm{mA} \, \mathrm{cm^{-2}}$ in both ROI 1 and ROI 2. 

Changes in ROI 1 parameters are much more distinct than in ROI 2, as was also seen in field inclination and vertical field strength (which may be related to the size of the region). Thus, stronger currents appear in both regions before the flare occurs, dropping back to earlier background values after the flare. An increase in current density before the flare indicates an emergence or build-up of non-potentiality in the field, with these observed changes likely to be linked to energy build-up before the flare, and energy release during to the flare. 

%%%%%%%%%%%%%%%%%%%%%%%%%%%%%%%%%%%%%%%%%%%%%%%%%%%%%%%%%%%%%%%%%%%%%%%%%%%%%%%%%%%%%%%%%%%%%%%%%%%%%%%%%%%%%%%%%%%%%%%%%%%%%%%%%%%%%%%%%%%%%%%%%%%%%%%%%%%%%%%%%%

\section{Discussion}
\label{paper1:conclusions}

Siginficant changes in various photospheric magnetic field parameters (namely vertical field, inclination angle, and vertical current density) have been discovered over a number of hours before and after a flare event. Note that the changes observed had occurred by $\sim2.5$~hours before the flare onset, with no significant changes observed $\sim$~6.5\,--\,10 hours beforehand. \citet{Schrijver07} states that the energy build-up phase can last for as much as a day in an active region, so it is interesting to see such short time-scale changes. This research has shown for the first time that significant variations in the photospheric magnetic field of AR small flux elements can be observed in the hours leading up to a flare. Considering the small magnitude of the flare (GOES B1.0), investigations of larger events could show even stronger changes (see Chapter~\ref{chapter:NL} for the analysis of a GOES C9.7 flare event).

\begin{figure}[!t]
\centerline{\includegraphics[width=1.0\textwidth,clip]{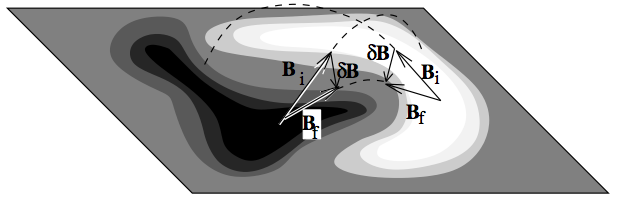}}
\caption[Ilustration by \citep{hudson08} predicting a decrease in inclination across the NL after a solar flare.]{Ilustration by \citep{hudson08} predicting a decrease in inclination across the NL due to a flare, which agrees with the inclination results in this chapter. Initial photospheric field vectors, $\mathbf{B_\mathrm{i}}$, tilt by $\delta \mathbf{B}$ towards the horizontal to final states, $\mathbf{B_\mathrm{f}}$. This tilt is a result of coronal restructuring during a flare/CME (denoted here by the dashed lines showing changes in the connectivity of the coronal field).}
\label{paper1:hudson}
\end{figure}

An $\sim8^{\circ}$ change in median field inclination towards the vertical was found leading up to the flare,  with a $\sim7^{\circ}$ return towards the horizontal afterwards. Although no previous work has looked at these particular changes over hours before a flare, some studies have also reported changes in field orientation after a flare. \citet{li09} found an inclination angle change of $\sim5^{\circ}$ towards the horizontal in a region of enhanced G-band intensity after an X-class flare, and the inclination becoming more vertical by $\sim3^{\circ}$ in a region of diminished G-band intensity.  Although their study focuses on penumbral regions, the region becoming more horizontal after the flare is located close to the flaring NL, similar to the findings of this thesis. After the publication of this research, a study by \citet{wangs12} noted a rapid and permanent change in field inclination towards the horizontal by $\sim7^{\circ}$ within $\sim$~30 minutes after an X2.2 flare (however they observed no significant changes before the flare). This change occured in a compact region along the flaring NL. This concept is also mentioned in some theoretical studies, e.g., \citet{hudson08} predicted that the photospheric magnetic fields close to the NL would become more horizontal in a simple flare restructuring model (see Figure~\ref{paper1:hudson}). The restructuring found in this event seems to agree with the predictions of \citeauthor{hudson08}.

The coronal `implosion' model of \citeauthor{hudson08} describes initial sources of non-potential (stressed) coronal field that reconfigure due to flaring. An eruption can occur if the force balance between the upward magnetic pressure force of the stressed field, and the downward magnetic tension force of the overlying quasi-potential field, is disrupted. A decrease in coronal magnetic energy (which will be investigated in more detail in Chapter~\ref{chapter:3D}) should lead to a reduction of the upward magnetic pressure, which would inevitably result in the contraction of the overlying coronal magnetic field in an `implosion'. Given a change $\delta \mathbf{B}$ in the photospheric field across the flaring magnetic NL as shown in Figure~\ref{paper1:hudson}, a contraction in the Maxwell stress tensor (see Equation~\ref{maxwell_stress}) occurs which causes a change in the Lorentz force per unit area. 

\citeauthor{hudson08} notes that work done by the Lorentz force on the back reaction could supply enough energy to explain flare-driven seismic waves. However, of relevance for this research is the proposed change in field configuration to becoming more horizontal after restructuring. Previous observations of contracting flaring loops during the early phase of flares (manifested by converging conjugate footpoints and descending looptop emission) have been reported in X-ray, EUV, H$\alpha$, and microwave wavelengths. For example, \citet{liu09} observed EUV coronal loops overlying an eruptive filament pushing inward during a flare, which is associated with a converging motion of the conjugate HXR footpoints and the downward motion of the HXR looptop source. It is certainly interesting to also see evidence of this `implosion' scenario from the evolution of magnetic field inclination in this research. 

Examining previous findings of transverse field changes, \citet{wang02} used vector magnetogram observations to find an impulsive increase of the transverse-field strength and magnetic shear after three X-class flares. \citet{li09} found a transverse field increase of 20\% after an X3.4 flare, and \citet{wangs12} also found a permanent enhancement in transverse field of $\sim$~30\% of the pre-flare magnitude.  No significant changes in transverse-field strength were found here, either immediately before or after the flare (only slight decreases in the pre-flare average values, that correspond to the ramping up of the field vector). However, the difference in active regions must be noted, with the \citeauthor{wang02} and \citeauthor{li09} works focusing on larger-magnitude flares from $\delta$ sunspot groups, and \citet{wangs12} examining a whole $\beta\gamma$ region. The insignificant changes in the transverse-field strength found here are explained by the competing field strength and inclination changes before and after the flare. For example, a large increase in inclination angle for ROI 1 (Figure~\ref{paper1:figure_3}d) between the second and third scans is accompanied by a slight increase in field strength (Figure~\ref{paper1:figure_3}a), giving approximately no change in the horizontal field (Figure~\ref{paper1:figure_3}b). The \citet{li09} result supports the reconnection picture of \citet{liu05}, whereby newly connected fields near the magnetic NL contributed to field inclination becoming more horizontal. This picture suggests that the field lines after the flare in this study become newly-reconnected, low-lying, more horizontal field lines near the flaring NL. 

Vertical-field magnitude was found here to increase in both ROIs before the flare, and decrease by approximately the same amount afterwards. \citet{wang02} examined LOS magnetograms as well as vector data, finding an increase in magnetic flux of the leading polarity in six X-class flares. \citet{sudolharvey05} used longitudinal magnetogram data from the \textit{Global Oscillation Network Group} to find abrupt and permanent changes in the LOS magnetic field after 15 X-class flares. They found decreases in vertical field twice as often as increases, and in 75\% of cases the magnetic-field change occurred in less than ten minutes. \citeauthor{sudolharvey05} quote median LOS field changes of 90 G and found that the strongest field changes typically occur in penumbrae. This behaviour of decreasing vertical field is reflected in these findings, although larger changes of $\sim330$~G are observed here in a region outside of the penumbra. Higher resolution observations are used in this research compared to the previous studies mentioned, which may explain the larger changes found.

The observations obtained for this event show an increase in negative vertical current density within $\sim6.5$\,--\,2.5 hours before the flare, with a decrease towards the initial pre-flare values after the flare. Strong emerging currents have often been linked with flare triggers, e.g., \citet{su09} observed the current density for the same active region three days later when a C8.5 flare occurred, finding strong currents along the field lines. \citet{canou10} also examined the vertical current density for the same active region using a different extrapolation method than \citet{su09}. The extrapolation found footpoints of the twisted flux ropes to be anchored in a region of significant vertical current (i.e., in the core of the flux region rather than along the field lines). They observed the breakdown of the force-free assumption along the NL due to non-zero vertical current density and suggested that this could be due to the emergence of the twisted flux ropes, or perhaps the presence of non-null magnetic forces. They also determined that enough free magnetic energy existed to power the C8.5 flare studied by \citet{su09} and a C4.2 flare a few days later. Similar mechanisms could possibly be at work to cause the earlier lower-magnitude flare examined in this study. 

\citet{regnierpriest07aa} made a number of observations regarding AR configurations, whilst noting the discrepancies that exist between using different extrapolation methods (which will be investigated further in Chapter~\ref{chapter:3D}). They found that strong currents present in the magnetic configuration were responsible for highly twisted and sheared field lines in a decaying active region. In contrast, weak currents existed in a newly emerged active region. They also suggest a strong dependance of vertical current density on the nature of the active region, e.g., the stage of the regions evolution or the distribution of the sources of magnetic field.  Most previous work has focused on considerably more complex active regions that produce M- or X-class flares, so it is important to note that distinct changes in the magnetic field were still observed for this B-class flare.
 
It is worth mentioning that \citet{okamoto09} observed converging motions in Ca \textsc{ii} H movies of the same active region studied in this chapter, which they describe as driven by moat flows from the sunspot towards the trailing plage NL (i.e., near the locations of ROI 1 and ROI 2).  \citet{Schrijver00} mention typical spatial scales of moat flow regions of  $\sim10 - 20$~Mm measured from the outer edge of the penumbra. This suggests that ROI 1 and 2 lie on the outer edge of the moat flow region, and perhaps a moving magnetic feature was being driven towards the plage region. This driving would cause the field near the NL to become more vertical before the flare, as per the results found, and might explain the pre-flare energy build-up phase. The field would then relax and become more horizontal after the energy release, as is found here. This driving can also be compared to converging motions towards a NL that are highlighted in a number of eruptive MHD simulations \citep[e.g., ][]{amari03,amari11}. \citeauthor{amari03} mention a three part magnetic structure associated with their model's disruption phase, with a twisted flux rope running through a global arcade and above small loops. The newly formed small loops, described as due to reconnection, are perhaps indicative of the more horizontally inclined post-flare field of this study compared to pre-flare build up values.

The research described in Chapter~\ref{chapter:3D} aims to clarify the connectivity of the ROIs and changes in the 3D topology. The resulting disambiguated field vector will be used as an input to magnetic field extrapolations to determine various topology measures. However, it is interesting to see such clear changes in field vector characteristics (such as inclination, vertical field strength, and vertical current density) leading up to and after the flare, before making higher order calculations of 3D field topology. The results of this chapter have shown significant pre-flare changes on a shorter timescale (i.e., hours) than has previously been found. These forms of field orientation changes could prove to be useful precursors for flare forecasting in the future. The field inclination evolution is of particular interest, as the results in this chapter confirm the concept of field configuration changes due to flaring highlighted by previous theories. It is also interesting to find that the magnetic NL is the clear location of flaring, as has been found previously (and discussed in Chapter~\ref{chapter:introduction}). The research described in the next chapter will investigate the field inclination of a different AR in more detail, examining the spatial variation as a function of distance from the NL as well as temporal evolution over the course of a flare.

%%%%%%%%%%%%%%%%%%%%%%%%%%%%%%%%%%%%%%%%%%%%%%%%%%%%%%%%%%%%%%%%%%%%%%%%%%%%%%%%%%%%%%%%%%%%%%%%%%%%%%%%%%%%%%%%%%%%%%%%%%%%%%%%%%%%%%%%%%%%%%%%%%%%%%%%%		% research 1
	
%\include{7/group}		% group chapater
		
% this file is called up by thesis.tex
% content in this file will be fed into the main document

%: ----------------------- name of chapter  -------------------------
\chapter{Field Inclination Changes near a Neutral Line} % top level followed by section, subsection
\label{chapter:NL}

\ifpdf
    \graphicspath{{5/figures/PNG/}{5/figures/PDF/}{5/figures/}}
\else
    \graphicspath{{5/figures/EPS/}{5/figures/}}
\fi

\hrule height 1mm
\vspace{0.5mm}
\hrule height 0.4mm 
\noindent 
\\ {\it As briefly discussed in Section~\ref{intro:complexity} and throughout Chapter~\ref{chapter:paper1}, magnetic neutral lines (NLs) are often found to be associated with flare observations. Thorough investigations of the magnetic field configurations across these regions must be made in order to fully understand the processes involved in flare events and flare triggers. In this chapter, the magnetic field evolution along a NL during a flare observation period is investigated. Specifically, the evolution of magnetic field inclination in a plage region near a sunspot group is examined over a $\sim21$~hour period before and after a solar flare. The research topic is first briefly introduced in Section~\ref{NL:intro}, and observations described in Section~\ref{NL:obs}. The research in this chapter is currently in preparation for publication in\apjl.
}
\\ 
\hrule height 0.4mm
\vspace{0.5mm}
\hrule height 1mm 

\newpage

%%%%%%%%%%%%%%%%%%%%%%%%%%%%%%%%%%%%%%%%%%%%%%%%%%%%%%%%%%%%%%%%%%%%%%%%%%%%%%%%%%%%%%%%%%%%%%%%%%%%%%

\section{Introduction}
\label{NL:intro}

Early observations of magnetic NLs established that flaring is likely to occur near the NL, as by definition these divide areas of opposite polarity magnetic field \citep{severny58, zirin73}. Neutral lines can play a central role in magnetic reconnection (being a likely setting for the actual reconnection process), and this mechanism is widely assumed to drive the initial impulsive phase of solar flares \citep{demoulin93}. Investigating the area across the NL is thus paramount for understanding the flaring process, as well as possible flare triggers. The magnetic field inclination is of particular interest, as numerous flare models and observational studies have proposed field configuration changes before and after a solar flare (e.g., \citealp{liu05}; \citealp{hudson08}; \citealp{fisher12}). The results of observations discussed in Chapter~\ref{chapter:paper1} confirmed some of these proposals, and highlighted the need to study field inclination across NL regions in more detail.

\begin{figure}[!t]
\centering
\includegraphics[width=.6\textwidth]{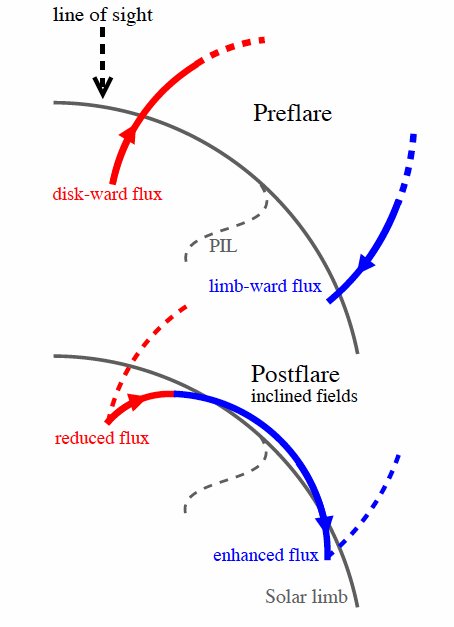}
\caption[Illustration by \citet{wang10} demonstrating their observation of LOS field lines becoming more horizontal with respect to the solar surface after a flare, with the limbward flux increasing and diskward flux decreasing.]{Illustration by \citet{wang10} demonstrating their observation of LOS field lines becoming more horizontal with respect to the solar surface after a flare, with the limbward flux increasing and diskward flux decreasing. The polarity inversion line (PIL; i.e., NL) is shown by the dashed grey line. Their results agree well with the predictions of \citet{hudson08}.}
\label{NL:wang10}
\end{figure}

Previous work has been conducted on the spatial variation of active region magnetic field inclination across NLs \citep[e.g.,][]{solanki03a,wang12}. \citet{wang12} used \emph{Hinode} SOT/SP vector magnetograms to study the relationship between intensity and magnetic inclination of detailed penumbral structure, examining the averaged inclination of bright and dark penumbral fibrils as a function of distance across the penumbra. Other studies have focused on the temporal evolution (rather than spatial) of field inclination near the NL after flaring. For example \citet{wang10} studied, amongst other events, a GOES M8.7 flare event using BBSO vector magnetograms. They examined time profiles of transverse field, inclination angle, and Lorentz force per unit area within a defined compact region at the magnetic NL. Their Figure 3 is reproduced in Figure~\ref{NL:wang10}, illustrating the changes they observed in LOS fields after the flare.

Most studies focus solely on spatial or temporal changes in the field inclination, but this does not give a complete picture of the field configuration over a flaring period. A recent study by \citet{gosain12} briefly combined both areas of focus, using \emph{SDO}/HMI observations to study photospheric magnetic field changes in an active region. They examined the field inclination difference between two times before and after an X-class flare for a region across the NL. However, no studies as yet have examined the complete evolution of an active region's field inclination leading up to and after a flare, which is the aim of the research in this chapter.

In this Chapter, the evolution of magnetic field inclination across a magnetic NL is examined during a 21-hour period of observation in which a C-class solar flare occurred. The variation of inclination is investigated over $\sim14.5$~hours prior the flare, to search for possible pre-flare build-up, and over $\sim5$~hours after the flare ended, to test currently proposed field reconfiguration models due to flaring. The inclination is also examined spatially across the identified region of interest (as a function of distance from the NL), to pinpoint the exact location of greatest changes.

%%%%%%%%%%%%%%%%%%%%%%%%%%%%%%%%%%%%%%%%%%%%%%%%%%%%%%%%%%%%%%%%%%%%%%%%%%%%%%%%%%%%%%%%%%%%%%%%%%%%%%

\section{Observations and Data Analysis}
\label{NL:obs}

\begin{figure}[!t]
\centerline{\includegraphics[width=0.8\textwidth]{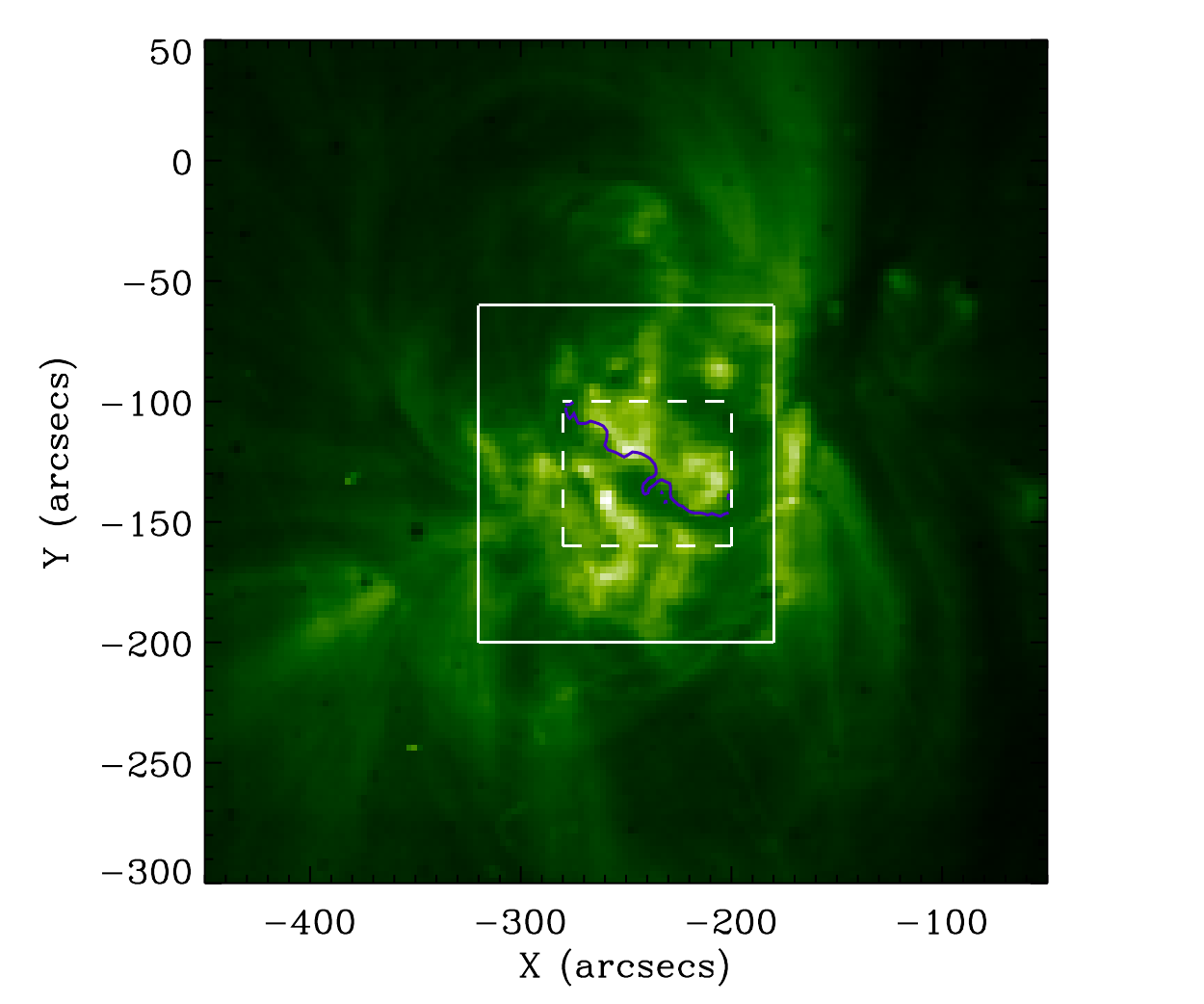}}

\

\centerline{\includegraphics[width=0.95\textwidth]{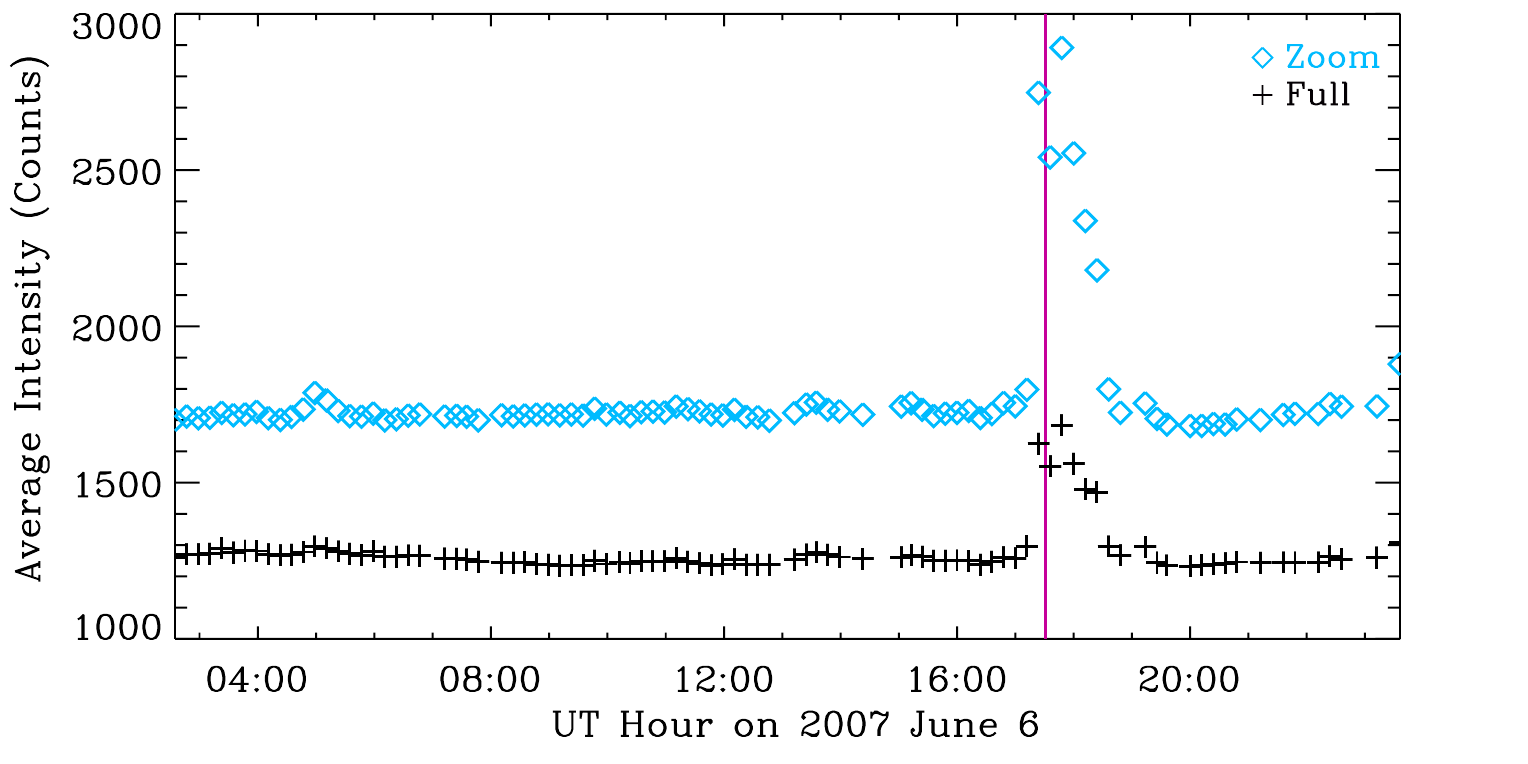}}
\caption[Upper panel shows SOHO/EIT 195\AA\ image of NOAA 10960 at the time of flaring, while lower panel shows corresponding lightcurves of the AR over a 21-hour period of observation.]{Upper panel shows SOHO/EIT 195~\AA\ image of NOAA 10960 at the time of flare peak, with the solid white box outlining the whole AR FOV, and dashed white box the zoomed-in FOV. Purple contours indicate SOHO/MDI LOS magnetic field strength at 0~G (i.e., the magnetic NL). Lower panel shows corresponding average intensity lightcurves for the full FOV (black crosses) and zoomed-in FOV (blue diamonds). The purple vertical line indicates the time of flare peak.}
\label{paper3:lightcurve}
\end{figure}

Active region NOAA 10960 crossed the solar disk from 2 -- 13 June 2007. A $\beta\gamma\delta$ Hale classification region, it was the source of several large solar flares, including a GOES C9.7 flare when it was close to disk centre on 2007 June 6 (peaking at 17:25~UT). Fast map observations from \emph{Hinode}/SOT-SP were obtained of the region, covering a 21-hour period before and after the flare (with no other flares during that time). To confirm no other flares occurred during this time period, SOHO/EIT lightcurves were obtained similar to Chapter~\ref{chapter:paper1}, and Figure~\ref{paper3:lightcurve} shows the results of this analysis. The upper panel shows a 195~\AA\ image from 17:11~UT (during the flare), with two FOVs outlined from which the lightcurves were obtained. The lower panel shows average values of 195~\AA\ intensity for the whole AR FOV (black crosses), as well as a zoomed-in FOV across the magnetic NL (blue diamonds) that will be used for analysis later. A significant increase in average intensity is observed at the time of flaring (purple vertical line), with no other large increases observed throughout the 21-hour period. The slight increase in average intensity observed at $\sim 05:00$~UT is due to instrumental noise.

\begin{figure}[!t]
\centerline{\includegraphics[width=0.9\textwidth]{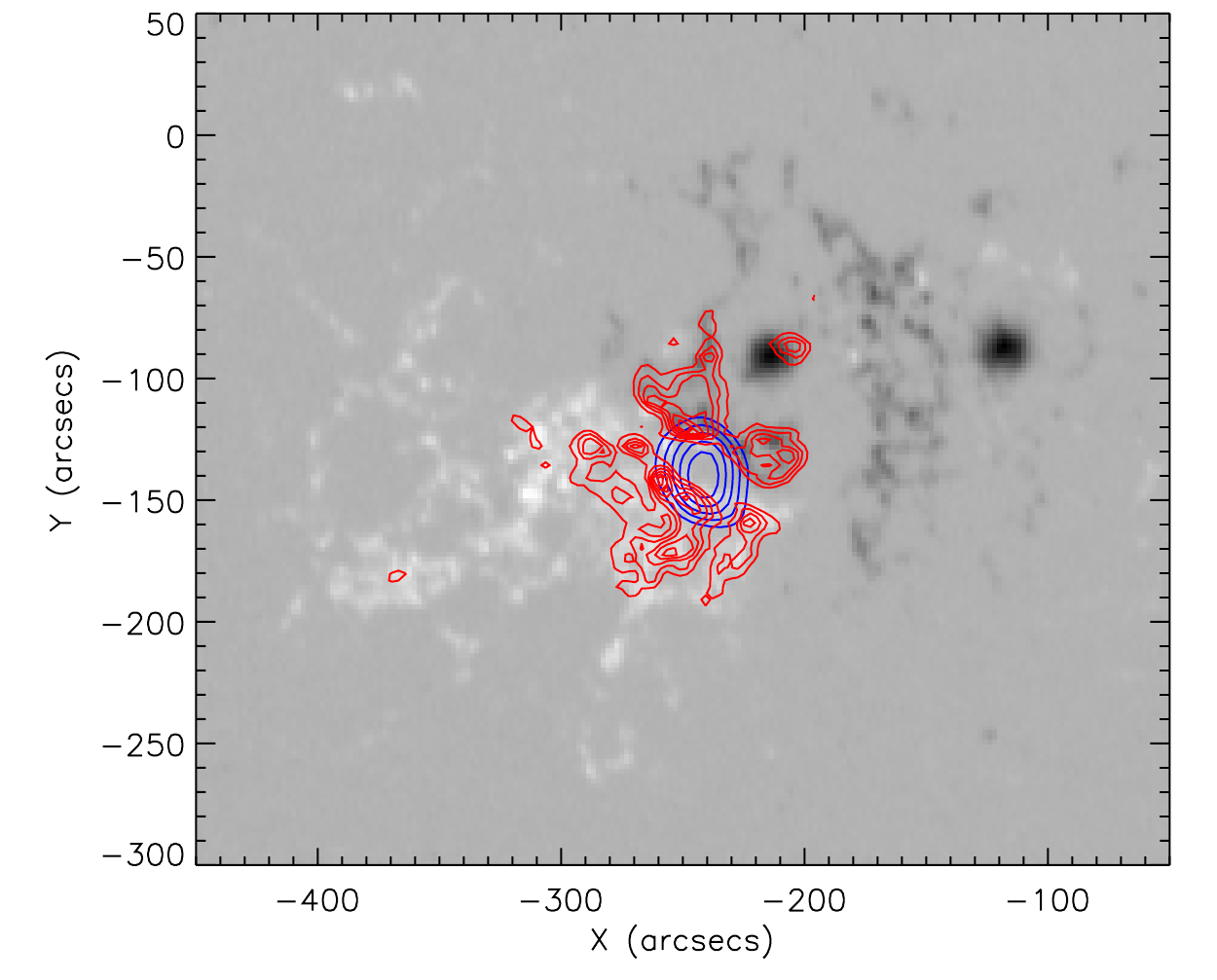}}
\caption[\emph{SOHO}/MDI magnetogram of the AR on 2007 June 6, with reconstructed RHESSI image and 195~\AA\ EUV peak contours overlayed.]{\emph{SOHO}/MDI magnetogram of the AR at 17:34~UT on 2007 June 6. Blue contours show the peak intensity of a reconstructed RHESSI image in the $6 - 12$~keV energy range. Red contours show the peak intensity of a SOHO/EIT 195~\AA\ image.}
\label{nl:rhessi}
\end{figure}

There were three SOT-SP scans before the flare and two afterwards, with each scan taking $\sim60$~minutes to record. Table~\ref{NL:scan_times} outlines the flare and scan times. The flare location outlined in Table~\ref{NL:scan_times} corresponds to a reconstructed RHESSI image peak, obtained using the RHESSI Quicklook Browser Interface. Figure~\ref{nl:rhessi} shows the peak intensity contours of the reconstructed RHESSI image and a SOHO/EIT 195~\AA\ image overlayed on a \emph{SOHO}/MDI magnetogram of the AR at 17:34~UT. The EUV brightening is concentrated both above and below the magnetic NL, with the RHESSI contours located along the NL itself.

\begin{table}
\centering
\caption{Summary of SOT-SP scan times on  2007 June 6.}
\begin{tabular}{ c c c c c }
\vspace{0.1cm} \\
\hline\hline
  Scan Number & Begin Time 	& End Time 	& Centre of FOV \\
  			& (UT)			&  (UT)		&    (Solar X, Solar Y)           \\
  \hline
  1                        & 02:39     	& 03:41	   	&  -374$''$, -125$''$\\ 
  2                        & 07:34     	& 08:37	   	&  -331$''$, -125$''$\\ 
  3 			& 12:30     	& 13:32	  	&  -292$''$, -125$''$\\ 
Flare	& 16:55	 	& 17:35		&   -$236''$, -$140''$\\ 
  4 			& 19:04	 	& 20:06	 	&  -234$''$, -126$''$\\
  5 			& 22:22     	& 23:23	  	&  -207$''$, -126$''$\\ 
  \hline\hline
  \end{tabular}
  \label{NL:scan_times}
\end{table}

Photospheric vector magnetic field information was obtained using the \helix\ atmospheric inversion code. A one-magnetic-component model atmosphere was used similar to Section~\ref{section:paper1}, with a local straylight component included. Calculated values of polarisation threshold are outlined in Table~\ref{nl:polarisation} (see Section~\ref{paper1:observations} for a description), with an average quiet region value of $P \sim3.5 \times 10^{-3}~{I_c}$, i.e., units of continuum intensity. The 180$^\circ$ azimuthal ambiguity was resolved using the \textsf{AMBIG} code, using a transverse field strength threshold of 150~G. All resulting magnetic field parameters were converted from the image plane to the heliographic plane using the method described in Section~\ref{rad_trans:coords}, and co-aligned to the third pre-flare scan for further analysis.

\begin{table}[!t]
\centering
\caption[Values of total polarisation in quiet and active (umbra) regions for all five SOT-SP scans.]{Values of total polarisation in quiet and active (umbra) regions for all five SOT-SP scans. For a slice along the third scan 600$^\mathrm{{th}}$ pixel on the x axis, the 260$^\mathrm{{th}}$ and 490$^\mathrm{{th}}$ pixels along the y axis were chosen for the active and quiet region samples, respectively. }
\begin{tabular}{ c c c c c }
\vspace{0.1cm} \\
\hline\hline
  Scan Number 	& $P_{quiet}$ 		& $P_{active}$ 	 \\
  				& $\times 10 ^{-3}$~$I_c$			&  $\times 10 ^{-3}$~$I_c$			           \\
  \hline
  1                        & 4.06    		& 85.83	   	\\ 
  2                        & 2.38    		& 88.30	   	\\ 
  3 			& 4.18     		& 83.49	  	\\ 
  4 			& 2.81	 	& 67.98	 	\\
  5 			& 3.96	 	& 81.82	 	\\
  \hline\hline
  \end{tabular}
  \label{nl:polarisation}
\end{table}

\begin{sidewaysfigure}
\centering
\includegraphics[width=\textwidth]{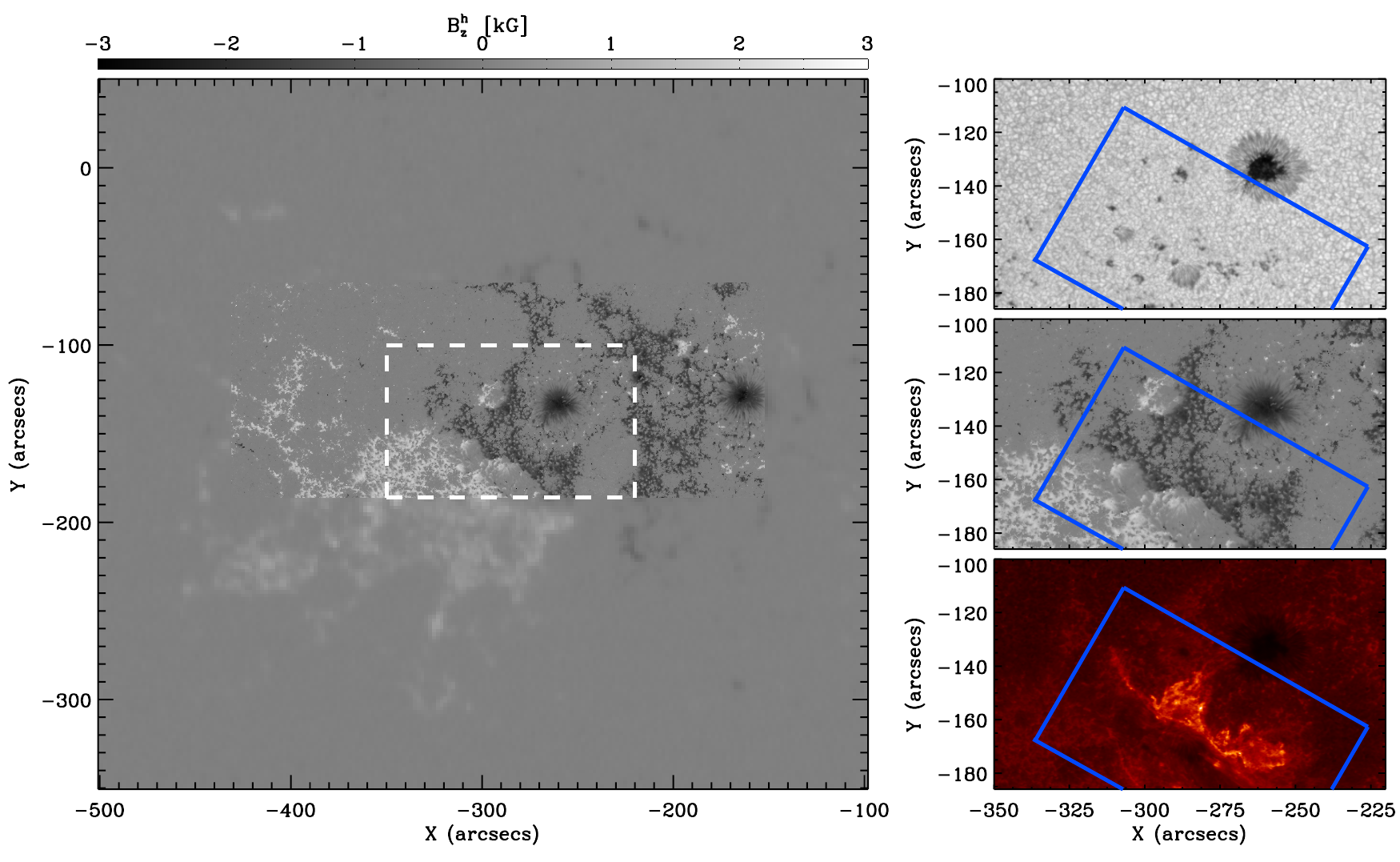}
\caption[Context image showing the active region at the time of the third scan (and  Ca~\textsc{ii}~H intensity at flare peak.)]{Context image showing the active region at the time of the third scan (and  Ca~\textsc{ii}~H intensity at flare peak, 17:24~UT). The left column shows MDI magnetogram, with \emph{Hinode} vector magnetogram overlayed. The dashed white box outlines the FOV shown in the right column: continuum intensity (upper row), vertical field strength (middle) and Ca~\textsc{ii}~H intensity (lower). The blue boxes outline the FOV used for further analysis in the area of greatest chromospheric flare brightening.}
\label{nl:context}
\end{sidewaysfigure}

The left panel of Figure~\ref{nl:context} shows a SOHO/MDI magnetogram of the active region at the time of the third scan, with the corresponding Hinode vector magnetogram overlayed. Ca~\textsc{ii}~H images (3968~\AA) of the active region at the time of flare peak were obtained from the SOT-BFI. A large amount of chromospheric flare brightening was identified at flare peak in an area of negative polarity to the solar south east of the left sunspot, near the magnetic NL with positive polarity plage below it.  A white dashed box outlines the general location where this brightening occurs, and can be seen clearer in a zoomed-in Ca~\textsc{ii}~H image of this area in the lower right panel of Figure~\ref{nl:context}. Zoom-ins of continuum intensity (upper right panel) and vertical field (middle right panel) are also presented. 

\begin{figure}
\centerline{\includegraphics[width=\textwidth]{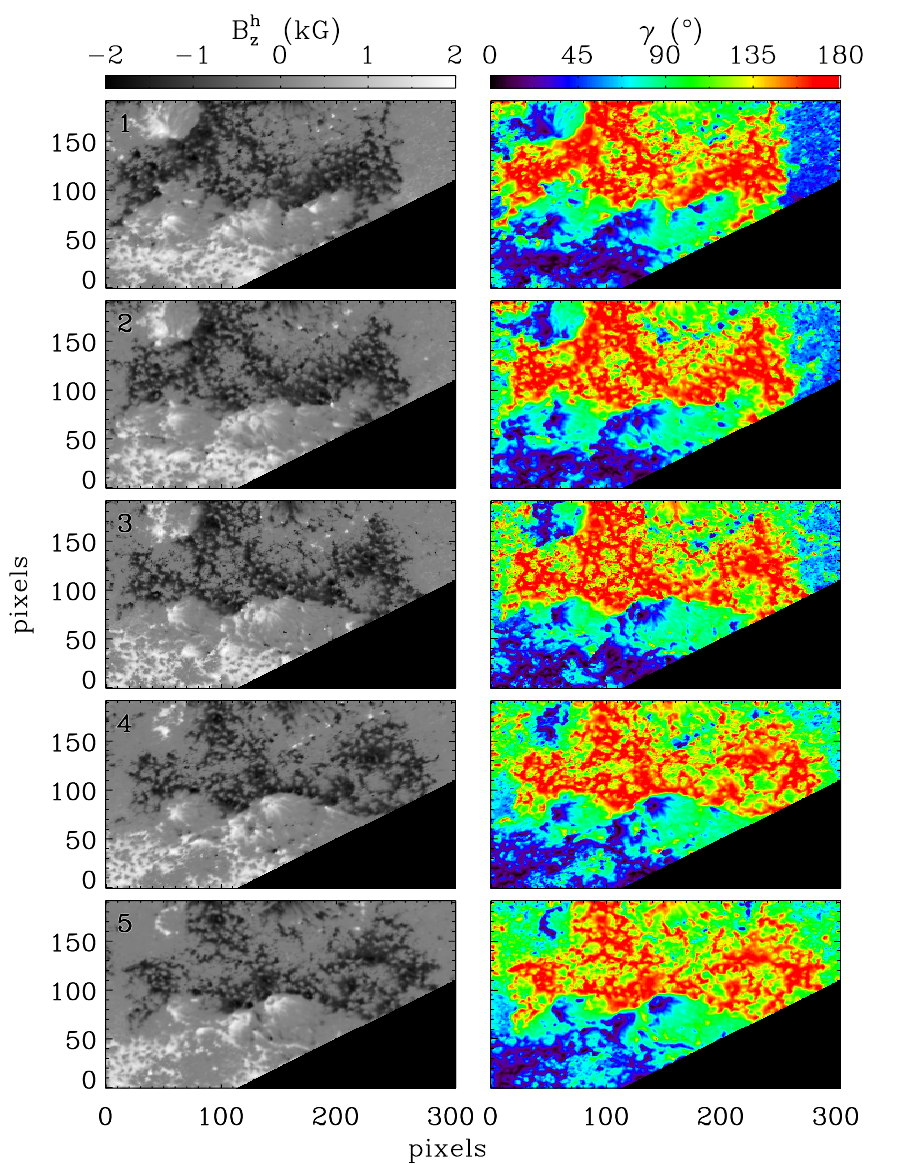}}
\caption[Zoomed-in rotated FOV to be used for analysis: vertical field ($\mathrm{B_{z}^{h}}$) is shown in the left column, and inclination angle ($\gamma$) in the right column.]{Zoomed-in rotated FOV to be used for analysis: vertical field ($\mathrm{B_{z}^{h}}$) is shown in the left column, and inclination angle ($\gamma$) in the right column. Upper to lower rows show all five scan times in order (see labels in left column). The black areas in the lower right of the panels indicate no data.}
\label{nl:scan_image}
\end{figure}

This region of flare brightening was deemed worthy of further investigation; the blue boxes in the right column of Figure~\ref{nl:context} identify a more specific zoomed-in FOV that is analysed further. Note that the main sunspot was avoided. Figure~\ref{nl:scan_image} shows a sample of the zoomed-in rotated FOV for vertical field (left column) and inclination (right column), for all five scan times. The main negative polarity region (black colour in left column) dominates the FOV, with vertically inclined field (red colour in the right column). The positive polarity plage region (white colour in left column) in the bottom left of the FOV has the opposite inclination (blue colour in right column), as would be expected from an opposite polarity. Unfortunately the positive polarity region is not shown fully due to a lack of SOT-SP data here. There is also a smaller region of positive polarity to the top right of the FOV. The magnetic NL lies horizontally in these rotated images, at a location of about the 80$^{\mathrm{th}}$ pixel on the y-axis. It is hard to determine by eye any clear evolution of these parameters over time, therefore the region was investigated further using a number of analysis techniques that will be presented in the next section.

%%%%%%%%%%%%%%%%%%%%%%%%%%%%%%%%%%%%%%%%%%%%%%%%%%%%%%%%%%%%%%%%%%%%%%%%%%%%%%%%%%%%%%%%%%%%%%%%%%%%%%

\section{Results}
\label{NL:res}

\begin{sidewaysfigure}
\centering
\includegraphics[width=\textwidth]{figure_1.pdf}
\caption[In the left column, the chosen FOV from third scan for vertical field (upper row), inclination angle (middle row), and horizontal field (lower row) is shown. Corresponding macropixels are shown in the middle column, and standard deviation in the right column.]{In the left column, the chosen FOV from third scan for vertical field (upper row), inclination angle (middle row), and horizontal field (lower row) is shown. Corresponding macropixels are shown in the middle column, and standard deviation in the right column. The horizontal lines indicate the NL zone. The green contour in the upper left panel outlines the significant Ca~\textsc{ii}~H flare brightening (at the 600~DN level) observed at 17:24~UT.}
\label{nl:figure_1}
\end{sidewaysfigure}

\begin{figure}[!t]
\centerline{\includegraphics[width=\textwidth]{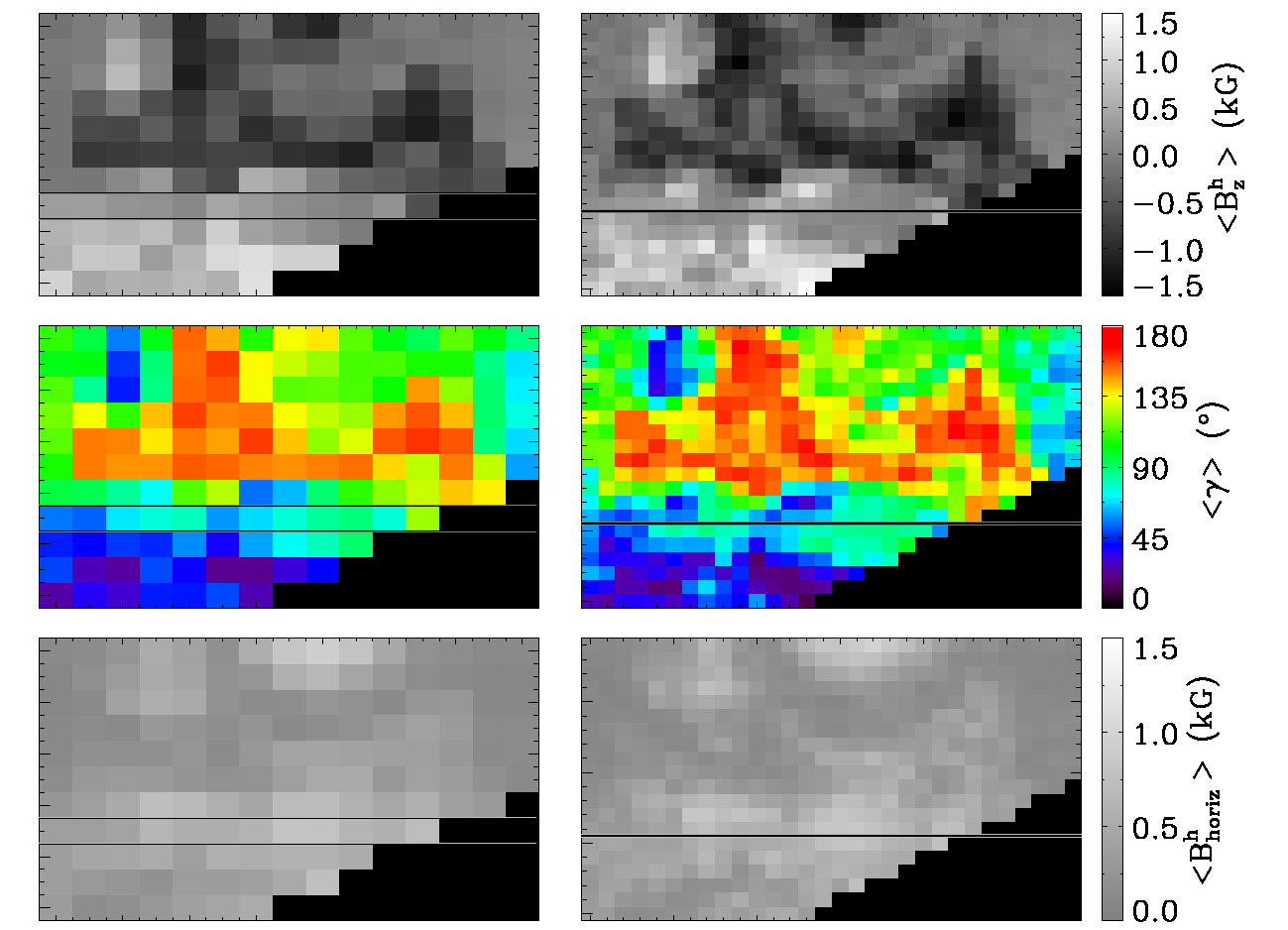}}
\caption[Comparison macropixel size for the third pre-flare scan in vertical field, inclination angle, and horizontal field.]{Comparison of macropixel size for the third pre-flare scan in vertical field (upper row), inclination angle (middle row), and horizontal field (lower row). Macropixels of $20\times20$~pixels in size are shown in the left column, and macropixels of $10\times10$~pixels in size are shown in the right column.}
\label{nl:macropixels}
\end{figure}

The FOV selected for further study was manually rotated by $30^\circ$ anti-clockwise for analysis purposes, as shown in the left column of Figure~\ref{nl:figure_1} (with vertical field, inclination angle, and horizontal field shown for the third scan). Although all five scans are co-aligned to the third pre-flare scan, the negative and positive polarity regions both exhibit small features that evolve spatially with time (as can be seen in Figure~\ref{nl:scan_image}). This evolution presents a temporal pixel-by-pixel comparison, so to examine differences over time larger areas are needed. Thus, macropixels of $16\times16$~pixels in size were created, with each pixel containing the average value of the parameter in that $16\times16$~pixel area, as is shown in the second column of Figure~\ref{nl:figure_1}. Note that other macropixel sizes were examined before this particular size was selected, and Figure~\ref{nl:macropixels} shows examples of macropixels of $20\times20$~pixels (left column) and $10\times10$~pixels (right column) in size. However, after careful examination of the evolution of the region in these three sizes, $20\times20$~pixels was deemed too coarse, and $10\times10$~pixels too fine, for proper analysis, thus a size of $16\times16$~pixels was selected. A zone for the NL was also defined, by finding the area that closest represents zero vertical field along the x-direction (delineated by the horizontal bars in Figure~\ref{nl:figure_1}). Inclination angle was selected to be investigated further over the course of the observation period, the results of which are discussed in this Section.

\begin{figure}[!t]
\centering
\includegraphics[width=\textwidth]{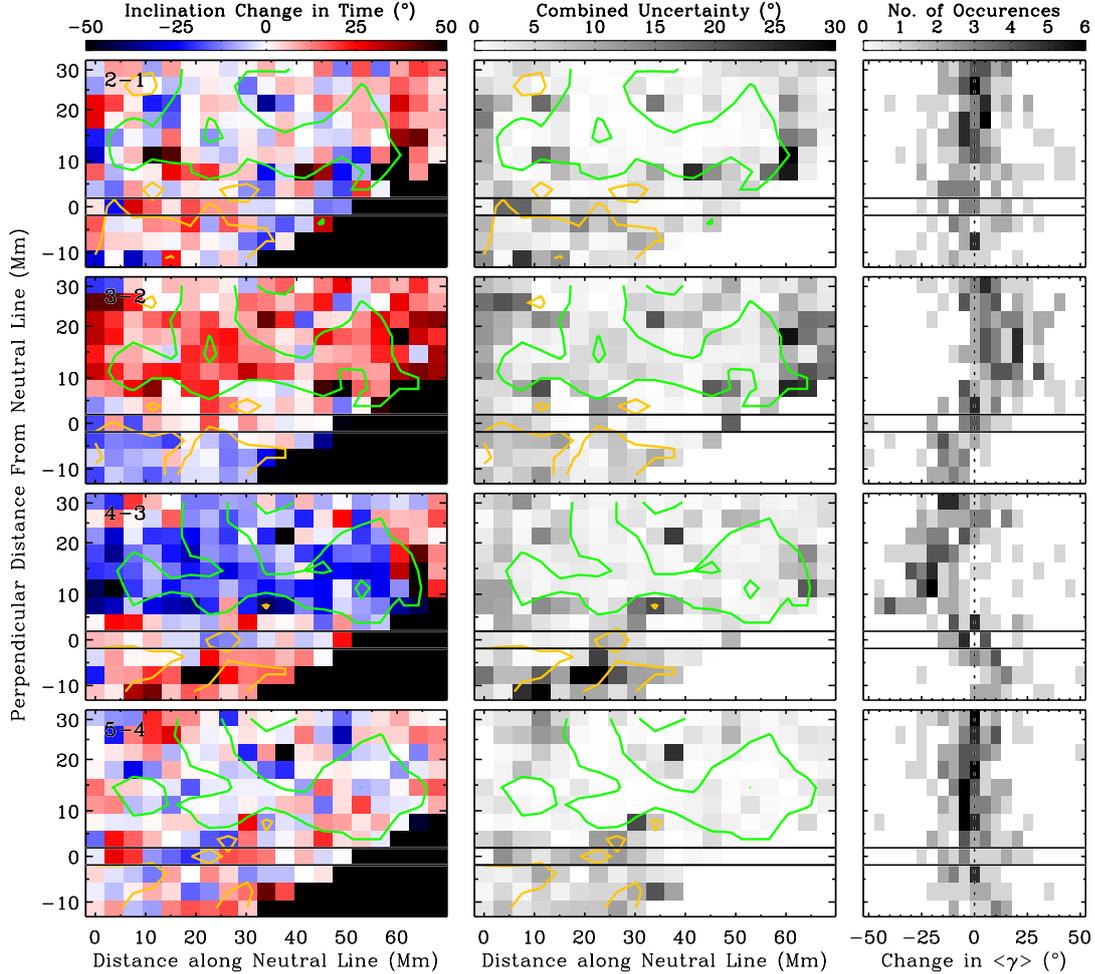}
\caption[Macropixel average inclination angle difference 2D spatio temporal evolution is shown in the left column, with corresponding combined uncertainty in the middle column. Also shown is the 1D spatial distribution evolution (at varying NL distance separation) in the right column.]{Macropixel average inclination angle difference 2D spatio temporal evolution is shown in the left column, with corresponding combined uncertainty in the middle column. Also shown is the 1D spatial distribution evolution (at varying NL distance separation) in the right column. Green and orange contours show macropixel vertical field strength at $-~500$~G and $+~500$~G respectively, and horizontal lines delineate the NL zone. Note in the left column that the red (blue) colour above the NL indicates the field is more vertical (horizontal) in the leading scan, the opposite being the case below the NL.}
\label{nl:figure_2}
\end{figure}

\subsection{Inclination Angle Differences}
In order to investigate the evolution of the field inclination across the observation period, the results for each scan must be differenced. The left column of Figure~\ref{nl:figure_2} shows the calculated inclination angle differences, i.e., macropixel average inclination in scan 1 subtracted from macropixel average inclination in scan~2, etc. The combined uncertainties corresponding to the combined macropixel standard deviations are shown in the middle column. Note that the red (positive values) and blue (negative values) colours above the NL indicate the field becoming more vertical and horizontal respectively (with the opposite being the case below the NL). Histograms are presented in the right column, showing the spread of inclination angle values along the x-direction for each macropixel value along the y-axis. The upper row shows inclination difference for Scans [2 $-$ 1], i.e., a quiet pre-flare period. There is no clear trend found in the plots, with perhaps a slight tendency of field orientation towards more vertical values in the histogram. Similar results are observed in the quiet post-flare plots of Scans [5 $-$ 4] (lower row), with an even a lower spread of inclination differences both above and below the NL.

Larger magnitude changes are found in the immediate pre- and post- flare rows, as compared to the quiet period rows. The calculated inclination difference for Scans [3 $-$ 2] (second row of Figure~\ref{nl:figure_2}) clearly shows the field configuration becoming more vertical between $\sim8.5-4.5$~hours before the flare (see left column; mainly red above NL zone, blue below). Note the largest uncertainty values are located towards the right of the FOV, where there is little negative polarity flux. The corresponding histogram confirms this change towards the vertical, with the clearest change in macropixels above the NL. The Scans [4 $-$ 3] plots (third row of Figure~\ref{nl:figure_2}) show a change in field configuration towards the horizontal (left column; mainly blue above the NL, and somewhat red below), confirmed by the trend shown by the corresponding histogram. The spurious positive values in the histogram above the NL mostly originate in the region of increased uncertainty to the right of the FOV. There is also an increased uncertainty found below the NL (see middle column). Note that scan 3 was recorded $\sim4.5-3.5$~hours before the flare, and scan 4 was recorded $\sim1.5-2.5$~hours after the flare. Therefore it can be said that the more horizontal configuration may be due to the flare itself (within the $\sim 2$ hour time range).

\begin{figure}[!t]
\centerline{\includegraphics[width=0.75\textwidth]{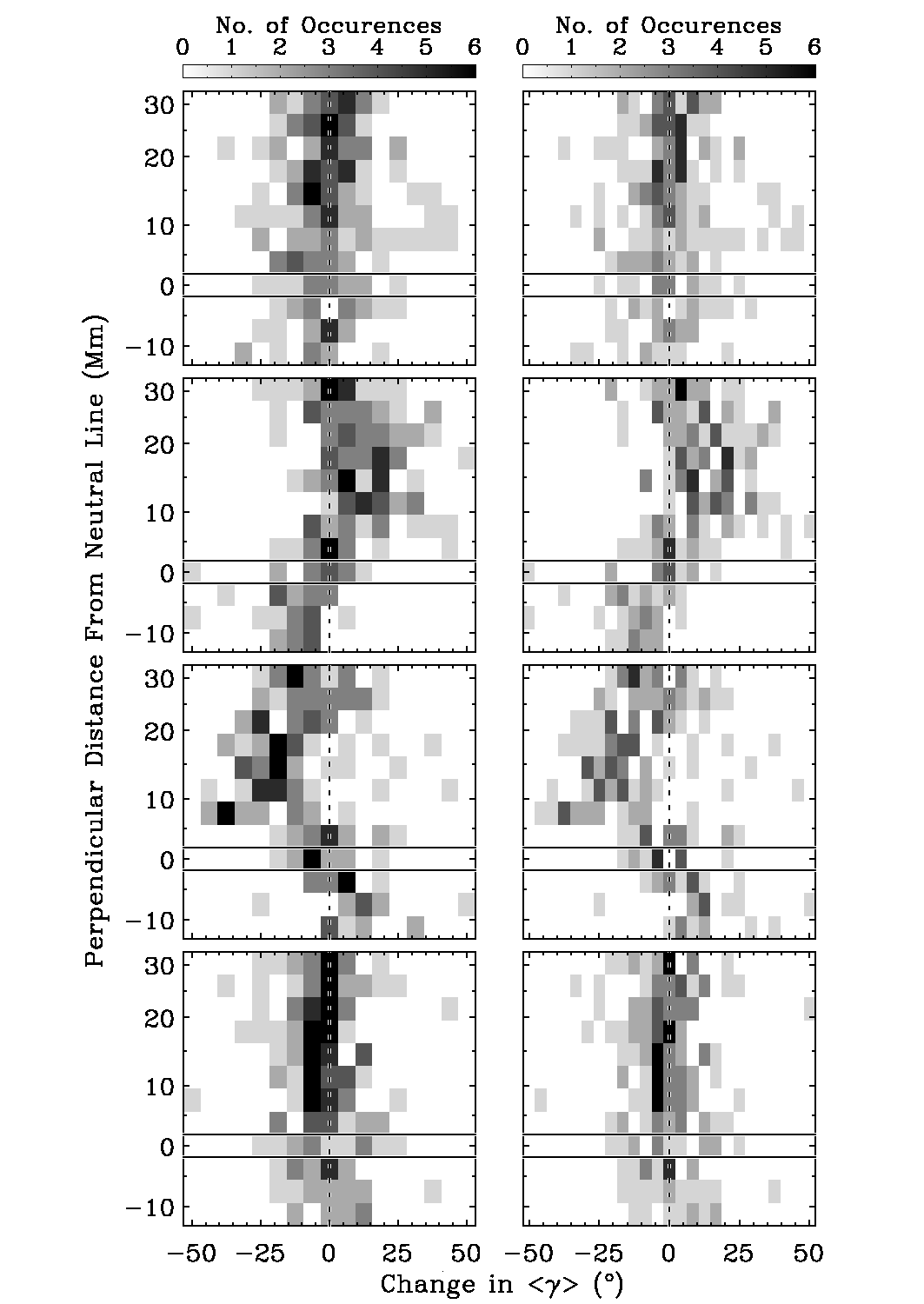}}
\caption[Histograms showing the spread of inclination angle values along the x-direction for varying bin sizes.]{Histograms showing the spread of inclination angle values along the x-direction for each macropixel value along the y-axis. A bin size of 6$^\circ$ was used in the left column, and of 4$^\circ$ in the right column.}
\label{nl:histograms}
\end{figure}

Note that a bin size of 5$^\circ$ was selected for the histograms shown in the right column of Figure~\ref{nl:figure_2}, as the changes in inclination over time are most clearly observed at this bin size. However, for consistency, Figure~\ref{nl:histograms} shows the same results at two other bin sizes: 6$^\circ$ (left column) and 4$^\circ$ (right column). Similar trends are observed for both of these bin sizes compared to the bin size of 5$^\circ$.  However, outlying values are more apparent with a bin size of 6$^\circ$, especially in the Scans [2 - 1] plots, showing a stronger tendency towards the vertical here. Nevertheless, there is a clear change in field configuration to a more vertical orientation before the flare (particularly above the NL), and a more horizontal configuration is observed after the flare has ended.

\subsection{Spatial Variation near the Magnetic NL}
\begin{figure}[!t]
\centering
\includegraphics[width=0.9\textwidth]{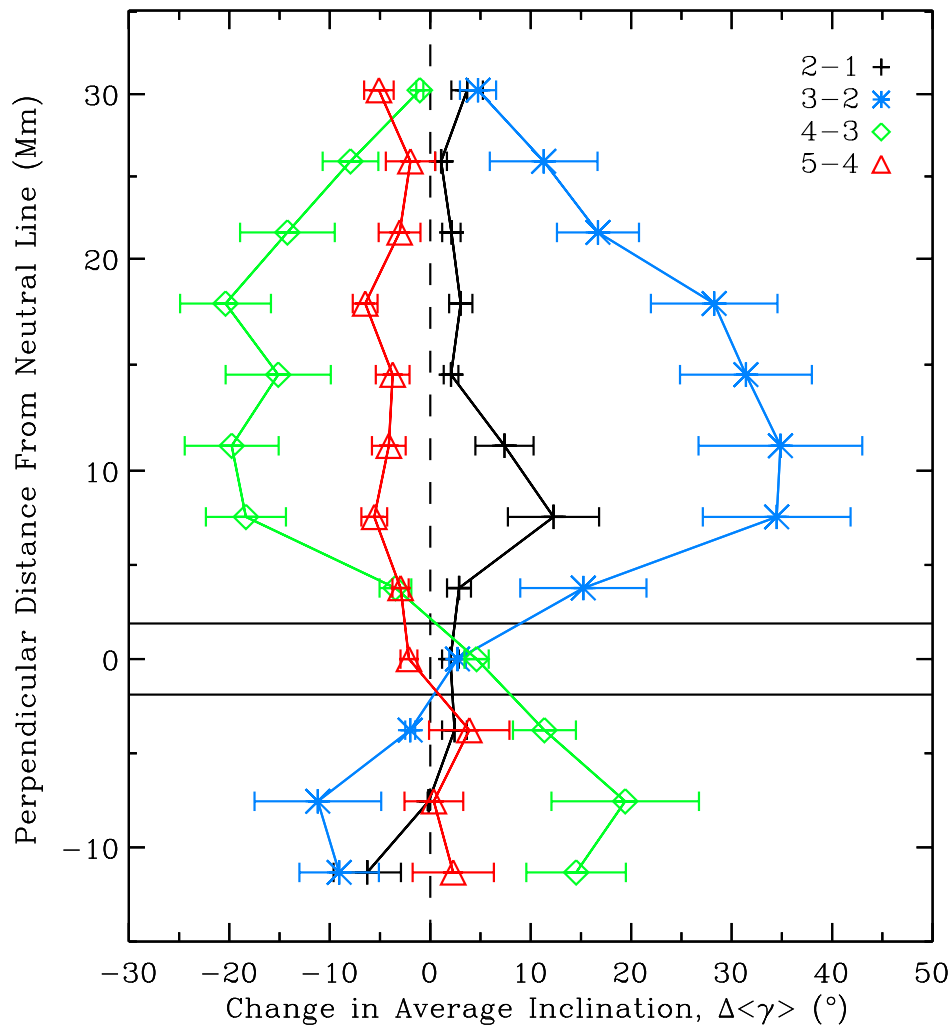}
\caption[Difference in average inclination of all pixels along the spatial x-axis direction (along NL) for each y-axis location in Figure~\ref{nl:figure_2}.]{Difference in average inclination of all pixels along the spatial x-axis direction (along NL) for each y-axis location in Figure~\ref{nl:figure_2}. Scan time differences are coloured as per the legend. The vertical dashed line indicates zero average inclination difference, and horizontal bars outline the borders of the NL zone. Error bars indicate the standard deviation of the values in the spatial x-direction along the NL.}
\label{nl:figure_3}
\end{figure}

In order to examine the changes found as a function of distance from the NL in more detail, all inclination angle difference values along the x direction were averaged for each y-axis location, the results of which are shown in Figure~\ref{nl:figure_3}.  Here, error bars indicate the standard deviation of values in spatial direction along the NL. This figure confirms the field configuration changes observed in previous figures, with stronger changes clearly occurring in macropixels above the NL. The quiet pre- and post- flare curves are close to the dashed vertical line at $0^\circ$, with only a slight increase towards the vertical above the NL between scans 1 and 2. The field becomes much more vertical before the flare (i.e., between scans 2 and 3) by $\sim30^\circ$ above the NL at its peak, and by $\sim10^\circ$ below the NL. In contrast, the field becomes more horizontal between scans 3 and 4 by $\sim20^\circ$ above the NL, and by $\sim10^\circ$ below the NL. Note that these changes are from macropixels that do contain some component of field evolution and horizontal motion. As a result, absolute inclination changes for discrete fields may be smaller in magnitude.

\begin{figure}[!t]
\centering
\includegraphics[width=0.65\textwidth]{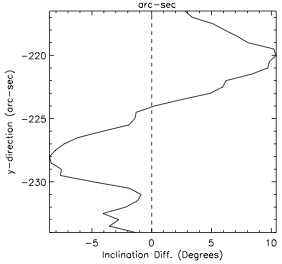}
\caption[A mean profile of the change in inclination angle a region near a polarity inversion line \citep{gosain12}.]{A mean profile of the change in inclination angle within a region near a polarity inversion line (averaged along the abscissa, x-direction) \citep{gosain12}.
}
\label{nl:gosain}
\end{figure}

Note that Figure~\ref{nl:figure_3} is similar to Figure 6 of \citet{gosain12}, which is reproduced in Figure~\ref{nl:gosain}. This shows the mean profile of the change in the inclination angle within a region across the NL. They similarly find the field configuration close to the magnetic NL becoming more horizontal due to a solar flare. However, \citeauthor{gosain12} only differences two times before and after a flare, rather than the more complete evolution shown in this chapter. Considering that the changes in inclination found due to the flare in this research are similar to previous work, the strong changes observed only hours before the flare are also likely to be accurate. 

It is also worth noting that there is a clear trend in Figure~\ref{nl:figure_3} of larger differences in average inclination closer to the NL. The magnitude of average inclination increase and decrease reaches a peak within $\sim 5-10$~Mm of the NL zone, where the negative polarity region begins (and where chromospheric flare brightening is observed in Figure~\ref{nl:figure_1}). The magnitude of the inclination change with time then drops off at larger perpendicular distances from the NL, which would be expected if the flare source is located close to the NL.

%%%%%%%%%%%%%%%%%%%%%%%%%%%%%%%%%%%%%%%%%%%%%%%%%%%%%%%%%%%%%%%%%%%%%%%%%%%%%%%%%%%%%%%%%%%%%%%%%%%%%%

\section{Discussion}
\label{NL:concs}

The results of this chapter have shown a complete picture of the change in field inclination over a 21-hour period before and after a C-class flare event. Only slight changes are found in the periods between $\sim13.5-8.5$~hours before (i.e., between scans 1 and 2) and $\sim2.5-5$~hours after (i.e., between scans 4 and 5) the flare, as is to be expected from a quiet period with no flare activity. There is a slight `bump' of the field close to the NL orienting towards the vertical between scan 1 and 2 (see Figure~\ref{nl:figure_3}), suggesting that some of the pre-flare build-up may begin around this time. The inclination differences between scan 4 and scan 5 closely resemble those between scan 1 and scan 2, being situated near zero in average inclination difference. This suggests that, within $\sim 2.5-5$~hours after flaring, the field configuration has already returned to a `quiet' state.

Large changes in inclination angle across the magnetic NL have been found on short timescales during the period of flaring studied in this research. Previous work has also found the field becoming more horizontal after flaring, e.g., \citet{wang12} found that the mean transverse field strength increases by $\sim90$~G about 1 hour after an M-class flare, with a decrease in inclination angle of $\sim3^{\circ}$ towards the horizontal. \citet{gosain12} found an inward collapse of field lines toward the neutral line by $\sim10^{\circ}$ within $\sim15$~minutes of the start time of an X-class flare. Here, we find a greater change in inclination, with the field becoming more horizontal by $\sim20^\circ$ (maximum) between $\sim1.5-2.5$~hours after a smaller C-class flare event. It is worth noting that the results found by \citeauthor{wang12} and \citeauthor{gosain12}, and as those observed in this chapter, are all consistent with the `coronal implosion' scenario of \citet{hudson08}. As mentioned in Chapter~\ref{chapter:paper1}, \citeauthor{hudson08} predicted that photospheric magnetic fields close to the NL would become more horizontal in a simple flare-restructuring model.

In addition to finding the field becoming more horizontal after the flare, the field was also observed to become more vertical beforehand. The change in inclination angle towards vertical occurs on a relatively short timescale of sometime between $\sim8.5-4.5$~hours before the GOES-defined beginning of the flare. A short timescale of pre-flare inclination changes on the order of hours was previously observed in the results of Chapter~\ref{section:paper1} for a lower-magnitude (B-class) flare. However, most previous work has focused on more longer-term changes before flaring, on the order of days \citep{Schrijver07}. The large increase of $\sim30^\circ$ found here shows the importance of observing magnetic field parameters on these shorter timescales before flaring, as they could be useful for flare forecasting methods in the future. 

It is interesting to note that the changes in inclination angle before and after the flare are greater than those observed for the event studied in Chapter~\ref{section:paper1}. A change of $\sim8^\circ$ before and due the flare was found previously, compared to $\sim20^\circ$ before and $\sim30^\circ$ due to the flare studied in this Chapter. It is worth noting the difference in events studied that could explain these magnitude differences. The event studied in Chapter~\ref{section:paper1} is a small B-class flare occurring outside a simple $\beta$ region consisting of a main spot with opposite trailing plage. Here, a larger FOV for a C-class event is studied in a more complex $\beta\gamma\delta$ sunspot region. It is likely the field configurations of the two different regions differ, with stronger changes occurring in the more complex AR before and due to a larger magnitude flare. Although the field structures likely differ, it is promising that the same kind of re-configuration is observed, with the field inclination becoming more vertical before both flare events, and then becoming more horizontal due to the flares themselves.

It is clear from Figure~\ref{nl:figure_3} that the greatest changes both before and after the flare can be found closest to the NL, with the magnitude of change decreasing as a function of perpendicular distance from the NL. This is to be expected considering the widely held belief that the source of a flare is likely located close to a magnetic NL \citep{moore85,solanki03}. Magnetic reconnection is favourable across a NL location, as was mentioned in Section~\ref{intro:flaring_locations}. However, with chromospheric flare brightening occuring both above and below the NL (as seen in the upper left panel of Figure~\ref{nl:figure_1}), it is unfortunate there is a lack of SOT-SP data for the positive polarity plage region below the NL. The changes found below the NL could in fact be greater than observed, and any results here for macropixels below the NL are likely an underestimation (have a larger uncertainty). However, the region of Ca~\textsc{ii}~H flare brightening located just above the NL matches closely with the largest change in values observed between $\sim10-25~\mathrm{Mm}$ from the NL. 

The research in this chapter has shown the first results of the full evolution of field inclination over 21 hours as a function of spatial location from the NL over the course of a solar flare. So far in this thesis, only the photospheric magnetic field has been studied. The next step is to examine the coronal magnetic field, in order to gain a deeper understanding of short-term evolution of AR magnetic fields during flaring periods. In the next chapter, magnetic field extrapolations are used in order to explore any changes in the 3D coronal magnetic field before and after the flare studied in Chapter~\ref{chapter:paper1}. The event studied in this chapter was not used due to its limited FOV.

		% reserch 3

%\include{1_introduction/mhd}	% mhd chapter

% this file is called up by thesis.tex
% content in this file will be fed into the main document

%: ----------------------- name of chapter  -------------------------
\chapter{Coronal Magnetic Geometry and Energy Evolution} % top level followed by section, subsection
\label{chapter:3D}

\ifpdf
    \graphicspath{{6/figures/PNG/}{6/figures/PDF/}{6/figures/}}
\else
    \graphicspath{{6/figures/.pdf/}{6/figures/}}
\fi
\hrule height 1mm
\vspace{0.5mm}
\hrule height 0.4mm 
\noindent 
\\ {\it In this chapter, the same event examined in Chapter~\ref{chapter:paper1} is also looked at here, i.e., the evolution of the magnetic field in NOAA region 10953 leading up to and after a GOES B1.0 flare. Chapter~\ref{chapter:paper1} described pre- and post- flare changes in photospheric vector magnetic field parameters of flux elements outside the primary sunspot. Here, 3D geometry is investigated using potential, linear force-free, and non-linear force-free field extrapolations in order to fully understand the evolution of the field lines. The deviation of the non-linear field configuration from that of the linear and potential configurations is investigated, as well as a study of the free energy available leading up to and after a flare. The research described in this chapter has been published in Murray, et al, \aap, 2013. 
\\  
}
\hrule height 1mm
\vspace{0.5mm}
\hrule height 0.4mm 

\newpage

%%%%%%%%%%%%%%%%%%%%%%%%%%%%%%%%%%%%%%%%%%%%%%%%%%%%%%%%%%%%%%%%%%%%%%%%%%%%%%%%%%%%%%%%%%%%%%%%%%%%%%%%%%%%%%%%%%%%%%%%%%%%%%%%%%%%%%%%%%%%%%%%

\section{Introduction}
\label{3D:intro}

As described in Chapter~\ref{chapter:introduction}, solar flares occur when energy stored in AR magnetic fields is suddenly released, with the stored magnetic energy being converted to kinetic energy of energetic particles, mass motions, and radiation emitted across the entire electromagnetic spectrum \citep[see, e.g.,][]{fletcher11}. Energies of large flares can be as high as $\sim10^{32}$~ergs, over short time scales of tens of minutes. Section~\ref{intro:complexity} mentioned that the field configuration of ARs is believed to be linked to the likelihood of flaring, and that flare triggers are often associated with flux emergence and increased shear and twist in the field \citep[see review by][and references therein]{priest02}. However, the processes involved in magnetic energy storage and release within ARs are still not fully understood, and in this chapter the aim is to study these processes in more detail. 

Energy release is expected to occur in the corona, but magnetic field observations are usually based in the photosphere. In order to investigate the coronal magnetic field, photospheric measurements are used as a starting point for 3D magnetic field extrapolations \citep[Chapter~\ref{chapter:theory} here, and][]{gary89}. Although there are inaccuracies involved with the various assumptions that must be made to obtain 3D extrapolations, much progress has been made in recent years with high-resolution photospheric magnetic field measurements now available from spacecraft such as $Hinode$ and \emph{SDO}.

In this chapter, the evolution of the 3D coronal magnetic field is investigated in an AR using the three types of extrapolation procedure described in Section~\ref{mhd:extraps}: potential, linear force free (LFF), and non-linear force free (NLFF). The corona is generally considered to be force free, and all three types of extrapolation assume the force-free approximation (Equation~\ref{3D:force_free_eqn}). The three generalised forms of this relation (Equations~\ref{potential},~\ref{lff},~\ref{nlff}) describe the three types of extrapolation. To summarise these equations, for a potential field the force-free parameter $\alpha$ is zero (no currents), for a LFF field $\alpha$ is non-zero but constant throughout a given volume, and for a NLFF field $\alpha$ is allowed to vary spatially (differing from field line to field line, but constant along one field line). Vector-magnetic field information was obtained in Chapter~\ref{chapter:paper1}, and thus all three types of extrapolation can be investigated here (as previously mentioned, vector-magnetic fields are required for NLFF extrapolations).

The underlying principle of force-free coronal fields has previously been extensively studied. Initially, \citet{woltjer58} theoretically examined a magnetic field configuration over the course of magnetic energy release, proposing that force-free fields with constant $\alpha$ (LFF) represent the state of lowest magnetic energy in a closed system. \citet{taylor86} applied this theory to laboratory plasma experiments, suggesting that the total magnetic helicity of a flux system is invariant during the relaxation process to this minimum-energy state. This theory and experimental evidence led to the concept that the free magnetic energy that may be released during field relaxation in a solar AR is the excess energy above the LFF field with the same magnetic helicity \citep{heyvaerts84}. This relaxation process will henceforth be referred to as Taylor relaxation throughout the rest of this thesis. 

\citeauthor{taylor86} considered a plasma with low $\beta$, enclosed in a simply-connected, perfectly conducting surface, surrounding a volume $V$. Supposing the plasma is initially not in equilibrium, (possibly undergoing magnetic reconnection), it is assumed all the magnetic energy in this isolated configuration is free energy, meaning energy that can be extracted from the system as work \citep{raastad09}. When the plasma configuration expels all its free energy then it is at the point of minimum energy,   i.e., the Taylor state. \citeauthor{taylor86}'s derivation began with the equation for perfectly conducting fluid variations in a magnetic field, $\partial \mathbf{B}/ \partial t - \nabla \times (\mathbf{v} \times \mathbf{B}) = 0$, where $\mathbf{v}$ is the fluid velocity. \citeauthor{taylor86} then derived helicity as a quantity that needs to be conserved \citep{bellan00},
\begin{equation}
\label{paper2:helic_def}
H~=~\int_{V} \mathbf{A} \cdot \mathbf{B}~ dV ~,
\end{equation}
where $\mathbf{A}$ is the vector potential of $\mathbf{B}$, such that $\mathbf{B} = \nabla \times \mathbf{A}$. A differential equation was then derived using global helicity to minimise energy inside the system,
\begin{equation}
\nabla \times \mathbf{B}~=~\lambda \mathbf{B}~,
\end{equation}
where $\lambda$ is an eigenvalue that determines the exact force-free configuration of the plasma \citep{taylor74}, and can be conceptually thought of as $E/H$ of the isolated configuration \citep{raastad09}, where $E$ is magnetic energy, and $H$ helicity as defined in Equation~\ref{paper2:helic_def}. Several theoretical, numerical, and observational studies have investigated whether Taylor relaxation occurs during solar flares, with some agreeing to its presence (e.g., \citealp{nandy03}; \citealp{browning08}) and others disagreeing \citep[e.g.,][]{amari00, bleybel02}. These studies will be discussed further in Section~\ref{3D:concs}.

Investigating the magnetic and free magnetic energy in ARs is essential to study the physical processes occurring during solar flares. Pre-flare studies of magnetic energy evolution are rare, with previous work mainly focused on long timescale changes. For example, \citet{thalmannwiegelmann08} observe a gradual increase in magnetic energy in an AR over a day before a solar flare. Changes in energy values due to flaring have been studied more extensively, with previous work reporting decreasing magnetic energy after flares. For example, \citet{bleybel02} and \citet{regniercanfield06} use NLFF field extrapolations to estimate the magnetic energy budget before and after a flare, both finding that the magnetic energy usually decreases over the course of the flare. 
Also, energy budgets for combined flare-CME events are studied in detail by \citet{emslie05}, finding $\sim$~20--30\% of the available free magnetic energy is required to power the events examined. \citet{emslie12} expand this study to 38 more solar eruptive events, finding available non-potential magnetic energy of the order $\sim 10^{33}$~erg being sufficient to power a CME, flare accelerated particles, and hot thermal plasma. Approximately 30\% of this available energy is released in the eruptive event, with the remainder staying in the AR as stored magnetic energy. 

The aim of the research in this chapter is to examine AR magnetic energy evolution before and after a flare (on much shorter timescales than previously studied), in order to gain a better understanding of the processes involved in energy release. Comparisons will be made between the results of three different forms of 3D magnetic field extrapolation by examining the evolution of the AR. In Section~\ref{3D:obs}, the observations and analysis techniques employed are briefly discussed. Section~\ref{3D:res} presents the main results, in particular geometrical differences in traced field-line solutions (Section~\ref{3D:interpolation}), orientation differences in field-line footpoint solutions (Section~\ref{3D:orientation}), and the evolution of magnetic energy and free magnetic energy (Section~\ref{3D:energy}). Finally, some conclusions are discussed in Section~\ref{3D:concs}.

%%%%%%%%%%%%%%%%%%%%%%%%%%%%%%%%%%%%%%%%%%%%%%%%%%%%%%%%%%%%%%%%%%%%%%%%%%%%%%%%%%%%%%%%%%%%%%%%%%%%%%%%%%%%%%%%%%%%%%%%%%%%%%%%%%%%%%%%%%%%%%%%

\section{Observations and Data Analysis}
\label{3D:obs}

Active region NOAA 10953, which was studied in Chapter~\ref{chapter:paper1}, is again studied here. Details of the complete analysis procedure applied to the data are contained in Chapters~\ref{chapter:instrumentation} and~\ref{chapter:paper1}, including the atmospheric inversion procedure, 180$^\circ$~azimuth disambiguation, and heliographic coordinate conversion. Four SOT-SP scans are examined from 2007~April~29 covering a period of 12~hours, leading up to and after a \emph{GOES} B1.0 solar flare. The scan times were summarised in Table~\ref{paper1:table_obs}.

\begin{figure}[!t]
\centerline{\includegraphics[width=\textwidth]{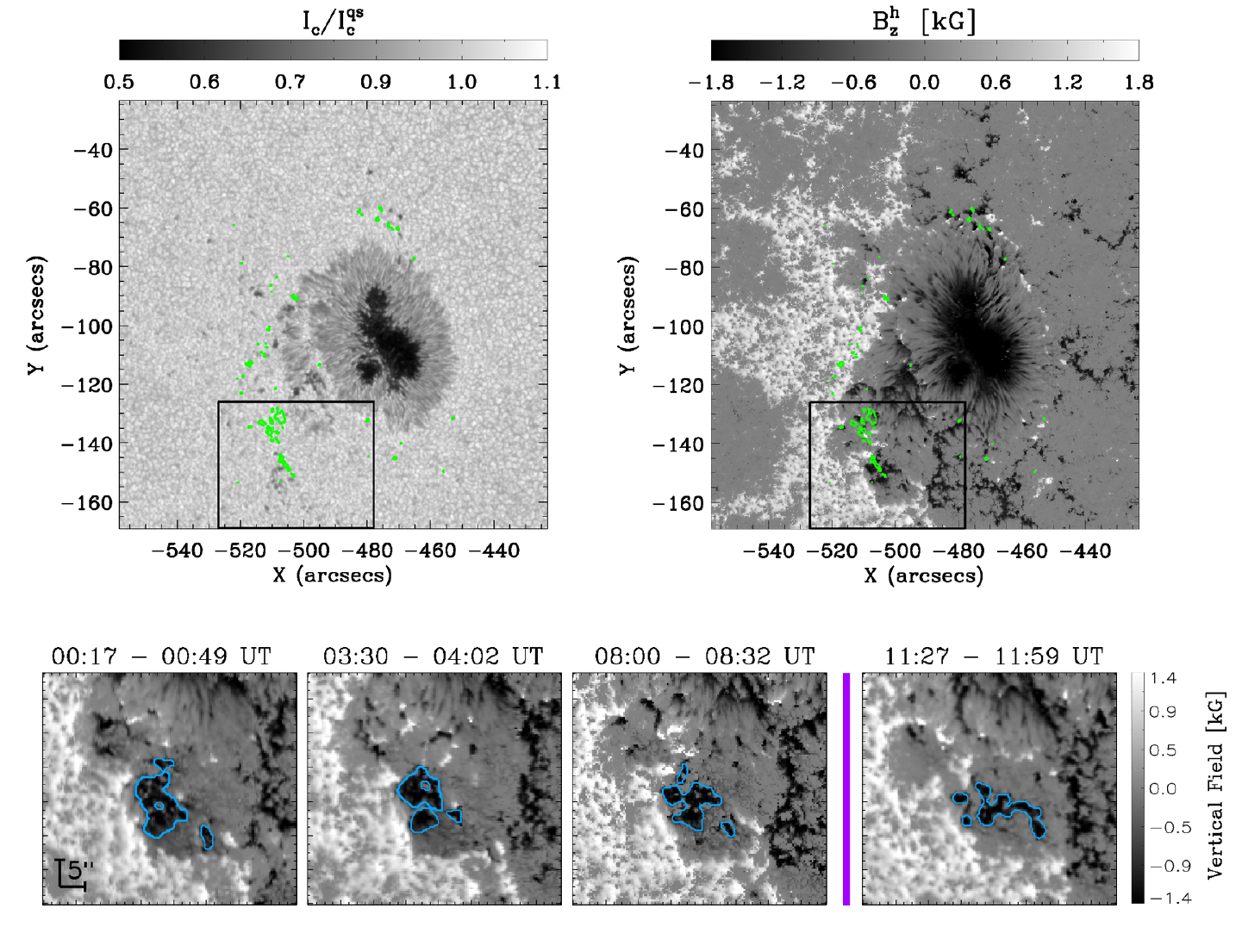}}
\caption[Upper row shows the active region pre-flare state in continuum intensity (left) and vertical field strength (right). The sub-region indicated by the box is shown in the lower row as preprocessed vertical field strength, at increasing scan time from left to right.]{Upper row shows the active region pre-flare state in continuum intensity (left) and vertical field strength (right). Green contours overlaid on both images indicate Ca\,\textsc{ii}\,H flare brightening observed at flare peak. The sub-region indicated by the box is shown in the lower row as preprocessed vertical field strength, at increasing scan time from left to right. A region of interest, thresholded at $-$750~G, is defined by the blue contours, and the purple line indicates the time of flaring between scans 3 and 4.}
\label{3D:context_image}
\end{figure}

The top row of Figure~\ref{3D:context_image} serves as a reminder of Figure~\ref{paper1:figure_1}, showing the specific area of the active region which is further investigated in this chapter. These panels contain the continuum intensity (upper left) and surface vertical field strength (upper right) of the third scan (i.e., the scan prior to the flare). Previously in Chapter~\ref{chapter:paper1}, two ROIs were identified (see Figure~\ref{paper1:figure_2}) in an area of increased Ca\,\textsc{ii}\,H flare emission. These regions are contained within the box in both upper panels of Figure~\ref{3D:context_image}, with green contours showing the Ca\,\textsc{ii}\,H intensity at the flare peak. This zoomed-in box is used later for a portion of the analysis in Section~\ref{3D:energy}. The surface magnetic field evolution of this region was previously considered in Chapter~\ref{chapter:paper1}, finding significant changes in vector field parameters in the ROIs during the observation period. Both ROIs showed a change in field inclination towards the vertical before the flare, with the field returning towards the horizontal afterwards. This variation in inclination agrees with inclination changes across the neutral line predicted by \citet{hudson08}. However, 3D coronal extrapolations are necessary to fully understand the evolution of the field, which is the specific purpose of this research.

In this chapter the vector-magnetic field data analysed in Chapter~\ref{chapter:paper1} is subjected to further analysis. The previous heliographic planar vector field information (i.e., $B_{x}^{\mathrm{h}}$, $B_{y}^{\mathrm{h}}$ and $B_{z}^{\mathrm{h}}$) were used as inputs to the three forms of 3D magnetic field extrapolation presented in Section~\ref{mhd:extraps}. The {\tt LINFF} code was used to calculate the potential and LFF fields, as described in Section~\ref{mhd:pot_lff}. To obtain the potential extrapolation using {\tt LINFF}, $\alpha$ is simply set to zero. For the case of the LFF extrapolations, the required values of constant $\alpha$ were calculated by fitting $J_{z}^{\mathrm{h}}$ vs. $B_{z}^{\mathrm{h}}$ from the results of Chapter~\ref{chapter:paper1} \citep[see][]{hahn05}. The values of $\alpha$ obtained for the full FOV of each scan (i.e., $455 \times 455$~pixels$^2$) are outlined in Table~\ref{3D:alpha}. Note that this table also includes values for a zoomed-in region that will be described at the end of this Section. The values for the full FOV are reasonably similar, showing a slight decrease over time before the flare, with a marginal increase afterwards. 

\begin{table}[!]
\centering
\caption[Values of $\alpha$ calculated for each of the four SOT-SP scans.]{Values of $\alpha$ calculated for each of the four SOT-SP scans for the full and zoomed-in FOVs.}
\begin{tabular}{ c c c c c }
\vspace{0.1cm} \\
\hline\hline
  Scan Number 		& Full FOV  	 & Zoomed-in FOV  							 \\
  					& $\alpha$ ($\mathrm{Mm}^{-1}$)  & $\alpha$ ($\mathrm{Mm}^{-1}$)\\ %10^{-7}\mathrm{m}^{-1}$			 \\
  \hline
  1                        		& 0.30 & 0.23\\ %3.0    									\\ 
  2                        		& 0.29 & 0.26\\ %2.9     									\\ 
  3 					& 0.27 & 0.22\\ %2.7     									\\ 
  4 					& 0.28 & 0.17\\ %2.8	 								\\
  \hline\hline
  \end{tabular}
  \label{3D:alpha}
\end{table}

The chosen form of NLFF extrapolation is the weighted optimization method described in Section~\ref{mhd:nlff}. As previously mentioned, the code is currently one of the most accurate NLFF procedures available (see reviews by \citealp{schrijver06}; \citealp{metcalf08}; \citealp{derosa09}). It is worth noting that this NLFF code directly minimises the force-balance equation, which avoids the explicit computation of $\alpha$. The photospheric magnetic field data was also preprocessed before its use as a boundary condition for 3D extrapolations, as explained in Section~\ref{mhd:nlff}. Comparison of the flux balance, net force, and net torque terms defined in Equations~\ref{mhd:force},~\ref{mhd:torque}, and ~\ref{mhd:flux}, is a useful consistency check for the preprocessing method. To serve as a suitable lower-boundary condition, vector magnetograms must be approximately flux balanced, and the net force and torque should vanish. Thus, the smaller the calculated values (i.e., values $\ll$ unity), the closer the data is to a force-free state. The results for the data set used in this chapter are promising. For example, the sum of net force and net torque before preprocessing was on average $\sim0.4790$, and after preprocessing this value drops to $\sim0.0016$. See Table~\ref{3D:prepro} for these parameters calculated before and after preprocessing. Thus, the preprocessing method has resulted in a data set that is considerably more consistent with the assumption of a force-free photospheric, and hence coronal, magnetic field. The bottom row of Figure~\ref{3D:context_image} shows the evolution of the preprocessed surface vertical field strength in the area of increased Ca\,\textsc{ii}\,H intensity, from the zoom-in region delineated by the box in the upper panels of Figure~\ref{3D:context_image}.

\begin{table}
\centering
\caption{Values of flux, torque and force before and after preprocessing. Note these values are dimensionless, normalised to the magnetic pressure.}
\begin{tabular}{  c c  c c c}
\hline
\hline
Scan Number & $\epsilon_{flux}$  &   $\epsilon_{force}$ &   $\epsilon_{torque} $&   $\epsilon_{force}+\epsilon_{torque}$ \\
\hline
Before preprocesssing	&  &  & & \\
1  	&  0.0248 &  0.2716 & 0.1151 & 0.3868 \\
2	&  0.0261 &  0.2831 & 0.1404 & 0.4236 \\
3	&  0.0281 &  0.3454 & 0.2158 & 0.5612 \\
4	&  0.0292 &  0.3417 & 0.2027 & 0.5444 \\
 \hline
After preprocessing	&  &  & & \\
1	&  0.0245 &  0.0009 & 0.0007 & 0.0015 \\
2	&  0.0257 &  0.0009 & 0.0007 & 0.0016 \\
3 	&  0.0275 &  0.0010 & 0.0007 & 0.0016 \\
4 	&  0.0288 &  0.0009 & 0.0006 & 0.0015 \\
 \hline
 \hline
  \label{3D:prepro}
  \end{tabular}
\end{table}

A single ROI in the area of flare brightening was selected for further investigation. This was identified by thresholding the preprocessed surface vertical field strength at $-$750~G (depicted by contours in the lower panels of Figure~\ref{3D:context_image}). This ROI corresponds to ROI 2 investigated in Chapter~\ref{chapter:paper1}, but is not completely identical due to the data preprocessing (i.e., effective smoothing). ROI 1 was not investigated, as it was deemed too small and fragmented after the preprocessing to observe any statistically meaningful variations in the extrapolated field. Extrapolation solutions were calculated for the full FOV of each of the four scans, with computational volumes comprising of $455 \times 455 \times 228$~pixels$^3$ (i.e., $105\times105\times52$~Mm$^3$). It should be noted that the zoomed-in region for further consideration in Section~\ref{3D:energy} makes use of the full height of the computational volume, covering $165\times135\times228$~pixels$^3$ (i.e., $38\times31\times52$~Mm$^3$).

%%%%%%%%%%%%%%%%%%%%%%%%%%%%%%%%%%%%%%%%%%%%%%%%%%%%%%%%%%%%%%%%%%%%%%%%%%%%%%%%%%%%%%%%%%%%%%%%%%%%%%%%%%%%%%%%%%%%%%%%%%%%%%%%%%%%%%%%%%%%%%%%

\section{Results}
\label{3D:res}

\begin{figure}[!t]
\centerline{\includegraphics[width=\textwidth]{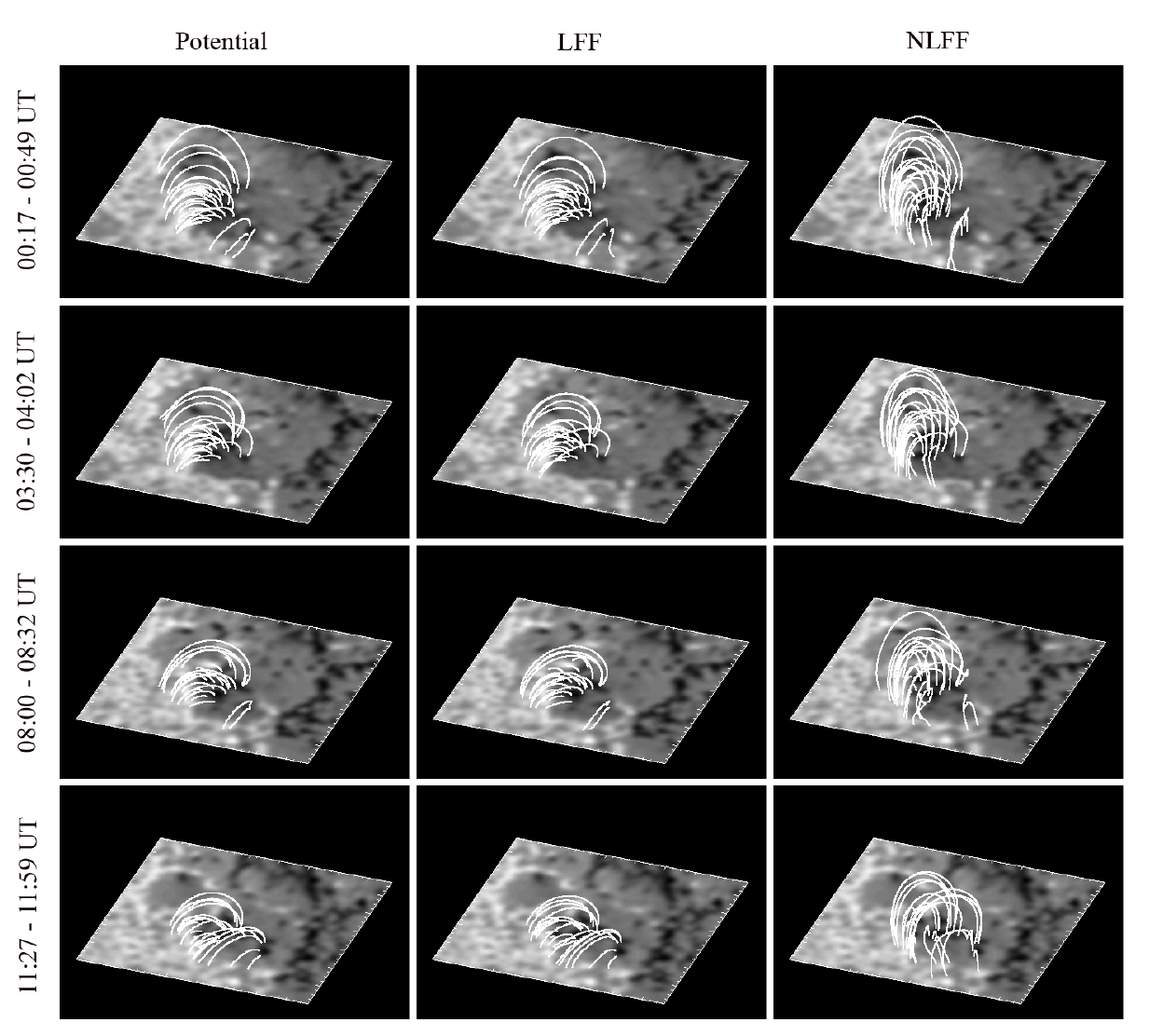}}
\caption{Zoomed-in FOV of extrapolated volume (see Figure~\ref{3D:context_image}) at increasing scan time from left to right. Upper to lower row: potential field extrapolation, LFF extrapolation, and NLFF extrapolation.}
\label{3D:figure_extraps}
\end{figure}

Figure~\ref{3D:figure_extraps} shows zoom-ins on the results of the three types of extrapolation performed on the full FOV, with increasing scan time from left to right (the flare occurs between scans 3 and 4). Field-line traces for every 30$^\mathrm{th}$ pixel in the ROI identified in the bottom panel of Figure~\ref{3D:context_image} are shown. The potential and LFF field-line solutions look similar, while the NLFF field lines reach greater heights and connect further south.  

\begin{figure}[!t]
\centerline{\includegraphics[width=.7\textwidth]{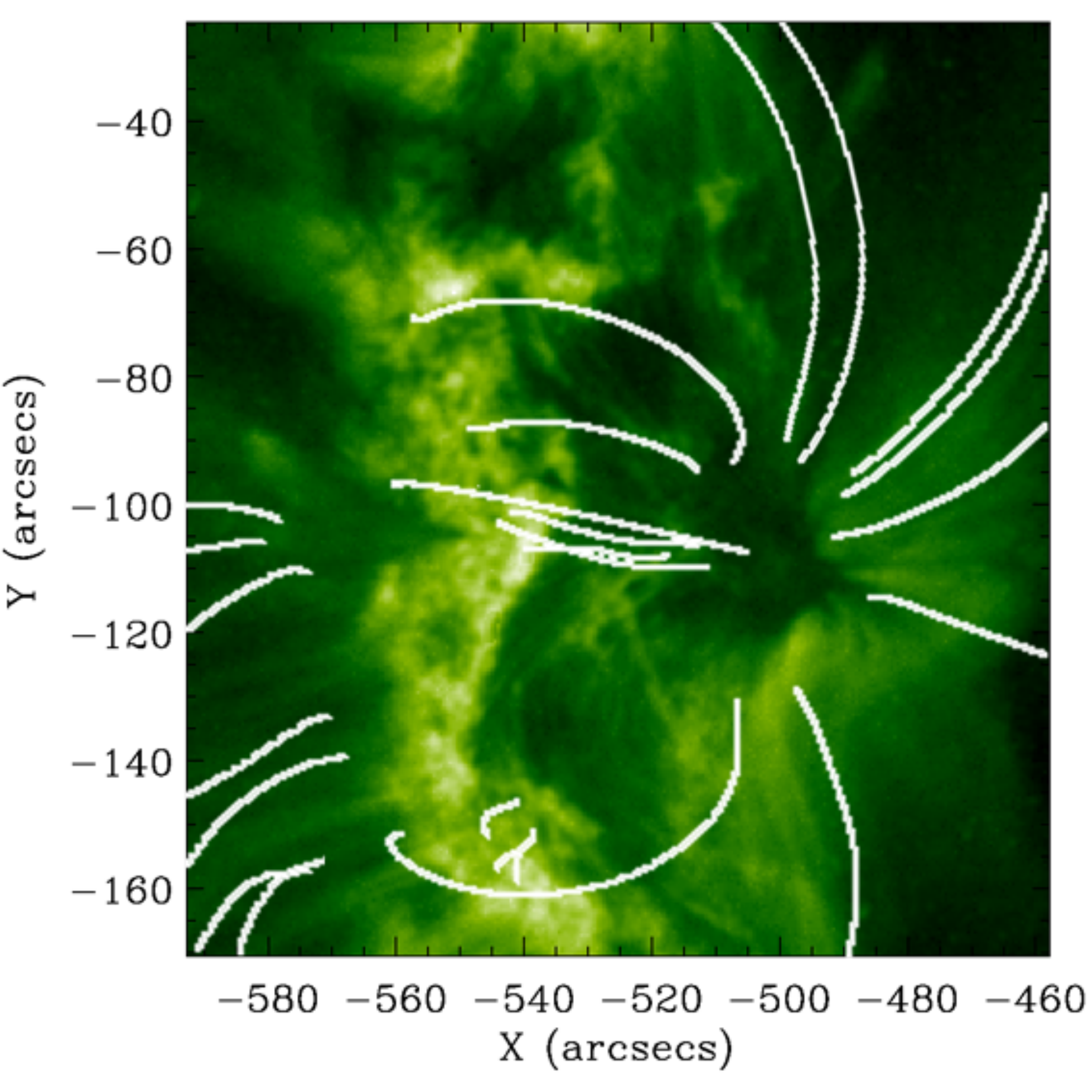}}
\caption{TRACE 195~\AA\ image of NOAA active region 10953 at 03:45~UT on 2007 April 29. The overplotted white lines represent selected field lines from NLFF extrapolation results of scan 2.}
\label{3D:trace}
\end{figure}

As mentioned in \citet{derosa09} and \citet{wiegelmann12}, it is useful to compare the extrapolated field with coronal-loop observations as a consistency check. This quantifies the extent to which the extrapolation correctly reproduces the coronal magnetic field configuration. Unfortunately, very few high-resolution coronal observations were available for this particular event. Observations were obtained from the \emph{Transition Region and Coronal Explorer} (\emph{TRACE}) space telescope for comparison; a 195~\AA\ image was recorded at 03:45~UT on 2007 April 29 (i.e., during scan 2). Figure~\ref{3D:trace} shows select NLFF field lines traced over the image. Note that only the NLFF extrapolation is examined, as it is considered to be the most accurate representation of the coronal field (compared to the potential or LFF extrapolations). It is unfortunately difficult to identify coronal loops in this 195~\AA\ wavelength image (171~\AA\ for example would have been more useful), but the extrapolated field lines seem to reasonably agree with the observed loops, with some deviations also existing. Agreement is particularly clear in the loops traced to the northeast of the image.

\begin{figure}[!t]
\centerline{\includegraphics[width=\textwidth]{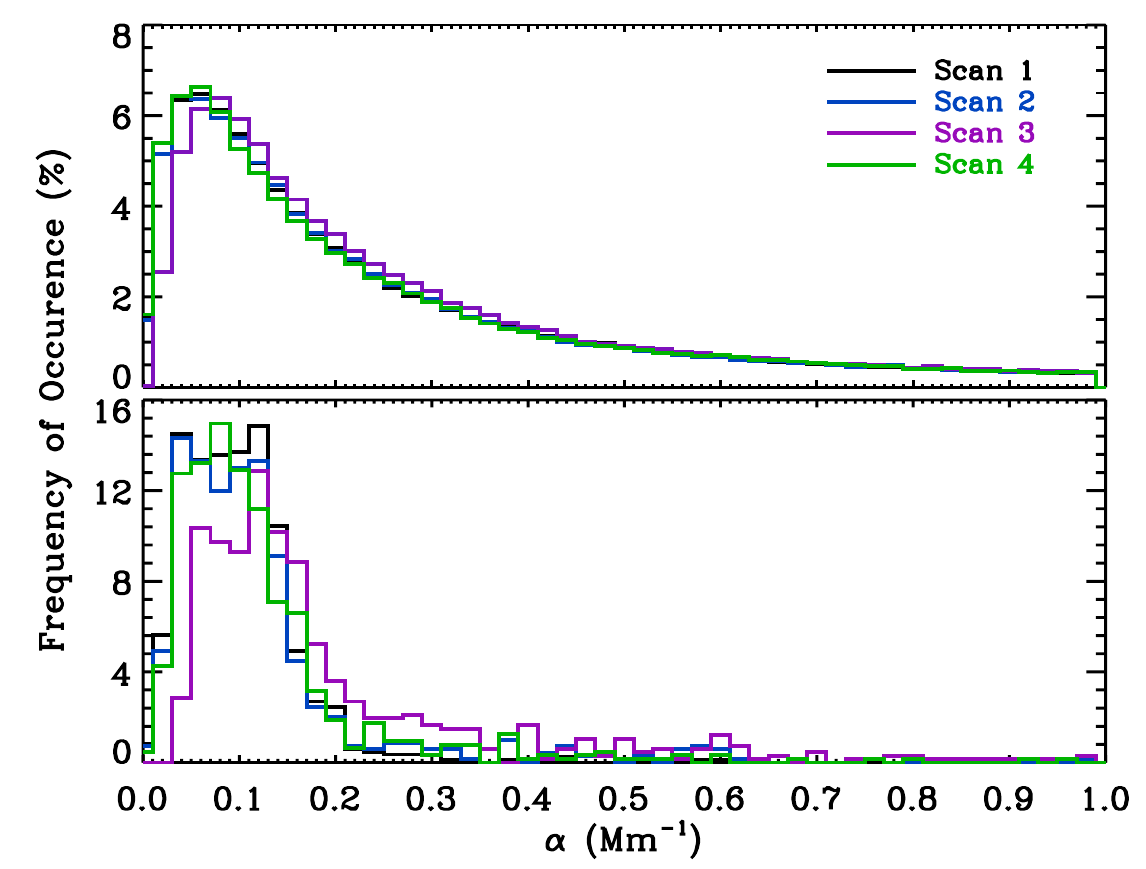}}
\caption[Distribution of calculated NLFF $\alpha$ values for traced field lines from all pixels from all pixels within the full computational volume (upper) and ROI (lower).]{Distribution of calculated NLFF $\alpha$ values for traced field lines from all pixels from all pixels within the full computational volume (upper) and ROI (lower). A bin size of 0.02~Mm$^{-1}$ was used. Y-axes show percentage occurrences, while the x-axis indicates the value of $\alpha$ in Mm$^{-1}$. Each scan distribution is coloured as per the legend.}
\label{3D:alpha_hists}
\end{figure}

In order to study relaxation of the coronal field, it is necessary to determine differences in the field structure between the three forms of extrapolation solution. 
As noted in Section~\ref{3D:obs}, the NLFF code used here does not compute $\alpha$. An estimate of $\alpha$ for all traced field lines within the computational volume can be obtained using Equation~\ref{nlff} ($\nabla \times \mathbf{B} = \alpha \mathbf{B}$) with the results of the NLFF extrapolation procedure. Average values for $\alpha$ were calculated for each traced field line within the full computational volume as well as in the ROI, and histograms of the results are shown in Figure~\ref{3D:alpha_hists}. For NLFF extrapolations $\alpha$ is assumed constant along the entire field line, and this allows examining the traced field line averages from the source pixels within a region. All histograms throughout this chapter have been normalised to the total number of pixels of a ROI or FOV, so as to show the percentage occurrence.  A consistency check was carried out on full field lines traced from the source pixels, and $\alpha$ was found to be constant within $\pm~0.02$~Mm$^{-1}$. Figure~\ref{3D:alpha_hists} shows $\alpha$ to be very similar for all scans in the full computational volume, with the distributions peaking between $\sim0.03 - 0.09$~Mm$^{-1}$, and a slight variation in scan 3 towards larger values. For the ROI, $\alpha$ also seems to be similar between scans but a larger variation is observed, and distributions peak between $\sim0.08 - 0.13$~Mm$^{-1}$. The distribution for the third pre-flare scan differs the most from the other scans, being shifted towards larger values of $\alpha$. This indicates an increase in the amount of twist in the field in the hours leading up to the flare. The fourth scan distribution indicates a decrease in $\alpha$, and hence twist, after the flare has occurred.

Although the changes in $\alpha$ observed in the ROI over the four scans are not large, it is promising to find expected increases (decreases) before (after) the flare. A number of other 3D magnetic field vector parameters will be examined here, showing more significant changes over the observation period. In particular, geometrical differences will be investigated between field-line traces that are caused by the differing presence of currents in each extrapolation solution. Section~\ref{3D:interpolation} presents an analysis of 3D spatial offsets between the extrapolation solution field lines, while differences in field-line footpoint locations are presented in Section~\ref{3D:orientation}. Finally, the total magnetic and possible free magnetic energies are calculated in Section~\ref{3D:energy}.

%%%%%%%%%%%%%%%%%%%%%%%%%%%%%%%%%%%%%%%%%%%%%%%%%%%%%%%%%%%%%%%%%%%%%%%%%%%%%%%%%%%%%%%%%%%%%%%%

\subsection{Spatial Differences in Field-line Traces} 
\label{3D:interpolation}
To obtain an indirect indication of how $\alpha$ may be varying in the NLFF extrapolation solution, a first step is to determine to what degree the extrapolation solutions differ when considering field-line traces from the same starting pixel. This can be quantified in terms of offsets in the computational $x$, $y$, and $z$ spatial coordinates at specific points along the length of a field-line trace. It should be noted that field lines traced from the same source pixel (i.e., from the ROI) may have different path lengths in each extrapolation solution through 3D space before they return to the photospheric boundary. In order to calculate 3D displacements between comparable locations along each field line, the variable length arrays of ($x$, $y$, $z$) coordinates for each field line trace were interpolated to 11 points. These correspond to relative locations separated by 1/10 of the total path length. Displacements were calculated in each of the $x$, $y$, and $z$ spatial coordinates between the same relative locations (i.e., 1/10 along the potential trace minus 1/10 along the LFF and NLFF traces, etc\ldots) with the displacement values along one entire field-line trace being averaged. The larger this displacement value is, the further apart the LFF and NLFF traces are from the potential case and the more current there is in the system.

\begin{sidewaysfigure}
\centering
\includegraphics[width=\textwidth]{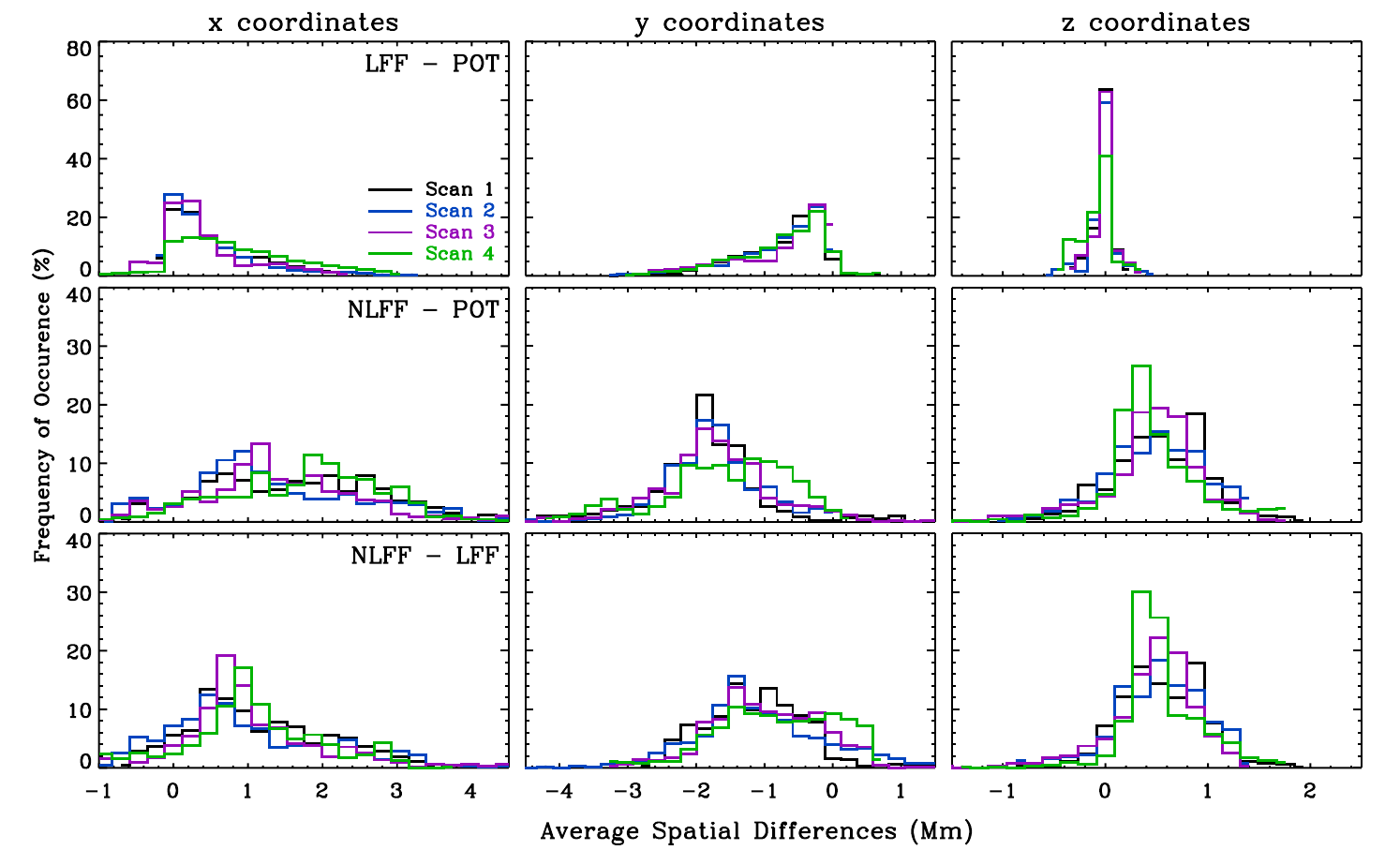}
\caption[Differences in $x$ (left), $y$ (centre) and $z$ (right) coordinates of field-line traces between LFF and potential (upper), NLFF and potential (middle), and NLFF and LFF extrapolations (lower).]{Differences in $x$ (left), $y$ (centre) and $z$ (right) coordinates of field-line traces between LFF and potential (upper), NLFF and potential (middle), and NLFF and LFF extrapolations (lower). Each scan distribution is coloured as per the legend, while bin sizes of 1.0 were used for all $x$ and $y$ coordinates, 0.5 for (LFF $-$ POT) $z$ coordinates, and 0.75 for (NLFF $-$ POT) and (NLFF $-$ LFF) $z$ coordinates. }
\label{3D:figure_interpolation}
\end{sidewaysfigure}

Figure~\ref{3D:figure_interpolation} shows the results of differencing and averaging these arrays of $x$, $y$, and $z$ interpolated coordinates for each ROI originating field-line trace. Histogram values for all four scans are overlaid for comparison. In terms of $x$ coordinates, the mainly positive distributions in the upper left panel show the potential traces generally reach further east than the LFF traces (i.e, reaching smaller $x$ pixel values). The distributions for the (NLFF -- POT) and (NLFF -- LFF) cases are broader (due to the NLFF solution effectively having a distribution of $\alpha$ values), with the NLFF field lines not reaching as far east as the LFF field lines. In terms of $y$ coordinates, the LFF field generally reaches further south than the potential configuration (shown by the mainly negative $y$-coordinate distributions in the upper middle panel). The central and lower-middle panels again show broader distributions, with the NLFF field-line traces reaching much further south than the potential and LFF traces.

\begin{figure}[!t]
\centering
\includegraphics[width=0.8\columnwidth]{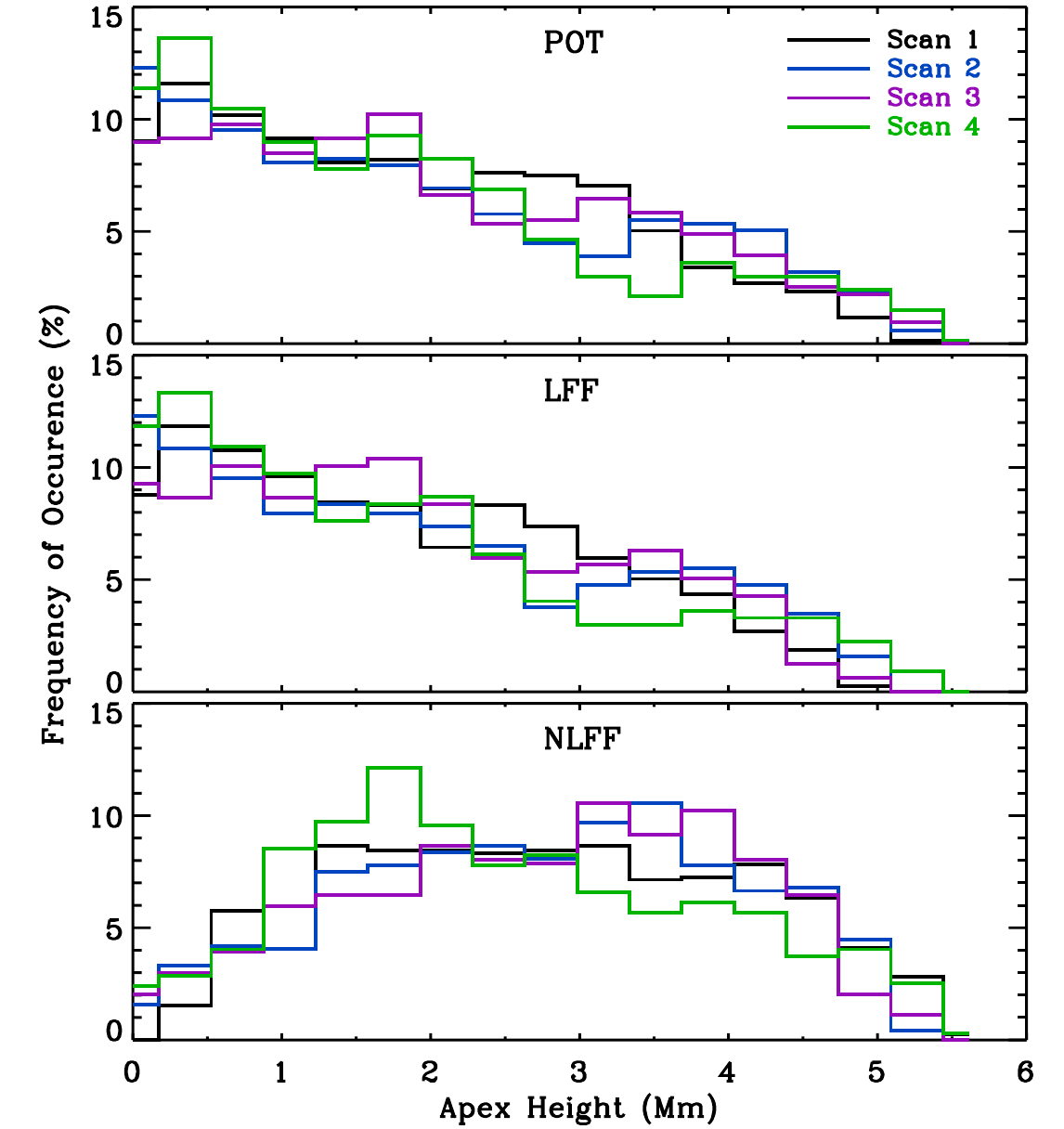}
\caption[Distribution of apex heights from all ROI pixels for potential (upper), LFF (middle), and NLFF (lower) extrapolations. ]{Distribution of apex heights from all ROI pixels for potential (upper), LFF (middle), and NLFF (lower) extrapolations. Y-axes show percentage occurrences, while the x-axis indicates the height in Mm. Each scan distribution is coloured as per the legend.}
\label{apex_height}
\end{figure}

\begin{figure}[!t]
\centering
\includegraphics[width=0.65\textwidth]{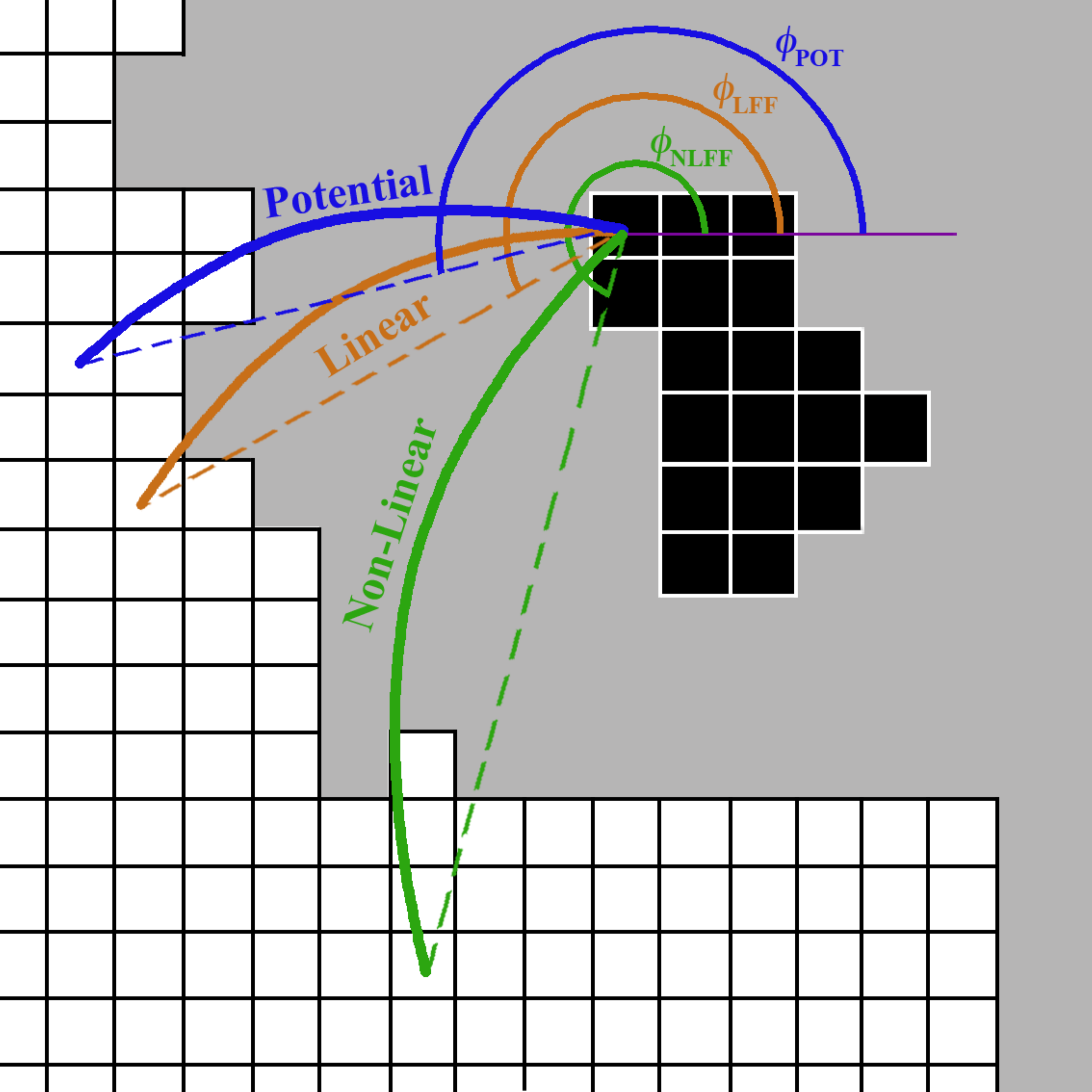}
\caption[2D cartoon of typical extrapolated field line traces.]{2D cartoon of typical extrapolated field line traces. The black area indicates the negative polarity ROI, and white area the plage region. The grids outline single pixels. Field-line traces and position angles are indicated in blue (potential), orange (LFF), and green (NLFF).}
\label{3D:figure_cartoon}
\end{figure}

The $z$-coordinate distributions for (LFF -- POT) are much narrower than those of the (NLFF -- LFF) and (NLFF -- POT) distributions, with a greater percentage of values close to zero. The height differences between the NLFF and both the potential and LFF field-line traces indicate that the NLFF field lines reach much larger heights within the computational volume. The z-coordinate heights were investigated further, by calculating the maximum field-line height for all pixels within the ROI. The result of this calculation is shown in Figure~\ref{apex_height}. No significant changes are observed with time for the potential or LFF extrapolations. The largest changes are observed in the NLFF results, with the apex height increasing before the flare in scan 3, and then decreasing afterwards in scan 4. Note that NLFF fields reaching greater heights has been previously found,  e.g., \citet{regnierpriest07aa} concluded for NOAA region 8151 that the NLFF fields were statistically higher, longer and had a stronger magnetic field than the potential ones. \citet{liu11} found that, for closed field lines, the NLFF lines can reach higher altitudes that those of the potential field (and are also more vertical and stretched). The characteristics derived from Figure~\ref{3D:figure_interpolation} are summarised in Figure~\ref{3D:figure_cartoon}, which illustrates typical field-line traces from an ROI pixel where the NLFF field has a larger magnitude of $\alpha$ (and hence a greater degree of twist) than that of the LFF field.

Considering the variation of Figure~\ref{3D:figure_interpolation} across time, none of the coordinate difference distributions vary significantly from scans 1 to 3, whereas larger changes are observed in scan 4 (i.e., after the flare). This is clearest for the $y$ coordinates when considering the central and lower-middle panels. A portion of each of these distributions have shifted towards smaller values, with a greater percentage of the (NLFF -- LFF) distribution having values of zero compared to the (NLFF -- POT) distribution. Thus, after the flare some of the NLFF field lines lie closer in 3D space to the LFF field lines than to the potential ones. This confirms that the LFF field contains current (as it deviates from the current-free potential field) but that the NLFF field has stronger currents (as it deviates further). Smaller differences between the NLFF and LFF fields in the post-flare scan indicates a decrease in NLFF currents relative to those in the LFF field. However, some of this is due to the increased LFF $\alpha$ in scan 4 (Section~\ref{3D:obs}). Even though portions of the extrapolation solutions are similar after the flare, the NLFF field contains stronger currents than the LFF field.

%%%%%%%%%%%%%%%%%%%%%%%%%%%%%%%%%%%%%%%%%%%%%%%%%%%%%%%%%%%%%%%%%%%%%%%%%%%%%%%%%%%%%%%%%%%%%%%%

\subsection{Orientation Differences Between Field-line Footpoints}
\label{3D:orientation}

As indicated in the previous subsection, the amount of current present in the coronal magnetic field affects the direction of field-line traces and ultimately the location of their footpoints (resulting from the twist that currents introduce in the magnetic field). In order to get a better picture of the effective distribution of $\alpha$ values in the NLFF solutions, the footpoint position angle, $\phi$, of a field-line trace is defined as,
\begin{equation}
\phi~=~tan^{-1}\left(\frac{y_{\mathrm{f}}-y_{\mathrm{i}}}{x_{\mathrm{f}}-x_{\mathrm{i}}} \right) \ ,
\end{equation}
where $x$ and $y$ are pixel coordinates on the solar surface, subscript `i' denotes initial coordinate, and subscript `f' denotes final coordinate. The initial coordinates are taken as each pixel within the negative polarity ROI, such that the footpoint position angle measures the counter-clockwise angle from solar west to the traced field-line end footpoint in the positive polarity plage (see Figure~\ref{3D:figure_cartoon} for an illustration).

\begin{figure}[!t]
\centering
\includegraphics[width=0.8\textwidth]{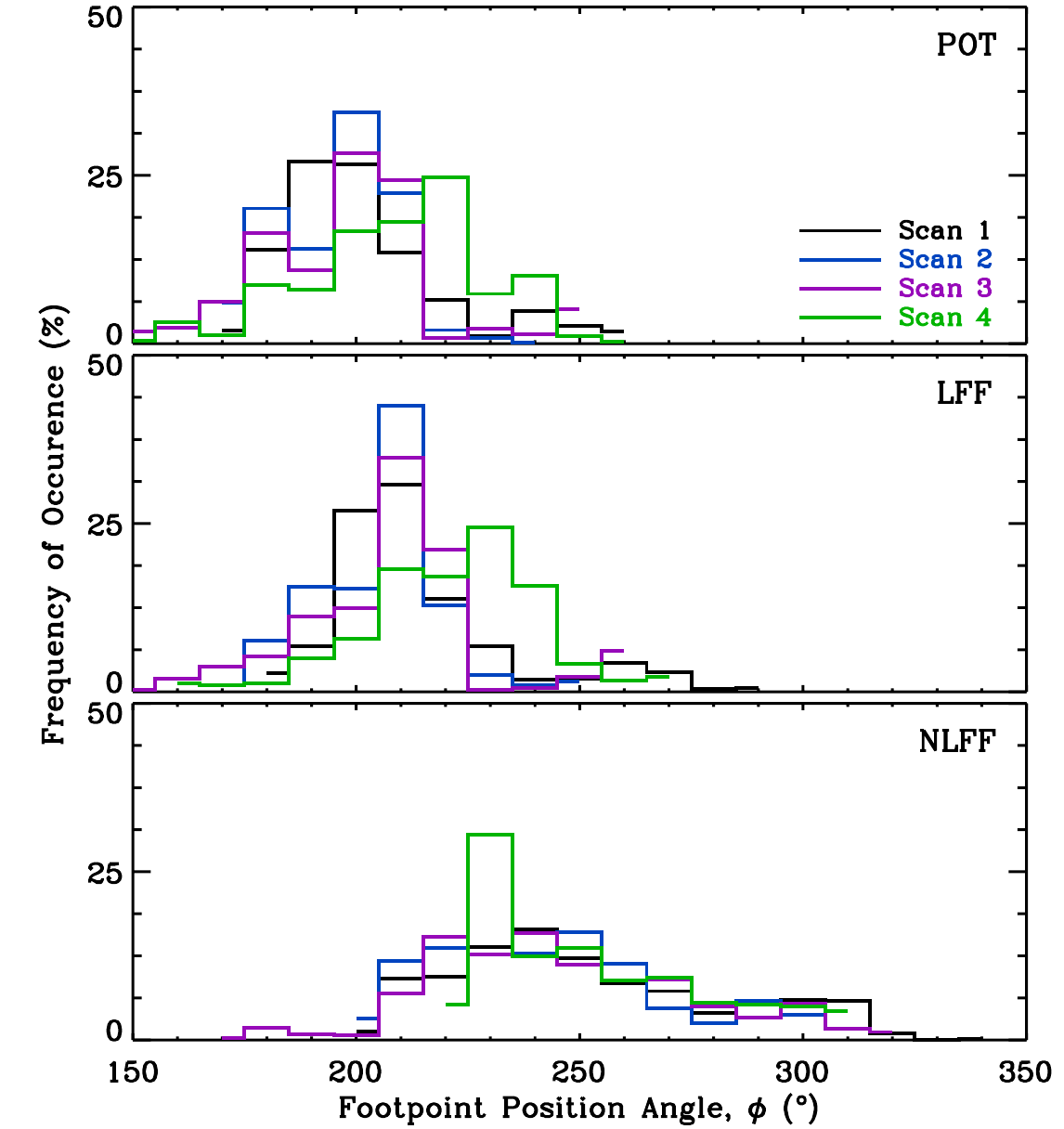}
\caption[Distribution of footpoint position angles from all ROI pixels for potential (upper), LFF (middle), and NLFF (lower) extrapolations.]{Distribution of footpoint position angles from all ROI pixels for potential (upper), LFF (middle), and NLFF (lower) extrapolations. Y-axes show percentage occurrences, while the x-axis gives the end footpoint position angle in degrees counter-clockwise from solar west. Each scan distribution is coloured as per the legend, while bin sizes of 10$^\circ$ were used throughout.}
\label{3D:figure_azimuth}
\end{figure}

Figure~\ref{3D:figure_azimuth} contains the results of this calculation for each of the extrapolation formats from all four scans. The position angle distribution for the potential case ($\phi^\mathrm{POT}$; upper panel) does not vary much over the first three scans, but a greater portion of the distribution occurs at larger angles after the flare. The position angle distribution for the LFF case ($\phi^\mathrm{LFF}$; middle panel) is very similar to $\phi^\mathrm{POT}$, but shifted to larger angles by $\sim$~10--15$^{\circ}$ in all four scans. This is expected with the introduction of twist in the field from the presence of currents that are equally distributed throughout space for the LFF case. Finally, the position angle distributions for the NLFF case ($\phi^\mathrm{NLFF}$; lower panel) are shifted further and broadened. This is also expected, with the NLFF case having a distribution of effective $\alpha$ values throughout space (and hence a distribution of currents and twist in the field). The three pre-flare scans exhibit very similar $\phi^\mathrm{NLFF}$ distributions. However, in scan 4 after the flare the $\phi^\mathrm{NLFF}$ distribution has decreased occurrence of lower angle values and shows a stronger peak at a similar position angle to the peak of the $\phi^\mathrm{LFF}$ distribution. It should be noted that significant portions of the $\phi^\mathrm{LFF}$ and $\phi^\mathrm{NLFF}$ distributions do not overlap.

\begin{sidewaysfigure}
\centering
\includegraphics[width=1\textwidth]{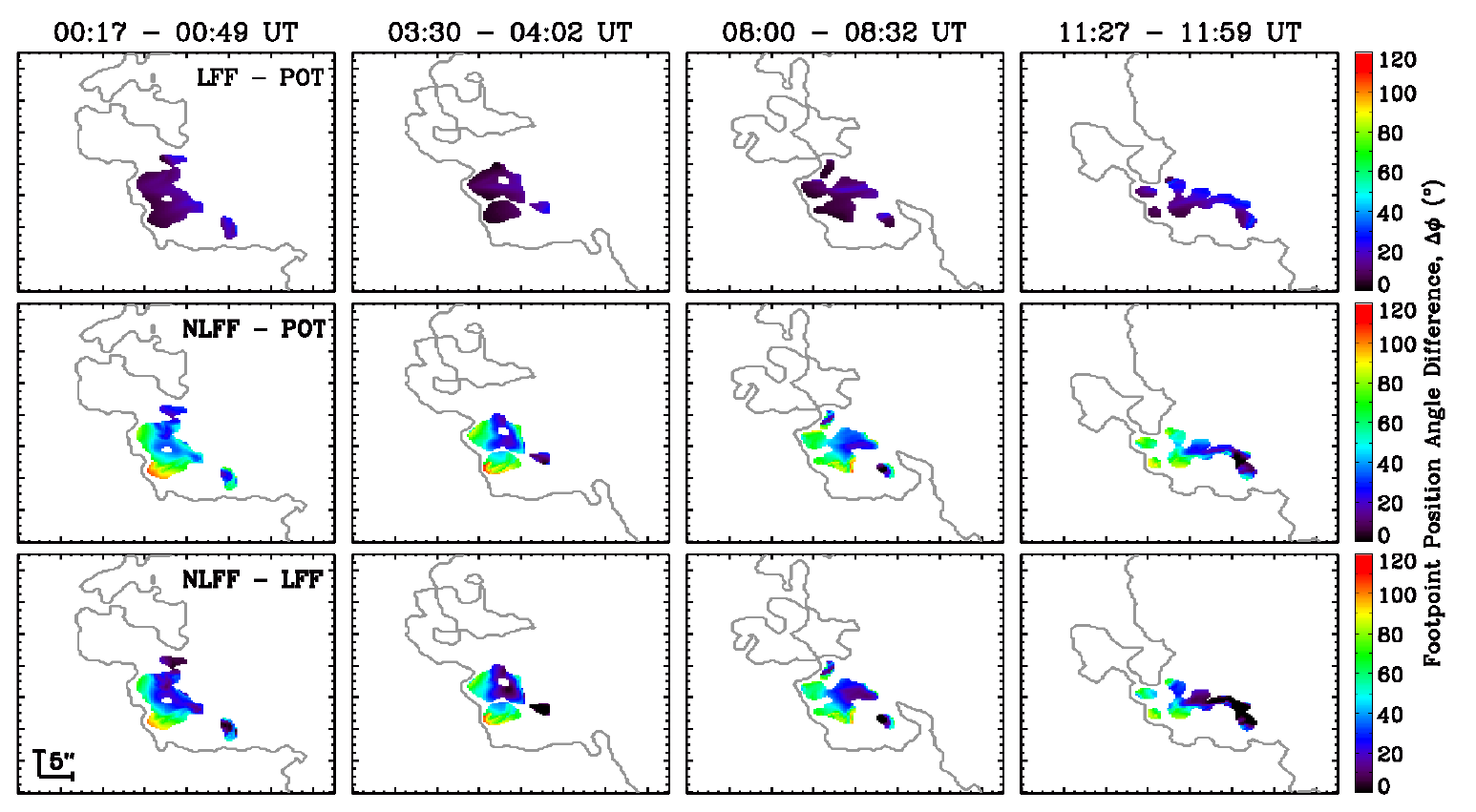}
\caption[Spatial variation of footpoint position angle difference, $\Delta\phi$, from originating ROI pixel locations (see Figure~\ref{3D:figure_hist} for value distributions).]{Spatial variation of footpoint position angle difference, $\Delta\phi$, from originating ROI pixel locations (see Figure~\ref{3D:figure_hist} for value distributions). Scan time increases from left to right, while rows depict values of $\Delta\phi^{\mathrm{LFF - POT}}$ (upper), $\Delta\phi^{\mathrm{NLFF - POT}}$ (middle), and $\Delta\phi^{\mathrm{NLFF - LFF}}$ (lower). A neutral line is plotted in grey for context, as seen in the lower row of Figure~\ref{3D:context_image}.}
\label{3D:figure_mask}
\end{sidewaysfigure}

The footpoint orientations and temporal variation reported so far correspond to non-localised characteristics of the magnetic field. The spatial distribution of differences between footpoint orientations from the different extrapolation solutions contains information on what portions of the ROI harbour strong currents (and hence magnetic energy). For each ROI source pixel, footpoint position angle values for the three extrapolation traced field-line solutions were subtracted from one another to give the footpoint position angle difference, $\Delta\phi$. This calculated quantity is presented in Figure~\ref{3D:figure_mask}, where larger values correspond to greater deviation in footpoint position angle between the selected pair of extrapolation solutions (i.e., greater twist in the field). The $\Delta\phi^\mathrm{LFF - POT}$ values are generally small over all scans, with a slight increase in scan 4. This again confirms the known increase of the LFF constant $\alpha$ value in that scan (Section~\ref{3D:obs}). Larger quantities are observed for $\Delta\phi^\mathrm{NLFF- POT}$ and $\Delta\phi^\mathrm{NLFF- LFF}$, with the largest located in the south east of the ROI (i.e., nearest to the magnetic NL with the positive plage). This indicates that the portion of the ROI containing the strongest currents (thus magnetic energy) is that closest to the NL involved in the flare.

\begin{sidewaysfigure}
\centering
\includegraphics[width=1\textwidth]{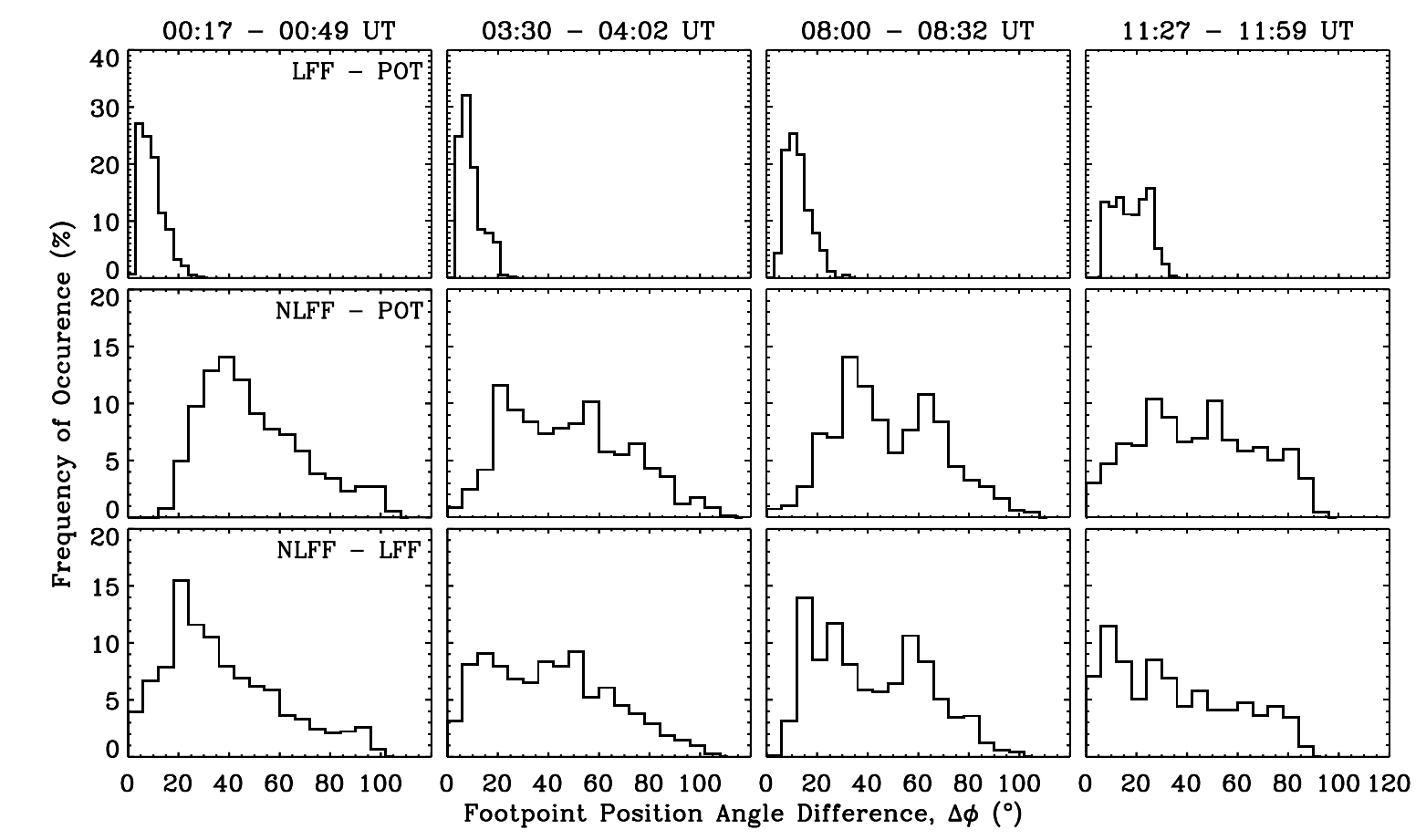}
\caption[Distributions of footpoint position angle difference, $\Delta\phi$ (see Figure~\ref{3D:figure_mask} for spatial representation).]{Distributions of footpoint position angle difference, $\Delta\phi$ (see Figure~\ref{3D:figure_mask} for spatial representation). Scan time increases from left to right, while rows depict values of $\Delta\phi^{\mathrm{LFF - POT}}$ (upper; bin size 3$^\circ$), $\Delta\phi^{\mathrm{NLFF - POT}}$ (middle; bin size 6$^\circ$), and $\Delta\phi^{\mathrm{NLFF - LFF}}$ (lower; bin size 6$^\circ$).}
\label{3D:figure_hist}
\end{sidewaysfigure}

Figure~\ref{3D:figure_hist} represents the footpoint position angle differences as histograms to aid in a more quantitative comparison. The upper row clearly shows the fairly constant, narrow distribution of $\Delta\phi^\mathrm{LFF - POT}$ over scans 1 to 3, which broadens and shifts to larger difference values after the flare (as indicated in Figure~\ref{3D:figure_mask}). Both the $\Delta\phi^\mathrm{NLFF - POT}$ (middle row) and $\Delta\phi^\mathrm{NLFF - LFF}$ (bottom row) distributions are considerably broader, showing some minor variation over the three pre-flare scans. However, both distributions have a greater percentage of ROI pixels with angle differences close to zero in scan 4 after the flare. The offset observed between the LFF and potential cases in Figure~\ref{3D:figure_azimuth} results in the $\Delta\phi^\mathrm{NLFF - LFF}$ distribution being shifted to lower values by $\sim$~10--15$^\circ$ from the $\Delta\phi^\mathrm{NLFF - POT}$ distribution in all scans. This follows from the $\phi^{LFF}$ distribution being shifted by $\sim$~10--15$^\circ$ relative to $\phi^{POT}$ (see above). Combining both of these results, the NLFF and LFF post-flare fields appear to be closer in configuration than the NLFF and LFF pre-flare fields were.

%%%%%%%%%%%%%%%%%%%%%%%%%%%%%%%%%%%%%%%%%%%%%%%%%%%%%%%%%%%%%%%%%%%%%%%%%%%%%%%%%%%%%%%%%%%%%%%%

\subsection{Magnetic and Free Magnetic Energies}
\label{3D:energy}

It is important to investigate the amount of energy stored in a magnetic configuration, as some of this stored energy is released in the flaring process. The analysis up to now has been concerned with studying geometrical differences between field-line traces in the three extrapolation solutions. However, the effects that the potential, LFF, and NLFF values of $\alpha$ have on the coronal energy content can be calculated directly from the extrapolation solutions. The magnetic energy, $E_\mathrm{m}$, contained in the field is \citep{schmidt64},
\begin{equation}
\label{mag_en}
E_\mathrm{m}~=~\int_{V}\frac{B^{2}}{8\pi}dV ~\ ,
\end{equation}
where $V$ is the 3D computational volume under consideration and $B$ is the magnitude of the magnetic field vector at every point in computational ($x$, $y$, $z$) space. However, the values of $E_\mathrm{m}$ for each of the extrapolation solutions do not correspond to the total energy available for flaring \citep{aly84}. Instead, two forms of free magnetic energy, $\Delta E_{\mathrm{m}}$, can be calculated, 
\begin{eqnarray}
\Delta E_{\mathrm{m}}^{\mathrm{NLFF-POT}} & = & E_\mathrm{m}^{\mathrm{NLFF}}~-~E_\mathrm{m}^{\mathrm{POT}}~  \ , \label{nlff-pot} \\
\Delta E_\mathrm{m}^{\mathrm{NLFF-LFF}} & = & E_\mathrm{m}^{\mathrm{NLFF}}~-~E_\mathrm{m}^{\mathrm{LFF}} ~ \ , \label{nlff-lff}
\end{eqnarray}
which correspond to using either the potential (Eqn.~\ref{nlff-pot}) or the LFF (Eqn.~\ref{nlff-lff}) field configuration as the minimum energy state. Note that there is no free magnetic energy in a potential field configuration, but the LFF and NLFF configurations have free energy due to the presence of currents. Values for $\Delta E_\mathrm{m}^{\mathrm{LFF-POT}}$ are not computed here since the ultimate aim of this research is to study the relaxation of the NLFF field over the course of a flare (as it is considered the most accurate representation of the field for small-scale changes).

Two different volumes are used for the energy calculations: the whole active region volume ($455\times455\times228$~pixels$^3$), and the surface zoom-in represented by the boxes in Figures~\ref{3D:context_image} and \ref{3D:figure_extraps} (i.e., $155\times135\times228$~pixels$^3$). Note that the zoomed-in volume is a subset of the full FOV extrapolation solution and uses the whole height of the extrapolation volume. The zoomed-in box was chosen as it corresponds to the location of greatest flare emission in Ca\,\textsc{ii}\,H. The box extent was determined by first choosing the furthest footpoint coordinates of all ROI field-line traces from the four scans as boundary points, but a number of small magnetic flux elements enter and leave this region over the course of the observations. In order to avoid any spurious effects these transient flux elements might have on the calculated energy evolution, the box was enlarged by $\sim$~15~pixels on each side. This results in minimal amounts of flux entering or leaving the box over the four scans (Figure~\ref{3D:context_image}, lower panels). Note, the zoomed-in region excludes the entire sunspot umbra and penumbra.

\begin{sidewaysfigure}
\centering{\includegraphics[width=\textwidth]{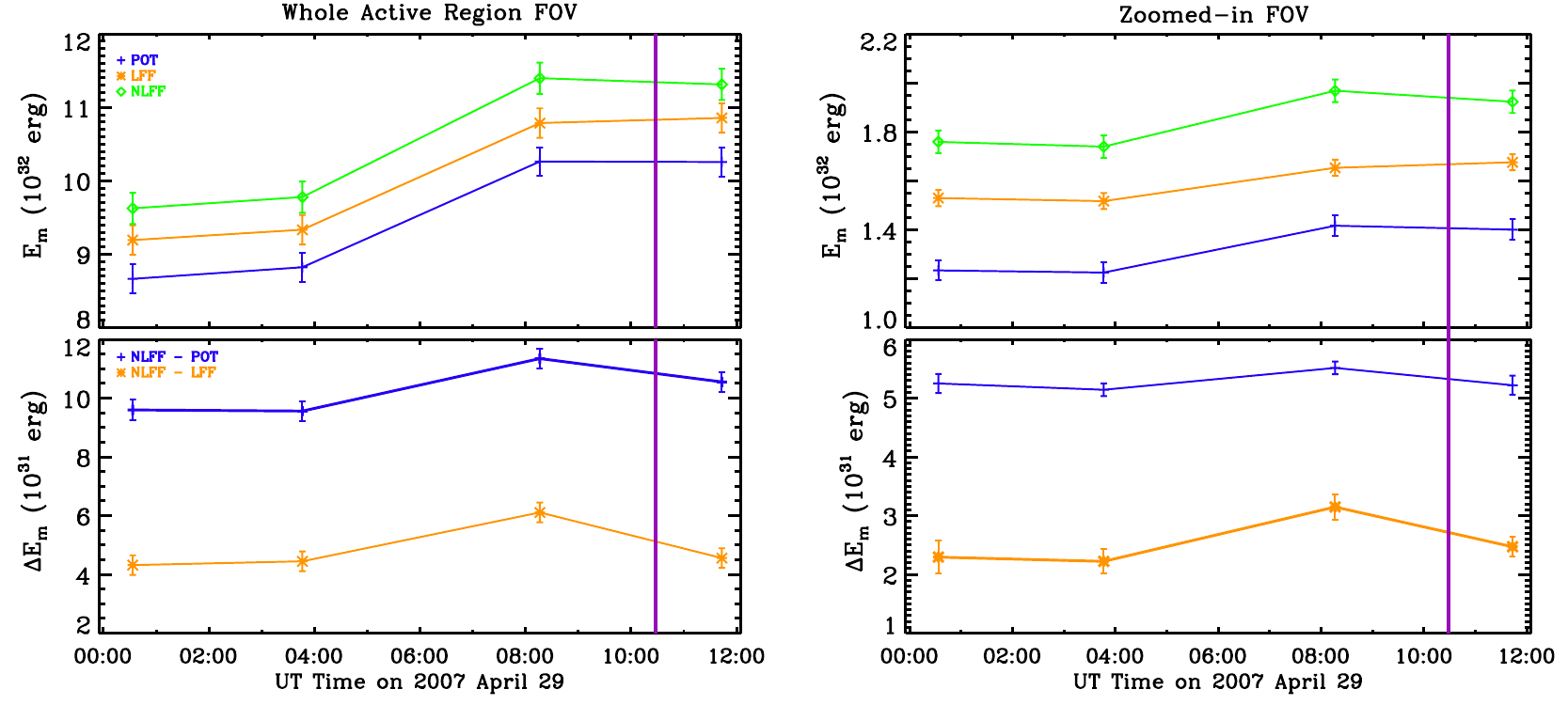}}
\caption[Magnetic energy ($E_\mathrm{m}$, upper) and free magnetic energy ($\Delta E_\mathrm{m}$, lower) for each extrapolation type.]{Magnetic energy ($E_\mathrm{m}$, upper) and free magnetic energy ($\Delta E_\mathrm{m}$, lower) for each extrapolation type, as identified in the legends. Energies for the whole active region volume are shown on the left with those for the smaller zoomed-in volume shown on the right, while vertical bars denote errors. The purple vertical lines between the scans 3 and 4 mark the flare peak time.}
\label{3D:figure_energy}
\end{sidewaysfigure}

The temporal variation of the calculated $E_\mathrm{m}$ values are displayed in the upper panels of Figure~\ref{3D:figure_energy}, with results from the whole active region volume presented in the left column and the zoomed-in volume in the right column. Understandably, values of $E_\mathrm{m}$ from the whole active region volume are much greater than those from the zoomed-in volume. In addition, $E_\mathrm{m}^{\mathrm{NLFF}}$ is consistently larger than $E_\mathrm{m}^{\mathrm{LFF}}$, which is consistently larger than $E_\mathrm{m}^{\mathrm{POT}}$. This corroborates the findings of Sects.~\ref{3D:interpolation} and \ref{3D:orientation}, such that for all scans the NLFF extrapolations contain more twist in the field (i.e., stronger currents and greater magnetic energies) than the LFF extrapolations. For both volumes under consideration, $E_\mathrm{m}$ varies little between the first two scans, but increases in magnitude by scan 3 (i.e., prior to the flare). After the flare, most of the magnetic energies decrease marginally. However, none return to the earlier pre-flare `quiet' values. Differences in $E_{\mathrm{m}}^{\mathrm{NLFF}}$ between scans 3 and 4 contain changes in total magnetic energy due to the combined effect of the flare and evolution of the region over a $\sim$~3 -- 4~hour period. Between these scans the whole active region volume shows a decrease in $E_{\mathrm{m}}^{\mathrm{NLFF}}$ of $(9\pm2) \times 10^{30}$~erg, with the zoomed-in volume decreasing by $(5\pm1) \times 10^{30}$~erg. Both of these changes are likely to be dominated by the flare, given the essentially constant level of total potential magnetic energy observed over the flare.

The results of the free magnetic energy calculations are presented in the lower panels of Figure~\ref{3D:figure_energy}. Similar to the total magnetic energies, no significant changes are observed for either volume between the first two scans. Both $\Delta E_\mathrm{m}^{\mathrm{NLFF-POT}}$ and $\Delta E_\mathrm{m}^{\mathrm{NLFF-LFF}}$ increase from scan 2 to 3 (i.e., just prior to the flare). After the flare, $\Delta E_\mathrm{m}^{\mathrm{NLFF-LFF}}$ decreases towards pre-flare `quiet' values, remaining slightly higher than the level observed in scans 1 and 2. $\Delta E_\mathrm{m}^{\mathrm{NLFF-POT}}$ also decreases, but not as much as $\Delta E_\mathrm{m}^{\mathrm{NLFF-LFF}}$. Table~\ref{3D:table_energies} shows the differences in NLFF energy values between the third and fourth scan, i.e., how the energies changed due to the flare and evolution of the region in the $\sim3 - 4$~hour window. In terms of absolute changes over the flare, $\Delta E_\mathrm{m}^{\mathrm{NLFF-LFF}}$ decreases by $(1.6 \pm 0.3)\times10^{31}$~erg in the whole active region volume, while the zoomed-in volume decreases by $(7\pm2)\times10^{30}$~erg. Caution should be taken when interpreting changes in $\Delta E_\mathrm{m}^{\mathrm{NLFF-LFF}}$ over time (e.g., it is known here from Section~\ref{3D:obs} that the value of LFF $\alpha$ is different before and after the flare). However, it is at least encouraging that the changes in free magnetic energy above the LFF state are within a factor of 2 of the absolute changes in total NLFF magnetic energy. This indicates that $\sim$~15 -- 25\% of the free magnetic energy that existed above the LFF state before the flare has been lost from the system over the course of the flare.

\begin{table}
\centering
\caption[Difference in selected magnetic energy values before and after the \emph{GOES} B1.0 flare on 2007 April 29.]{Difference in selected magnetic energy values before and after the \emph{GOES} B1.0 flare on 2007 April 29, summarising the changes shown in Figure~\ref{3D:figure_energy}.}
\begin{tabular}{  l c  c }
\hline
\hline
FOV & Energy Value  &   Change from Scan 3 to Scan 4 \\
&   &    ($\times 10^{31}$ergs) \\
\hline
Whole Active Region \dotfill 	& $E_\mathrm{m}^{\mathrm{NLFF}}$ & $- 0.9 \pm 0.2 $\\
\dotfill  	& $\Delta E_\mathrm{m}^{\mathrm{NLFF-POT}}$	& $- 0.8 \pm 0.2$ \\
 \dotfill 	& $\Delta E_\mathrm{m}^{\mathrm{NLFF-LFF}}$	&  $- 1.6 \pm 0.3$ \\
 \hline
Zoomed-in  \dotfill 	& $E_\mathrm{m}^{\mathrm{NLFF}}$		&  $- 0.5 \pm 0.1$\\
\dotfill 	& $\Delta E_\mathrm{m}^{\mathrm{NLFF-POT}}$	& $- 0.3 \pm 0.1$ \\
\dotfill 	& $\Delta E_\mathrm{m}^{\mathrm{NLFF-LFF}}$ 	&  $- 0.7 \pm 0.2$ \\
 \hline
 \hline
  \label{3D:table_energies}
  \end{tabular}
\end{table}

%%%%%%%%%%%%%%%%%%%%%%%%%%%%%%%%%%%%%%%%%%%%%%%%%%%%%%%%%%%%%%%%%%%%%%%%%%%%%%%%%%%%%%%%%%%%%%%%
%%%%%%%%%%%%%%%%%%%%%%%%%%%%%%%%%%%%%%%%%%%%%%%%

\section{Discussion}
\label{3D:concs}

The magnetic field configuration of a ROI within an active region has been studied over the course of a \emph{GOES} B1.0 magnitude flare. Geometrical differences between field-line traces through potential, LFF, and NLFF extrapolation solutions have been analysed, as well as their resulting magnetic energies. It is found that the general orientation of ROI field-line footpoints do not change significantly in the hours leading up to the flare, despite the ROI showing an increase in total magnetic energies. However, there are signatures of field redistribution after the flare that indicate incomplete Taylor relaxation: a portion (i.e., not all) of the NLFF field configuration becomes similar to that of the LFF field. Consideration of the magnetic and free magnetic energies after the flare indicate that the region still has energy available for further flaring. It is worth noting that a \emph{GOES} B1.2 flare occurs in the same spatial region on 2007~April~29 at 14:35~UT, less than 3~hours after the final SOT-SP scan analysed here.

A number of previous works have debated the relevance of Taylor relaxation to the flaring process. \citet{amari00} use the presence of non-linearities in the post-relaxation state of numerical simulations to suggest that Taylor's theory does not apply to flares and CMEs. \citet{bleybel02} reach the same conclusion using observations, finding that the relaxed post-eruption state is inconsistent with a LFF state. However, a large amount of helicity was ejected from the region studied. In relation to this, \citet{regnierpriest07apj} state that if helicity is not conserved (e.g., during a CME) then the minimum energy state can be a potential one. It is also not yet known whether helicity conservation is the only constraint on relaxation \citep{pontin11}. Although helicity is not calculated here due to time constraints, it should be noted that the studied flare event has no associated CME.

In contrast, other studies have demonstrated the presence of Taylor relaxation. Numerical simulations of energy release in a coronal loop by \citet{browning08} indicate that the relaxed equilibrium state corresponds closely to a constant $\alpha$ field. \citet{nandy03} report the observational detection of a process akin to partial Taylor relaxation in flare-productive active regions that never achieve completely LFF states within their observation periods. It is worth noting that \citeauthor{nandy03} find that the relaxation process occurs on of the order of a week. The 12-hour period studied here is a considerably shorter time scale, which may contribute to only partial Taylor relaxation being observed. It has also been suggested that partial relaxation is perhaps to be expected from a magnetically complex system with flux emergence and cancellation to be accounted for \citep{pontin11}.

In the build up to the flare, a clear increase is observed in all magnetic energies and free magnetic energies, occurring $\sim$~6.5 -- 2.5 hours prior to the start time of the event. On average, in the zoomed-in region volume the magnetic energy increases by $\sim7 \times10^{30}$~erg and free magnetic energy increases by $\sim2 \times10^{31}$~erg. In the whole active region volume, the magnetic energy increases by $\sim20 \times10^{30}$~erg and free magnetic energy increases by $\sim15 \times10^{31}$~erg. This may indicate a shorter time scale for flare energy input than has been previously observed \citep[e.g., a gradual increase in magnetic energy over the course of a day before an M-class flare is reported by][]{thalmannwiegelmann08}, but could be related to the low magnitude of the event studied here. It is worth noting, in studying \emph{Hinode} data associated with X-class flares, \citet{jing10} found no clear and consistent pre-flare pattern in the temporal variation of free magnetic energy above the potential state.

Considering changes over the flare, a marginal decrease is observed in most of the magnetic energies. Previous authors have been primarily concerned with reporting changes in the free magnetic energy above the potential state over solar flares. For example, \citet{sun12} find a decrease in $\Delta E_\mathrm{m}^{\mathrm{NLFF-POT}}$ of $\sim$~$3 \times 10^{31}$~erg within 1~hour of an X2.2 flare (which they believe to be an underestimation), \citet{thalmannwiegelmann08} find a decrease of $\sim$~$5 \times 10^{32}$~erg over an M6.1 flare, and \citet{thalmannwiegelmann08a} find a decrease of $\sim$~$2 \times 10^{31}$~erg after a C1.0 flare ($\sim$~40\% of the available free magnetic energy). The B1.0 flare-related changes of (3 or 8)\,$\times~10^{30}$~erg in $\Delta E_\mathrm{m}^{\mathrm{NLFF-POT}}$ (for the zoomed-in and whole active region volumes considered here, respectively) agree with the scaling of the flare event magnitudes. However, \citet{regnierpriest07apj} suggest that $\Delta E_\mathrm{m}^{\mathrm{NLFF-POT}}$ gives an upper limit for the energy that can be released during large flares, while $\Delta E_\mathrm{m}^{\mathrm{NLFF-LFF}}$ is a good estimate of the energy available for small flares. In this research the value of $\Delta E_\mathrm{m}^{\mathrm{NLFF-LFF}}$ prior to the flare is $\sim$~(3 or 6)\,$\times~10^{31}$~erg (for the zoomed-in and whole active region volumes, respectively) with a corresponding decrease over the flare of $\sim$~(1 or 2)\,$\times~10^{31}$~erg. This indicates that $\sim$~20~--~30\% of the free magnetic energy above the LFF state is removed during the course of the flare.

In summary, it seems that the magnetic configuration of active region NOAA 10953 did not fully relax to either a potential or LFF state after the B1.0 flare on 2007~April~29. In addition, the energy budget remained sufficient to trigger another flare within 4~hours. Finding an AR in a partially relaxed state after previous flaring may then be a good indicator of impending flare activity. Also, the increase of magnetic energy on short time scales before a flare could be useful for near-realtime forecasting. However, the methods used to obtain these quantities are too computationally intensive to be currently applied in near-realtime.

		% research 2

% this file is called up by thesis.tex
% content in this file will be fed into the main document

\chapter{Conclusions and Future work} % top level followed by section, subsection
\label{chapter:concs}

% ----------------------- paths to graphics ------------------------

% change according to folder and file names
\ifpdf
    \graphicspath{{8/figures/PNG/}{8/figures/PDF/}{8/figures/}}
\else
    \graphicspath{{8/figures/EPS/}{8/figures/}}
\fi

% ----------------------- contents from here ------------------------

\newpage

The research throughout this thesis has investigated the magnetic field evolution of flaring sunspot regions in detail. Any changes found in the lead-up to a flare event could be useful precursors for flare forecasting, and changes observed due to the flare itself test currently proposed flaring AR field configurations. The collection of results described throughout this thesis have succeeded in finding interesting changes both before and after flares in locations of chromospheric flare brightening. Temporal changes in photospheric and coronal magnetic field parameters have been discovered, as well as spatial changes across a magnetic neutral line. In this chapter, the main results obtained are summarised (Section~\ref{conc:principal}), as well as noting some outstanding issues (Section~\ref{conc:problem}), and outlining directions for future research (Section~\ref{conc:future}).

%%%%%%%%%%%%%%%%%%%%%%%%%%%%%%%%%%%%%%%%%%%%%%%%%%%%%%%%%%%%%%%%%%%%%%%%%%%%%%%%%%%%%%%%%%%%%%%%%%%%%%%%

\section{Principal Results}
\label{conc:principal}
Observations primarily obtained from \emph{Hinode}/SOT have been used in this study to examine the photospheric vector magnetic field of flaring ARs. In addition, coronal magnetic field geometry and energy changes were analysed using multiple extrapolation techniques. The main results arising from these studies can be summarised as follows.

\subsection{Photospheric Magnetic Field (Murray et al, 2011, 2012)}
\begin{itemize}

\item {\bf Temporal Field Inclination Changes~~~} Interesting changes in the solar surface magnetic field vector have been detected in the events studied in this thesis. Significant increases in negative vertical field strength, negative vertical current density, and field inclination angle towards the vertical before a B-class flare have been discovered using vector magnetic field observations of small flux elements. A change in field inclination towards the vertical has also been discovered in a region across a magnetic NL before a C-class flare. 
Vertical field strength and vertical current density are also observed to decrease in magnitude after the B-class flare, and a change in field orientation towards the horizontal is observed over the course of the two flares studied. This change in field inclination (from more vertical before a flare to more horizontal afterwards) is reminiscent of the field re-configuration predictions outlined by the `coronal implosion model' of \citet{hudson08} (see Figure~\ref{paper1:hudson}).

\item {\bf Spatial Field Inclination Changes~~~} The research presented in Chapter~\ref{chapter:NL} has shown, for the first time, the full evolution of field inclination across a flaring magnetic NL, with both temporal and spatial changes discovered before and after a C-class flare. In terms of the spatial variation of field inclination, a clear distinction in changes at various locations across  the NL has been observed. The strongest changes are observed close to the NL itself, with the magnitude of changes dropping off with perpendicular distance from the NL. 

\item {\bf Flare Locations~~~} The chromospheric observations obtained of all events studied indicated flare brightening occurring near or along magnetic neutral line locations. Significant changes in the directly measured photospheric and inferred coronal magnetic field parameters in Chapters~\ref{chapter:paper1} and \ref{chapter:3D} were observed at these NL locations (see Section~\ref{conc:coronal}). This has confirmed previous theoretical and observational studies suggesting NLs to be typical flaring locations (\citealp{hagyard84}; \citealp{demoulin93}; \citealp{fisher12}).
\end{itemize}

\subsection{Coronal Magnetic Field (Murray et al, 2013a)}
\label{conc:coronal}
\begin{itemize}
\item {\bf Extrapolation Differences~~~} After comparison of potential, LFF, and NLFF extrapolation methods for the 2007 April 29 observations, it is clear that significant geometrical differences exist between the different field configurations obtained. The field lines obtained by the NLFF extrapolation stretch to higher heights than the potential or LFF field lines. The NLFF configuration also has consistently larger magnetic energy values than the potential or LFF configurations, leading to the presence of free magnetic energy. From comparison with coronal loops, and analysis of magnetic geometry and energy, it can be said that the NLFF field configuration appears to be the most accurate representation of the 3D magnetic field structure in the region studied.

\item {\bf 3D Geometrical Analysis~~~} Although clear distinctions are observed between the three different extrapolated geometrical configurations, the differences between them do not clearly vary with time before the flare. No significant differences in geometry are observed, perhaps suggesting that 3D geometry is not the best focus when investigating useful prediction parameters for flaring, considering the distinct changes observed in 2D photospheric vector magnetic field parameters before the flare. Perhaps for now, until 3D extrapolation techniques become more accurate and less computationally intensive, photospheric magnetic field information may be a better flare-forecasting tool.

\item {\bf Magnetic Energy Changes~~~} However, (free) magnetic energy may be a 3D parameter of interest, with large changes in (free) magnetic energy detected before and after the B-class flare event studied. Similar to the changes observed in the photospheric magnetic field, (free) magnetic energy values significantly increase before the flare. Magnetic energy and free magnetic energy values are also observed to decrease after this flare event.
\end{itemize}

\begin{figure}[!t]
\centerline{\includegraphics[width=0.9\textwidth]{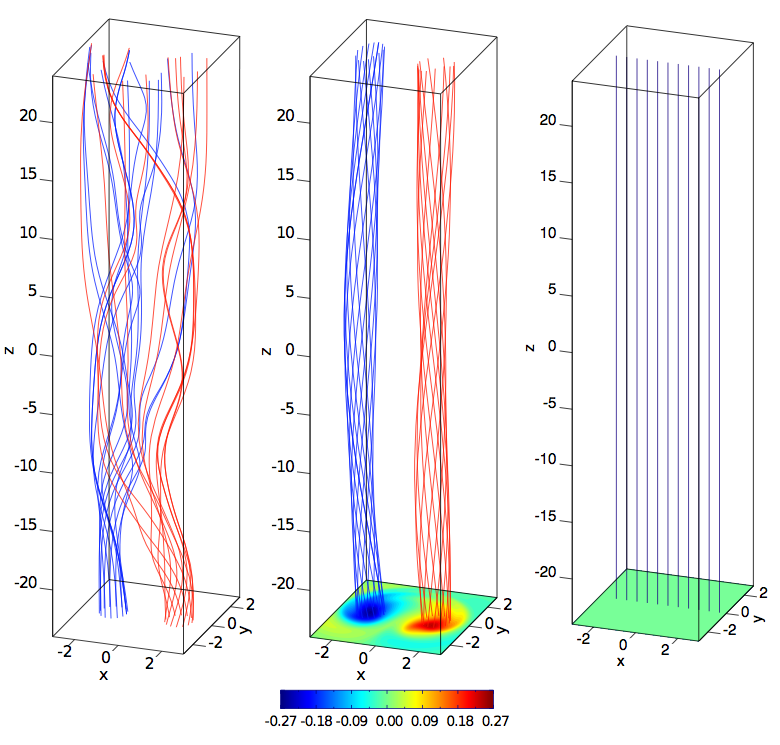}}
\caption[Magnetic field lines in a braided magnetic field (left), NLFF final state (middle), and ideal Taylor state (right) \citep{pontin11}.]{Magnetic field lines in a braided magnetic field (left) which undergoes a resistive relaxation. The final state (centre) is a NLFF field. Contours of the force-free parameter $\alpha$ are shown on the lower boundary. A simple Taylor relaxation would predict a uniform field (right) with $\alpha = 0$ as the end state of relaxation \citep{pontin11}.}
\label{conc:pontin_taylor}
\end{figure} 

\begin{itemize}
\item {\bf Partial Relaxation~~~} Evidence of partial Taylor relaxation has been detected due to the B-class flare studied in this thesis (Chapter~\ref{chapter:3D}). The magnetic field configuration does not return to a completely potential state, retaining some NLFF complexity. Evidence has been presented for the NLFF field configuration becoming more similar, but not equal, to the LFF configuration than to the potential configuration after the flare, hence why only partial relaxation is noted. This observation agrees particularly well with expectations of field topology changes due to a flare with no associated CME  (\citealp{regnierpriest07apj}; \citealp{browning08}). \citet{pontin11} investigated relaxation in a braided flux tube designed to model a coronal loop, and found the field to relax into two separate LFF flux tubes of different helicity sign, separated by a region of non-constant $\alpha$. Figure~\ref{conc:pontin_taylor} shows the initial, final, and ideal Taylor states found. Their results are quite distinct to that of a simple Taylor relaxation which would predict a uniform vertical field in this case. However helicity was conserved, and full relaxation did occur, suggesting something currently unknown, and perhaps more fundamental, prevented the field from reaching a LFF configuration \citep{wilmotsmith10}. Most studies up to now have looked at more complex regions with large flares (as mentioned above). In this thesis, it is interesting to see that the field does not completely relax after a small flare in a very simply-structured sunspot region, highlighting the need to study these `simpler' events in greater detail.
\end{itemize}

\subsection{General Comments}
\begin{itemize}
\item {\bf Field Change Magnitudes~~~} The various results obtained in this thesis come from studying magnetic field evolution over the course of B- and C- class flare events. The flare class also correlates with the size of the region showing chromospheric flaring brightening, with small flux elements investigated for the B-class event (FOV analysed $\sim$~30~Mm$^2$), and a larger area examined for the C-class ($\sim70 \times 45$~Mm area) event. 
Many previous studies of this field of research have focused on large-scale events (generally M- or X- class) in whole ARs (\citealp{wang02}; \citealp{sudolharvey05}), as greater changes are expected from larger more complex sunspot groups. It is interesting to observe quite large variations in magnetic field parameters, such as field strength, inclination, current density, and magnetic energy in even the smallest regions of relatively simple sunspot structures. Such large changes before and after a small B-class flare have not been previously observed.

\item {\bf Field Change Timescales~~~} It is worth noting that the changes observed in the various observation periods studied in this thesis are on short time scales of just hours before and after the flares. The research presented in Chapter~\ref{section:paper1} is the first study in which these changes on hour scales before a small flare event have been observed. Many previous studies examined magnetic field evolution on longer timescales over the course of a flare event than investigated here, especially when looking at whole AR changes. For example, \citet{thalmannwiegelmann08} compare a whole AR magnetic field evolution over three separate days before a flare event to the AR field on the day after. Other studies of smaller regions have focused only on the response of the magnetic field to a flare over a period of hours \citep{wangs12,gosain12}. The research in this thesis has provided full evolution of the regions under investigation, finding changes on much shorter timescales than previously expected.
\end{itemize}

%%%%%%%%%%%%%%%%%%%%%%%%%%%%%%%%%%%%%%%%%%%%%%%%%%%%%%%%%%%%%%%%%%%%%%%%%%%%%%%%%%%%%%%%%%%%%%%%%%%%%%%%

\section{Outstanding Issues and Questions}
\label{conc:problem}

Although the analysis procedures used this thesis have been shown to enhance our understanding of flaring AR evolution, they are not without their limits. The following points outline some of the main issues that exist, as well as some open questions remaining.

\subsection{Photospheric Magnetic Field}
\begin{figure}[!t]
\centerline{\includegraphics[width=0.9\textwidth]{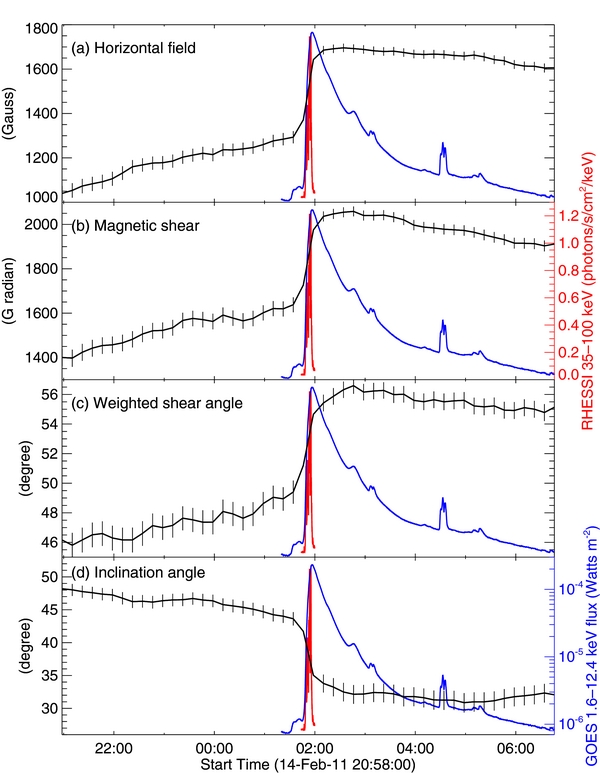}}
\caption[Temporal evolution of various magnetic properties in a region near a magnetic NL \citep{wangs12}.]{Temporal evolution of variousmagnetic properties in a region near a NL: (a) horizontal field, (b) magnetic shear, (c) weighted shear angle, (d) inclination angle. Light curves of RHESSI HXR flux in the $35-100$~keV energy range (red) and GOES flux in $1-8$~\AA\ (blue) are overplotted. The vertical error bars indicate a $3\sigma$ level \citep{wangs12}.}
\label{conc:wangs}
\end{figure}

\begin{itemize}
\label{conc:problem_photo}

\item {\bf Stokes Inversion Parameters~~~} The errors involved with using \emph{Hinode}/SOT-SP data with the \helix atmospheric inversion code are not well-known. An extensive statistical analysis would have to be carried out on the \helix code to substantiate these errors, however unfortunately a lack of time and resources prevented this to be carried out for this thesis. The errors involved with using the \textsf{AMBIG} disambiguation code are also still un-substantiated. \citet{wiegelmann12} state that until an exact error computation becomes available from inversion and ambiguity removal of the photospheric magnetic field vector, a reasonable assumption is that the field is measured more accurately in strong field regions, and the error in the photospheric transverse field is at least one order of magnitude higher than the LOS component. It is worth noting however that preprocessing of the vector magnetic data does help remove uncertainties in the transverse field before use as inputs to 3D extrapolation procedures (see Section~\ref{3D:obs}).

\item {\bf Limited Vector Magnetic FOV~~~} Observations used in this thesis are limited by the FOV of the instrument used, as well as the cadence of the instrument. For example, the region of positive plage to the south of the \emph{Hinode}/SOT-SP FOV is missing in the 2007 June 6 studied in Chapter~\ref{chapter:NL}. There are also only four scans over a $\sim12$~hour period for the 2007 April 29 event, and five scans over $\sim21$~hours for the 2007 June 6. Joint observations can also be a problem after an event is chosen for analysis. For example, there were few coronal wavelength observations available for the 2007 April 29 event, with only TRACE 195\AA\ being able to be used for a consistency check (see Figure~\ref{3D:trace}). Full-disk LOS (and limited vector) magnetograms are now available every $\sim$~12 minutes from \emph{SDO}/HMI, with recent work using this data rather than older \emph{SOHO}/MDI data. For example, \citet{wangs12} used full-disk HMI vector  observations to study changes in horizontal field strength, magnetic shear, and inclination angle over the course of a X-class flare. See Figure~\ref{conc:wangs} for an example of their results. However the data available from \emph{SDO}/HMI are not as high resolution as the limited FOV \emph{Hinode}/SOT-SP instrument. A trade off has to be made between cadence, resolution, and viable events. For this thesis, it was more important to have high-resolution observations of a near-disk-centre AR (to avoid projection effects) that flared only once during the observation period (to avoid contamination of other events), which can be rare to find. Luckily the events found for study in this thesis had generally good coverage, and interesting results were obtained across all observation periods.
\end{itemize}

\subsection{Coronal Magnetic Field}

\begin{itemize}

\item {\bf NLFF Techniques~~~} There is an on-going issue with accuracy of 3D magnetic field extrapolation methods, especially regarding boundary conditions for NLFF extrapolations, as mentioned in Section~\ref{mhd:nlff}. However the accuracy of these techniques is continuously improving, with work such as that outlined in \citet{wiegelmann10} investigating these issues. Preprocessing the data, and using the most accurate NLFF extrapolation currently available is the best that could be done in this thesis to mitigate these issues. Hopefully the accuracy of 3D extrapolation techniques will continue to improve in the future, as our understanding of the solar corona improves.

\item {\bf Magnetic Topology~~~} One open question that remains after this thesis study is what the in-depth 3D field topology of the regions studied looks like, i.e., determining null points, separatrices etc. In 2D, magnetic reconnection occurs at X-points, but in 3D it can occur at null points, along separatrix surfaces and along separators where strong electric current sheets can be created. Unfortunately time constraints did not allow the 3D topology of the AR on 2007 April 29 to be investigated further than magnetic geometry and energy changes, however 3D skeleton analysis could be useful to pinpoint energy storage sites for possible magnetic reconnection. Further work could include determining null point locations, numbers, curvature, height, and other related properties.

\item {\bf Field Relaxation~~~} Another open question that remains is whether full Taylor relaxation is ever achieved in NOAA AR 10953 after the 2007 April 29 event, and if so, when? Does the region return to a completely potential state after several other small B-class flares that day, or continue to become more like a LFF field with constant $\alpha$? Or perhaps it remains in a NLFF configuration, as only other small B-class events occur from NOAA 10953 over the next few days. Larger B-class flares and a C8.5 flare do not occur in the AR until 2007 May 2 (with possible associated CMEs according to the \emph{SOHO}/LASCO CME Catalog\footnote[1]{\url{http://cdaw.gsfc.nasa.gov/CME_list/}}). The observation period available for this event unfortunately limits the ability to answer these questions in this thesis. However studying similar sunspot structures in the future with similar small-flare events, over longer observation periods, will help provide more concrete answers to these questions.

\item {\bf Magnetic Helicity~~~} Although a wide range of magnetic field parameters in various sunspot regions have been studied here, other parameters remain unstudied. For example, one open question is how does helicity evolve in the regions studied? Knowing how the helicity changes with time before/after a flare event may help to quantify various aspects of the magnetic field structure \citep[see][]{berger99}. This would be particularly useful for the 2007 April 29 event, as it would compliment the 3D geometry and energy values observed. Discovering whether helicity is conserved or not in the AR studied would help place the partial Taylor relaxation observed into a wider context. 

\end{itemize}

Nevertheless, the results obtained in this thesis show magnetic fields in even the smallest of flux regions evolving significantly with time during periods of flare observation. This confirms the importance of analysing high resolution vector magnetic field observations when investigating flare mechanisms.

%%%%%%%%%%%%%%%%%%%%%%%%%%%%%%%%%%%%%%%%%%%%%%%%%%%%%%%%%%%%%%%%%%%%%%%%%%%%%%%%%%%%%%%%%%%%%%%%%%%%%%%%

\section{Future Work}
\label{conc:future}

This research has discovered numerous interesting insights into the evolution of the photospheric and coronal magnetic field during periods of flare activity. However, the results are certainly not exhaustive, with open questions remaining regarding flare triggers and processes, and field topology changes involved with flaring. Further work is required before a complete understanding of AR evolution can be achieved. Besides the open questions discussed in the previous Section, the following points outline some future research that could be useful for expanding upon the results obtained in this thesis.

\begin{itemize}

\item {\bf Vector Magnetic Uncertainties~~~} As briefly mentioned in Section~\ref{conc:problem_photo}, a full statistical analysis of the atmospheric inversion code used in this thesis would be extremely beneficial for estimating uncertainties. \citet{deltoro10} examined the reliability of the \textsf{MILOS} inversion code on a synthesised data set, but no-one has as yet calculated the error involved in running \emph{Hinode}/SOT-SP data through the \helix code. \textsf{PIKAIA} (see Section~\ref{rad_trans:helixmethod}) is based on random number generators and not dependent on starting values, so each pixel can be inverted several times and the spread of the parameter values obtained gives a sense of reliability. This is in contrast to the older Levenberg-Marquadt method often used with inversions, which always finds the same answer for a given set of starting values and is not guaranteed to find global minima. The inversion must be repeated numerous times, which would take at least several months of multi-processor computational time. The mean, variance, skewness and kurtosis can then be explored for every pixel on the map. Running such a statistical analysis on the data obtained will clearly show which parameters are relatively stable and how their stability depends on various polarisation measures.

\item {\bf Multi-Instrument Observations~~~} The results of this thesis have outlined that joint observations from both \emph{SDO}/HMI and \emph{Hinode}/SOT-SP could be extremely beneficial to analysis. Embedding the higher resolution SOT-SP vector magnetograms in full-disk HMI vector magnetograms would certainly help any FOV issues. However, \citet{cheng10} and \citet{guo10} note that applying the preprocessing procedure of \citet{wiegelmann06} to the magnetic field in the original FOV vector magnetogram is better than embedding it in a larger FOV LOS magnetogram (where the transverse components of the magnetic field are unknown). This mainly relates to the lower resolution \emph{SOHO}/MDI data which was only available for this research (and hence embedding was not used here), but transverse field measurements may be more accurate for the higher resolution HMI data. This would enable analysis of a whole AR, including any spurious flux elements that may be lying outside the SOT-SP FOV, which could cause issues in 3D extrapolations if missing. The improved cadence of HMI would also be beneficial for analysis of various magnetic field parameters. Using both HMI and SP data sets together will fill in some of the gaps that currently exist in, for example, Figure~\ref{paper1:figure_3} or ~\ref{3D:figure_energy}. A comparison of the results obtained by both sets of vector magnetograms has also not yet been carried out, and would be of definite interest for future work. One of the hopes of this thesis was to run a comparison of these two data sets for a particular flare event, but HMI Stokes profiles were unfortunately not publicly released in time.

\item {\bf Event Statistics~~~} Although two separate flare events in two different ARs were studied in this thesis, with similar results for each, a wider range of events studied would certainly be desirable. No further events were studied in this thesis due to the time-consuming nature of the analysis process. Note that currently a typical run-time of the \helix code on a  $512 \times 512$ pixels$^2$ region takes $\sim3$ days on a 7-core server, with a NLFF extrapolation run taking $\sim$ 10 days on one 8GB core. However, looking for temporal and spatial changes in additional regions of varying complexity and flare class would certainly aid the investigation into flaring AR magnetic field evolution. It would also be beneficial to expand the 3D field topology analysis to the C-class event studied here, as well as other new events, as the 3D topology of only the 2007 April 29 AR event was examined in this thesis due to time constraints.

\item {\bf Energy Budgets~~~} The investigation of (free) magnetic energy in this thesis has proven beneficial for investigations of magnetic field evolution in flaring regions, and (free) magnetic energy values could be used as flare precursors in the future. Some previous studies have tried to estimate energy budgets of solar flares (e.g., \citealp{bleybel02}; \citealp{emslie05,emslie12}), and it would be interesting to try to determine maximum values of (free) magnetic energy in flare events. \citet{schrijver12} estimated the frequency of extremely energetic solar events, finding at most a 10\% chance of a flare larger than X30 occurring in the next 30 years. Estimating when these large events may occur is increasingly important due to the impact space weather has on current technologies. \citeauthor{schrijver12} evaluated the probability of large-energy flares by combining flare observations with an ensemble of stellar flare observations. Using high resolution 3D vector magnetic field observations such as those studied in this thesis could improve current estimates. By studying Equation~\ref{mag_en}, a maximum value of magnetic energy can give an idea of maximum magnetic field strengths and length scales of the flaring sunspot region. \citeauthor{schrijver12} note that a flare energy of $10^{34}$~erg would require a sunspot that is 20 times larger than the historically observed maximum. Extremely large magnetic field strengths may also not be compatible with solar rotation rates. Solar flares with magnitudes above the  observational maximum of $\sim10^{33}$~ergs is perhaps unlikely. Nonetheless, calculating maximum values could be extremely useful in determining thresholds for flaring, which could be incorporated into flare forecasting models. 

\item {\bf Chromospheric Magnetic Field~~~} Chromospheric observations have been used throughout this thesis in order to find regions of interest to study (where flare-peak brightening occurs). A deeper investigation of chromospheric observations could be an interesting area of future research. Chromospheric vector magnetograms rather than photospheric vector magnetograms would also be more accurate inputs to 3D extrapolation procedures, since preprocessing basically tries to emulate chromospheric conditions in order to reach a state closer to a force-free state. Although chromospheric magnetic field data is presently not widely available, the launch of the Solar-C spacecraft in 2018 will provide these measurements at much higher spatial resolution and polarimetric accuracy than currently available from \emph{Hinode} (Solar-B). The SUVIT instrument will record photospheric and chromospheric vector magnetic field data at better than 0.1$"$ spatial resolution (compared to 0.32" for SOT-SP), and $10^{-4}$ polarimetric accuracy (compared to $10^{-3}$ for SOT-SP).

\item {\bf Active Region Simulations~~~} One aspect of analysis that has not been investigated in this thesis is the running of simulations of AR magnetic fields. There are currently a number of quite accurate MHD models of sunspots available \citep[see review by][]{rempel11}. 
Also, recent studies have compared observations obtained of ARs to MHD simulations of the same AR (e.g., \citealp{li11}; \citealp{kusano12}). \citeauthor{kusano12} survey nonlinear dynamics caused by a wide variety of magnetic structures in terms of 3D MHD simulations, and find two types of small magnetic structures that favour solar eruption onset. Their Figure 2 is reproduced in Figure~\ref{conc:mhd_kusano} showing their `opposite polarity' scenario, the other being a `reverse shear' scenario. It could be interesting to compare the ARs studied here with simulations, to see whether the changes in photospheric and coronal magnetic field parameters observed can be re-created with the models. The analysis procedure could also be improved using non-LTE or full MHD atmospheric models rather than the LTE Milne-Eddington inversion code used here. These methods are currently extremely computationally intensive, and thus have not been looked at in this thesis, however could be worthwhile future work as computational processing power continues to improves.
\end{itemize}

\begin{figure}
\centerline{\includegraphics[width=0.85\textwidth]{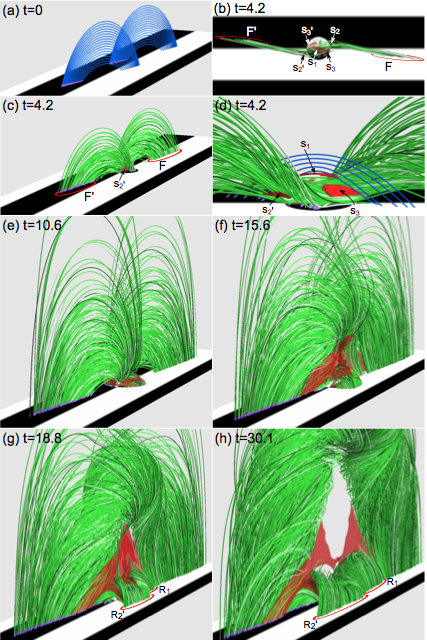}}
\caption[MHD simulation result showing an opposite polarity type scenario causing a solar eruption \citep{kusano12}.]{3D MHD Simulation result in which the `opposite polarity' type of emerging fluxes causes solar eruption-inducing reconnection dynamics. Each subset represent a birds eye view (a, c, e-h), top view (b), and enlarged side view (d) of the magnetic field at different times. Green tubes represent magnetic field lines with connectivity that differs from the initial state, with blue tubes indicating initial fields or those returning to an initial state \citep{kusano12}.}
\label{conc:mhd_kusano}
\end{figure} 

\begin{itemize}
\item {\bf Computational Optimisation~~~} Ultimately, the full use of high performance computing facilities will become necessary for analysis of increasingly higher resolution data sets in the future. This will improve the run-time of the procedures, as well as allowing multiple data sets to be analysed simultaneously, which will be extremely useful when analysing $\sim 90$~s cadence \emph{SDO} vector magnetograms. With improved computing power, changes in parameters such as field strength, field inclination, current density, or (free) magnetic energy could be monitored in near real-time. This will eventually allow incorporation into a space weather suite such as \url{SolarMonitor.org}. 
\end{itemize}

The research presented in this thesis has given an excellent insight into photospheric and coronal magnetic field evolution of flaring regions. For the first time, vector magnetic field changes have been observed mere hours before small magnitude flares. Also, vector magnetic field changes observed after the flare events studied have confirmed current models of field configuration variations due to flares. Any future research expanding upon the first detections presented here will be extremely beneficial in furthering our understanding of the magnetic nature of active regions and solar flare processes. 		% conclusions

%\include{8/materials_methods}        % description of lab methods

% --------------------------------------------------------------
%:                  BACK MATTER: appendices, refs,..
% --------------------------------------------------------------

% the back matter: appendix and references close the thesis

%: ----------------------- bibliography ------------------------

% The section below defines how references are listed and formatted
% The default below is 2 columns, small font, complete author names.
% Entries are also linked back to the page number in the text and to external URL if provided in the BibTex file.

% PhDbiblio-url2 = names small caps, title bold & hyperlinked, link to page 
%\begin{multicols}{2} % \begin{multicols}{ # columns}[ header text][ space]
\begin{small} % tiny(5) < scriptsize(7) < footnotesize(8) < small (9)

\bibliographystyle{Latex/Classes/jmb} % Title is link if provided
\renewcommand{\bibname}{References} % changes the header; default: Bibliography

\bibliography{9_backmatter/bibliography} % adjust this to fit your BibTex file

\end{small}
%\end{multicols}

% --------------------------------------------------------------
% Various bibliography styles exit. Replace above style as desired.

% in-text refs: (1) (1; 2)
% ref list: alphabetical; author(s) in small caps; initials last name; page(s)
%\bibliographystyle{Latex/Classes/PhDbiblio-case} % title forced lower case
%\bibliographystyle{Latex/Classes/PhDbiblio-bold} % title as in bibtex but bold
%\bibliographystyle{Latex/Classes/PhDbiblio-url} % bold + www link if provided

%\bibliographystyle{Latex/Classes/jmb} % calls style file jmb.bst
% in-text refs: author (year) without brackets
% ref list: alphabetical; author(s) in normal font; last name, initials; page(s)

%\bibliographystyle{plainnat} % calls style file plainnat.bst
% in-text refs: author (year) without brackets
% (this works with package natbib)

% --------------------------------------------------------------

% according to Dresden med fac summary has to be at the end
%\include{0_frontmatter/abstract}

\end{document}